\documentclass[iop,revtex4.1,numberedappendix,appendixfloats]{emulateapj}

\usepackage{natbib}
\usepackage{amsmath}
\usepackage{color}
\usepackage{url}
\usepackage{ulem}
\usepackage{multirow}
\usepackage{amsmath}
\usepackage{wasysym}

\usepackage{rotating, graphicx}

\bibpunct{(}{)}{;}{a}{}{,}

\def\apjl{ApJ}%
\def\apjs{ApJS}%
\def\ao{Appl.~Opt.}%
\def\aap{A\&A}%
\shorttitle{The 450\,days X-ray monitoring of the changing-look AGN 1ES\,1927+654}
\shortauthors{Ricci et al.}

\begin{document}

\title{The 450\,days X-ray monitoring of the changing-look AGN 1ES\,1927+654}

\author{C. Ricci\altaffilmark{1,2,*}, M. Loewenstein\altaffilmark{3,4}, E. Kara\altaffilmark{5}, R. Remillard\altaffilmark{5}, B. Trakhtenbrot\altaffilmark{6}, I. Arcavi\altaffilmark{6,7}, K. C. Gendreau\altaffilmark{3}, Z. Arzoumanian\altaffilmark{3}, A. C. Fabian\altaffilmark{8}, R. Li\altaffilmark{2},  L. C. Ho\altaffilmark{2,9}, C. L. MacLeod\altaffilmark{10}, E. Cackett\altaffilmark{11}, D. Altamirano\altaffilmark{12}, P. Gandhi\altaffilmark{12}, P. Kosec\altaffilmark{8,5}, D. Pasham\altaffilmark{5}, J. Steiner\altaffilmark{5}, C.-H. Chan\altaffilmark{13,6} }

\altaffiltext{1}{N\'ucleo de Astronom\'ia de la Facultad de Ingenier\'ia, Universidad Diego Portales, Av. Ej\'ercito Libertador 441, Santiago, Chile} 
\altaffiltext{2}{Kavli Institute for Astronomy and Astrophysics, Peking University, Beijing 100871, China}
\altaffiltext{3}{Astrophysics Science Division, NASA Goddard Space Flight Center, 8800 Greenbelt Road, Greenbelt, MD 20771, USA}
\altaffiltext{4}{Department of Astronomy, University of Maryland, College Park, MD 20742, USA}
\altaffiltext{5}{MIT Kavli Institute for Astrophysics and Space Research, 70 Vassar Street, Cambridge, MA 02139, USA}
\altaffiltext{6}{School of Physics and Astronomy, Tel Aviv University, Tel Aviv 69978, Israel}
\altaffiltext{7}{CIFAR Azrieli Global Scholars program, CIFAR, Toronto, Canada}
\altaffiltext{8}{Institute of Astronomy, University of Cambridge, Madingley Road, CB3 0HA Cambridge, UK}
\altaffiltext{9}{Department of Astronomy, School of Physics, Peking University, Beijing 100871, China}
\altaffiltext{10}{Center for Astrophysics, Harvard \& Smithsonian, 60 Garden Street, Cambridge, MA 02138-1516, USA}
\altaffiltext{11}{Department of Physics \& Astronomy, Wayne State University, 666 West Hancock Street, Detroit, MI 48201, USA}
\altaffiltext{12}{Department of Physics \& Astronomy, University of Southampton, Southampton, Hampshire S017 1BJ, UK}
\altaffiltext{13}{Racah Institute of Physics, Hebrew University of Jerusalem, Jerusalem 91904, Israel}

\altaffiltext{*}{claudio.ricci@mail.udp.cl}

\begin{abstract}
1ES\,1927+654 is a nearby active galactic nucleus (AGN) which underwent a changing-look event in early 2018, developing prominent broad Balmer lines which were absent in previous observations. We have followed up this object in the X-rays with an ongoing campaign that started in May 2018, and that includes 265 {\it NICER} (for a total of 678\,ks) and 14 {\it Swift}/XRT (26\,ks) observations, as well as three simultaneous {\it XMM-Newton}/{\it NuSTAR} (158/169\,ks) exposures. In the X-rays, 1ES\,1927+654 shows a behaviour unlike any previously known AGN. The source is extremely variable both in spectral shape and flux, and does not show any correlation between X-ray and UV flux on timescales of hours or weeks/months. After the outburst the power-law component almost completely disappeared, and the source showed an extremely soft continuum dominated by a blackbody component. The temperature of the blackbody increases with the luminosity, going from $kT\sim 80$\,eV (for a 0.3--2\,keV luminosity of $L_{0.3-2}\sim 10^{41.5}\rm\,erg\,s^{-1}$) to $\sim 200$\,eV (for $L_{0.3-2}\sim 10^{44}\rm\,erg\,s^{-1}$). The spectra show evidence of ionized outflows, and of a prominent feature at $\sim 1$\,keV, which can be reproduced by a broad emission line. The unique characteristics of 1ES\,1927+654 in the X-ray band suggest that it belongs to a new type of changing-look AGN. Future X-ray surveys might detect several more objects with similar properties.

\end{abstract}


\setcounter{footnote}{0}

\section{Introduction}

\noindent Accretion onto supermassive black holes (SMBHs) is thought to play an important role in the evolution of galaxies (e.g., \citealp{Kormendy:2013uf}), and to be the mechanism allowing the SMBHs we observe in most massive galaxies (e.g., \citealp{Magorrian:1998tg}) to reach masses of $M_{\rm BH} \sim 10^6-10^{10}M_{\odot}$. While most active galactic nuclei (AGN) in the local Universe accrete at low values of the Eddington ratio\footnote{$\lambda_{\rm Edd}=L\sigma_{\rm T}/4\pi GM_{\rm BH}m_{\rm p}c$, where $L$ is the source bolometric luminosity, $\sigma_{\rm T}$ is the Thomson cross-section, G is the Gravitational constant, $m_{\rm p}$ is the proton mass, $c$ is the speed of light.} ($\lambda_{\rm Edd}\lesssim0.3$, e.g., \citealp{Koss:2017zk}), periods of very rapid accretion, possibly super-Eddington, are expected to be extremely important to explain the growth of the first SMBHs (e.g., \citealp{Trakhtenbrot:2011yq,Volonteri:2012hm}), and the vast majority of massive black holes might in fact gain most of their mass through these episodes \citep{King:2003ml}.

Several aspects of the accretion process are still highly debated such as, for example, the main mechanisms triggering accretion, and why some galaxies host AGN while others do not (e.g., \citealp{Alexander:2012kx}). Moreover some AGN have been found to show very strong variability (e.g., \citealp{Rumbaugh:2018zr,Trakhtenbrot:2019uq,Timlin:2020zw}), which could be associated with changes in the inflow of matter or with instabilities of the accretion disk (e.g., \citealp{Kawaguchi:1998qy}). AGN variability in the optical/UV and X-rays has been directly observed over a very wide range of time-scales (e.g., \citealp{Sartori:2018gf}), ranging from minutes and hours (e.g., \citealp{McHardy:2014tg}) up to several years (e.g., \citealp{MacLeod:2010ku}). One of the most interesting aspects of this strongly variable behaviour is associated with the so-called {\it changing-look AGN}. Those are sources in which the optical/UV broad emission lines, produced in the broad-line region (BLR), appear or disappear (e.g., \citealp{LaMassa:2015ne,Ruan:2016hb,Lawrence:2018my}). These objects are different from the {\it X-ray} changing-look AGN, which display strong variation of the line-of-sight column density ($N_{\rm H}$, e.g., \citealp{Risaliti:2009df,Bianchi:2012vs,Marinucci:2016ne,Ricci:2016nh}), possibly related to moving clouds in the broad-line region (e.g., \citealp{Maiolino:2010by}) or in the torus (e.g., \citealp{Markowitz:2014nm,Buchner:2019ty}).

So far about 20-30 changing-look AGN have been discovered (e.g., \citealp{McElroy:2016qy,Krumpe:2017ul,Parker:2016fr,Yang:2018qt,Katebi:2019ly,Oknyansky:2019qy,MacLeod:2019pm}), but the physical mechanism behind these events is still poorly understood. Optical changing-look AGN cannot be explained by variable obscuration, since the broad-line region is too large to be obscured by the clouds of the dusty torus, except maybe for a narrow range of inclination angles. Moreover, these objects do not typically show any sign of obscuration in their X-ray spectra after transitioning from type\,1 to type\,2 (e.g., \citealp{Denney:2014ez}). It has been argued that the appearance (disappearance) of the broad lines could be due to an increase (decrease) of the accretion rate (e.g., \citealp{Elitzur:2014rp}), possibly related to instabilities in the accretion flow (e.g., \citealp{Ross:2018vl,Stern:2018kr}). In a few objects it has been proposed that the increase in the AGN luminosity could be due to tidal disruption events (TDEs; e.g., \citealp{Merloni:2015on}, \citealp{Komossa:2015iz} and references therein).

1ES\,1927+654 is the first accreting SMBH that was detected in {\it the act of changing phase} \citep{Trakhtenbrot:2019qy}. This object is a rather well known nearby ($z=0.019422$) AGN, showing peculiar characteristics, completely unexpected within the framework of the AGN unification model (e.g., \citealp{Antonucci:1993fu,Urry:1995xy,Netzer:2015zn,Ramos-Almeida:2017qq,Hickox:2018mz}). While its optical spectrum suggested a type-2 classification \citep{Bauer:2000rp}, in the X-rays the AGN appears completely unobscured (i.e., $N_{\rm H}\simeq10^{20}\rm\,cm^{-2}$; \citealp{Boller:2003cq,Gallo:2013hq}), making it a member of the {\it true type 2} class (e.g., \citealp{Panessa:2002cv,Brightman:2008wo,Shi:2010fq,Bianchi:2012bq}). Consistent with this idea, \cite{Tran:2011kk} did not find any evidence of hidden polarized broad H$\alpha$ or direct broad Pa$\beta$ and Br$\gamma$ lines. 
1ES\,1927+654 was found to have increased its flux by at least two magnitudes in the optical V-band in March 2018 (AT2018zf/ASASSN-18el, \citealp{Nicholls:2018yo,Stanek:2018an}) by the All-Sky Automated Survey for Supernovae (ASAS-SN; \citealp{Shappee:2014mw}). From the analysis of data obtained by the Asteroid Terrestrial-impact Last Alert System (ATLAS; \citealp{Tonry:2018qy}), \citet{Trakhtenbrot:2019qy} found that the event started on December 23 2017. Our optical spectroscopic follow-up campaign showed the appearence of strong broad Balmer emission lines a few months after the optical flux increase \citep{Trakhtenbrot:2019qy}. This implies that 1ES1927+654 underwent a {\it changing-look} phase, in which the circumnuclear gas reacted to the strong increase of the optical/UV flux, illuminating the broad line region, and creating the broad lines. 
\smallskip

Our ongoing X-ray monitoring campaign started in May\,\,2018, after the detection of broad optical lines, and consists of three joint {\it XMM-Newton}/{\it NuSTAR} observations (158/169\,\,ks in total), carried out in June (PI: N. Schartel/C. Ricci), December 2018 (PI: N. Schartel/E. Kara) and on May 2019 (PI: C. Ricci), 14 {\it Swift}/XRT observations (PI: I. Arcavi, C. Ricci, 26\,\,ks), and 265 {\it NICER} observations (678\,ks; see \citealp{Kara:2018fc} for a summary of the first set of observations). In the X-rays the source shows an extremely peculiar behaviour, with the X-ray corona, which is typcally the source of most of the X-ray radiation in AGN \citep{Haardt:1991dq,Merloni:2001qy,Fabian:2009bz,Kara:2016ya}, disappearing after the optical/UV outburst \citep{Ricci:2020}.
In this paper we report the detailed analysis of all X-ray observations carried out in the first $\sim 450$\,days of our monitoring campaign of this puzzling object, which includes observations for a total exposure time of $> 1$\,\,Ms. 
The paper is structured as follows: in \S\ref{sect:Xrayobs} we describe all the X-ray data sets used and the data reduction procedures adopted. In \S\ref{sect:xmmobs_spec} and \S\ref{sec:NICERobs} we illustrate the X-ray spectral and timing analysis of {\it XMM-Newton/NuSTAR} and {\it NICER} data, respectively. In \S\ref{sect:discussion} we discuss our results, while in \S\ref{sect:conclusion} we summarize our findings and present our conclusions.
Throughout the paper we adopt standard cosmological parameters ($H_{0}=70\rm\,km\,s^{-1}\,Mpc^{-1}$, $\Omega_{\mathrm{m}}=0.3$, $\Omega_{\Lambda}=0.7$). Unless otherwise stated, uncertainties are quoted at the 90\% confidence level.

\section{X-ray observations}\label{sect:Xrayobs}

A large number of X-ray observations were performed during 2018 and 2019 to monitor 1ES\,1927+654 using  {\it XMM-Newton}, {\it NuSTAR}, {\it Swift}/XRT and {\it NICER}. The distribution of these observations is shown in Fig.\,\ref{fig:longtermLC}, while the observation log is reported in Table\,\,\ref{tab:obslog} in Appendix\,\ref{sect:obslog}. These observations show very strong variability, both on timescales of $<1$\,day ($\sim 1-2$ orders of magnitude) and of months (up to $\sim 4$ orders of magnitude, see Fig.\,1 of \citealp{Ricci:2020}). In the following we describe the procedure adopted for the data reduction of {\it XMM-Newton} (EPIC, RGS and OM; \S\ref{sect:xmm_datared}), {\it NuSTAR} (\S\ref{sect:nustar_datared}), and {\it NICER} (\S\ref{sect:nicer_datared}) observations. Given the large numbers of high-quality {\it NICER} observations, we did not focus on the short ($\sim 2$\,ks) {\it Swift}/XRT observations here, and we report their reduction and spectral analysis in Appendix\,\ref{appendix:XRT_spectra}.

\begin{figure*}[t!]
  \begin{center}
\includegraphics[width=\textwidth]{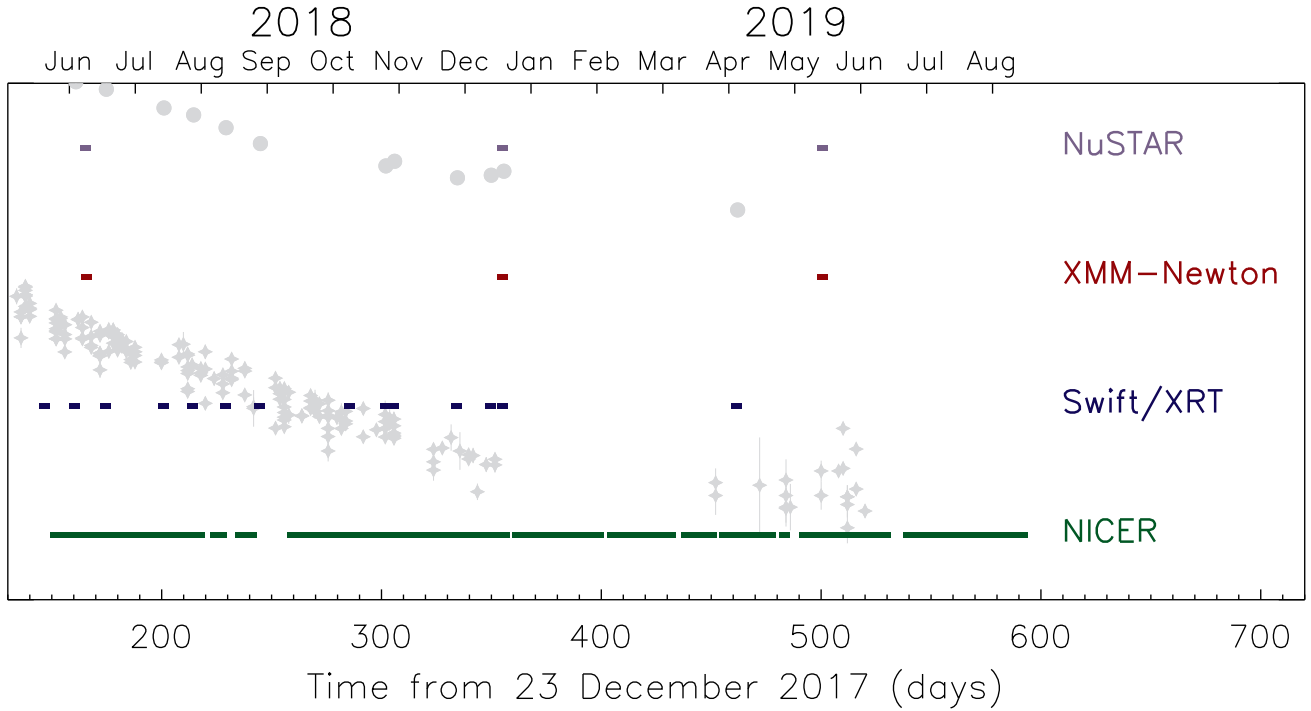}
    \caption{Schematic representation of the X-ray monitoring campaign of 1ES\,1927+654 reported here, covering the period between 17 May 2018 and 5 August 2019. The time of the detection here corresponds to that reported by Trakhtenbrot et al. 2019 (December 23 2017), based on observations from the Asteroid Terrestrial-impact Last Alert System (ATLAS; \citealp{Tonry:2018qy}). The light grey stars and circles are the ATLAS/o and {\it Swift}/UVM2 magnitudes (see \citealp{Trakhtenbrot:2019qy} and \citealp{Ricci:2020}). The complete X-ray observation log is reported in Table\,\,\ref{tab:obslog} in Appendix\,\ref{sect:obslog}.}
    \label{fig:longtermLC}
  \end{center}
\end{figure*}

\subsection{{\it XMM-Newton}}\label{sect:xmm_datared}

After the optical/UV outburst, {\it XMM-Newton} \citep{Jansen:2001ve} carried out three pointed observations of 1ES\,1927+654: on June\,5 2018 (46\,ks),  on December\,12 2018 (59\,ks), and on May 2019 (52\,ks). All observations were performed simultaneously with {\it NuSTAR}, and in all of them the source was clearly detected by the EPIC/PN \citep{Struder:2001uq} and MOS \citep{Turner:2001ec} cameras, as well as by the Reflection Grating Spectrometer (RGS; \citealp{den-Herder:2001bf}) and the optical monitor (OM, \citealp{Mason:2001mw}). 
All EPIC/PN observations were carried out in small-window mode, using a thin (June and December 2018) or a thick (May 2019) filter. The June and December EPIC/MOS observations were performed in large-window mode, while the one in May 2019 was done in small-window mode. A thin filter was used for all the EPIC/MOS exposures.
During the June 2018 observation the 0.3--10\,keV EPIC/PN (MOS) net count rate of the source was $4.2\rm\,ct\,s^{-1}$ ($0.72\rm\,ct\,s^{-1}$), while during the December 2018 and May 2019 observations it was significantly higher. In December\,\,2018 the net count rates were $18.9\rm\,ct\,s^{-1}$ ($1.4\rm\,ct\,s^{-1}$), and in May\,\,2019 they were $14\rm\,ct\,s^{-1}$ ($2.6\rm\,ct\,s^{-1}$). The EPIC/PN count rate of the May 2019 observation was lower than that of the December 2018 exposure, although the flux was higher, because of the different filter used.
For comparison, during the 2011 observation (see \citealp{Gallo:2013hq}), the EPIC/PN (thin filter) count rate was $5.2\rm\,ct\,s^{-1}$. 
 
The observation data files (ODFs) were reduced using the {\it XMM-Newton} Standard Analysis Software (SAS) version 17.0.0 \citep{Gabriel:2004fk}. The raw PN (MOS) data files were then processed using the \textsc{epchain} (\textsc{emchain}) task. The analysis of the background light curves in the 10--12\,keV band (EPIC/PN), and above 10\,keV (EPIC/MOS) of the June 2018 and May 2019 observations showed that no strong background activity was present in neither PN nor MOS, so that we were able to use the whole exposures. Some background activity was detected around the end of the December 2018 observation, with the count rates rising to $\sim 1\rm\,ct\,s^{-1}$ in EPIC/PN. However, due to the very high flux of the source during this observation, we did not filter out this last $\sim 6$\,ks. As a test, we compared the spectral parameters obtained by spectra with and without this period, and found them to be in very good agreement. In our data reduction only patterns that correspond to single and double events (PATTERN~$\leq 4$) were selected for PN, and patterns corresponding to single, double, triple and quadruple events were selected for MOS (PATTERN~$\leq 12$). 

For the June\,\,2018 observation, using circular regions centred on the object to extract the source spectrum we detected pile-up in all EPIC cameras, so that we adopted an annular region, with an inner (outer) radius of $6\arcsec$ ($40\arcsec$) and $5\arcsec$ ($30\arcsec$), for PN and MOS, respectively. The pile-up was considerably stronger in the December\,\,2018 and May\,\,2019 observations, and even using a large inner radius ($35\arcsec$) it was not possible to completely remove its effect on the MOS spectrum. Therefore, for these observations we used only the PN camera, extracting the spectrum using an inner (outer) radius of $15\arcsec$ ($40\arcsec$). The background spectra for the different observations were extracted from circular regions with a radius of $50\arcsec$  and $60\arcsec$ for PN and MOS, respectively. These regions were located on the same CCD as the source, where no other source was present.  The ARFs and RMFs were created using the \textsc{arfgen} and \textsc{rmfgen} tasks, respectively. For the June\,\,2018 observation the source and background spectra of the two MOS cameras, together with the RMF and ARF files, were merged using the \textsc{addascaspec} task. In the following we will refer to the combined spectra as the MOS spectrum. The EPIC spectra were binned to have at least $20$ counts per bin, and we applied the $\chi^2$ statistic.

{\it NICER} observed 1ES\,1927+654 for $\sim 7$\,ks during the December\,2018 {\it XMM-Newton} observation. This allowed us to compare the {\it XMM-Newton} and {\it NICER} spectra, and to look for the presence of any spectral artefacts related to pile-up and/or to X-ray loading in the EPIC/PN spectrum, since the observation was carried out when the source was extremely bright. In Appendix\,\ref{appendix:simultaneousNICER} we report the results of our analysis, showing that the spectral parameters obtained by fitting the two datasets are consistent (See Fig.\,\ref{fig:xmmobs2018lcnicer} and Table\,\ref{tab:fitXMMnicer_dec18}). This allowed us to exclude any significant impact of pile-up and/or to X-ray loading in the EPIC/PN spectrum.

The RGS data were reduced using the \textsc{rgsproc} task, and we used default source and background extraction regions. The spectra from the two detectors were combined into a single spectrum using the \textsc{rgscombine} task, for both the first and second order. The RGS first order count rates were $0.26\rm\,ct\,s^{-1}$, $1.6\rm\,ct\,s^{-1}$ and $1.9\rm\,ct\,s^{-1}$ for the June\,\,2018, December\,\,2018 and May\,\,2019 {\it XM-Newton} observations, respectively.
The RGS spectra were binned to have at least $1$ counts per bin, and we applied Cash statistics (C-stat, \citealp{Cash:1979ys,Kaastra:2017ef}).

The OM data were reduced using the standard guidelines. Details will be reported elsewhere (Li et al. in prep.).

\subsection{{\it NuSTAR}}\label{sect:nustar_datared}

The {\it Nuclear Spectroscopic Telescope Array} ({\it NuSTAR}, \citealp{Harrison:2013zr}) observed 1ES\,1927+654 simultaneously with {\it XMM-Newton}, with exposures of $46$\,ks (June\,\,2018), $65$\,ks (December\,\,2018) and $58$\,ks (May\,\,2019). {\it NuSTAR} data were processed using the {\it NuSTAR} Data Analysis Software \textsc{nustardas}\,v1.8.0 within Heasoft\,v6.24, using the latest calibration files \citep{Madsen:2015ee}. The source was not detected in the 3--24\,keV band during the June 2018 observation, and we recovered upper limits on the flux of $F_{3-5}\leq10^{-14}$, $F_{5-10}\leq6.3\times10^{-15}$ and $F_{10-24}\leq6.3\times10^{-15}\,\rm erg\,s^{-1}\,cm^{-2}$, for the 3--5\,keV, 5--10\,keV and 10--24\,keV bands, respectively. A clear detection up to 8\,keV (11\,keV) was obtained in December 2018 (May 2019), when the flux level of the source was considerably higher. Given the strong variability of the source, we tested for differences in the spectral shape and flux of the spectrum obtained from the whole observation and that extracted from a time interval coincident with the {\it XMM-Newton} observation. For the December 2018 observation we did not find any difference in the spectral shape (see Appendix\,\ref{sect:XMMvsNustar}), so that for our analysis we considered the spectrum obtained from the whole {\it NuSTAR} exposure.

The FPMA and FPMB spectra were extracted using the \textsc{nuproducts} task, selecting circular regions with a radius of $50\arcsec$. The background spectra and light-curves were obtained in a similar fashion, using a circular region of $60\arcsec$ radius located where no other source was detected. The 3--8\,keV net count rate of the source in the December\,\,2018 and May\,2019 observations was $2.3\times 10^{-2}\rm\,ct\,s^{-1}$ and $4\times 10^{-2}\rm\,ct\,s^{-1}$, respectively. The {\it NuSTAR} FPMA/FPMB spectra were merged, as done for the spectra of the two MOS cameras (in \S\ref{sect:xmmobs_spec_june}). The spectra were binned to have at least $20$ counts per bin, and $\chi^2$ statistic was used for the fitting.

\subsection{{\it NICER}}\label{sect:nicer_datared}

Our {\it Neutron Star Interior Composition Explorer} ({\it NICER},
\citealp{Gendreau:2012cr,Arzoumanian:2014qf,Gendreau:2016lq})
observational campaign commenced on May 22, 2018 and is currently still
ongoing. We report here on our analysis of data obtained through
August 5, 2019 (see Table\,\,\ref{tab:obslog} in
Appendix\,\ref{sect:obslog} for the full observation log). The median
exposure during this interval was $\sim 2$\,ks, with a maximum of 20.5\,ks. The typical observation cadence was once per day, with the time
spacing between successive observation sequences varying between 5\,hours
and 8.7\,days. The unfiltered {\it NICER} events were first reprocessed to
assure that the November 2018 energy scale update was applied to all
{\it NICER} observation sequences, and data from {\it NICER} focal plane detector
modules 14 and 34 that are known to undergo episodes of increased
noise were excluded. An ARF tailored to exclude
these was utilized in the spectral analysis. Filtered event files were
constructed by applying the NICERDAS Version 5.0 standard cleaning
criteria that excludes times within the NICER-specific definition of
the SAA, and those with elevations with respect to the earth's limb
and bright earth below their NICERDAS Version 5.0 default
values.\footnote{https://heasarc.gsfc.nasa.gov/docs/nicer/data\_analysis
  \linebreak /nicer\_analysis\_guide.html} Additional
background-related filtering, described below, was also applied.

Although the standard cleaning eliminates much of {\it NICER} internal
background, spectral analysis requires the modelling of the inevitable
residual background. The {\it NICER} internal background, due primarily to
the presence of optical light at low energy and energetic charged
particles at high energy, depends on the orbital parameters and space
weather conditions at the time when the observation was conducted, and
is highly variable both in magnitude and spectral shape. In order to
model the background spectrum for each {\it NICER} observation sequence, we
utilize three background proxies extracted from the unfiltered event
list. One of these is the rate of focused background events in an
energy band (15--17\,keV) where the telescope effective area is
effectively negligible (so all events will be due to particle background), and another the rate of unfocused events
identified as having a ``ballistic deficit'' due to their having
originated at the edges of individual detectors. The latter metric
considers the same ratio of slow-chain PI to fast-chain PI values used
to exclude background in standard screening, to instead select
background events. This rate is tightly correlated with the cut-off
rigidity. A two-dimensional
library of spectra was constructed, indexed by the values of these two rates, and
extracted from the 2.7 Ms of {\it NICER} background field accumulated as of
January 28 2019. The background spectra for any time interval is
estimated by adopting the library spectrum corresponding to the
derived rates of these two proxies, and then rescaling by the ratio of
the 15--17 keV rate to the average rate for that library entry to
impose a ``soft landing'' at the edge of the bandpass. The third background proxy accounts for solar light-leak driven
optical light loading noise events, and is constructed from the
slow-chain count rate in the 0--0.2\,keV noise band that is below the
{\it NICER} nominal energy range for cleaned file events. The library
described above is restricted to ``nighttime'' conditions (noise band
rate $<200\rm\,ct\,s^{-1}$), and an additional library of ``daytime''
background residuals is constructed from the background fields indexed by
noise band rate for rates $>200\rm\,ct\,s^{-1}$.  The residual
spectrum for any time interval is estimated by adopting the library
daytime residual spectrum corresponding to the derived noised band
rate, and then rescaling by the ratio of the 0--0.2 keV rate to the
average such rate for that daytime library entry. The resulting
residual daytime spectrum is added to the nighttime spectra.

We subdivide the cleaned event file good time intervals into 120\,s
subexposures, discarding any leftover intervals shorter than 10\,s. Rates of the three background proxies are then calculated for each
new interval, and daytime and nighttime background spectra identified
accordingly. Intervals that have rates that fall outside of the range
in the nighttime library are eliminated from the list of time
intervals contributing to the source spectra. The final background
spectrum for each observation sequence is then the exposure-weighted
sum over these subexposures of the matched library spectra, with a
background exposure time that is similarly weighted. The highly
variable nature of both the source and background is reflected in the
fact that the {\it NICER} observations include sequences where the source is
not significantly detected above the background and those where the
0.3--3\,keV signal-to-noise ratio is $>1000$. The quality of the match
between the background spectrum estimated as described above and the
total observed spectrum at energies where the background is clearly
dominant varies. We eliminate some of the more poorly matched
intervals from consideration by excising those where the absolute
value of the net 13--15\,keV rate is $>0.10\rm\,ct\,s^{-1}$. Nevertheless,
the source and background spectra for each observation sequence must
be carefully scrutinized, and the analysis details (number of
components, bandpass used) tailored accordingly (\S
\ref{sec:NICERobs}). The final sum of the 265 source spectra exposure
times is $\sim 678$\,ks.

\section{{\it XMM-Newton}/{\it NuSTAR} observations}\label{sect:xmmobs_spec}

In the following we report the results obtained by the spectral analysis of the June 2018 (\S\,\ref{sect:xmmobs_spec_june}), December 2018 and May 2019 (\S\,\ref{sect:xmmobs_spec_december_may}) joint {\it XMM-Newton}/{\it NuSTAR} observations of 1ES\,1927+654. In \S\,\ref{sec:alternativemodelling} we report on the results obtained by applying partial covering and blurred reflection spectral models. We also discuss the X-ray and UV variability of the source (\S\ref{sec:xrayvariabilityXMM}) and the time-resolved spectroscopy (\S\ref{sec:timeresolved2018}) of the three observations, as well as the origin of the broad feature at $\sim 1$\,keV found in all {\it XMM-Newton}/{\it NuSTAR} observations (\S\ref{sect:broadline}).

\begin{figure}
  \begin{center}
\includegraphics[width=0.45\textwidth]{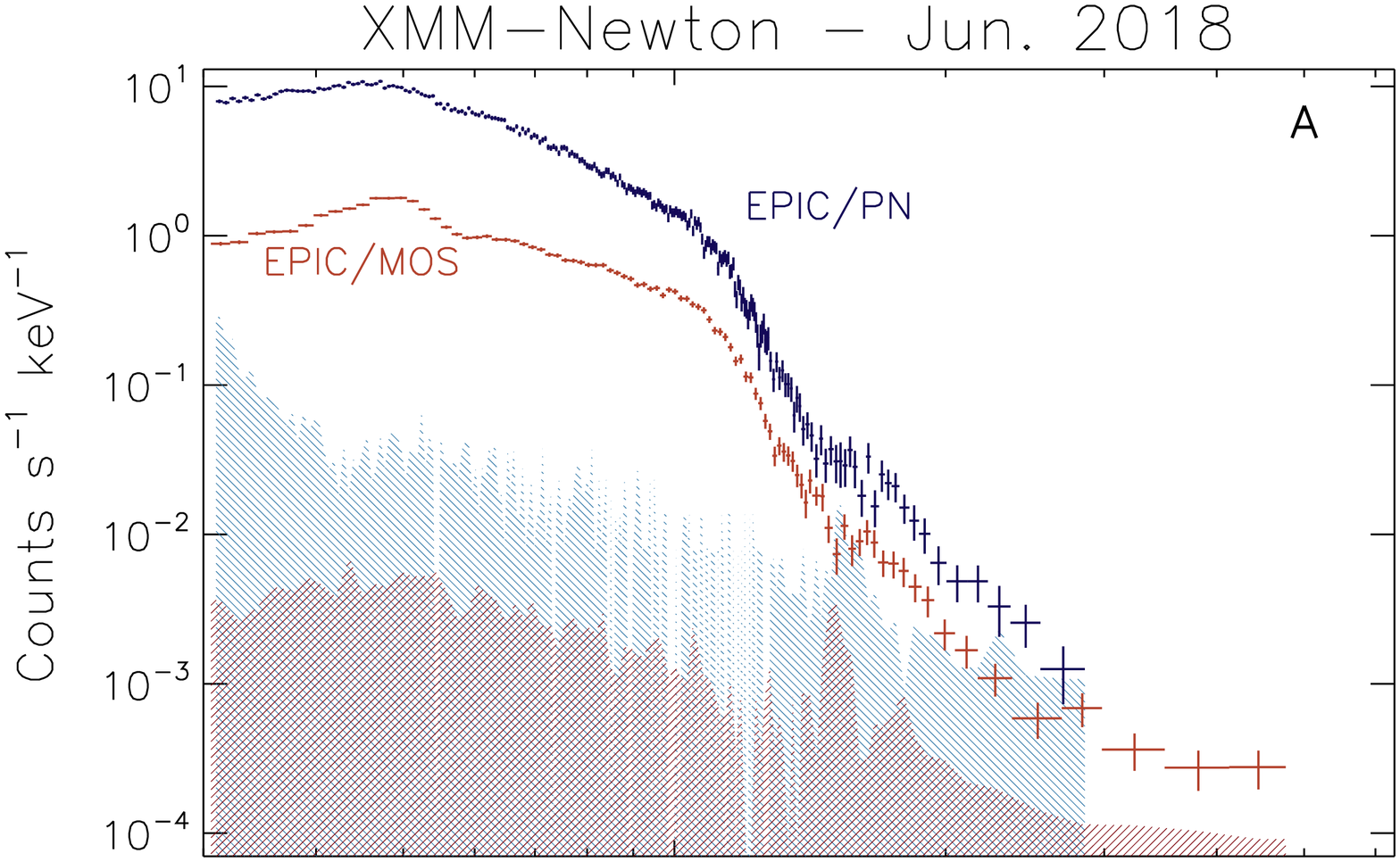}
\includegraphics[width=0.45\textwidth]{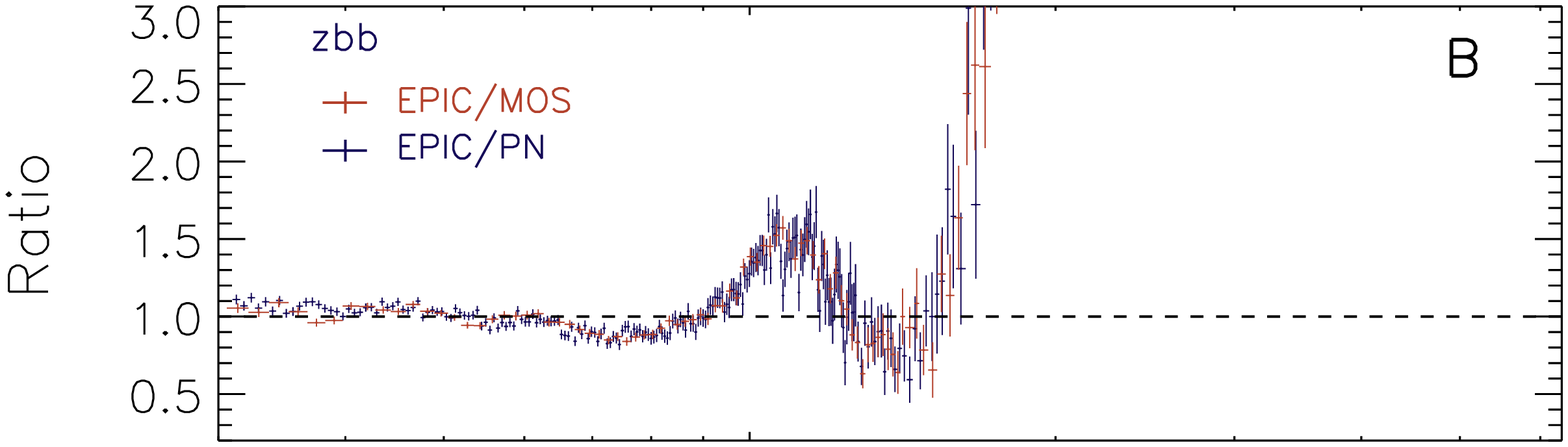}
\includegraphics[width=0.45\textwidth]{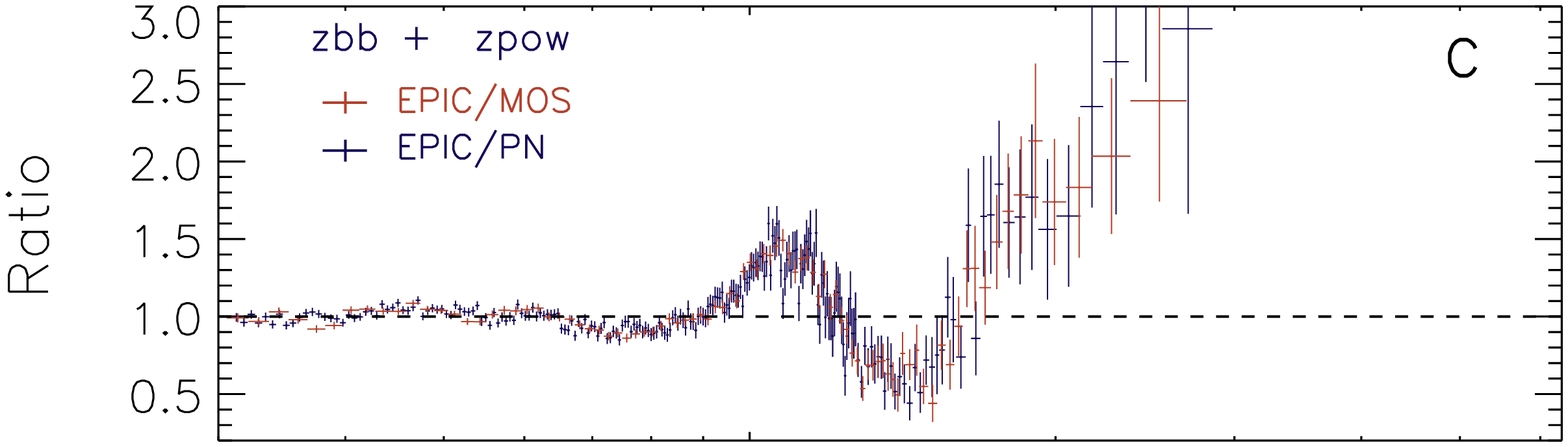}
\includegraphics[width=0.45\textwidth]{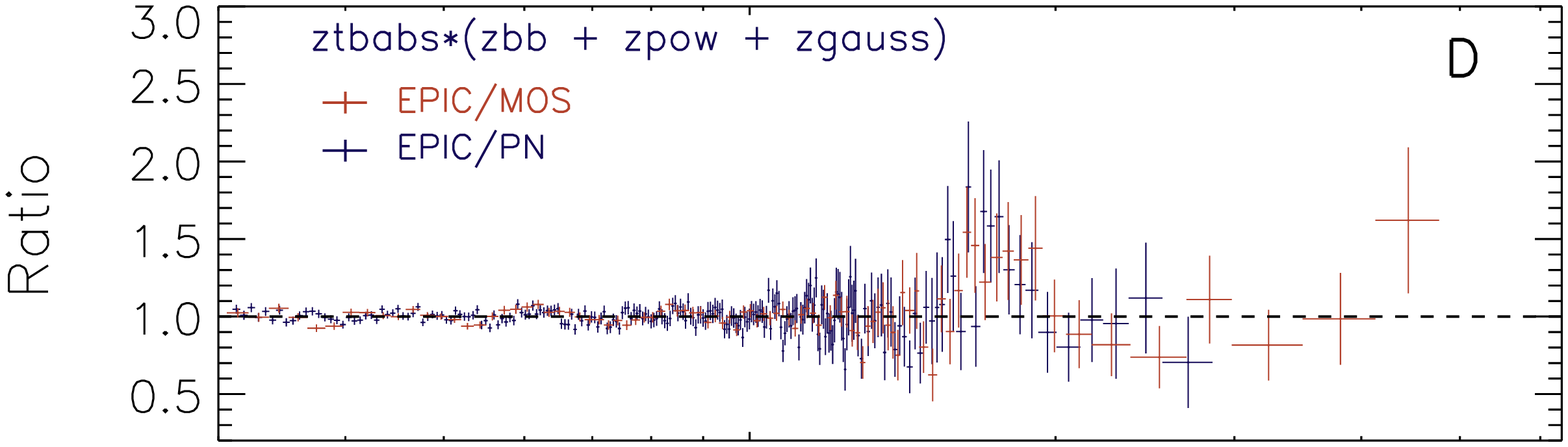}
\includegraphics[width=0.45\textwidth]{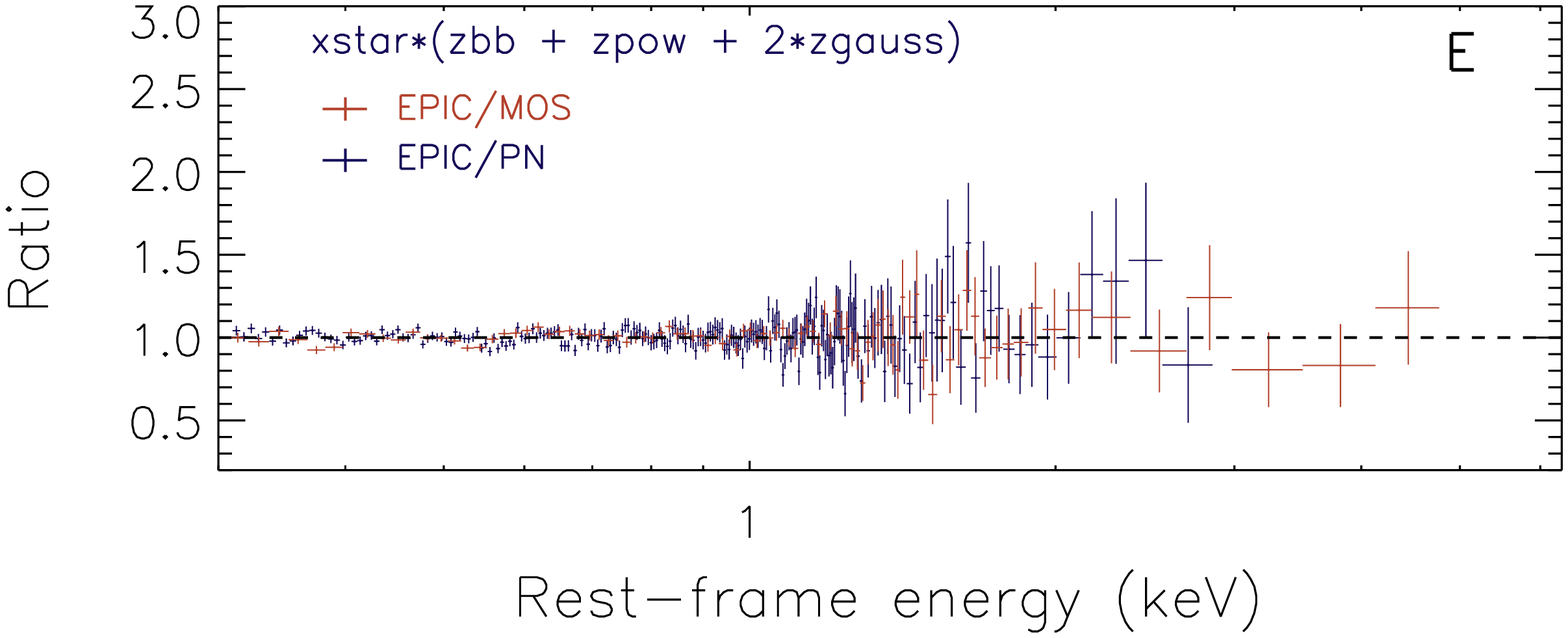}
    \caption{{\it Panel\,\,A}: June 2018 {\it XMM-Newton} EPIC/PN (blue) and MOS (red) spectra of 1ES 1927+654. The blue and red shaded area represent the background spectra of EPIC/PN and MOS, respectively. {\it Panels\,B--E}: ratio between models and the data, sorted from the least (B) to most complex (E) model. The different models used include a blackbody component (\textsc{zbb}), a power-law continuum (\textsc{zpo}), a neutral absorber (\textsc{ztbabs}), a ionized absorber (\textsc{xstar}) and Gaussian lines (\textsc{zgauss}). More details can be found in \S\ref{sect:xmmobs_spec_june}.}
    \label{fig:XMMspecJune18}
  \end{center}
\end{figure}

\tabletypesize{\normalsize}
\begin{deluxetable*}{llccc} 
\tablecaption{Spectral parameters obtained for the {\it XMM-Newton} June\,\,2018 (column\,2), December\,\,2018 (column\,3) and May\,\,2019 (column\,4) observations (EPIC+RGS). The model used included a blackbody component, a power law, two Gaussian absorption lines and two emission lines, a neutral and an ionized absorber.\label{tab:fitXMM}}
\tablehead{
%
 \colhead{ } & \colhead{(1) } & \colhead{(2)} &  \colhead{(3)} &  \colhead{(4)}  \\
\noalign{\smallskip}
 \colhead{ } & \colhead{ } & \colhead{June 2018} & \colhead{December 2018} &  \colhead{May 2019}  
}
\startdata 
\noalign{\smallskip}
a) & $N_{\rm H}$ ($10^{20}\rm\,cm^{-2}$)	&    \nodata           & $3.0\pm 0.1$&	$1.8_{-0.08}^{+0.06}$  \\
\noalign{\smallskip}
b) & $N_{\rm H}^W$ ($10^{20}\rm\,cm^{-2}$)	&      $ 2.2_{-0.5 }^{+0.6 }$         		& $ \leq 1.1$& $ \leq 1.1$ \\
\noalign{\smallskip}
c) &	$\log \xi$ ($\rm\,erg\,cm\,s^{-1}$)	&    $ 2.9_{-0.8 }^{+0.1 }$            & $  1.9_{-0.4 }^{+0.7 }$& $2.7^{+0.2}_{-1.1}$\\
\noalign{\smallskip}
d) & $z$	&         $-0.216 _{-0.018}^{+0.017 }$      	& $ - 0.212_{-0.003}^{+0.006 }$	&  $-0.249^{+0.06}_{-0.013}$ \\
\noalign{\smallskip}
e) & $\Gamma$	&        $1.9 _{-0.7}^{+0.5 }$       	& $  3.02\pm0.01$&  $2.98^{+0.08}_{-0.10}$  \\
\noalign{\smallskip}
f) & $E_{\rm cut}$	&        \nodata       	& $  2.76_{-0.04}^{+0.05 }$& $3.3\pm0.4$ \\
\noalign{\smallskip}
g) & $E_1$ (keV)	&       $ 1.01\pm 0.01$        	& $  1.074\pm 0.005$& $1.01^{+0.03}_{-0.05}$ \\
\noalign{\smallskip}
h) & $\sigma_1$ (eV)	&      $103_{-7}^{+6 }$          & $ 156\pm4$& $193^{+15}_{-25}$  \\
\noalign{\smallskip}
i) & $EW_1$ (eV)	&        $ 184_{-7 }^{+10 }$       	& $  94_{-22 }^{+23 }$& $79^{+30}_{-20}$ \\
\noalign{\smallskip}
j) & $E_2$ (keV)	&        $1.73\pm0.04$       	& $ 1.92_{-0.06}^{+0.05}$& $1.95^{+0.05}_{-0.06}$  \\
\noalign{\smallskip}
k) & $\sigma_2$ (eV)	&    $ 150_{-50 }^{+70 }$            & $  261_{-46 }^{+56 }$ & $300^{+NC}_{-52}$\\
\noalign{\smallskip}
l) & $EW_2$ (eV)	&      $ 460_{-142 }^{+228 }$         	& $  78_{-20 }^{+18 }$& $ 149^{+31}_{-44}$ \\
\noalign{\smallskip}
m) & $C_{\rm Nu}$	&         \nodata         	& $ 0.78\pm0.04$& $0.79^{+0.04}_{-0.05}$ \\
\noalign{\smallskip}
n) & $C_{\rm MOS}$	&         $ 1.05\pm0.01$         	& \nodata  &  \nodata \\
\noalign{\smallskip}
o) & $C_{\rm RGS}$	&          $ 0.92\pm0.01$     	& $ 0.916\pm0.004$& $0.921\pm 0.004$ \\
\noalign{\smallskip}
p) & $E_{\rm abs1}$ (keV)		&       \nodata        & $  0.396\pm0.002$& $0.376^{+0.07}_{-0.06}$\\
\noalign{\smallskip}
q) & $\sigma_{\rm abs1}$ (eV)	&     \nodata           & $ 0.2_{-0.1}^{+0.2}$& $25^{+7}_{-5}$ \\
\noalign{\smallskip}
r) & Strength ($10^{-3}$)	&      \nodata            & $ 1.4_{-0.6}^{+1.7}$& $11\pm3$ \\
\noalign{\smallskip}
s) & $E_{\rm abs2}$ (keV)	&         \nodata      	& $  0.527\pm0.001$& \nodata	\\
\noalign{\smallskip}
t) & $\sigma_{\rm abs2}$ (eV)	&    \nodata            & $ 0.5\pm0.1$& \nodata	 \\
\noalign{\smallskip}
u) & Strength ($10^{-3}$)	&      \nodata            & $ 3.0_{-0.8}^{+1.0}$&\nodata	 \\
\noalign{\smallskip}
v) & $kT$ (eV)	&            $ 102\pm1$   	& $  142\pm1$& $160^{+4}_{-6}$ \\
\noalign{\smallskip}
w) & $F_{0.3-10}$ ($\rm 10^{-11}\,\,erg\,s^{-1}\,cm^{-2}$) & $1.0 $ & $6.7$ & $7.9$\\
\noalign{\smallskip}
x)& $L_{0.3-10}$ ($\rm erg\,s^{-1}$) & $8.5\times 10^{42}$ & $5.7\times10^{43}$ & $6.7\times10^{43}$\\
\noalign{\smallskip}
y)& Stat/DOF	&        2943/2847 	        	& 4137/3712 &  3916/3549  
\enddata
\tablecomments{The table reports: the column density of the neutral absorber (a); the column density (b), ionization parameter (c) and redshift (d) of the ionized absorber; the photon index (e) and cutoff (f) of the power-law component; the energy (g), width (h) and equivalent width (i) of the first Gaussian emission line; the energy (j), width (k) and equivalent width (l) of the second Gaussian line; the cross-calibration constant of the {\it NuSTAR}/FPM (m), {\it XMM-Newton} MOS (n) and RGS (o) spectra; the energy (p), width (q) and strength (r) of the first Gaussian absorption line; the energy (s), width (t) and strength (u) of the second Gaussian absorption line; the temperature of the blackbody (v), the 0.3--10\,keV flux (w) and luminosity (x), and the value of the statistic and the number of degrees of freedom (y).}
\end{deluxetable*}

\subsection{The June 2018 observation}\label{sect:xmmobs_spec_june}

Our first joint {\it XMM-Newton}/{\it NuSTAR} X-ray observations were carried out on June 2018. The source an extremely soft continuum, very different from that observed in the previous {\it XMM-Newton} observation, carried out in May 2011, and from typical AGN. After the event in fact 1ES\,1927+654 showed only a very weak powerlaw component, which is found to typically dominate the X-ray emission of AGN (e.g., \citealp{Mushotzky:1993fj}). Because of its extreme softness, the source was not detected by {\it NuSTAR}, and the spectral analysis was carried out only using EPIC and RGS data. We used the EPIC PN (MOS) data in the 0.3--2.8\,keV (0.3--4.6\,keV) range, and RGS data in the 0.35--1.3\,\,keV interval.
The spectral analysis was carried out within \textsc{XSPEC}\,\,v12.10.0e \citep{Arnaud:1996kx}. In all models we took into account Galactic absorption at the position of the source ($N_{\rm H}=6.87\times 10^{20}\rm\,cm^{-2}$, \citealp{Kalberla:2005fk}) using the \textsc{tbabs} spectral model \citep{Wilms:2000vn}. A multiplicative constant (\textsc{cons} in XSPEC) was added to all models to take into account any cross-calibration offset between the different instruments.
\smallskip

The EPIC PN and MOS spectra of this observation are shown in panel\,\,A of Fig.\,\ref{fig:XMMspecJune18}. We started by fitting the X-ray spectrum with a simple blackbody model (\textsc{cons$\times$tbabs$\times$zbb}), which resulted in a chi-squared of $\chi^{2}=2483$ for 320 degrees of freedom (DOF), and left very strong residuals around 1\,keV and above $\simeq 1.5$\,keV (see panel\,\,B). The addition of a power-law component to the model [\textsc{cons$\times$tbabs$\times$(zbb+zpo)}] improved the fit ($\chi^{2}=1918$ for 318 DOF), still leaving significant residuals (panel\,\,C) between 1 and 2\,\,keV. Similarly to what was done by \citet{Gallo:2013hq} we also added a neutral absorber (\textsc{ztbabs} in \textsc{XSPEC}), possibly associated with gas in the host galaxy, leaving its column density free to vary [\textsc{cons$\times$tbabs$\times$ztbabs$\times$(zbb+zpo)}], which improved the fit ($\chi^{2}=1813$ for 317 DOF).
Applying this model to the RGS spectrum also shows significant residuals (see panel\,\,B of Fig.\,\ref{fig:RGS_spec_18} in Appendix\,\ref{appendix:RGS_spectra}), and in particular an excess around 1\,keV (clearly observed also in the second-order spectrum, see the top panels of Fig.\,\ref{fig:RGS_spec_order2}).

To take into account the strong residuals at $\sim$1\,keV, we added a Gaussian line, leaving its width and energy free to vary. This greatly improved the fit ($\chi^{2}=405$ for 314 DOF, panel\,\,D), leaving only some residuals at $\simeq 1.8$\,keV and around the absorption lines also observed in the RGS spectrum. We therefore added a second Gaussian, leaving both the energy and the width free to vary in the fit, which yielded a better fit ($\chi^{2}=375$ for 311 DOF). 
To model the absorption lines resolved in the RGS spectra, we then added a warm absorber using an \textsc{xstar} \citep{Kallman:2001ul,Bautista:2001bh} table built for a blackbody with a temperature of $kT=100$\,eV, fixing the turbulent velocity to $v_{\rm turb}=10^{4}\rm\,km\,s^{-1}$ [\textsc{cons$\times$tbabs$\times$ztbabs$\times$mtable\{xstar\}} \textsc{$\times$(zbb+zpo+zgauss+zgauss)}]. We left the redshift, ionization parameter and column density of the ionized absorber free to vary in the fit. This removed the need of a neutral absorber, which was not constrained by the fit ($N_{\rm H}\leq 2.3\times 10^{19}\rm\,cm^{-2}$), and resulted in a good fit ($\chi^{2}=346$ for 309 DOF, panel\,\,E). The ionized absorber provides a better fit than a simple neutral absorber, providing a difference in the chi-squared of $\Delta \chi^2=29$ for two additional DOF (i.e., ionization parameter and redshift).

\begin{figure}[ht!]    
 \begin{center}
    \includegraphics[width=0.48\textwidth]{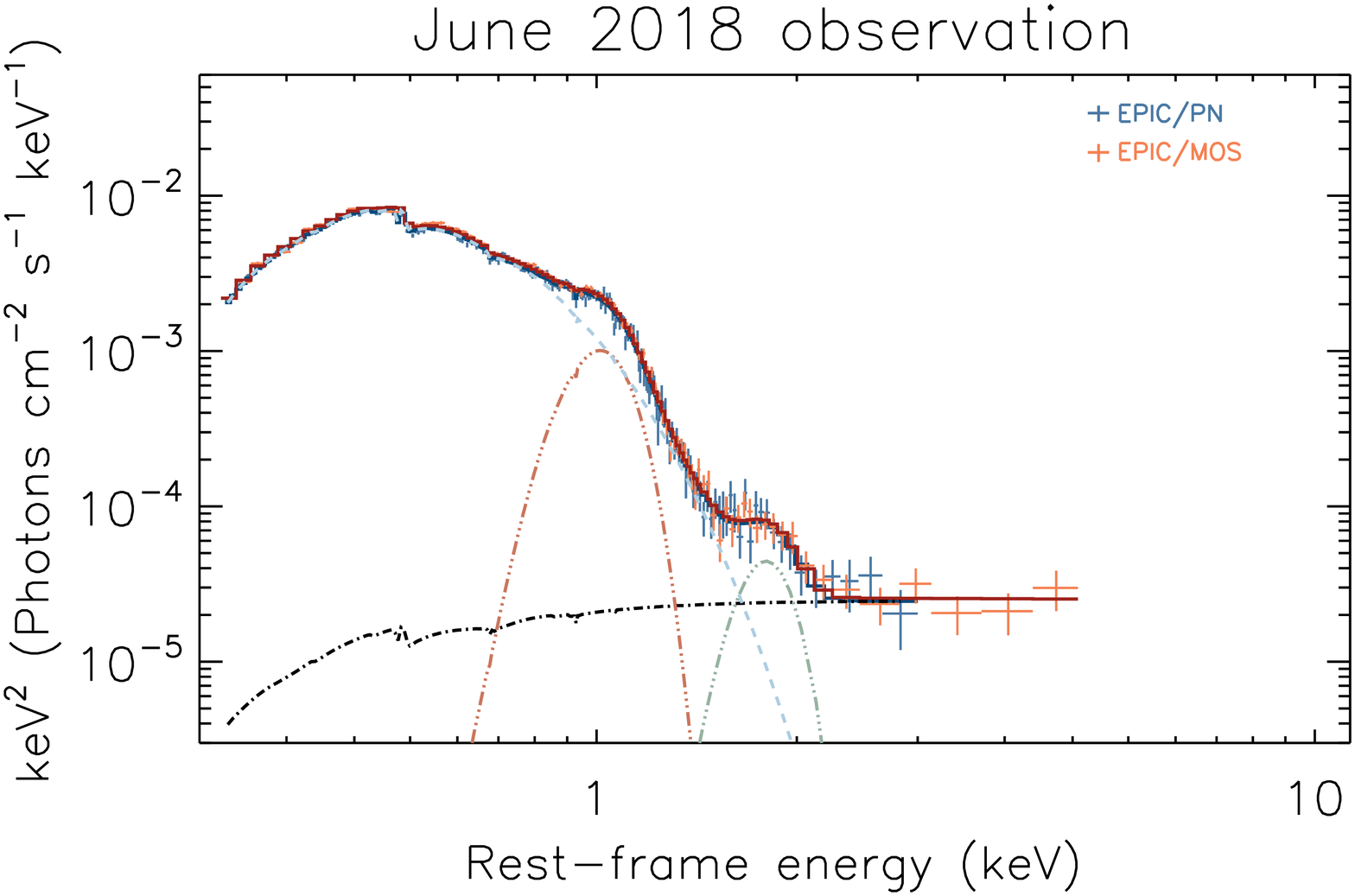}
    \par\medskip
    \includegraphics[width=0.48\textwidth]{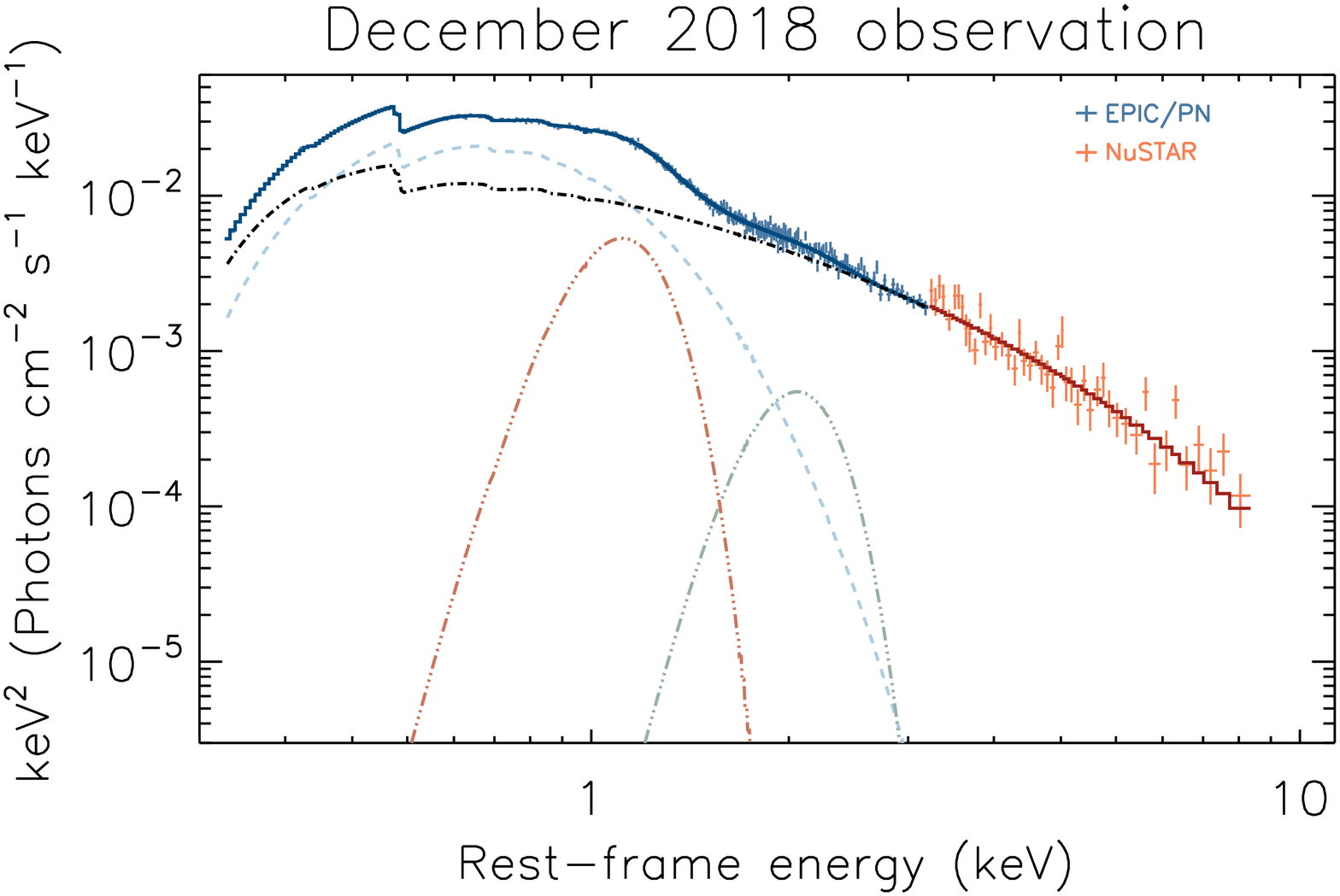}
    \par\medskip
    \includegraphics[width=0.48\textwidth]{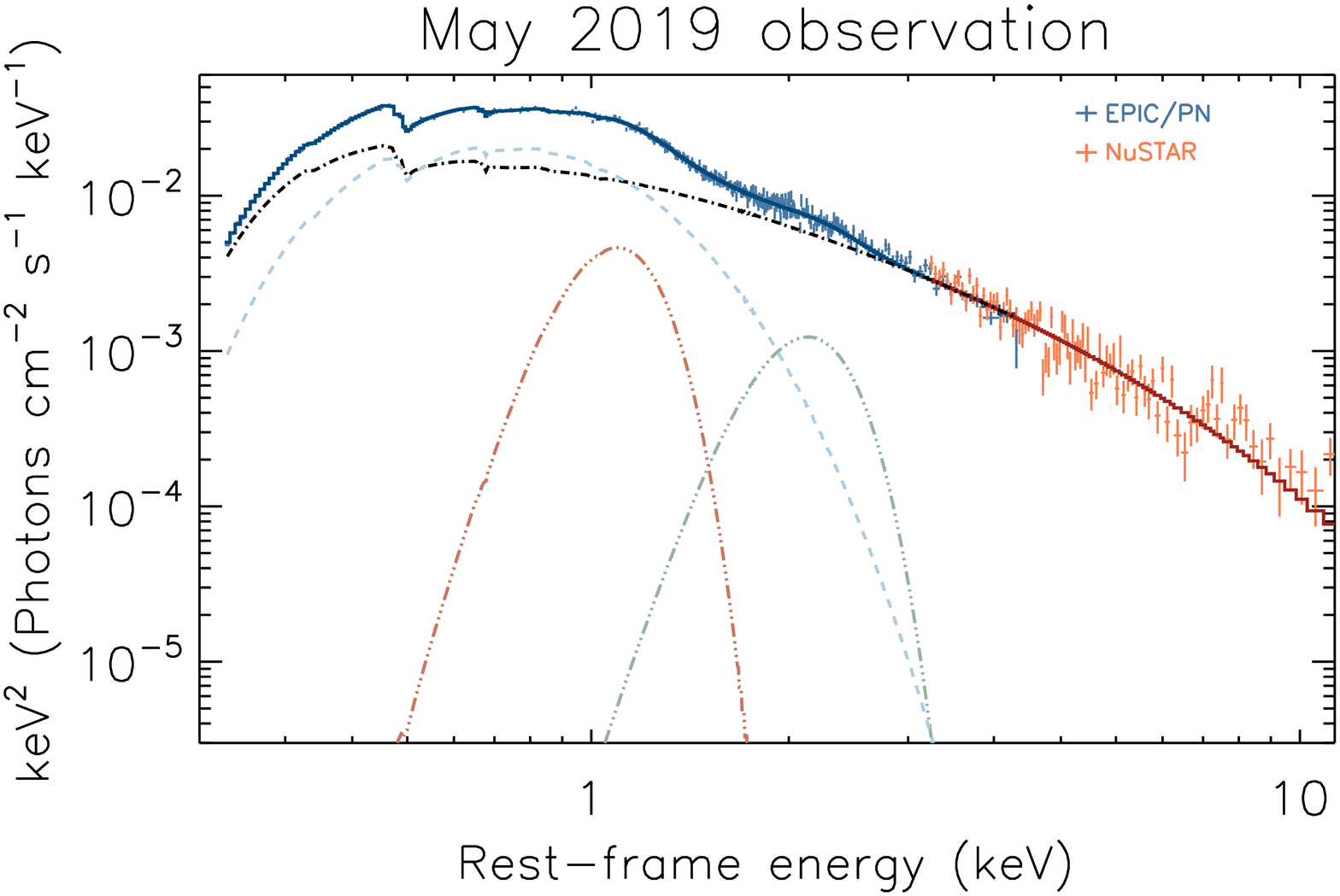}
 \end{center}
    \caption{{\it Top panel:} June 2018 {\it XMM-Newton} EPIC/PN (blue points) and MOS (red points) spectra of 1ES\,1927+654. The continuous line shows the best-fitting model, which includes a blackbody (cyan dashed line), a power-law (black dot-dashed line), and two Gaussian lines (dot-dot-dashed lines). These components are absorbed by low column density neutral and ionized material (see \S\ref{sect:xmmobs_spec_june} and column\,2 of Table\,\,\ref{tab:fitXMM} for details). The residuals of this fit are shown in panel\,\,E of Fig.\,\ref{fig:XMMspecJune18}. 
  {\it Middle planel:} same as the top panel for the December 2018 {\it XMM-Newton} EPIC/PN (blue points) and {\it NuSTAR} (red points) observations, using a cutoff power-law (black dot-dashed line). See \S\ref{sect:xmmobs_spec_december_may} and column\,3 of Table\,\,\ref{tab:fitXMM} for details. The residuals of this fit are shown in panel\,\,E (left) of Fig.\,\ref{fig:XMMspecDec18May19}.
   {\it Bottom planel:} same as the middle panel for the May 2019 {\it XMM-Newton} EPIC/PN (blue points) and {\it NuSTAR} (red points) observations. (see \S\ref{sect:xmmobs_spec_december_may} and column\,4 of Table\,\,\ref{tab:fitXMM} for details). The residuals of this fit are shown in panel\,\,E (right) of Fig.\,\ref{fig:XMMspecDec18May19}.\newline
 }
    \label{fig:xmmobseeufs}
\end{figure}

We then fitted the EPIC spectra together with RGS, using our best model, which included a blackbody, a power-law, two Gaussian lines and a layer of ionized material. This model can reproduce the data very well (Stat/DOF=2943/2847, where Stat includes both $\chi^2$ and C-stat; see panel\,\,C of Fig.\,\ref{fig:RGS_spec_18} in Appendix\,\ref{appendix:RGS_spectra}), and resulted in a temperature of the blackbody component of $kT=102\pm1$\,eV and a photon index of $\Gamma=1.9^{+0.5}_{-0.7}$, consistent with what we found using only EPIC data. The results of this final fit are reported in column\,2 of Table\,\,\ref{tab:fitXMM}, while the spectrum and the different components used are illustrated in the top panel of Fig.\,\ref{fig:xmmobseeufs}. We found that the ionized absorber has a redshift of $z\sim-0.22\pm0.02$, which implies a velocity of $v\sim0.26\rm\,c$ comparable to those observed in some AGN (e.g., \citealp{Tombesi:2010zr}) and Ultra-luminous X-ray sources (e.g., \citealp{Pinto:2016qn,Walton:2016ki}). We also created an absorption model with a lower turbulent velocity ($v_{\rm turb}=300\rm\,km\,s^{-1}$), considering a blackbody with a temperature of $kT=100$\,eV for the X-ray continuum.
We included this model in our fit to the EPIC and RGS spectra, and found that it  provides a worse fit (Stat/DOF=2963/2847) than that obtained by considering an absorber with very large turbulent velocity. While both the column density and the ionization parameter were consistent with those obtained assuming $v_{\rm turb}=10^{4}\rm\,km\,s^{-1}$, the redshift of the absorber was considerably lower ($z=-0.101\pm0.001$).

As a further test we replaced the \textsc{xstar} table with the \textsc{xstar} \textsc{warmabs} model, which is not based on pre-calculated grids, but calculates the spectrum on the fly. We produced specific population files using the observed X-ray spectrum of 1ES\,1927+654, rather than a single-component model. Fitting the X-ray spectrum with this model [\textsc{cons$\times$tbabs$\times$ztbabs$\times$warmabs$\times$} \textsc{(zbb+zpo+zgauss+zgauss)}] resulted in a good fit (Stat/DOF=3021/2846), and values of the temperature of the blackbody ($kT=102\pm1$\,eV) and photon index ($\Gamma=2.0^{+0.5}_{-0.7}$), consistent with what we found applying the XSTAR table. With this approach we could also get some constraints on the turbulent velocity of the absorber, which was found to be $v_{\rm turb}\geq 7418\rm\,km\,s^{-1}$.

We tested different X-ray continuum models to replace the blackbody component, including a multicolor blackbody model and a Comptonized plasma (see Appendix\,\ref{appendix:XMM_alternativecontinuummodels}), and we found that the data appear to be best fitted by the blackbody model.

\begin{figure*}
  \begin{center}
\includegraphics[width=0.45\textwidth]{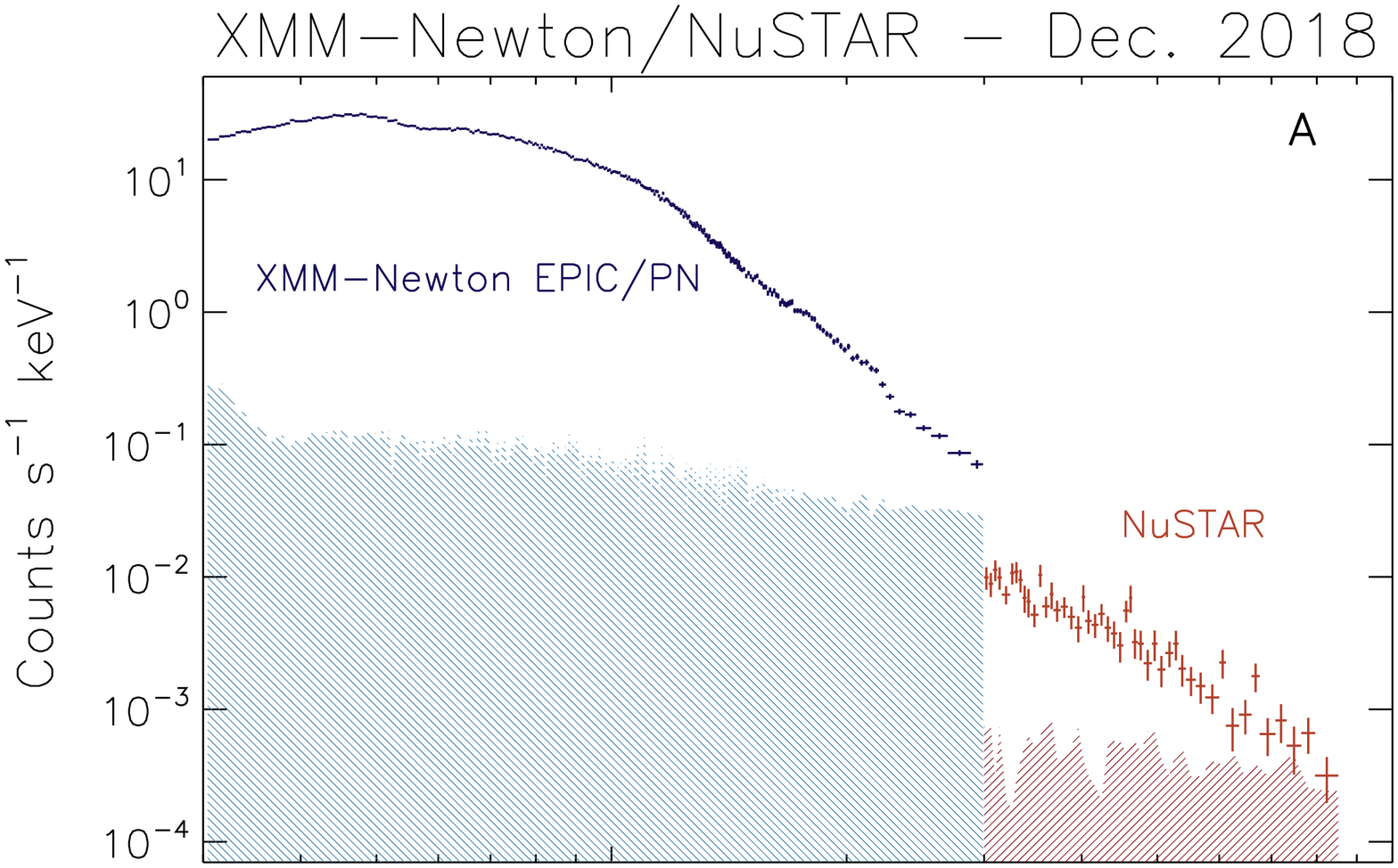}
\includegraphics[width=0.45\textwidth]{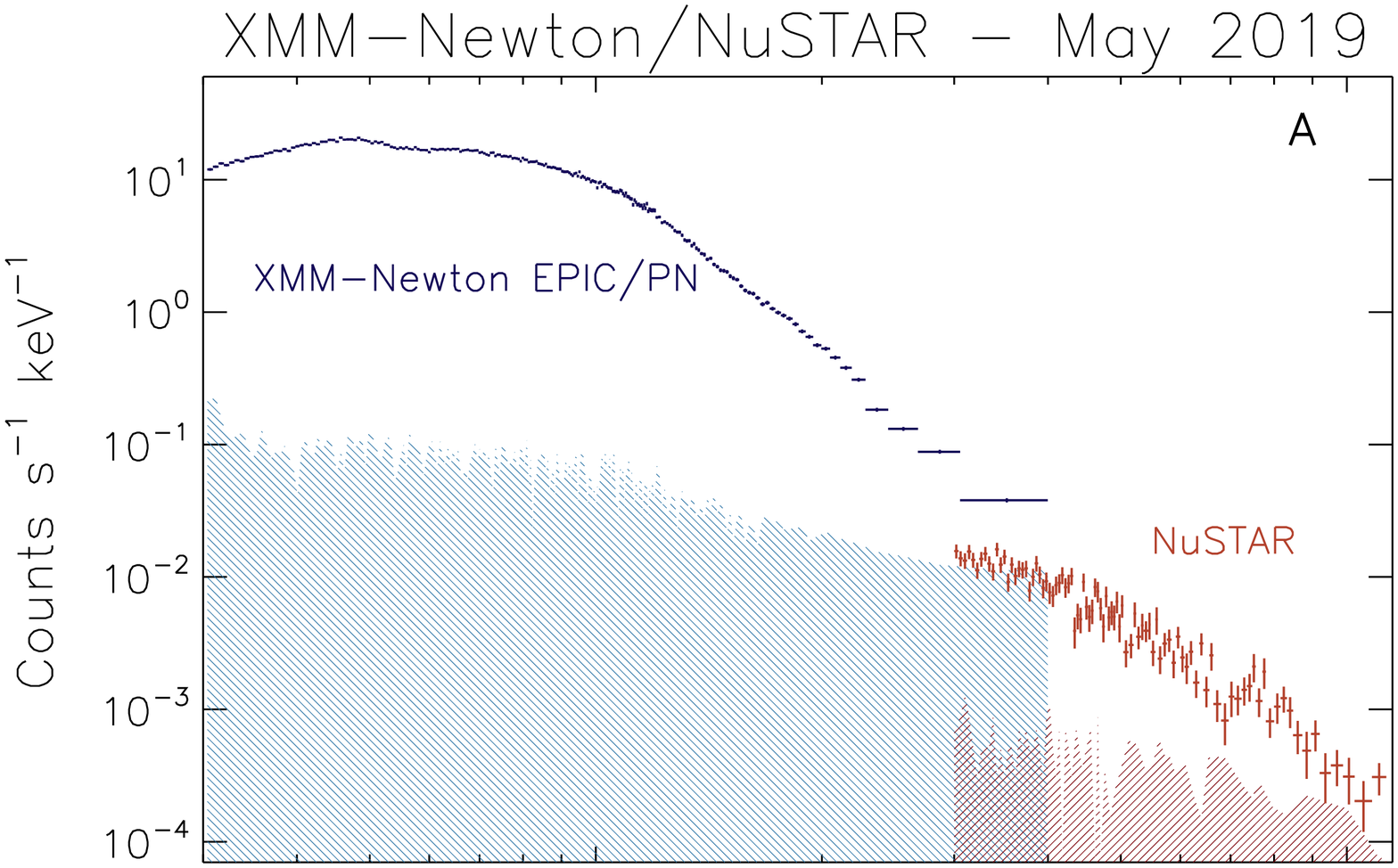}
\includegraphics[width=0.45\textwidth]{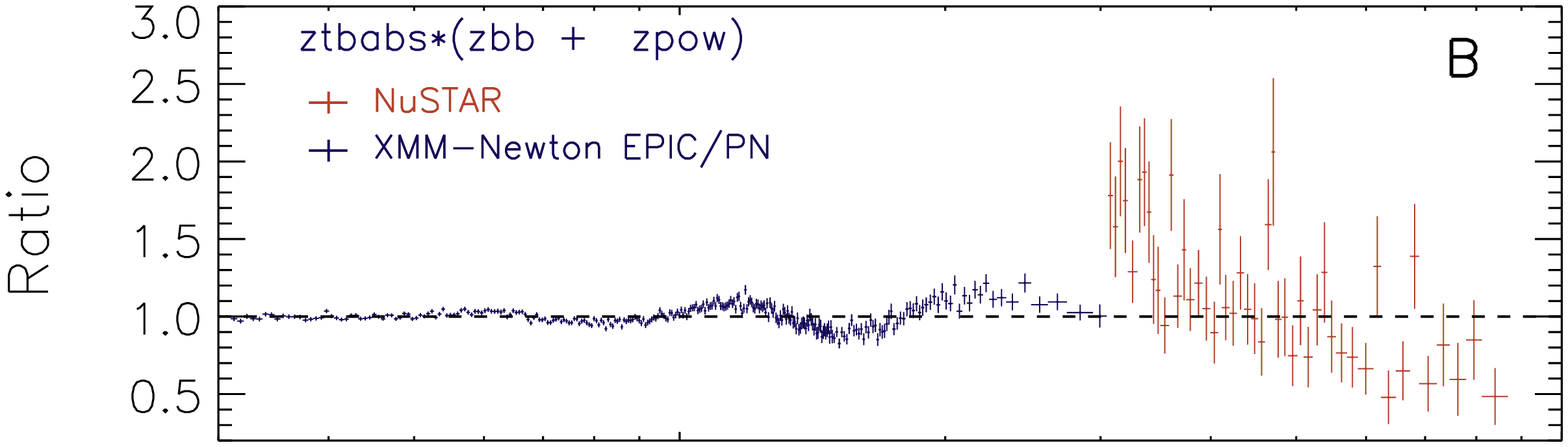}
\includegraphics[width=0.45\textwidth]{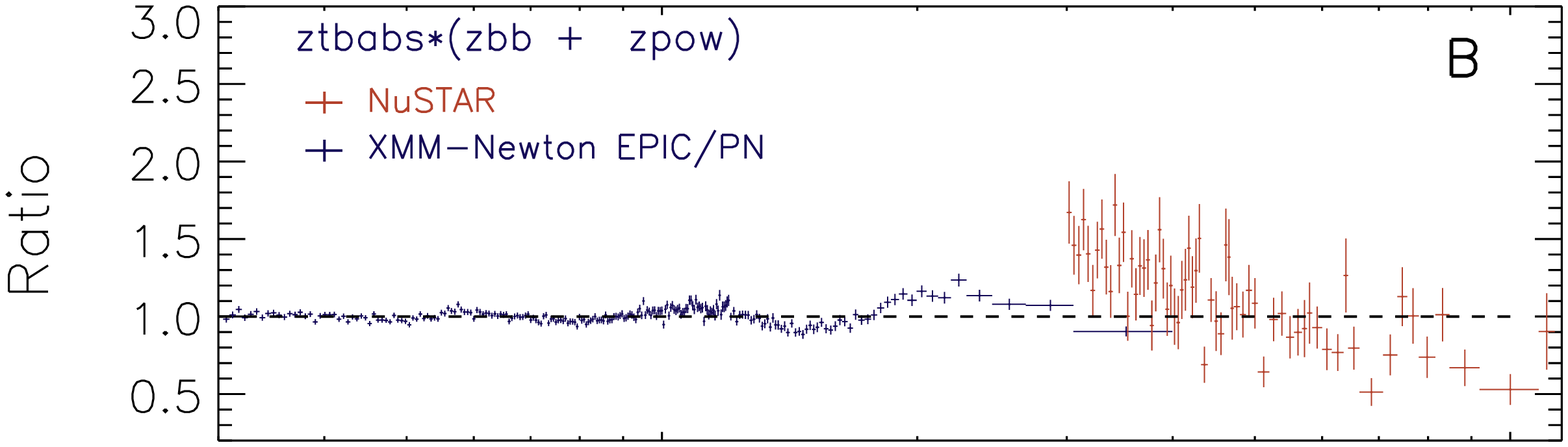}
\includegraphics[width=0.45\textwidth]{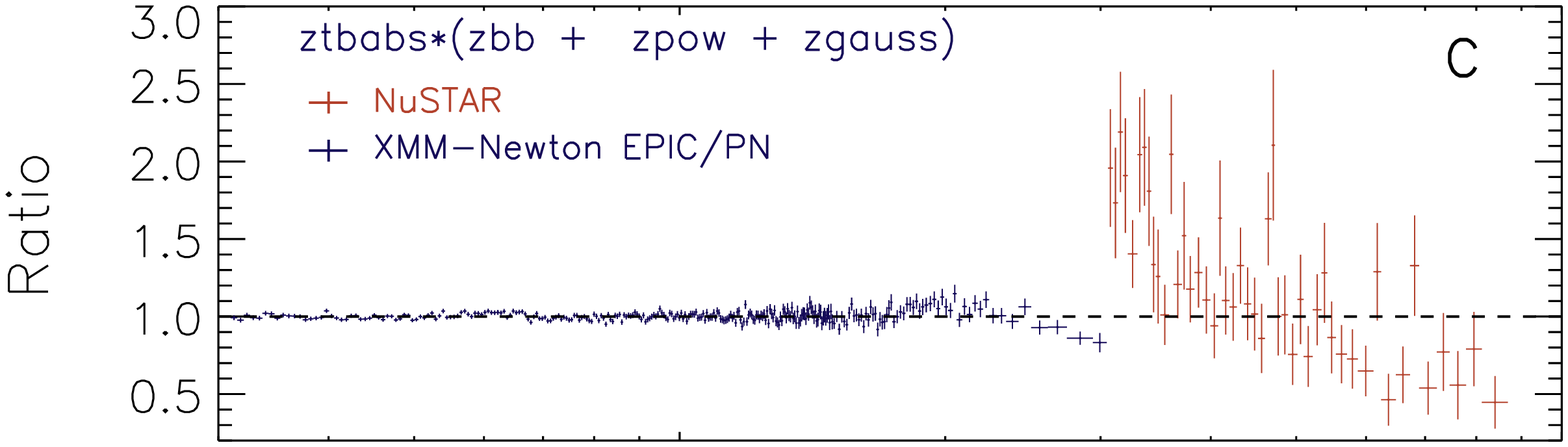}
\includegraphics[width=0.45\textwidth]{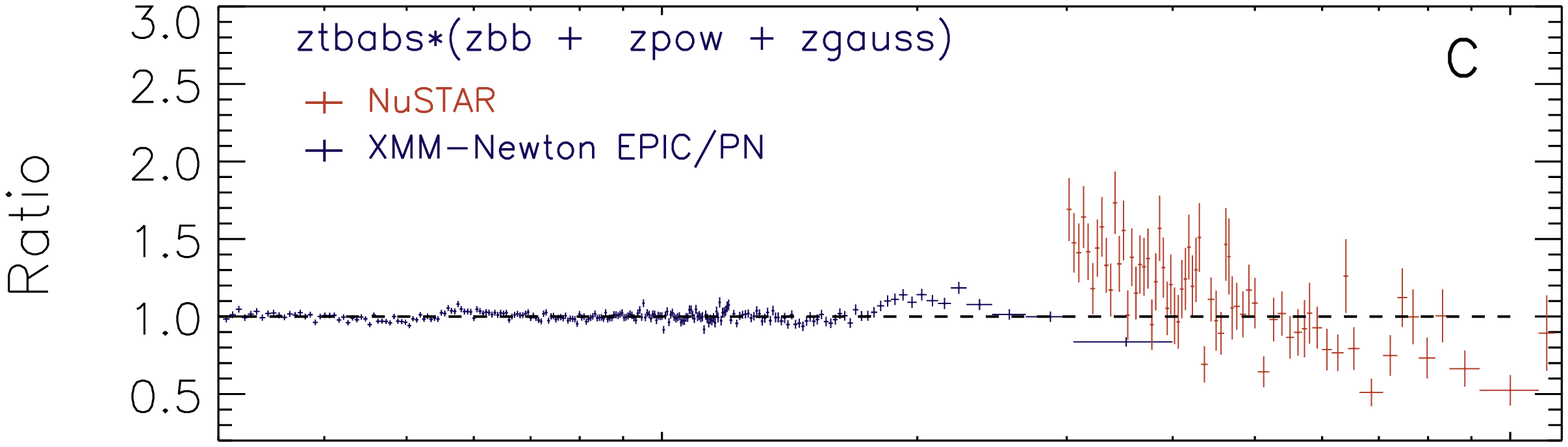}
\includegraphics[width=0.45\textwidth]{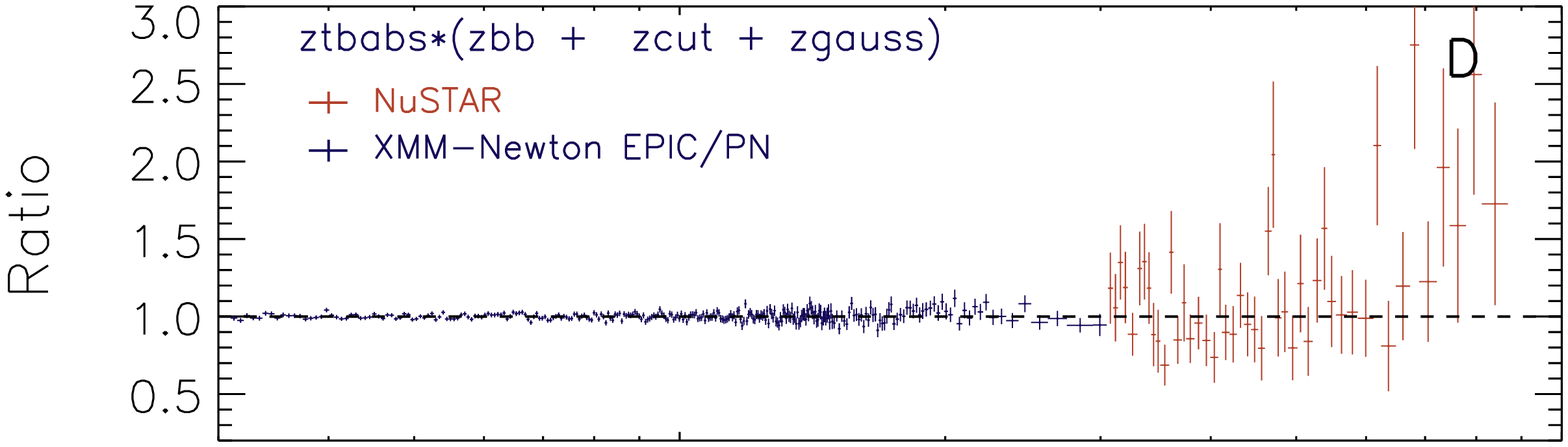}
\includegraphics[width=0.45\textwidth]{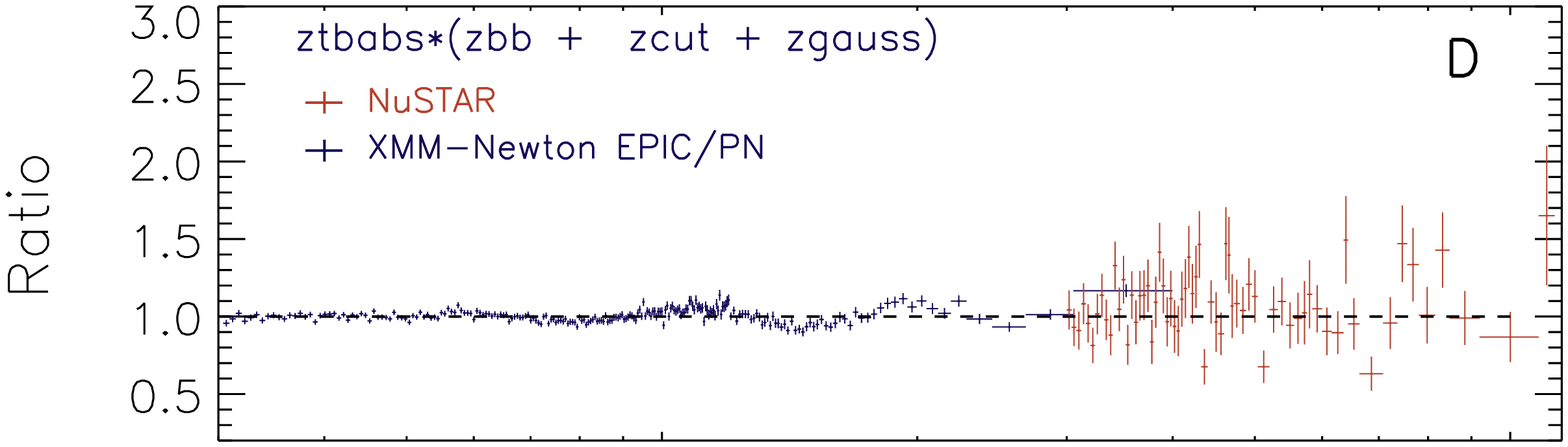}
\includegraphics[width=0.45\textwidth]{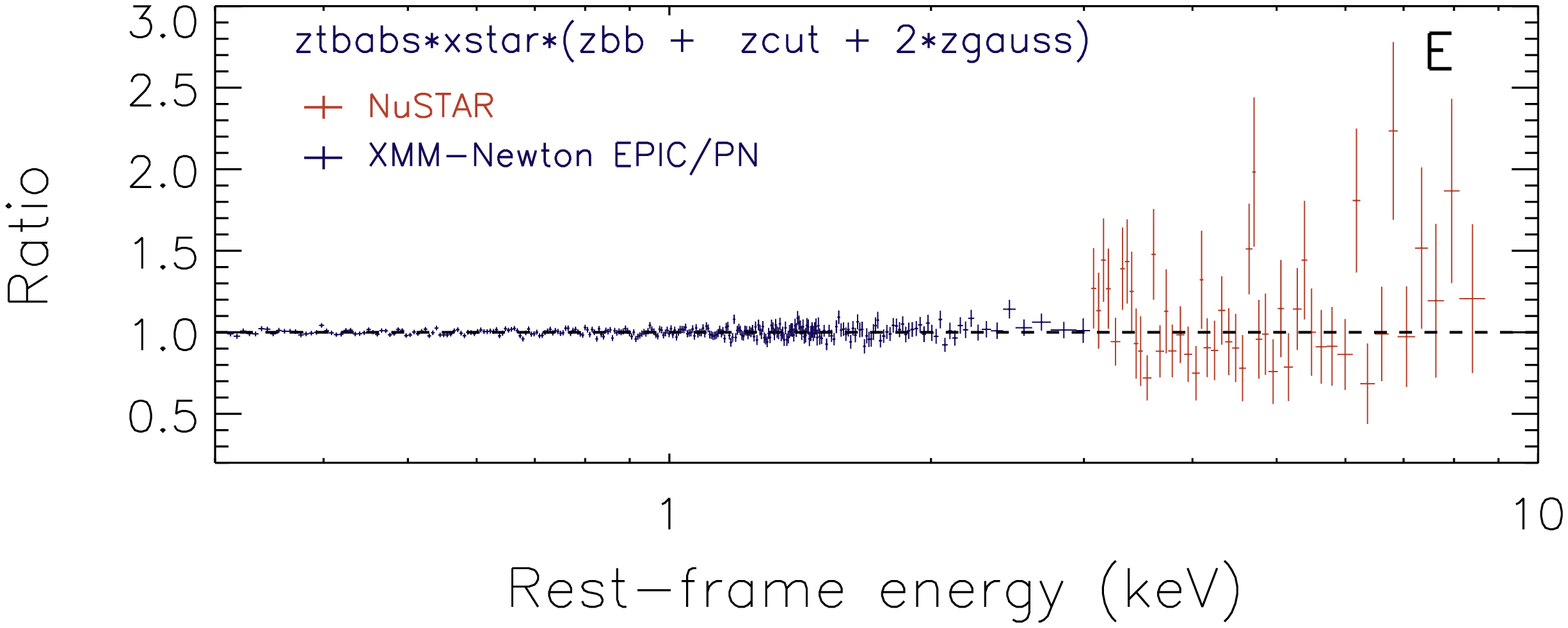}
\includegraphics[width=0.45\textwidth]{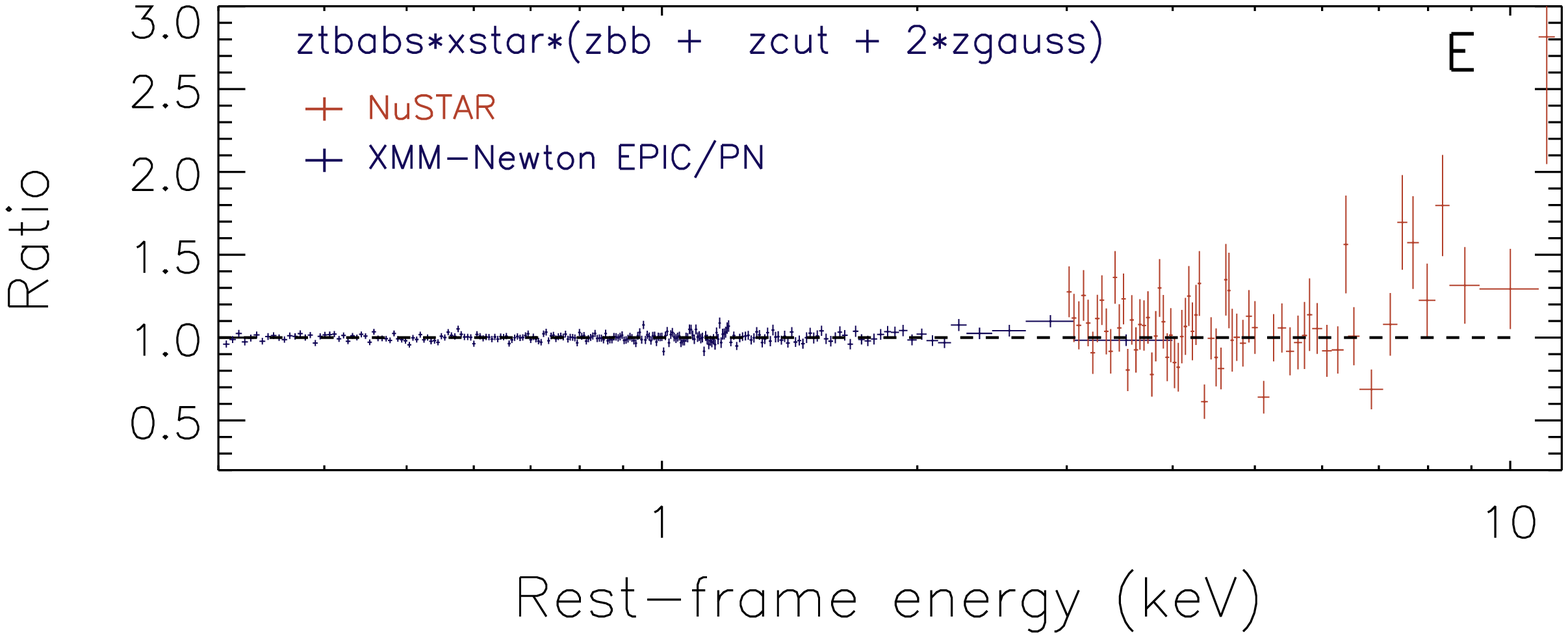}
    \caption{{\it Panel\,\,A}: December 2018 (left panels) and May 2019 (right panels) {\it XMM-Newton} EPIC/PN (blue) and {\it NuSTAR} (red) spectra of 1ES 1927+654. The blue and red shaded area are the background spectra of EPIC/PN and {\it NuSTAR}, respectively. {\it Panels\,B--E}: ratio between the models and the data, sorted from the least (B) to most complex (E) model. The different models used include a blackbody component (\textsc{zbb}), a power-law continuum (\textsc{zpo}), a neutral absorber (\textsc{ztbabs}), ionized absorbers (\textsc{xstar}) and Gaussian lines (\textsc{zgauss}). More details can be found in \S\ref{sect:xmmobs_spec_december_may}. }
    \label{fig:XMMspecDec18May19}
  \end{center}
\end{figure*}

\subsection{The December 2018 and May 2019 observations}\label{sect:xmmobs_spec_december_may}

Two additional joint {\it XMM-Newton}/{\it NuSTAR} observations of the source were carried out on December\,\,2018 and May\,\,2019 (panels\,A of Fig.\,\ref{fig:XMMspecDec18May19}). During these observations the source was $\simeq 7-8$ times brighter than in June, and the maximum luminosity was reached during the May\,\,2019 observation. While the spectra were found to be still extremely soft, the emission was significantly harder than in the previous {\it XMM-Newton} observation. This, combined with the higher overall flux, allowed {\it NuSTAR} to detect the source in both observations. For the December\,\,2018 (May\,\,2019) we used the EPIC/PN data in the 0.3--3\,keV (0.3--3\,keV) range, and RGS data in the 0.35--1.85\,keV (0.35--1.85\,keV) interval; {\it NuSTAR} data were used in the 3--8\,keV (3--11\,keV) range.

We started by applying a model that includes a blackbody plus a powerlaw component [\textsc{cons$\times$tbabs$\times$ztbabs$\times$(zbb+zpo)}]. This resulted in a poor fit ($\chi^2$/DOF=1637/567 and $\chi^2$/DOF=1340/667 for December\,\,2018 and May\,\,2019, respectively), with clear residuals in the 1--2\,keV region, similarly to what we found studying the June 2018 {\it XMM-Newton} observation (see $\S\ref{sect:xmmobs_spec_june}$). Moreover, the fit also fails to reproduce the spectrum above 3\,keV (panels\,B of Fig.\,\ref{fig:XMMspecDec18May19}).
Adding a Gaussian line [\textsc{cons$\times$tbabs$\times$ztbabs$\times$(zbb+zpo+zgauss)}], panels\,C] to this model removes the residuals at $\simeq 1$\,keV, leaving however very strong residuals above 3\,keV ($\chi^{2}$/DOF=820/564 and $\chi^{2}$/DOF=1079/664) and a cross-calibration constant for the {\it NuSTAR} spectrum that is rather unrealistic ($C_{\rm Nu}=0.41\pm 0.03$ and $C_{\rm Nu}=0.53\pm 0.03$), considering that the observations were carried out simultaneously.

\subsubsection{Cutoff powerlaw continuum}
Since a power-law continuum cannot reproduce the 3--10\,keV region, we used a cutoff power law model to account for the spectral curvature observed in {\it NuSTAR} data [\textsc{cons$\times$tbabs$\times$ztbabs$\times$(zbb+zcut+zgauss)}]. This reproduces much better the spectra ($\chi^{2}$/DOF=647/563 and $\chi^{2}$/DOF=1040/663, panels\,D).
An ionized absorber appears to be still present, as shown by the RGS spectra (see panels\,B of Fig.\,\ref{fig:RGS_spec_18} and Fig.\,\ref{fig:RGS_spec_19} in Appendix\,\ref{appendix:RGS_spectra}), with absorption features around $0.7-0.8$\,keV. We therefore used the same \textsc{xstar} table adopted for the June observation to take into account this absorber. 
Some residuals are also evident in the $1.7-2$\,keV region, consistent with what we found for the June\,\,2018 observation, so that we added also a second Gaussian component, which improved significantly the fit ($\chi^{2}$/DOF=621/557 and $\chi^{2}$/DOF=726/657, panels\,E).

As we did for the June\,\,2018 observation we used this model to fit the {\it NuSTAR}, {\it XMM-Newton} EPIC/PN and RGS data (Stat/DOF=4363/3718 and Stat/DOF=3949/3552). The RGS spectrum of the December\,\,2018 observation shows two additional absorption features at $\sim 0.4$\,keV and $\sim 0.53$\,keV that are not reproduced by the ionized absorber. Only the feature at $\sim 0.4$\,keV was found in the May\,\,2019 spectrum. We therefore added two (one) Gaussian absorption lines to the December\,\,2018 (May\,\,2019) spectrum, leaving both the energy and the width free to vary, which improved the fit (Stat/DOF=4137/3712 and Stat/DOF= 3916/3549, see panels\,C of Figs.\,\ref{fig:RGS_spec_18} and \ref{fig:RGS_spec_19}). The results of this fit are reported in columns\,\,3 and 4 of Table\,\,\ref{tab:fitXMM}, while the spectra and the different components used are illustrated in the middle and bottom panels of Fig.\,\ref{fig:xmmobseeufs}.

The temperature of the blackbody component is higher in December\,\,2018 (May\,\,2019) than it was in June\,\,2018 ($kT=142\pm1$\,eV and $kT=160^{+4}_{-6}$\,eV, respectively). The velocity of the outflow is slightly higher in May\,\,2019 than in the previous two observations. The photon index is consistent in the two observations, and the power-law component is significantly stronger with respect to the blackbody component than in December\,\,2018.

\subsubsection{Thermal Comptonization continuum}

In order to describe the hard X-ray component produced by the corona, which was only weakly detected in the first {\it XMM-Newton} observation, we have thus far used a cutoff power-law, consistent what is typically done for AGN (e.g., \citealp{Dadina:2007lc,Ricci:2017fj}). However, if the seed photons of the Comptonization process are the soft X-ray photons from the thermal component, this approach might not be physically self consistent. In order to test this scenario we used the \textsc{nthcomp} model \citep{Zdziarski:1996wp,Zycki:1999rc}, which provides an accurate description of the X-ray continuum arising from thermal Comptonisation. The parameters of this model are the temperature of the Comptonizing electrons ($kT_{\rm e}$), the temperature of the blackbody providing the seed photons ($kT_{\rm bb}$), the asymptotic power-law photon index ($\Gamma_{\rm nth}$) and the normalization. In \textsc{XSPEC} our model is \textsc{cons$\times$tbabs$\times$ztbabs$\times$mtable\{xstar\}} \textsc{$\times$(zbb+nthcomp+zgauss+zgauss)}. In our fits of the December\,\,2018 and May\,\,2019 {\it XMM-Newton} observations we substituted \textsc{nthcomp} to the cutoff power-law component, fixed the temperature of the seed photons to that of the thermal component ($kT_{\rm bb}=kT$), and let $kT_{\rm e}$ and $\Gamma_{\rm nth}$ free to vary.
In both cases this model provided a worse ($\chi^2/DOF=631.1/557$ for the December\,\,2018 observation) or similar ($\chi^2=727/657$ for May\,\,2019) fit to the cutoff power-law model. Both temperature of the electrons in the Comptonizing plasma and the photon index are higher in the May\,\,2019 observation ($kT_{\rm e}=4.4_{-1.5}^{+7.3}\rm\,keV$ and $\Gamma=4.4\pm0.1$) than in the December\,\,2018 one ($kT_{\rm e}=1.8_{-0.5}^{+1.0}\rm\,keV$ and $\Gamma=3.7\pm0.2$). Interestingly the blackbody temperature obtained is consistent between the two observations, and is significantly lower than when using a cutoff power-law model: $kT=124\pm1$\,eV and $kT=122\pm4$\,eV for the December\,\,2018 and May\,\,2019 observation, respectively. The temperature is however significantly higher than in our first {\it XMM-Newton} observation ($kT=102\pm1$\,eV).

\begin{figure}[t!]    
  \begin{center}
\includegraphics[width=0.48\textwidth]{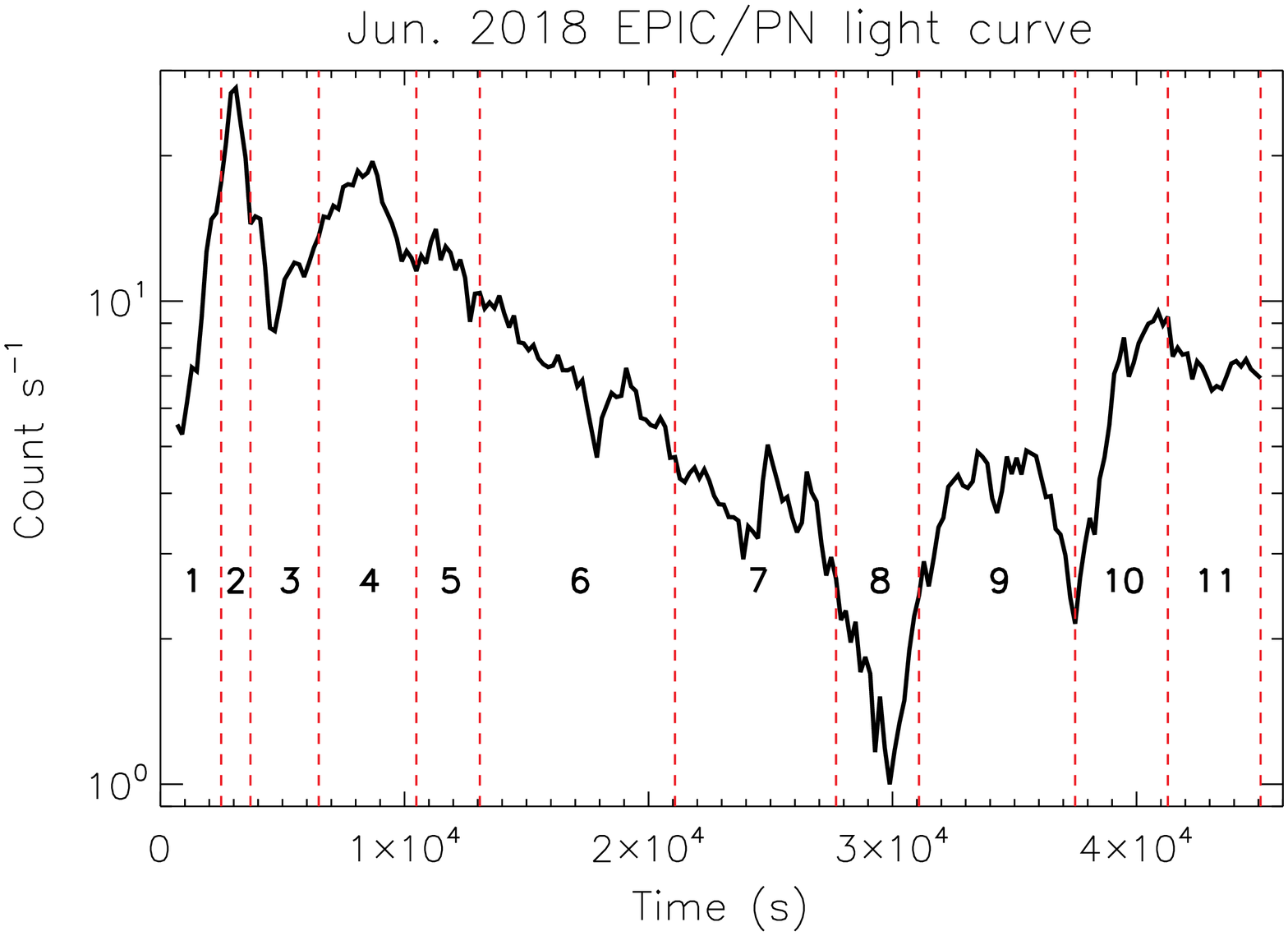}
\par\medskip
\par\medskip
\includegraphics[width=0.48\textwidth]{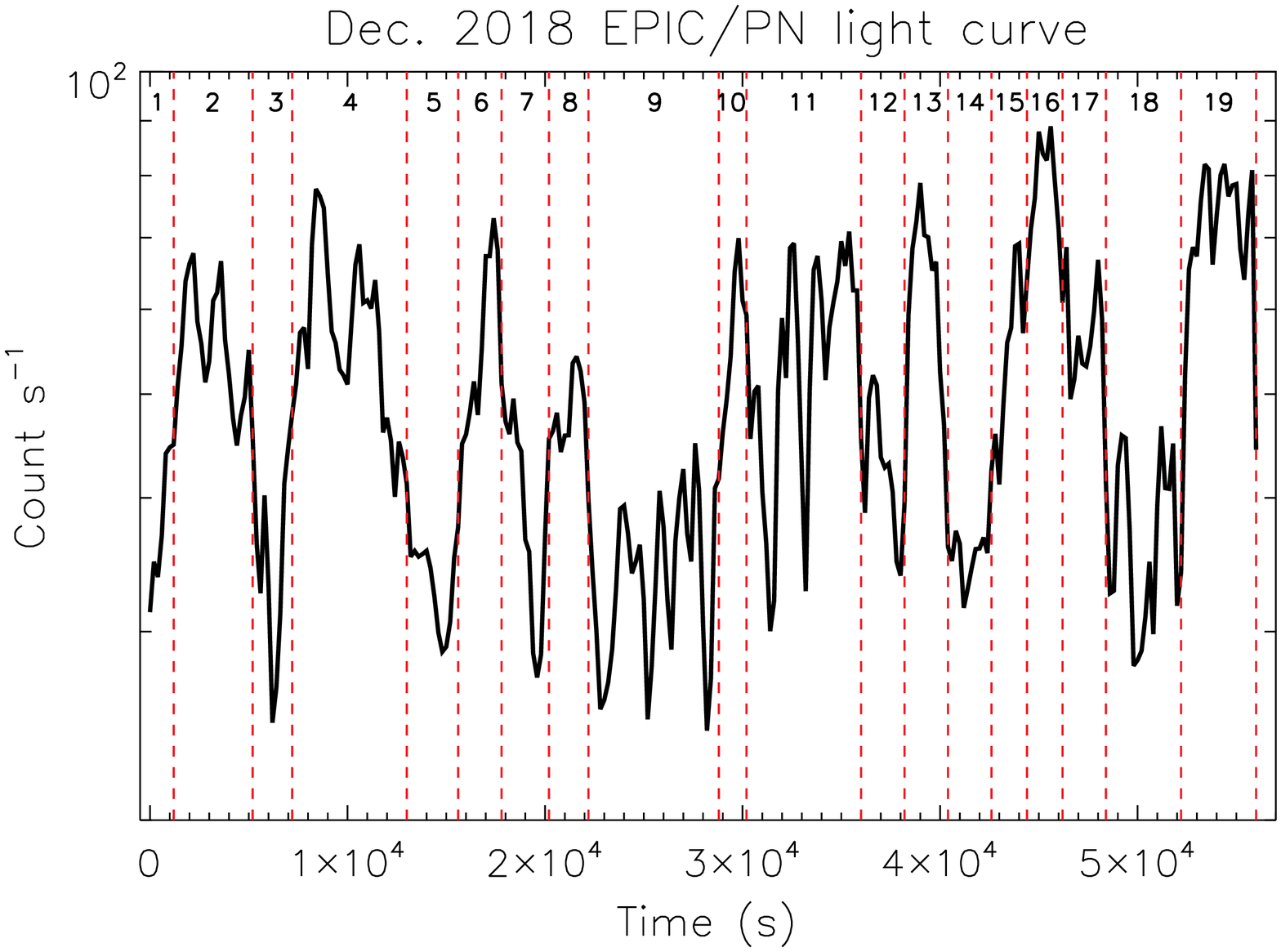}
\par\medskip
\includegraphics[width=0.48\textwidth]{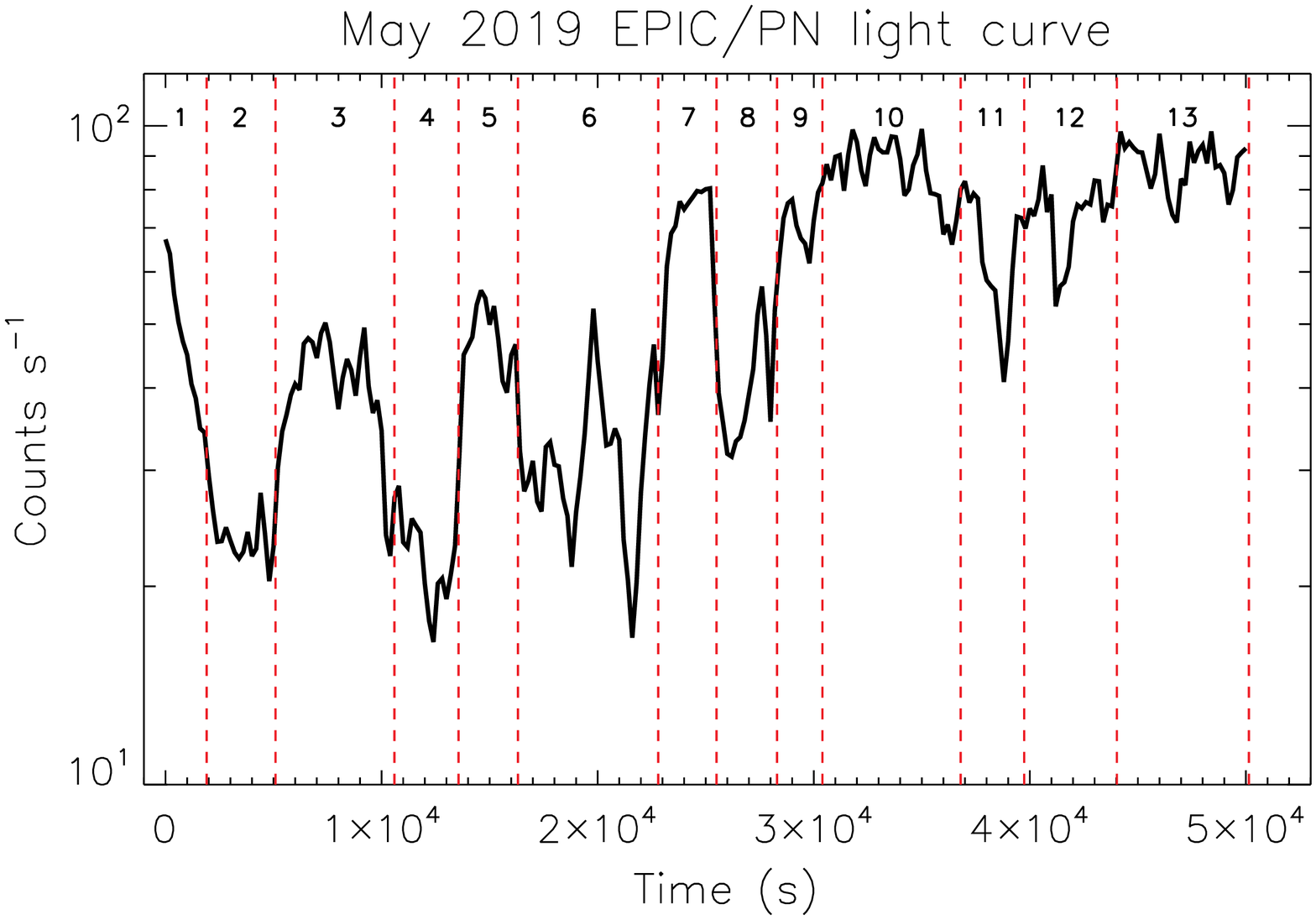}
    \caption{{\it XMM-Newton} EPIC/PN 0.3--10\,keV light curve of 1ES\,1927+654 for the June 2018 (top panel), December 2018 (middle panel) and May 2019 (bottom panel) observations (200\,s bins). The figure also shows the intervals used for the time-resolved spectroscopy, which are denoted by the vertical red dashed lines (see \S\ref{sec:timeresolved2018}).}
    \label{fig:XMMlc_18_bins}
  \end{center}
\end{figure}

\begin{figure*}
  \begin{center}
\includegraphics[width=0.68\textwidth]{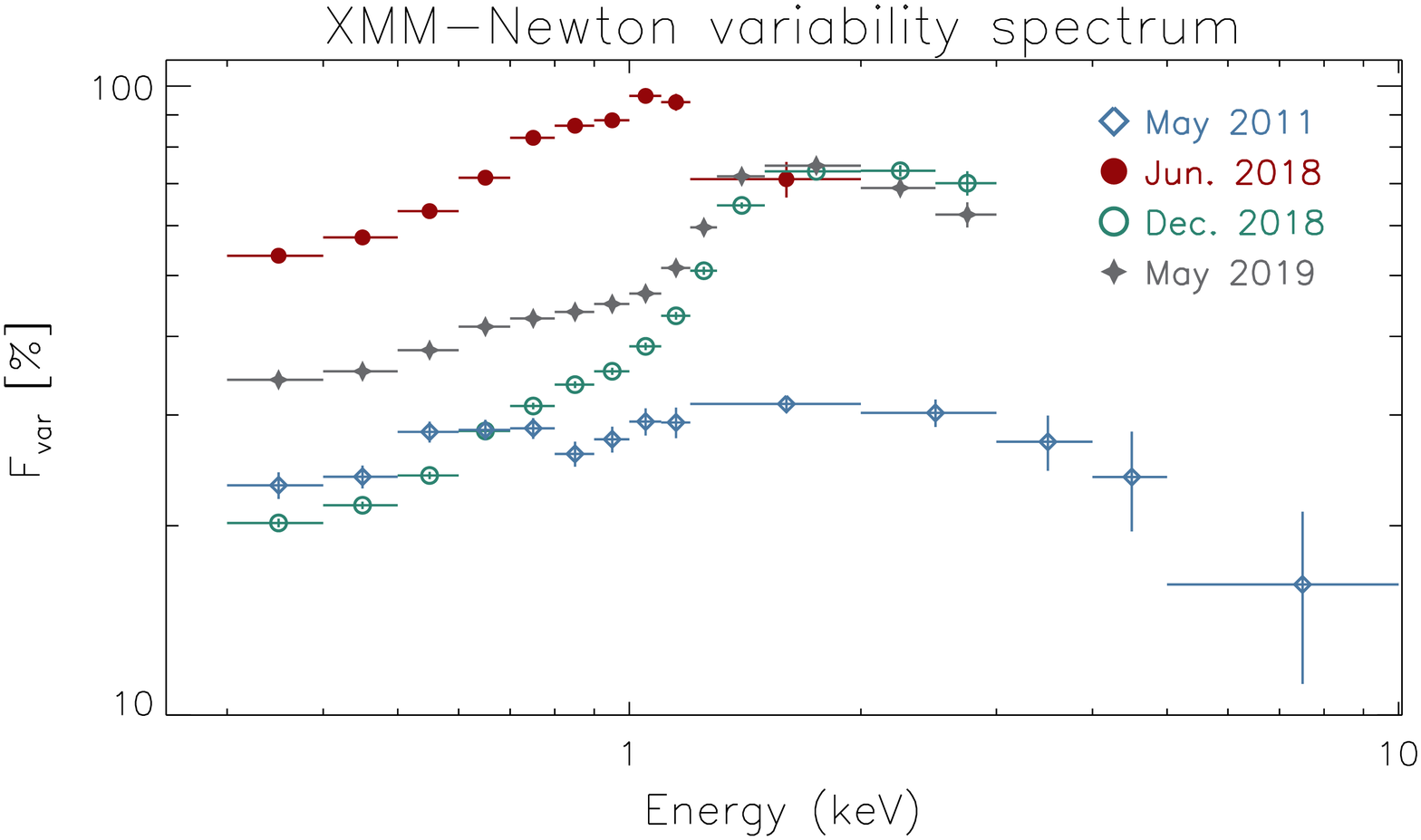}
    \caption{Variability spectra of the 2011 (blue filled diamonds), June\,\,2018 (red filled circles), December\,\,2018 (green empty circles) and May\,\,2019 (grey filled stars) {\it XMM-Newton}/EPIC PN observations on timescales of 200\,s. After the event, in June 2018 the source became much more variable at all the energies probed here. In December\,\,2018 and May\,\,2019, when the source luminosity was significantly higher the peak of the variability is found at higher energies.}
    \label{fig:XMMspecVarspec18}
  \end{center}
\end{figure*}

\subsection{Comparison with the 2011 {\it XMM-Newton} observation and alternative modelling}\label{sec:alternativemodelling}
The 2011 {\it XMM-Newton} observation of 1ES\,1927+654 was studied in detail by \cite{Gallo:2013hq}, who applied several different models, including a power-law absorbed by two partial covering neutral or ionized obscurers, as well as blurred ionized reflection. The authors found that all models could well reproduce the X-ray spectrum of the source, and they concluded that the X-ray emission in this source was consistent with being produced by a corona surrounding a standard accretion disc. In this section we test neutral and ionized partial covering models, as well as blurred ionized reflection, on the June 2018 and December 2018 {\it XMM-Newton}/{\it NuSTAR} observations. 

\subsubsection{Partial covering}
We first tested a model that includes a power law absorbed by two layers of partially covering neutral material (\textsc{tbabs$\times$ztbabs$\times$zpcfabs$\times$zpcfabs$\times$zpo}). This model leaves very strong residuals across most of the spectrum, and results in a very high reduced chi-squared ($\sim 39$ and $\sim 18$ for the June and December observations, respectively) and in a very steep power law ($\Gamma \sim 7.3$ and $\Gamma \sim 5.5$, respectively). Including a blackbody component [\textsc{tbabs$\times$ztbabs$\times$zpcfabs$\times$zpcfabs$\times$(zpo+zbb)}] improves the fit, but it still results in a poor reduced chi-squared ($\sim 4.5$ and $\sim 2$ for the June and December observations, respectively).
We then considered a model that includes partial covering ionized obscuration. We started by considering only a power-law continuum, as done by \citeauthor{Gallo:2013hq} (\citeyear{Gallo:2013hq}; \textsc{tbabs$\times$ztbabs$\times$zxipcf$\times$zxipcf$\times$zpo}). This cannot reproduce well the continuum, leaving strong residuals and resulting in a reduced $\chi^2$ of $\sim 1.4$ for both the June and December 2018 observations. Similarly, simultaneously fitting these two observations and the 2011 observation, leaving all the parameters of the absorbers free to vary while tying the photon index and normalization of the power law component, results in a poor fit ($\chi^2$\rm/DOF=6609/1550). This model fails to reproduce the spectrum even if we untie both the normalization and the photon index ($\chi^2$\rm/DOF=6539/1548 and $\chi^2$\rm/DOF=6098/1546, respectively).
The addition of a blackbody component and of a third partially covering ionized absorber [\textsc{tbabs$\times$ztbabs$\times$zxipcf$\times$zxipcf$\times$zxipcf$\times$(zpo+zbb)}] provides a good fit for both the June ($\chi^2\rm /DOF=346/306$) and the December ($\chi^2\rm /DOF=617/556$) observations. However, several of the parameters obtained from the fit are quite extreme: the photon indices would be $\Gamma\sim 5-6$, the velocity of the absorbers would be up to $\sim 0.55$c, while the unobscured luminosity of the June 2018 observation would be $\sim 2\times10^{47}\rm\,erg\,s^{-1}\,cm^{-2}$. This would make 1ES\,1927+654 the brightest AGN in the local Universe by several orders of magnitude. The strong change in the ratio between power-law and blackbody emission between the two observations (see \citealp{Ricci:2020}) would be even more extreme than what we found including two Gaussian lines (see Fig.\,\ref{fig:bb_po_ratio}), with the fractional contribution of the power-law component to the flux increasing of $\sim 2500$ times between the June and December\,\,2018 observations.

\subsubsection{Relativistic reflection}

We also tested several relativistic reflection models \citep{Dauser:2010qy,Dauser:2013kx,Garcia:2010yt,Garcia:2013dz,Garcia:2014kk}, such as \textsc{relxilllp} and \textsc{relxilllpD} (which considers a high-density accretion disk), and found that they also leave very strong residuals through the whole spectrum. The \textsc{relxilllpD} fit in particular results in a steep continuum ($\Gamma\sim 3.4$), a high inclination angle with respect to the disk ($i\sim89^{\circ}$, similarly to \citealp{Gallo:2013hq}), and in a reduced chi-squared of $\sim 2.7$.

\begin{figure}
\begin{center}
    \includegraphics[width=0.48\textwidth]{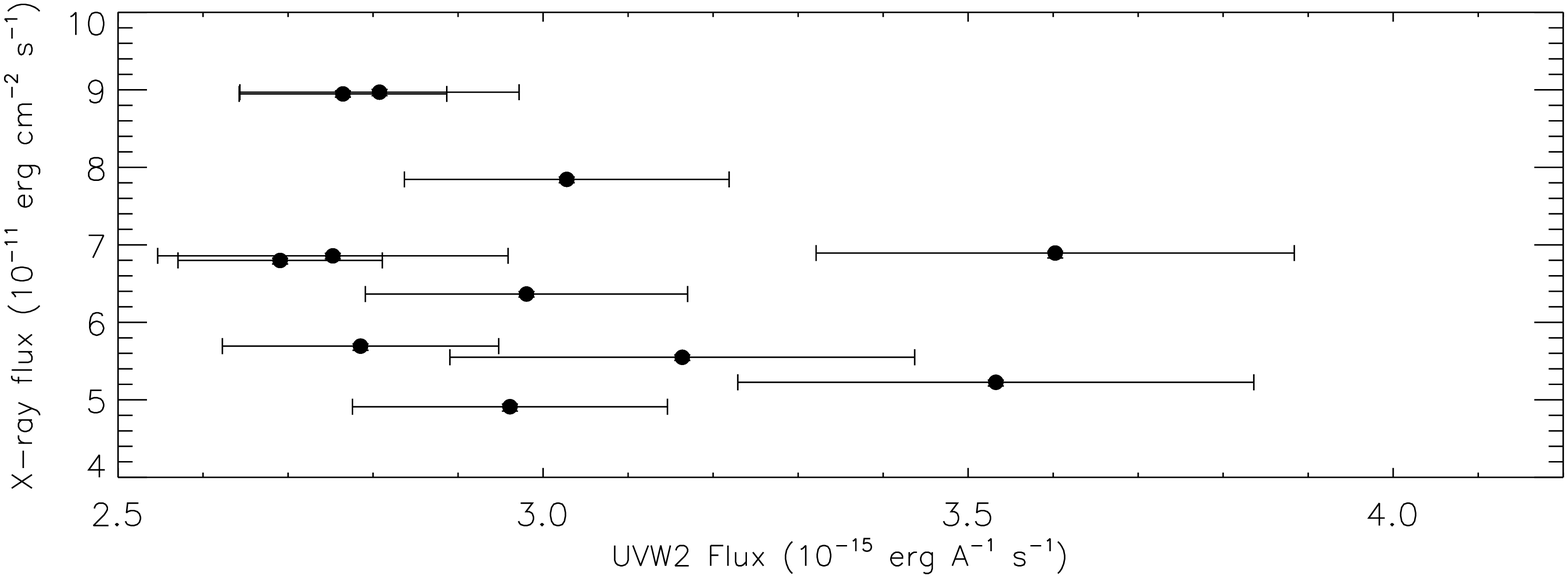}
    \includegraphics[width=0.48\textwidth]{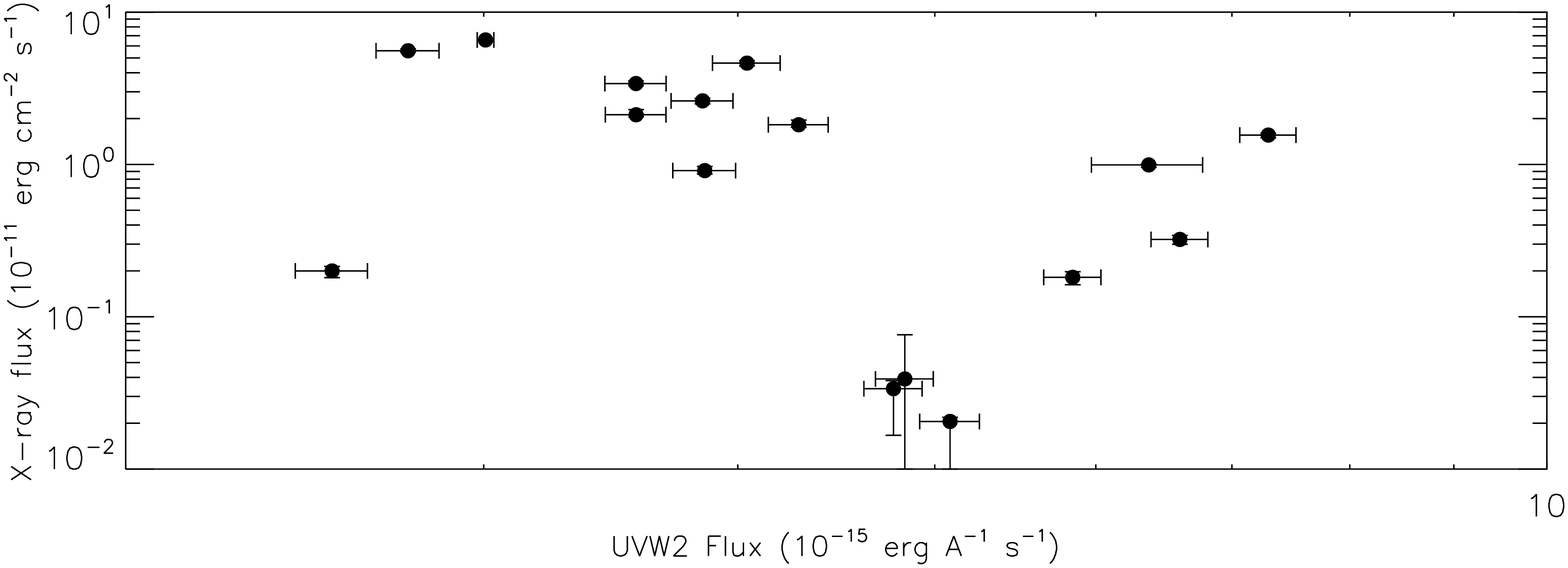}
     \caption{{\it Top panel:} X-ray flux (in the 0.3--3\,\,keV range) versus UVW2 flux (5.8\,\,eV) for the simultaneous EPIC/PN and OM exposures during the December\,\,2018 observation. {\it Bottom panel:} 0.3--3\,keV flux versus UVW2 flux for the {\it Swift}/XRT and UVOT observations, as well as for the June\,\,2018 and May\,\,2019 {\it XMM-Newton} observations.  }
   \label{fig:fluxXUV}
\end{center}
\end{figure}

\subsection{X-ray and UV variability}\label{sec:xrayvariabilityXMM}

\subsubsection{X-rays}\label{sect:xrayvar1}
1ES\,1927+654 shows very strong variability in the 0.3--10\,keV band during the three {\it XMM-Newton} observations (Fig.\,\ref{fig:XMMlc_18_bins}), though the variability characteristics are clearly different during each exposure. During the June\,\,2018 observation the AGN varies by a factor of 28 in $\sim27$\,ks (top panel), while the ratio between the minimum and maximum flux in December\,\,2018 and May\,\,2019 (middle and bottom panels, respectively) was only $\sim 4-5$, and the source varied more rapidly on shorter timescales.

In panels A and B of Fig.\,\ref{fig:XMMlc_18_bands} (Appendix\,\ref{appendix:XMM_lightcurves}) we show the variability in four different energy bands for the June\,\,2018 observation. The light curves show a very similar pattern, although with different ratios between the maximum and the minimum flux, ranging between $\sim 18$ for the 0.3--0.5\,keV band to a factor of $\sim 159$ in the $1-1.3$\,keV range. The light curves in six energy bands for the December\,\,2018 observation are illustrated in panels A, B and C of Fig.\,\ref{fig:XMMlc_18_december_bands}. Also in this case the variability is similar in all bands, but has a different ratio between the maximum and minimum flux, ranging from $\sim 3$ in the 0.3--0.5\,keV band to $\sim 70$ in the 2--3\,keV range. During the May\,\,2019 observation we also found that the variability is similar in all bands (panels A, B and C of Fig.\,\ref{fig:XMMlc_18_december_bands}), with the ratio between the maximum and minimum flux going from $\sim 4$ in the 0.3--0.5\,keV band to $\sim 59$ in the 1--1.3\,keV band. Interestingly, for all observations the hardness ratio varies following the flux of the source, suggesting that the radiation becomes harder when the flux increases (see panels C--E of Fig.\,\ref{fig:XMMlc_18_bands} and panels D--H of Fig.\, \ref{fig:XMMlc_18_december_bands}).

The {\it XMM-Newton} EPIC/PN variability spectrum of 1ES\,1927+654, calculated considering timescales of 200\,s, is shown in Fig.\,\ref{fig:XMMspecVarspec18} for the 2011 (blue diamonds, see \citealp{Gallo:2013hq}) observation, as well as for the June\,\,2018 (red filled circles), the December\,\,2018 (green empty circles) and the May\,\,2019 (cyan filled circles) observations. The fractional root mean square variability amplitude ($F_{\rm var}$) was calculated following \cite{Edelson:1990kx} and \citeauthor{Vaughan:2003qy} (\citeyear{Vaughan:2003qy}, see Eqs.\,10 and B2 in their paper). In the 2011 observation the source shows a peak at $\sim 1.5$\,keV, and a bell-shaped variability amplitude, similar to other well known unobscured AGN (e.g., \citealp{Gallo:2004bh,Gierlinski:2006ul}).
After the optical/UV outburst the source is much more variable than in 2011, and in particular in June 2018, only a few months after the detection of the optical/UV brightening, when $F_{\rm var}$ peaks at $96\pm2\%$ in the 1--1.1\,keV band, and then decreases to $71\pm5\%$ in the 1.2--2\,keV range. In December 2018 the peak of the variability is found at higher energies ($E\geq 1.5$\,keV), and is $F_{\rm var}=73\pm1\%$, a value comparable to what was observed in the same band six months before. Interestingly, in December\,\,2018 the variability below $\sim 0.7$\,keV is back to a level comparable with what was found in the May 2011 observation. During the May 2019 observations the shape of the variability spectrum of the source is similar to that observed in the December 2018 observation, with a peak in the $1.5-2$\,keV bin ($F_{\rm var}=75\pm1\%$). During this last observation the variability below 1\,keV is significantly higher than in December ($F_{\rm var}\simeq 30-50\%$). In 2018/2019 the value of $F_{\rm var}$ above $\sim 1.2$\,keV remains unchanged, and all of the changes occur at lower energies.

\subsubsection{UV}\label{sect:UV}

During the December\,\,2018 {\it XMM-Newton} observation OM carried out 11 exposures of the source in the UVW2 band (5.8\,\,eV). In order to compare the UV and X-ray flux we extracted EPIC/PN spectra using the same intervals adopted for the UVW2 exposures, and we fitted them using the same models we applied for the time-resolved spectroscopy (\S\ref{sec:timeresolved2018}).  
No clear correlation is found neither between the 0.3--3\,keV and the UVW2 flux (top panel of Fig.\,\ref{fig:fluxXUV}), nor between the UVW2 flux and the emission in different X-ray bands or cosidering different spectral components. Similarly, no clear trend is found when studying the relation between the UVW2 and the X-ray flux for the 14 {\it Swift} observations and the June\,\,2018 and May\,\,2019 {\it XMM-Newton} observations (bottom panel of Fig.\,\ref{fig:fluxXUV}).

\begin{figure*}
  \begin{center}
\includegraphics[width=0.32\textwidth]{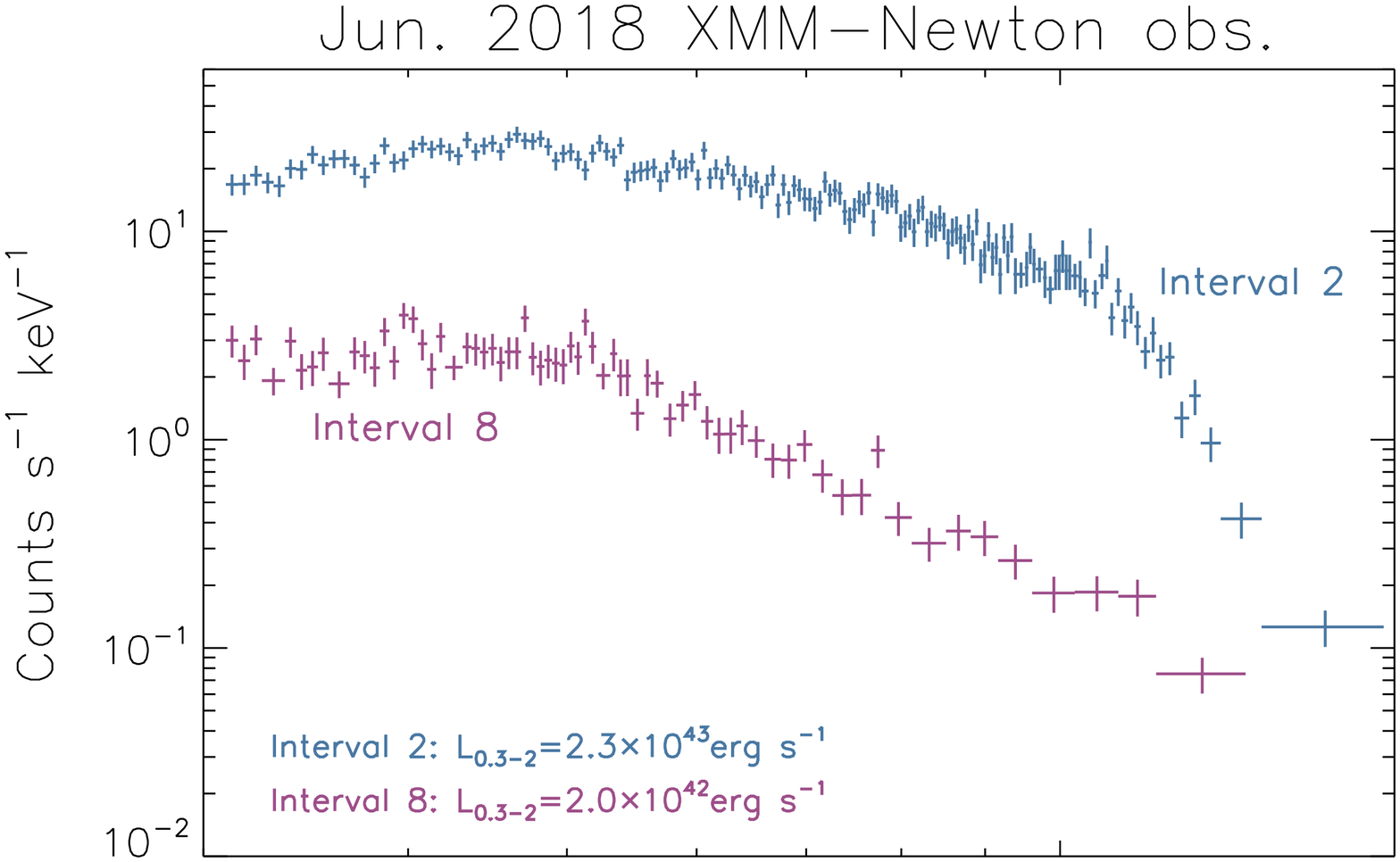}
\includegraphics[width=0.32\textwidth]{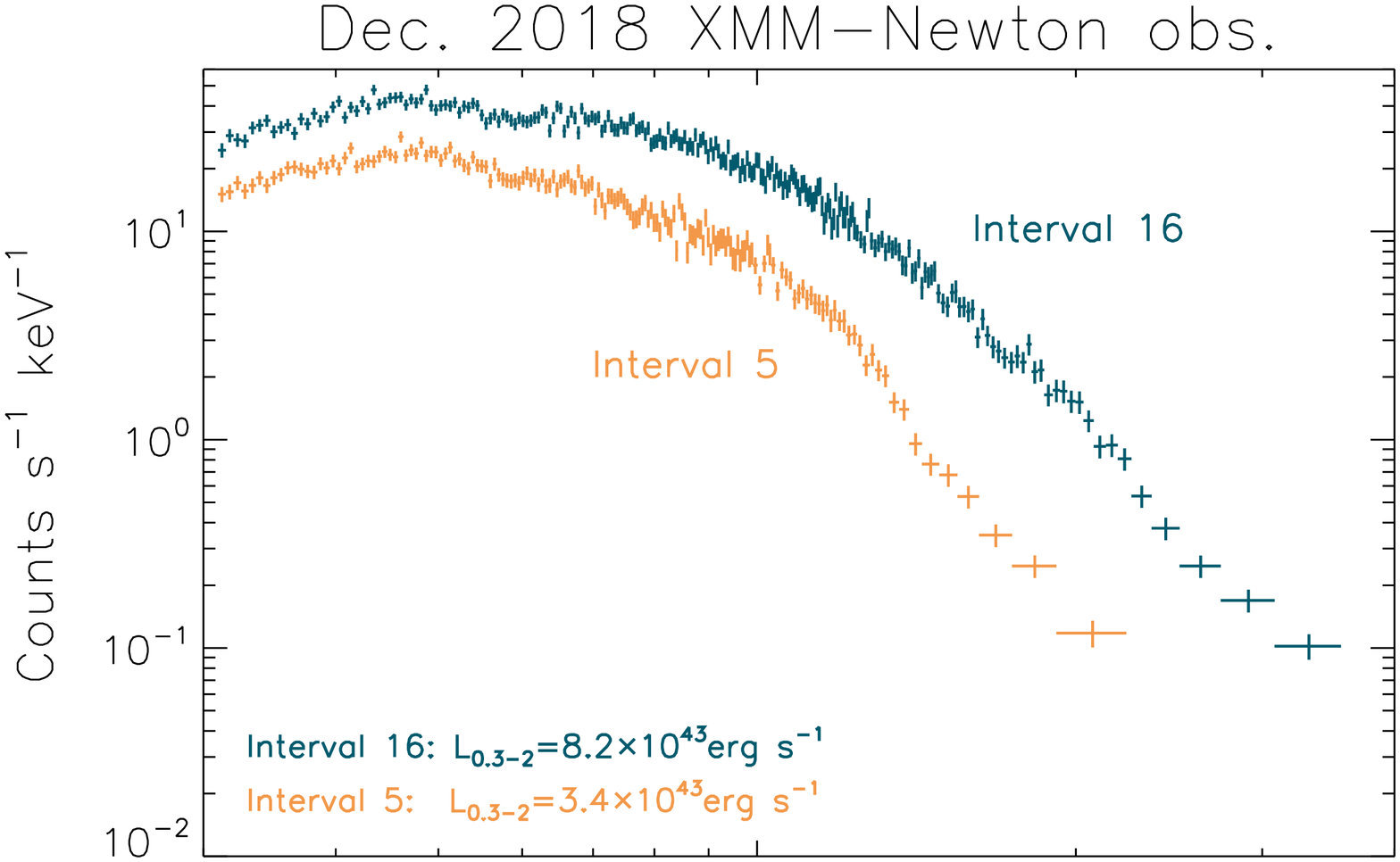}
\includegraphics[width=0.32\textwidth]{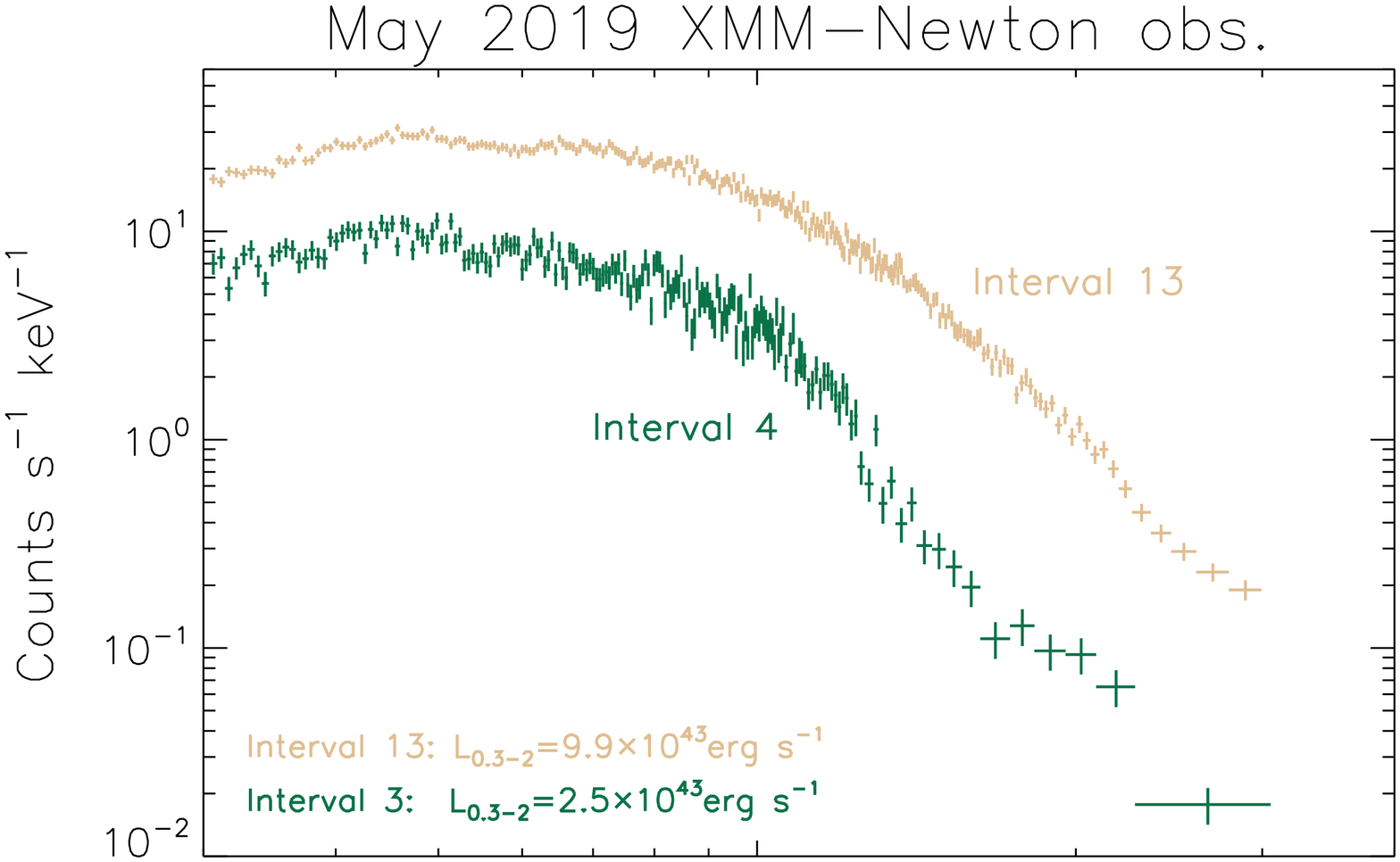}
\includegraphics[width=0.32\textwidth]{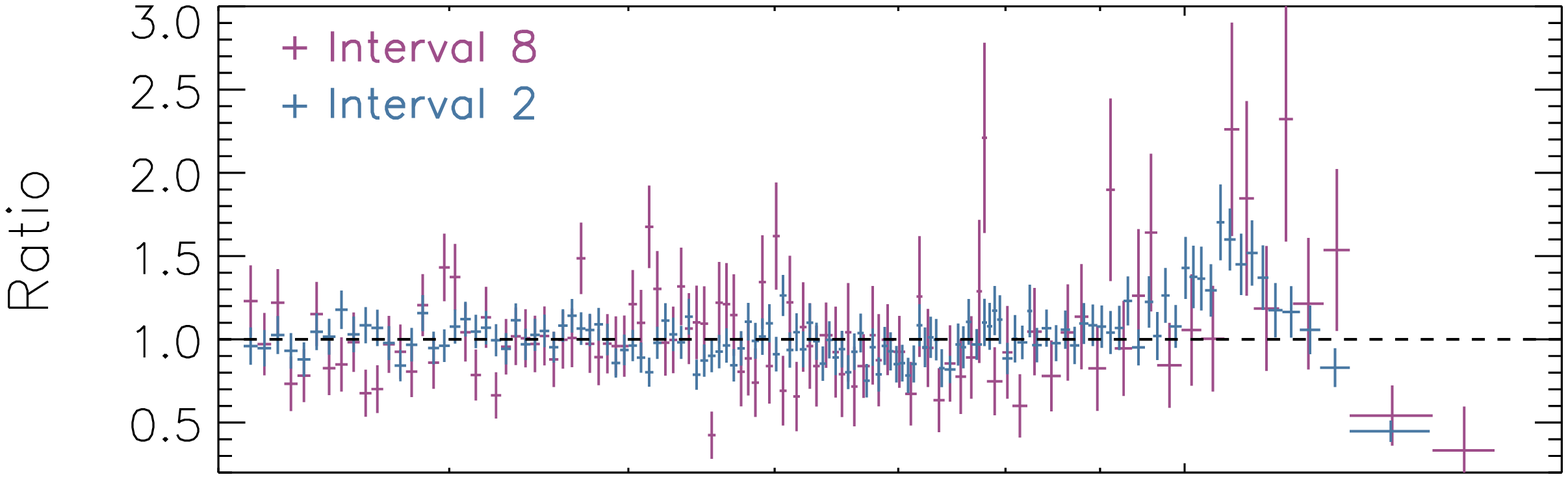}
\includegraphics[width=0.32\textwidth]{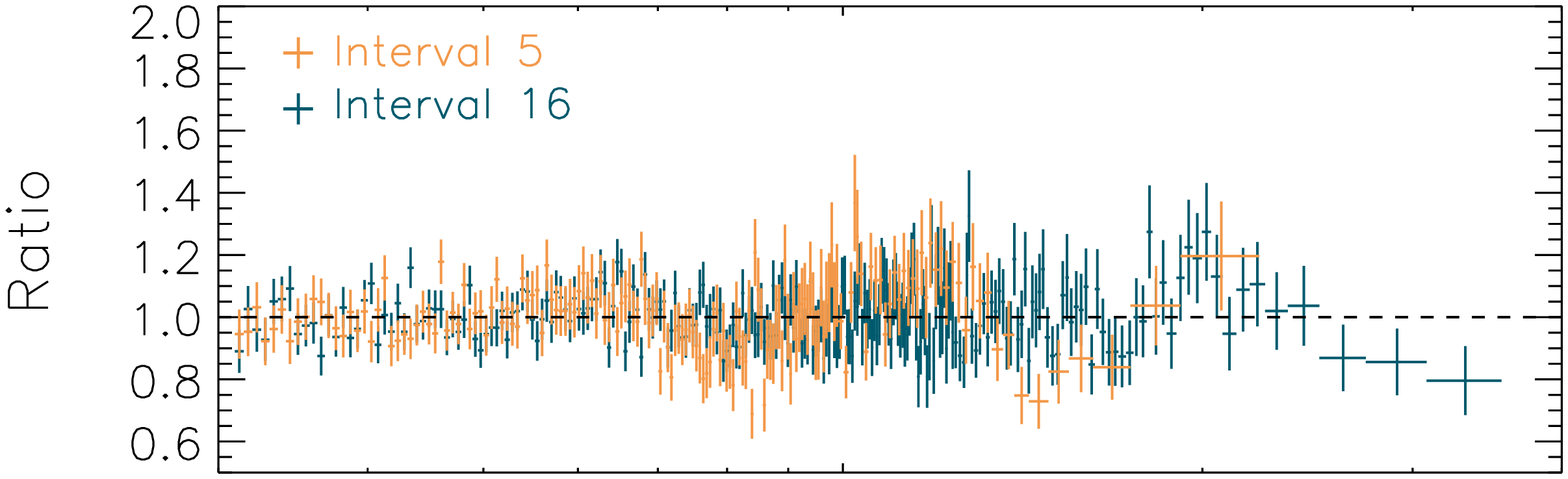}
\includegraphics[width=0.32\textwidth]{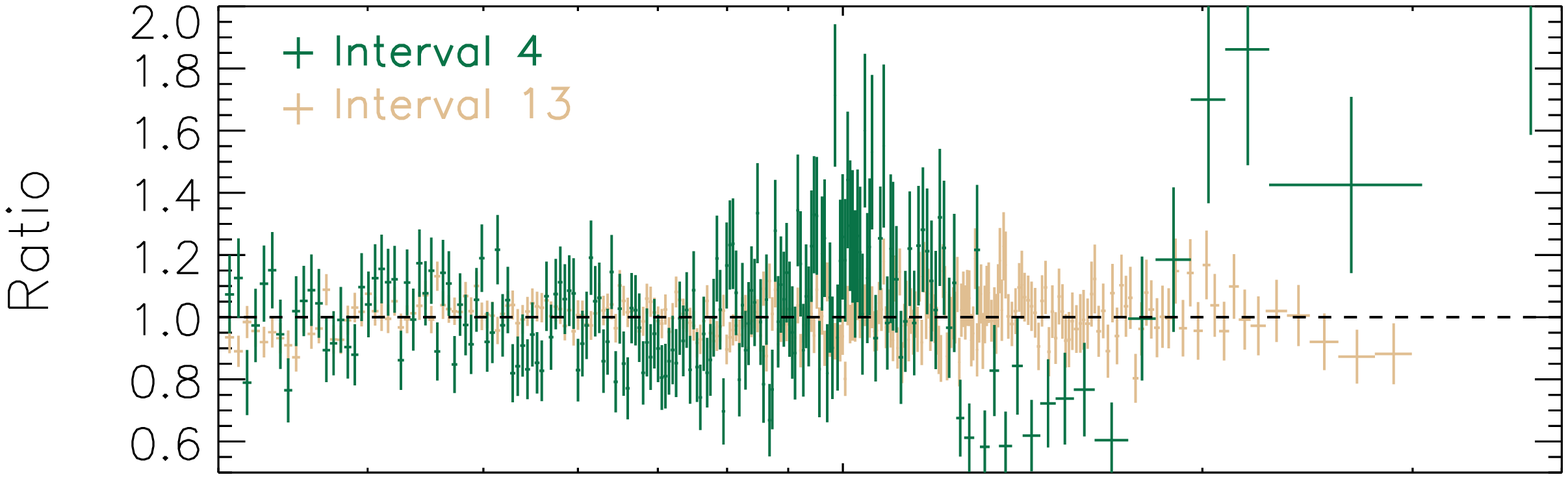}
\includegraphics[width=0.32\textwidth]{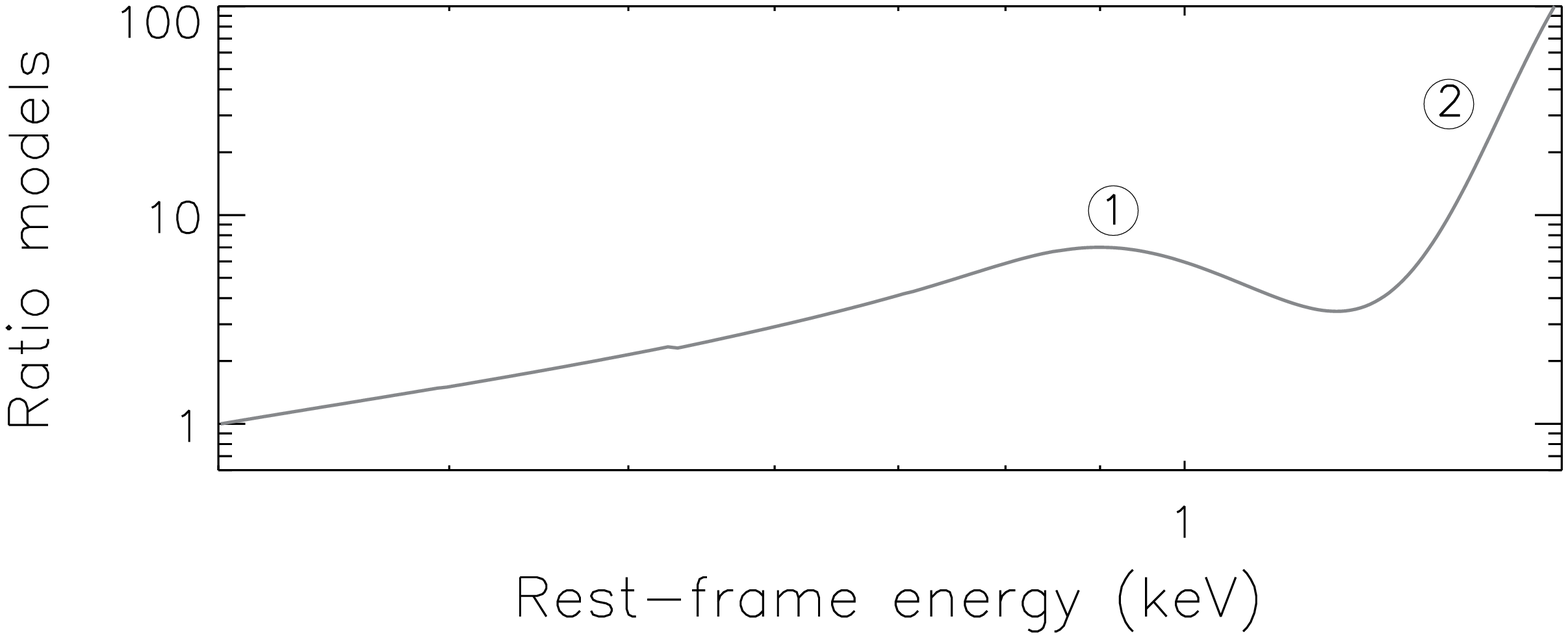}
\includegraphics[width=0.32\textwidth]{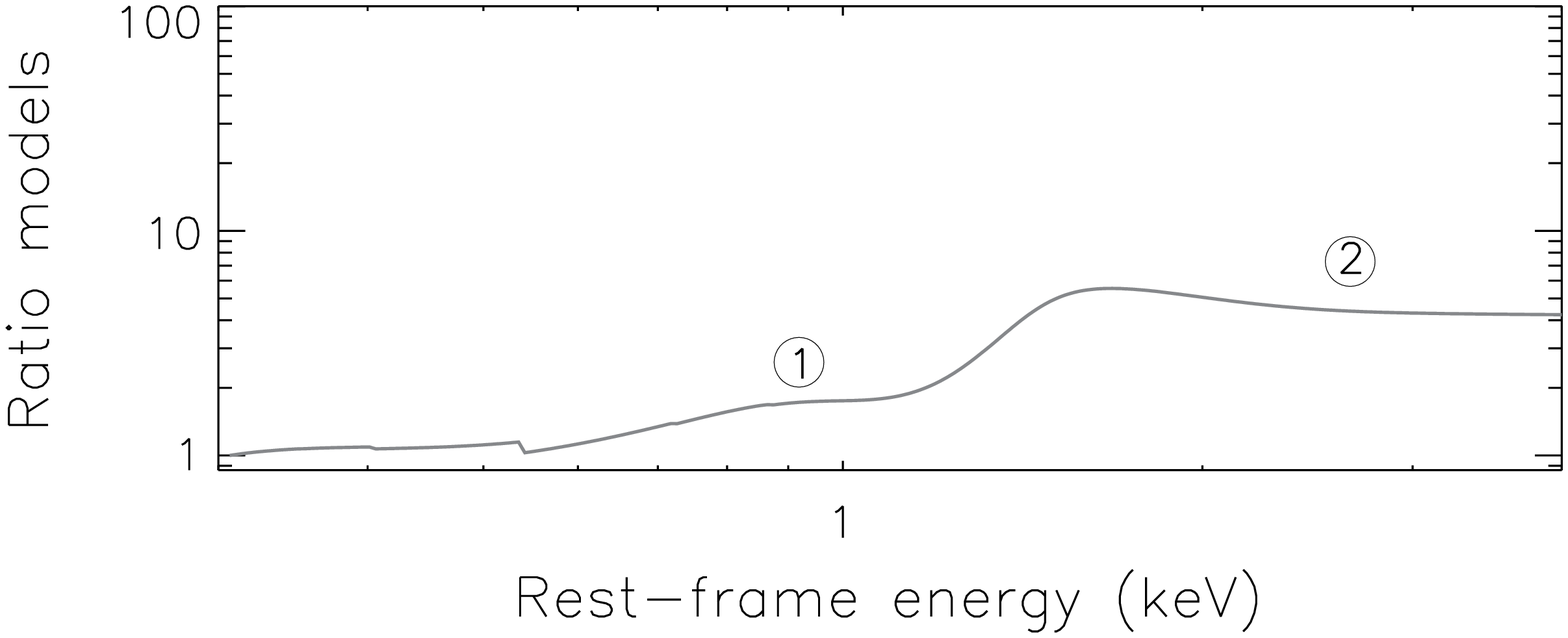}
\includegraphics[width=0.32\textwidth]{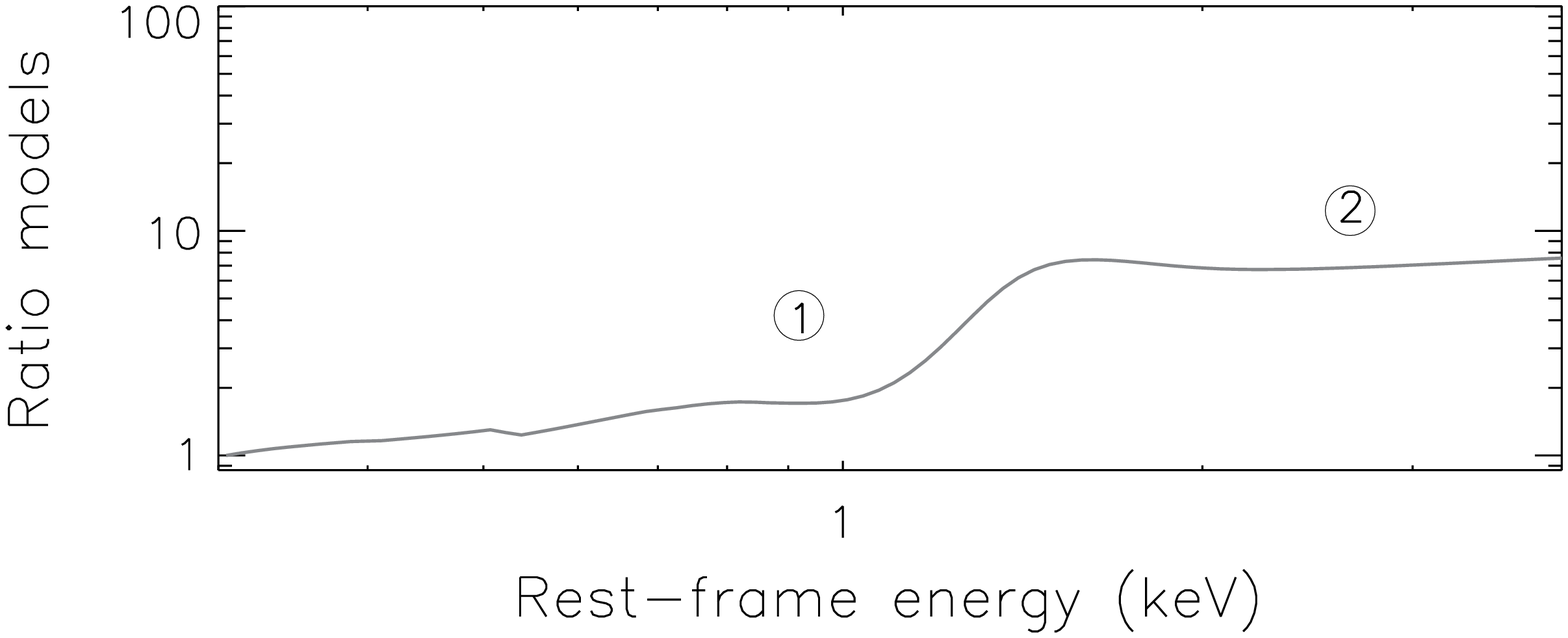}
    \caption{{\it Top panels:} {\it XMM-Newton} EPIC/PN spectra of the periods of the June\,\,2018 (left panel), December\,\, 2018 (middle panel) and May\,\,2019 (right panel) observations with the highest (intervals 2/16/13) and lowest (intervals 8/5/4) fluxes (see Fig.\,\ref{fig:XMMlc_18_bins}). The spectral results obtained for all intervals of these three observations are reported in Tables\,\,\ref{tab:fitXMM18epochs_Jun}--\ref{tab:fitXMM19epochs_May} of Appendix\,\ref{appendix:exposuresintervals}.  {\it Central panels:} ratio between the spectra and a powerlaw plus blackbody model, which shows the persistent presence of a feature at $\sim 1$\,keV, except at the highest luminosity (intervals 16 and 13 of the December\,\,2018 and May\,\,2019 observations, respectively). {\it Bottom panels:} ratio between the best-fitting spectral models of the highest and lowest flux intervals, after having normalized their fluxes at 0.3\,keV. The plots show that the difference arises both from a higher-temperature blackbody (1) and a stronger power-law component (2).}
    \label{fig:XMMspec18_twointervals}
  \end{center}
\end{figure*}

\subsection{Time-resolved spectroscopy}\label{sec:timeresolved2018}

Given the very strong X-ray variability of 1ES\,1927+654 on short ($\lesssim 5-10$\,ks) timescales, we studied its spectral evolution in different intervals using EPIC/PN for all  {\it XMM-Newton} observations (\S\ref{sec:timeresolved2018_2019}). This was done by considering different good time intervals for the spectral extraction. For the June 2018 observation we divided the exposure into 11 intervals, while for the December 2018 observation, given the higher flux and more rapid variability, we divided the observation into 19 periods. For the May 2019 observation we considered 13 periods. The intervals used for the spectral extraction are shown in Fig.\,\ref{fig:XMMlc_18_bins}. The exposures of the different intervals vary between 1.2\,ks and 8\,ks (see Appendix\,\ref{appendix:exposuresintervals}). To compare the spectral variations after the optical/UV outburst with the previous X-ray observations, we performed also time-resolved spectroscopy for the 2011 {\it XMM-Newton} observation (\S\,\ref{sec:timeresolvedMay2011}). Due to the low flux of some of the intervals, we fitted all spectra using Cash statistics, after binning the spectra to have at least one count per bin, to avoid possible issues related to empty bins in \textsc{XSPEC}.

\begin{figure}[t!]
  \begin{center}
\includegraphics[width=0.48\textwidth]{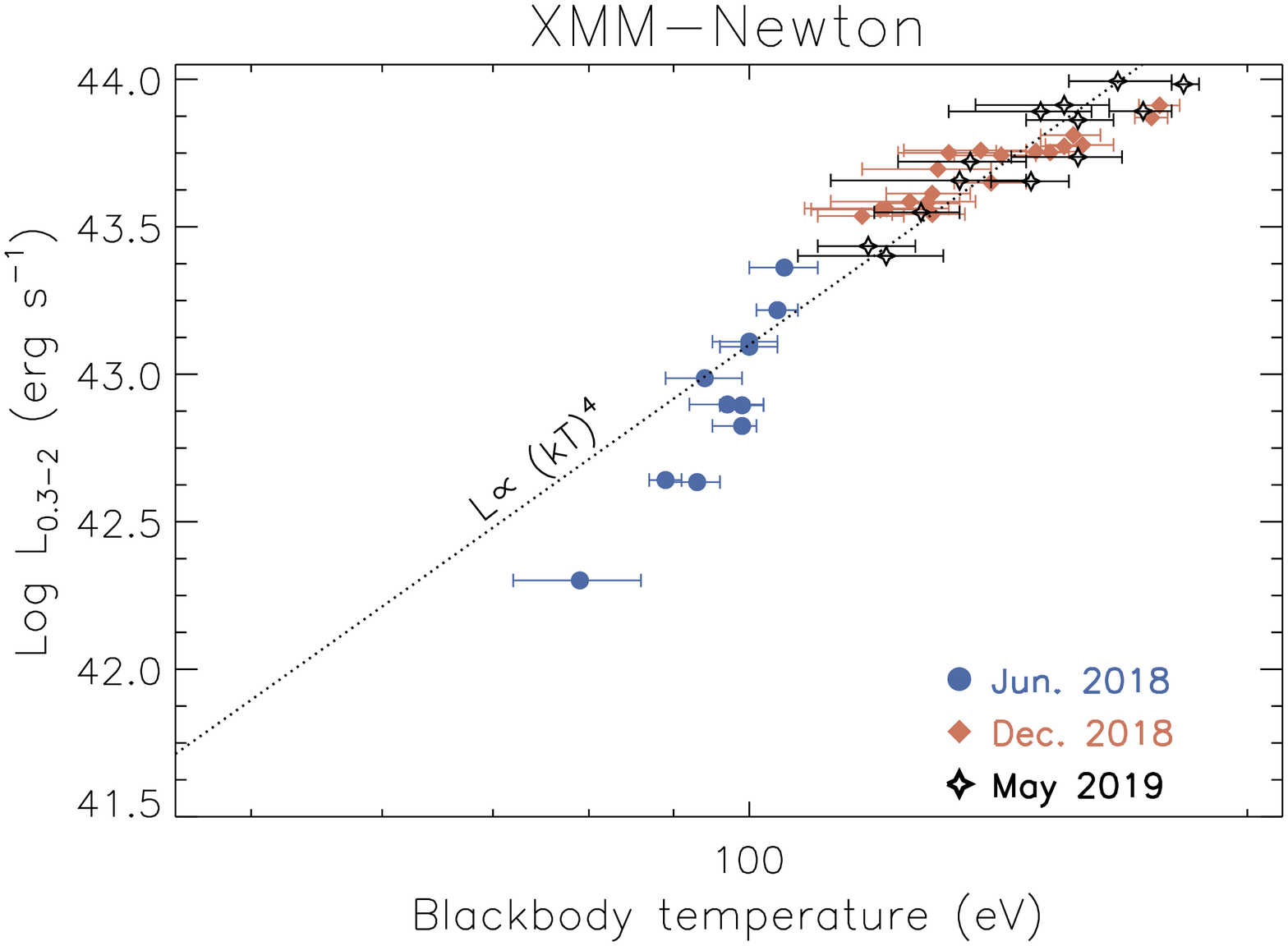}
\bigskip
\includegraphics[width=0.48\textwidth]{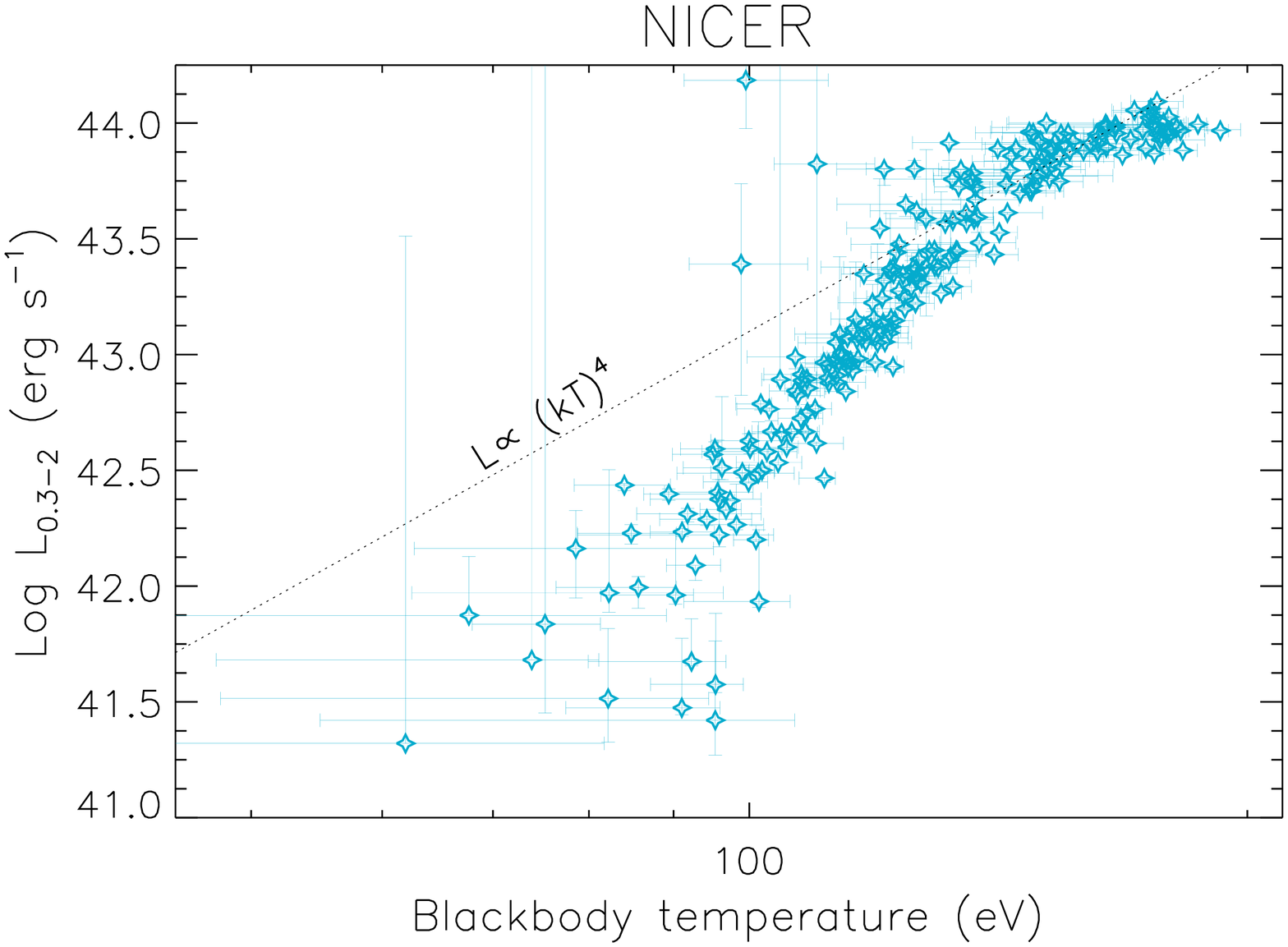}
\bigskip
\includegraphics[width=0.48\textwidth]{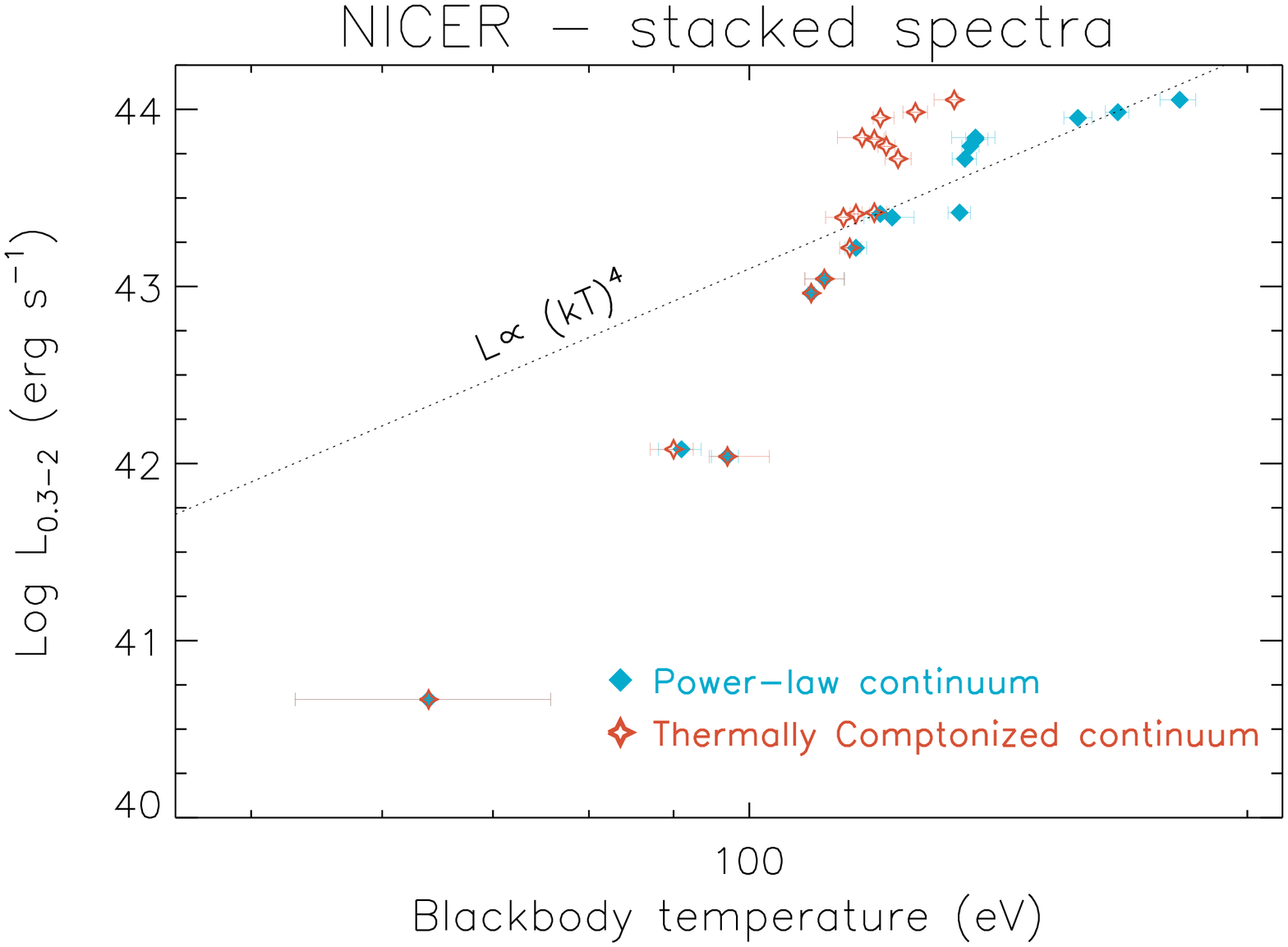}
    \caption{ {\it Top panel:} Temperature of the blackbody component versus the 0.3--2\,keV luminosity for the June 2018 (blue circles), December 2018 (red diamonds) and May 2019 (black stars) {\it XMM-Newton} observations (see \S\ref{sec:timeresolved2018}). The black dotted line represents the $L\propto(kT)^4$ relation. The X-ray spectral model used included a power-law, a blackbody component and a Gaussian line (see Appendix\,\ref{appendix:exposuresintervals}). {\it Middle panel:} Same as top panel for the individual {\it NICER} observations (see \S\ref{sect:NICERspectroscopy_individual}). {\it Bottom panel:} Same as top panel for the stacked {\it NICER} observations (see \S\ref{sect:NICERspectroscopy_stacked}). The cyan diamonds and the red stars are the values obtained by using a power law or a thermally Comptonized continuum, respectively.}
    \label{fig:kT_luminosity}
  \end{center}
\end{figure}

\subsubsection{June/December 2018 and May 2019}\label{sec:timeresolved2018_2019}

Strong spectral variability was detected in all {\it XMM-Newton} observations carried out after the optical/UV outburst.
We fitted all the spectra with a model that included neutral absorption, a power law and a blackbody component, plus a Gaussian line [\textsc{tbabs$\times$ztbabs$\times$(zpo+zbb+zgauss)}]. The spectral change between the highest and lowest flux intervals can be clearly seen in the top panels of Fig.\,\ref{fig:XMMspec18_twointervals}: the spectral shape changes significantly with flux, with the source becoming harder at higher fluxes, in good agreement with what is discussed in \S\ref{sect:xrayvar1}. The middle panels of the figure show the residuals obtained by fitting the spectra with a model that did not include an emission line [\textsc{tbabs$\times$ztbabs$\times$(zpo+zbb)}], highlighting the fact that the broad feature at 1\,keV is present both at the highest and the lowest flux level, with the exception of the highest luminosity intervals of the December\,\,2018 and May\,\,2019 observations. The bottom panels show the ratio between the best-fitting model of the highest and lowest-flux intervals, normalized at 0.3\,keV. The plot illustrates that the spectral difference arises due to both a higher-temperature blackbody (1) and a stronger power-law component (2) during the higher flux periods.

The results of the spectral fitting for all intervals using the emission line model are reported in Tables\,\,\ref{tab:fitXMM18epochs_Jun}--\ref{tab:fitXMM19epochs_May} in Appendix\,\ref{appendix:exposuresintervals}. The model reproduces the X-ray spectra of the source well in all the intervals. 
The temperature of the blackbody component ranges between $kT=79^{+7}_{-7}$\,eV and $kT=183^{+4}_{-3}$\,eV. A very tight relation between the flux level and the temperature of the blackbody is found  for all intervals (top panel of Fig.\,\ref{fig:kT_luminosity}), with $kT$ increasing as the source gets brighter. Fitting this trend with a function of the type $\log (L_{0.3-2}/\rm erg\,s^{-1})=a+b*\log(kT/\rm eV)$ we find a very steep slope ($b=7.9\pm1.1$) for the June\,2018 observation, while in December\,\,2018 and May\,\,2019 the trend is flatter ($b=2.0\pm0.2$ and $b=3.1\pm0.4$, respectively).  The median photon index is $\Gamma=3.52\pm0.13$, and spans a wide range ($\Gamma=0.8-5.5$) at the lowest luminosities. No clear relation is found between the photon index and the luminosity.

\begin{figure*}
  \begin{center}
\includegraphics[width=0.48\textwidth]{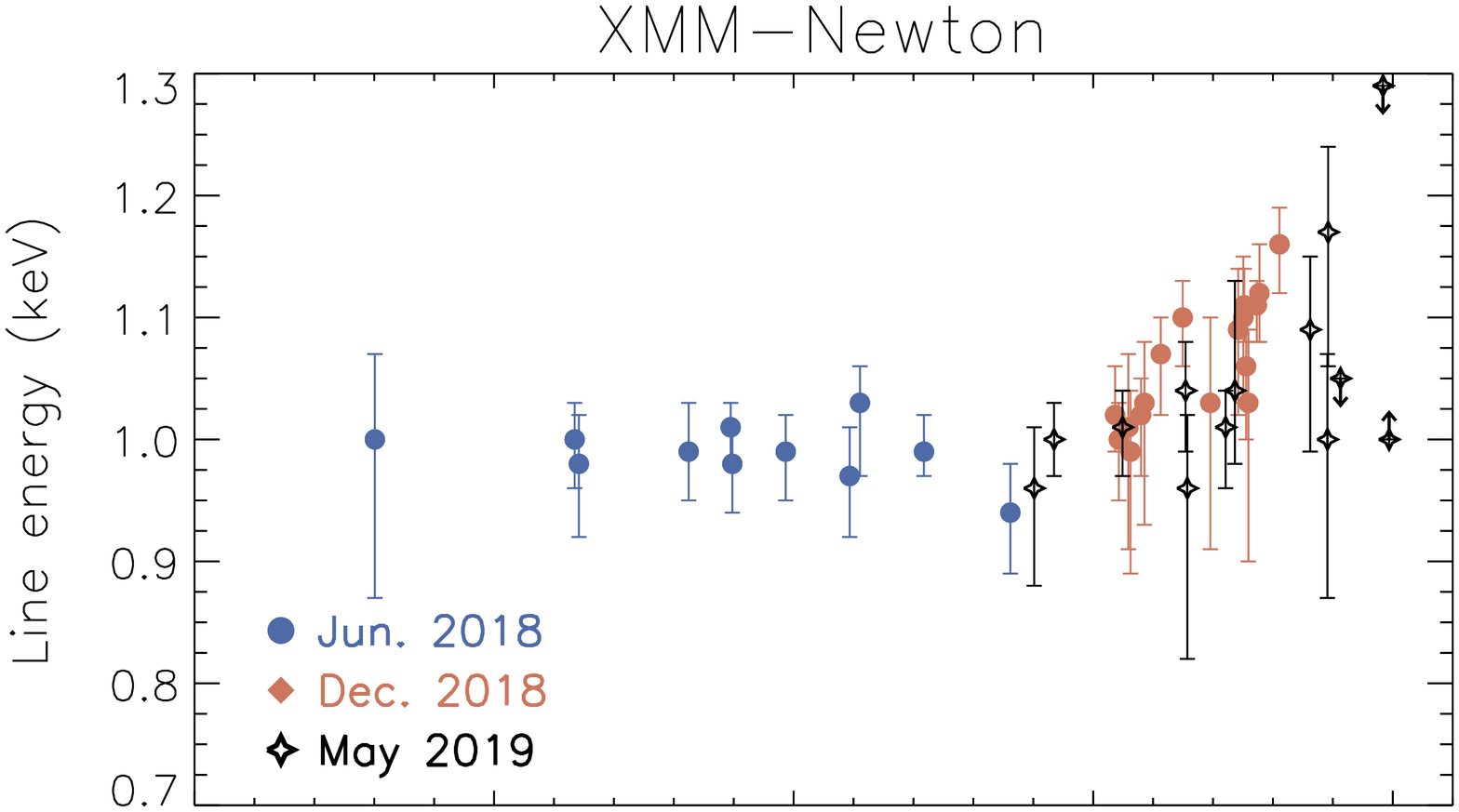}
\includegraphics[width=0.48\textwidth]{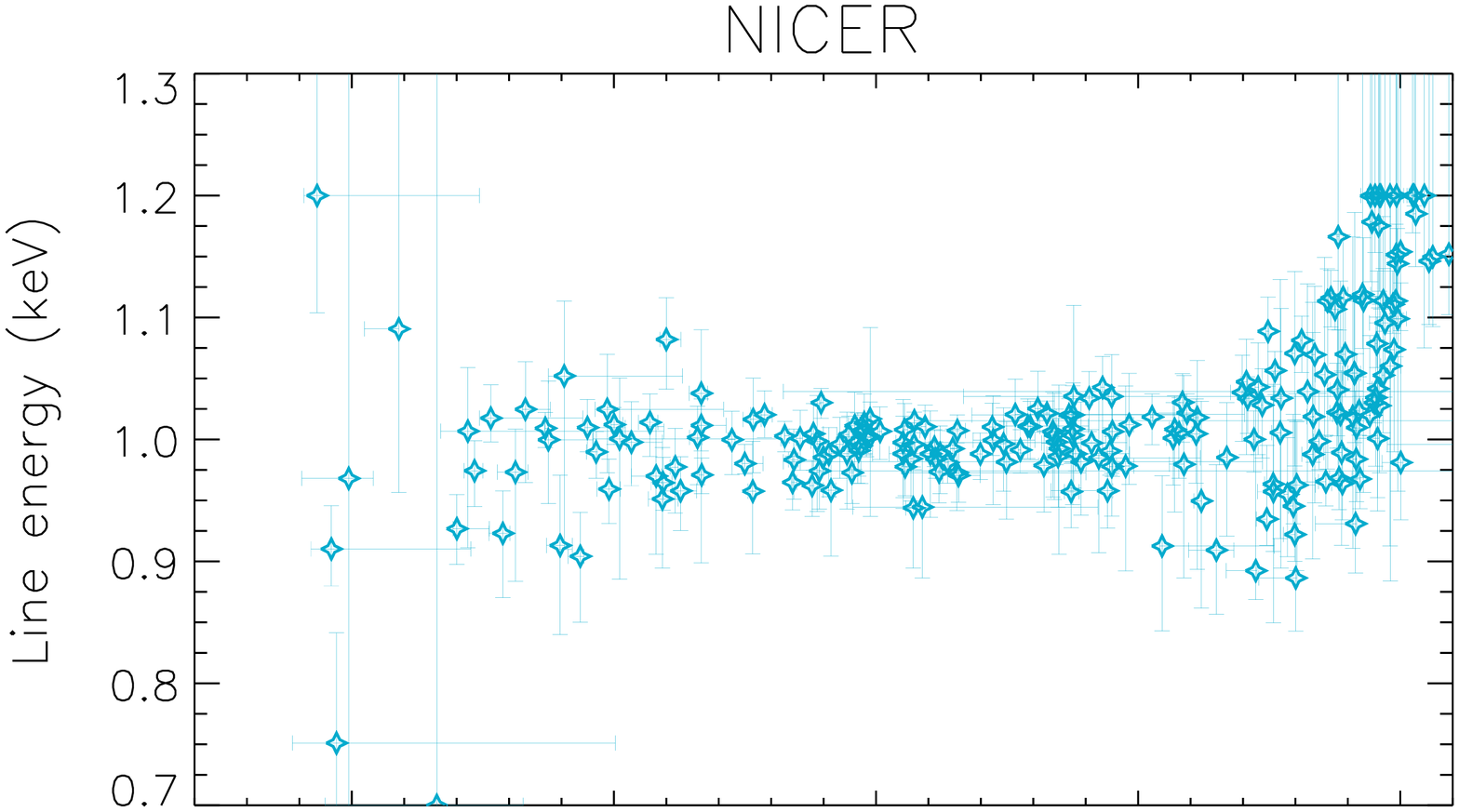}
\includegraphics[width=0.48\textwidth]{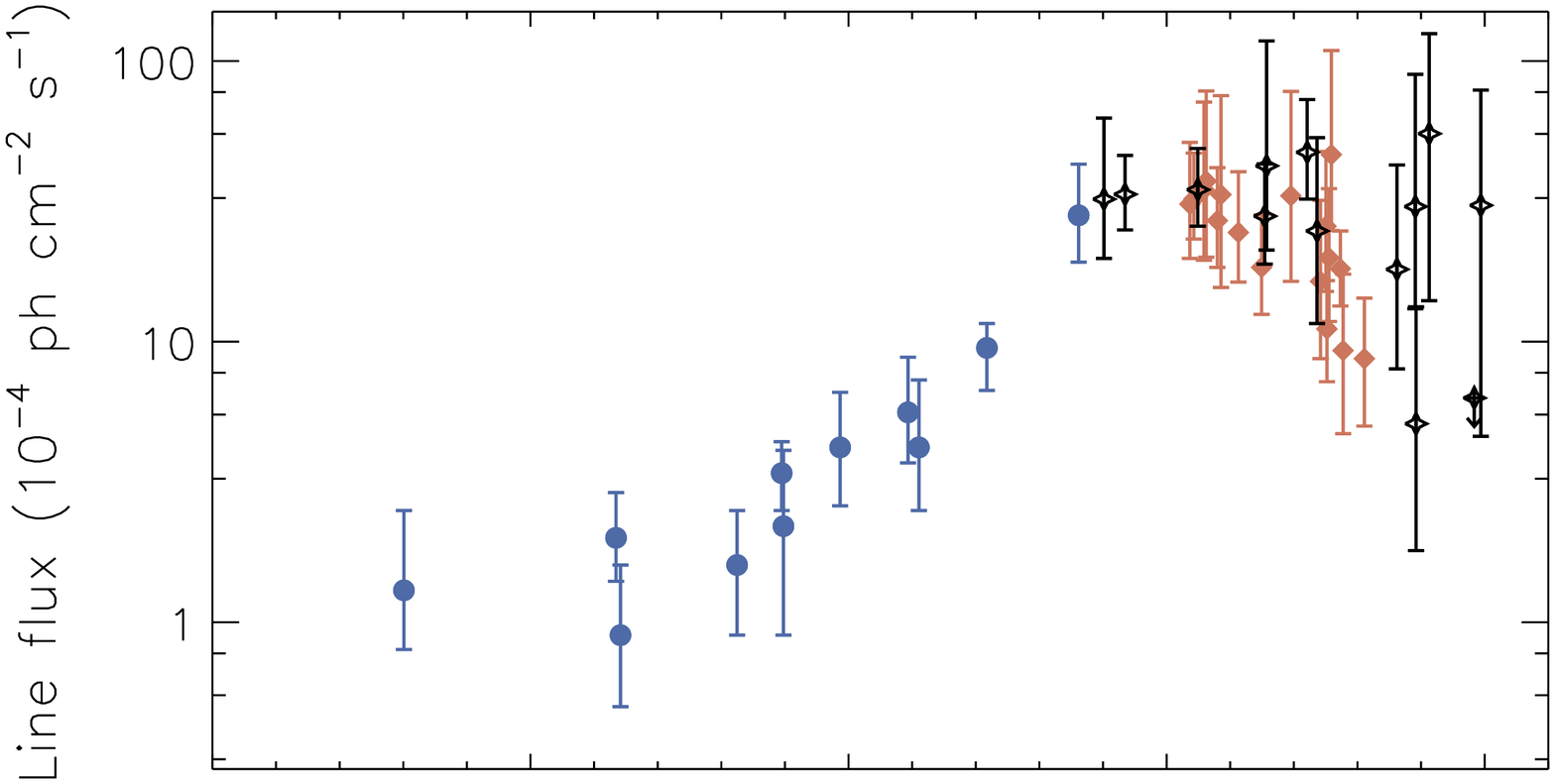}
\includegraphics[width=0.48\textwidth]{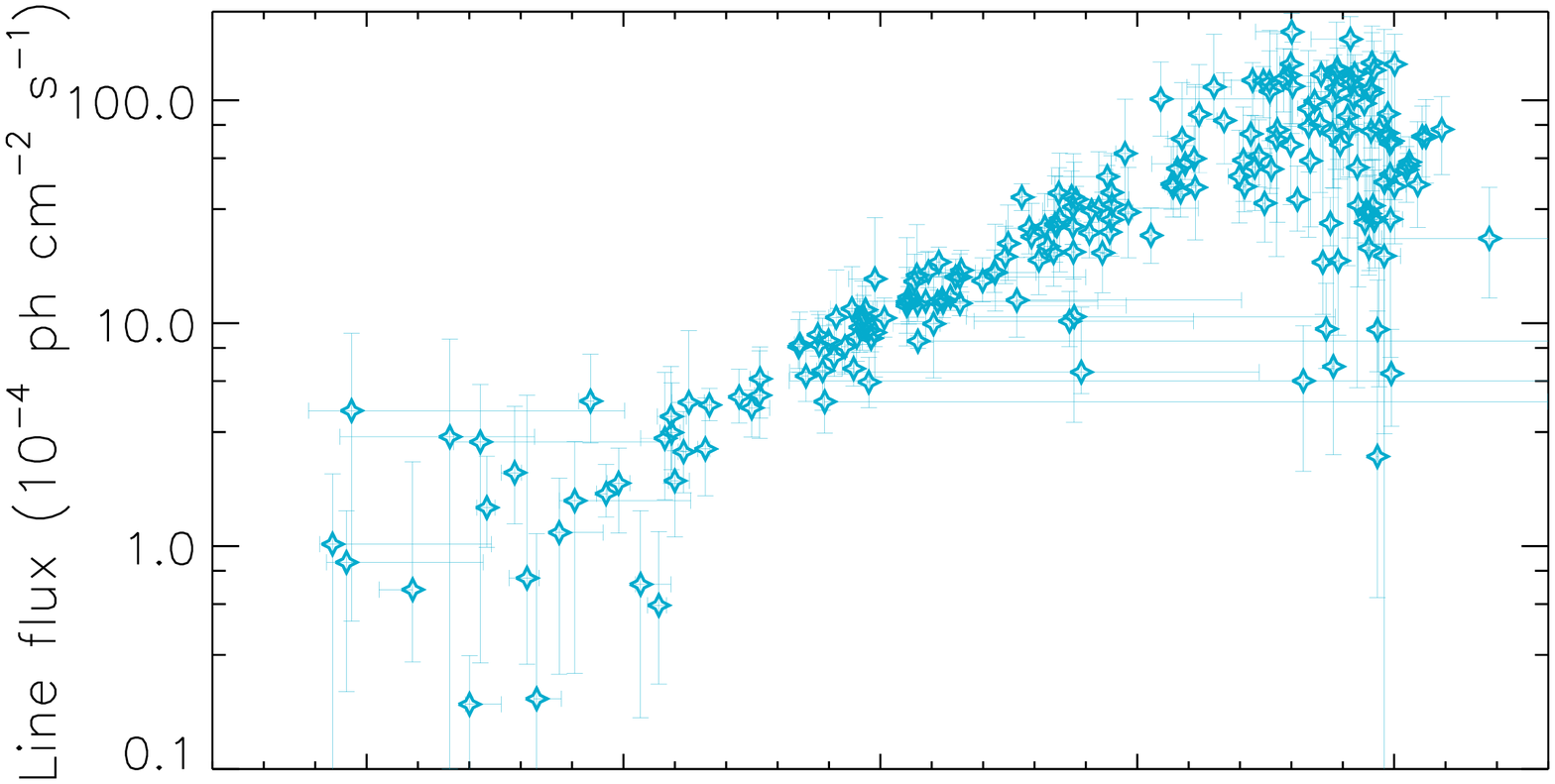}
\includegraphics[width=0.48\textwidth]{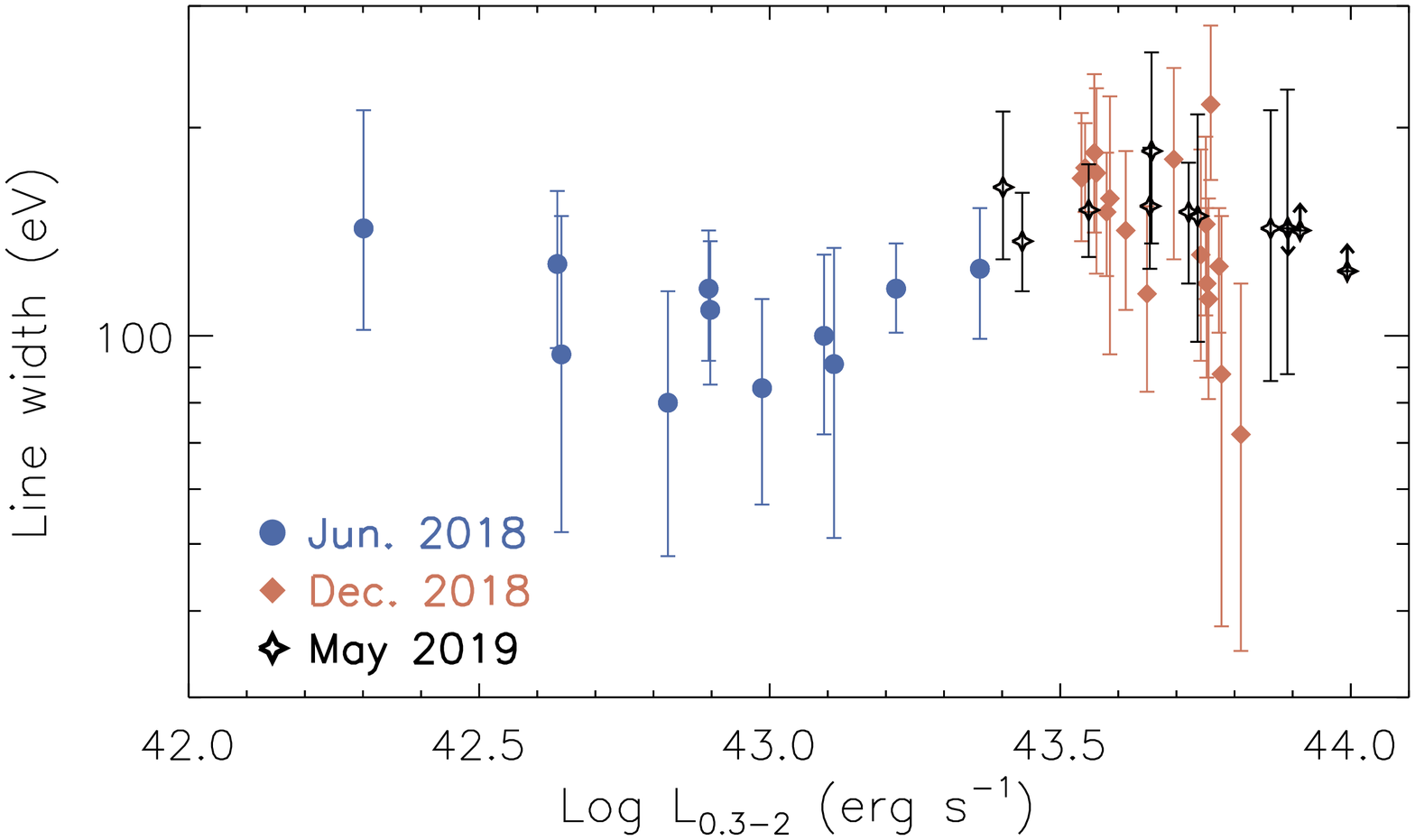}
\includegraphics[width=0.48\textwidth]{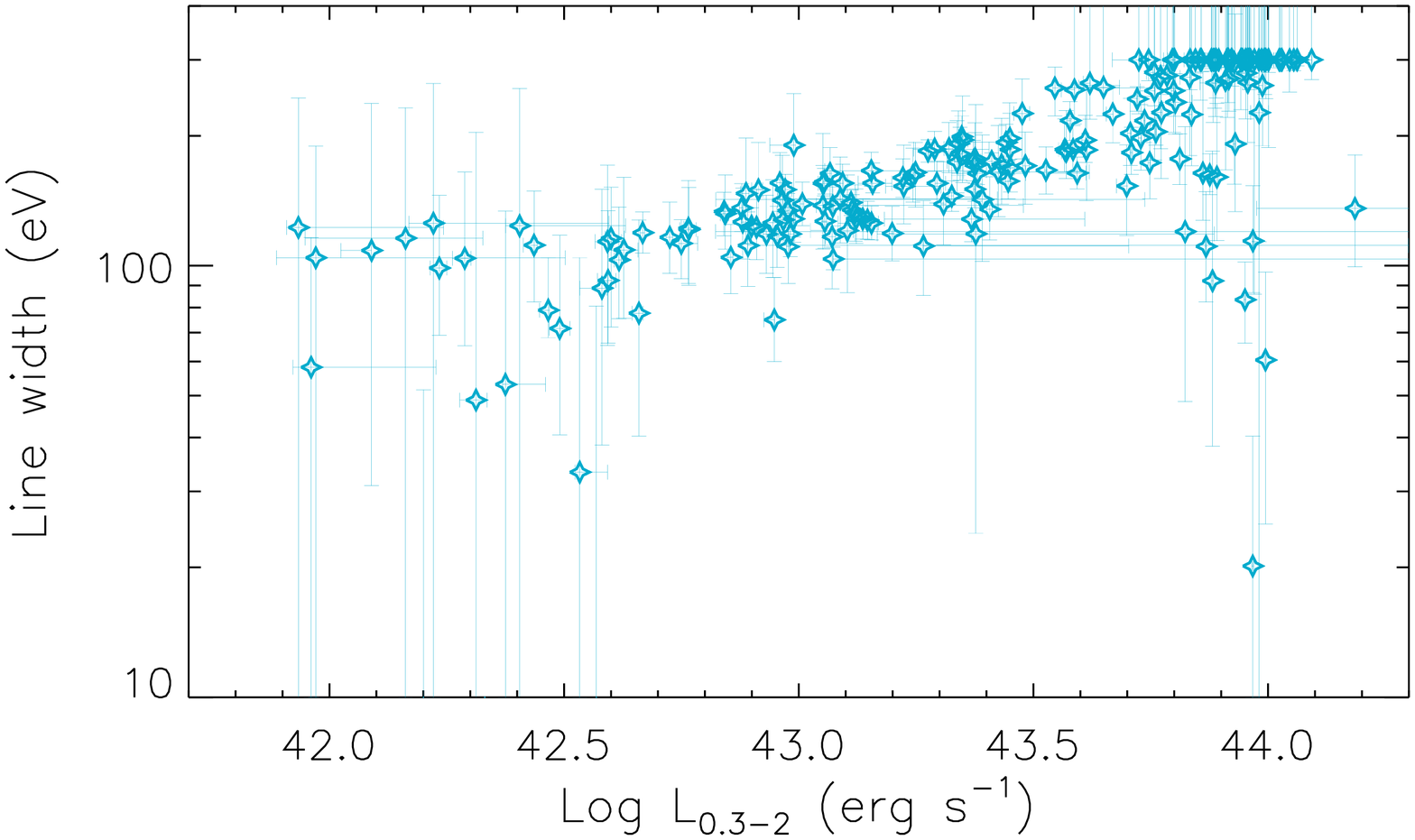}
    \caption{{\it Left panels:} Relation between the properties of the broad line at $\simeq 1$\,keV and the luminosity for the June 2018 (blue circles), December 2018 (red diamonds) and May 2019 (black stars) {\it XMM-Newton} observations (see \S\ref{sec:timeresolved2018}). {\it Right panel:} Same as left panels for the {\it NICER} observations (see \S\ref{sec:NICERobs}).}
    \label{fig:line_luminosity}
  \end{center}
\end{figure*}

The emission line is well constrained in all intervals except the most luminous ones (see central panels of Fig.\,\ref{fig:XMMspec18_twointervals}).
The energy of the line does not show any significant trend with luminosity for the June\,\,2018 observation, while in December\,\,2018, at the highest luminosities, the energy of the line appears to increase, up to $\sim 1.15$\,keV. No clear trend is observed in the May\,\,2019 observation.
During the June\,\,2018 observation the flux of the Gaussian component increases as the source gets brighter (blue filled circles in the middle left panel of Fig.\,\ref{fig:line_luminosity}), while in the December\,\,2018 and May\,\,2019 observations the line flux shows a flat trend with luminosity, and a possible decrease at the highest luminosities. A possible increase of the line width with luminosity is observed, with the Gaussian line being broader in December\,\,2018 and May\,\,2019 than in June\,\,2018.

To further examine the relation between spectral changes and luminosity, we also divided the June\,\,2018 RGS observation into two intervals, based on their flux. The high-flux interval covers the first 13\,ks of the observation (intervals 1--5 in Fig.\,\ref{fig:XMMlc_18_bins}), while the low-flux interval covers the 21--39\,ks range (intervals 7--9). We then fitted both intervals separately using our best model for the continuum, which includes a blackbody and a power law [\textsc{tbabs$\times$(zpo+zbb})]. The ratio between the model and the data is shown in Figure\,\ref{fig:RGS_highlowflux_june18} for the low (red squares) and high (black diamonds) flux intervals. No strong change is evident between the two spectra below 1.2\,keV, with the possible exception of the absorption feature at $E\simeq 0.7-0.8$\,keV, which appears to be broader in the low-flux interval.

\subsubsection{May 2011}\label{sec:timeresolvedMay2011}

In order to compare the relation between flux ratios in different bands and the 0.3--10\,keV flux with that before the outburst, we analyzed the light curves of the 2011 observations (see Appendix\,\ref{sect:2011xmmobs} for details). The ratios of different energy bands and the 0.3--0.5 keV flux versus the 0.3--10\,keV flux (in bins of 1\,ks) are shown in Fig.\,\ref{fig:XMMlc_11_bands_ratios} for the 0.75--1, 1--1.3, 1.3--2, 2--3, 3--5, 5--7 and 7--10\,keV bands. 
We fitted fourteen different intervals of this observation (see bottom panel of Fig.\,\ref{fig:xmmobs2011eeufs}) with a model that includes both a power-law and a blackbody component [\textsc{tbabs$\times$ztbabs$\times$(zpo+zbb})], leaving $\Gamma$, $N_{\rm H}$ and $kT$ free to vary. We did not find any significant relation between the blackbody temperature and the luminosity (Table\,\ref{tab:fitXMM11epochs}).

\subsection{The broad 1\,keV feature}\label{sect:broadline}

As discussed in \S\ref{sec:timeresolved2018_2019}, we found clear evidence in the {\it XMM-Newton} observations of a broad feature at 1\,keV, consistent with an emission line. The RGS spectra (see panels A and B of Figs.\,\ref{fig:RGS_spec_18} and \ref{fig:RGS_spec_19} in Appendix\,\ref{appendix:RGS_spectra}) show that the line might be asymmetric, with a shape that is reminiscent of the relativistically broadened profile typically observed in Fe K (e.g., \citealp{Fabian:2000if,Walton:2013ly}) and L (e.g., \citealp{Fabian:2009bz}) lines. We therefore fitted the line using the \textsc{relline} model \citep{Dauser:2010qy,Dauser:2013kx} instead of the symmetric Gaussian line. We fixed the inner radius in \textsc{relline} to $R_{\rm in}=-1$, so that it would self-consistently scale with the spin ($a$) of the SMBH, and the outer radius to $R_{\rm out}=400\rm\,r_{\rm g}$, where $\rm r_{\rm g}=GM_{\rm BH}/c^2$ is the gravitational radius. The two indices of the emissivity profile were set to have the same value and were left free to vary, while the distance from the SMBH at which the indices change was fixed to an arbitrary value. The normalization and energy of the line, as well as the spin of the SMBH and the inclination angle of the system, were also left free to vary. 
\smallskip

We studied the feature at $\sim 1$\,keV by fitting the two 2018 {\it XMM-Newton} observations simultaneously (PN, MOS and RGS for the June\,\,2018 observation; PN and RGS for the December\,\,2018 observation). We did not use the May\,\,2019 observations because the line was less prominent. We substituted the $\sim 1$\,keV Gaussian line with a relativistic line in our best model (see \S\ref{sect:xmmobs_spec_june} and \S\ref{sect:xmmobs_spec_december_may}), which includes a power-law and a blackbody component, as well as neutral absorption and a second Gaussian line at $\sim 1.8$\,keV [\textsc{cons$\times$tbabs$\times$ztbabs$\times$mtable\{xstar\}} \textsc{(zbb+zpo+relline+zgauss)}]. We also included two absorption lines for the December 2018 observation (see \S\,\ref{sect:xmmobs_spec_december_may}). We left all the parameters of the continuum, of the lines and of the warm absorbers free to vary between the two observations, setting instead the energy of the broad line, the spin of the SMBH and the inclination angle  to have the same values for both datasets. The normalization and the index of the emissivity profile was left free in both observations. The model fits the data well (Stat/DOF=7752/7021), and results in a spin of $a=0.41_{-0.05}^{+0.07}$, an inclination of $16.5_{-1.5}^{+2}$\,degrees, and an energy of the line of $1.37_{-0.03}^{+0.05}$\,keV, consistent with emission from Mg\,XI. During the first and second observation the index of the emissivity profile was $8.0_{-0.7}^{+1.2}$ and $4.6_{-0.3}^{+0.5}$, respectively. The temperature of the blackbody component obtained for the June observation is consistent with the one we found using a Gaussian line ($102\pm1$\,eV), while that of the December observation is slightly higher ($152\pm2$\,eV). It should be noted however that this fit does not result in a significant improvement of the chi-squared with respect to a model that includes a Gaussian line instead of the relativistic line (Stat/DOF=7754/7022).

\section{The {\it NICER} monitoring campaign}\label{sec:NICERobs}

{\it NICER} has been observing 1ES\,1927+654 for about 450\,\,days, starting from $\sim 150$\,\,days after the beginning of the event. This very intense monitoring, together with very large effective area of {\it NICER}, has allowed us to monitor the behaviour of the source within great detail.
In the following, we discuss the results obtained by studying the X-ray variability (\S\ref{sect:NICERvariability}) and spectroscopy (\S\ref{sect:NICERspectroscopy}) of 1ES\,1927+654 with {\it NICER} observations.

\subsection{X-ray Variability}\label{sect:NICERvariability}

The first {\it NICER} observations, carried out in May\,\,2018, $\sim 150$\,days after the beginning of the event, found 1ES\,1927+654 at a 0.5--10\,keV flux level similar to that observed in the 2011 {\it XMM-Newton} observation (see Fig.\,1 of \citealp{Ricci:2020}). However, the X-ray spectral shape of the source was at the time already very different from that of a typical AGN, with the continuum being dominated by a very soft component, as also seen in our {\it Swift}/XRT observations. 
Around the end of May\,2020 the flux of the source started to decline, reaching a minimum of $\sim 10^{40}\rm\,erg\,s^{-1}$ in July/August\,2018. In the following 100\,days the average X-ray emission increased rapidly, going even above the initial luminosity level and peaking at $\sim 10^{44}\rm\,erg\,s^{-1}$ in some intervals $\sim 300$\,days after the outburst. In the following 300\,days the median luminosity continued increasing. Between 300 and 400 days from the event it was $\log (L_{0.5-10}/\rm erg\,s^{-1})=43.27$, in the 400--500 days range it increased to $\log (L_{0.5-10}/\rm erg\,s^{-1})=43.89$ and finally in the last 100\,days of our monitoring campaign it reached $\log (L_{0.5-10}/\rm erg\,s^{-1})=44.25$. As discussed in \citealp{Ricci:2020} the X-ray luminosity appears to have an asymptote at $L_{0.5-10}/\rm erg\,s^{-1}\sim 44.25$. 
Interestingly, the very strong intraday X-ray variability that was observed in the first year after the event, when the source was found to vary of up to two orders of magnitude in a few hours, decreases strongly for $t\gtrsim400$\,days. In the last few months of our campaign 1ES\,1927+654 is in fact found to vary only of $\sim 0.3-0.5$\,dex in luminosity.

\begin{figure}
  \begin{center}
\includegraphics[width=0.45\textwidth]{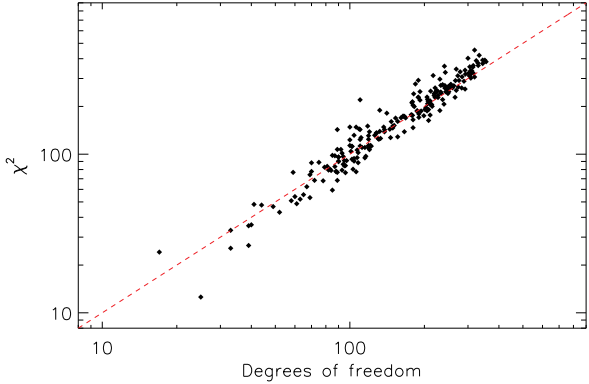}
\caption{Chi-squared versus degrees of freedom for the {\it NICER} spectra of our X-ray monitoring campaign. The red dashed line shows $\chi^2=\rm DOF$. }
    \label{fig:NICERfit}
  \end{center}
\end{figure}

\begin{figure*}
  \begin{center}
\includegraphics[width=0.45\textwidth]{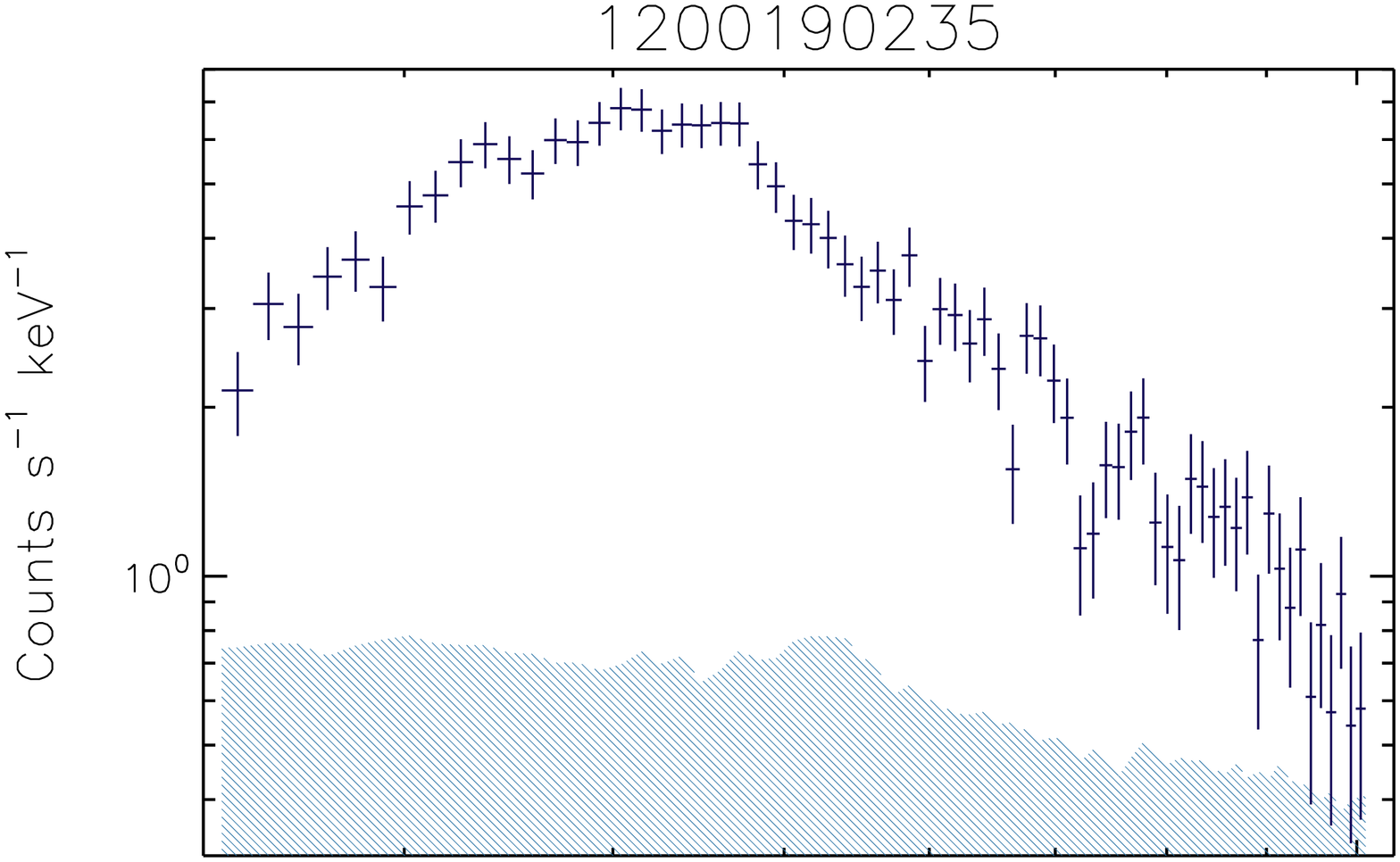}
\includegraphics[width=0.45\textwidth]{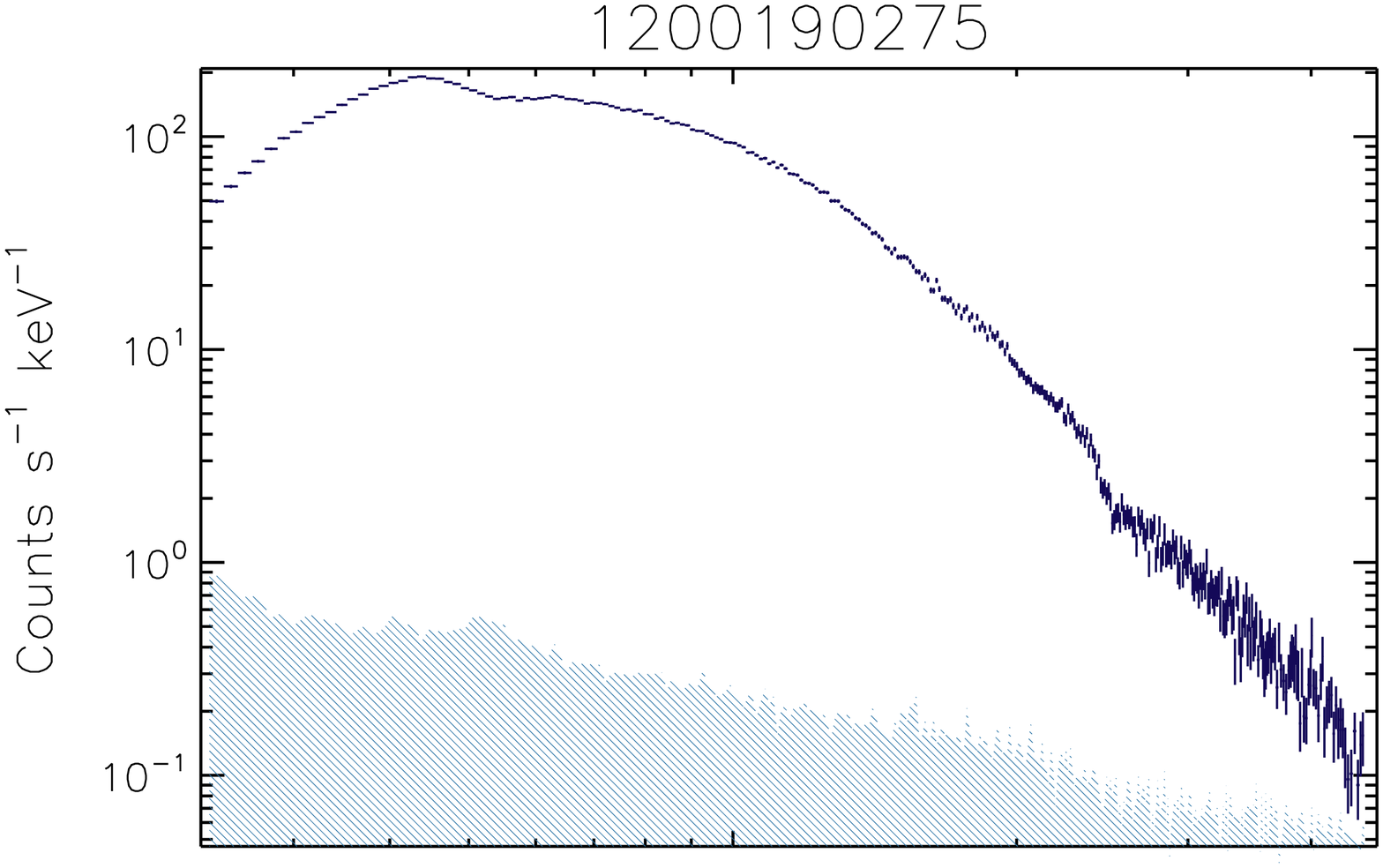}
\includegraphics[width=0.45\textwidth]{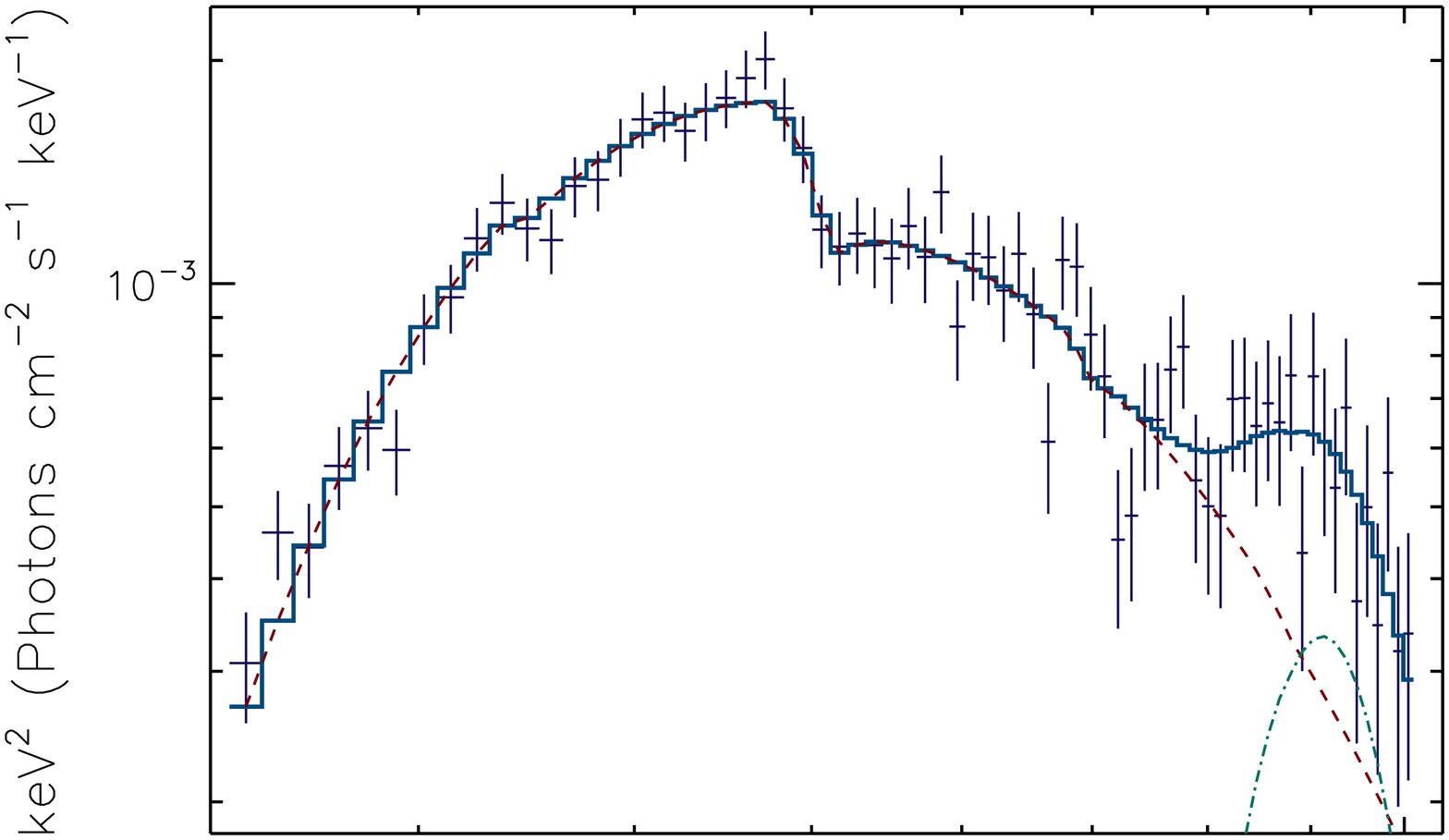}
\includegraphics[width=0.45\textwidth]{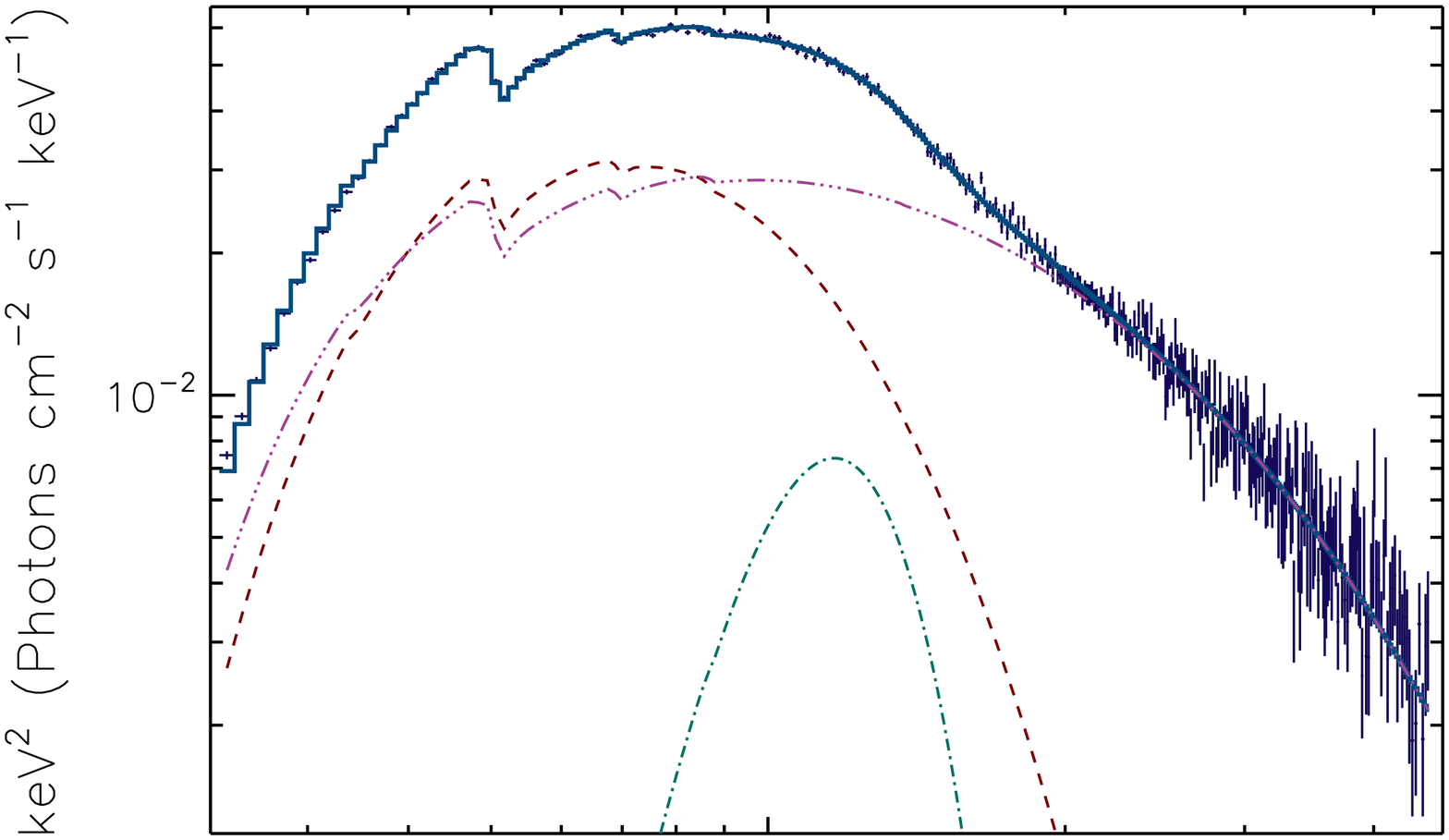}
\includegraphics[width=0.45\textwidth]{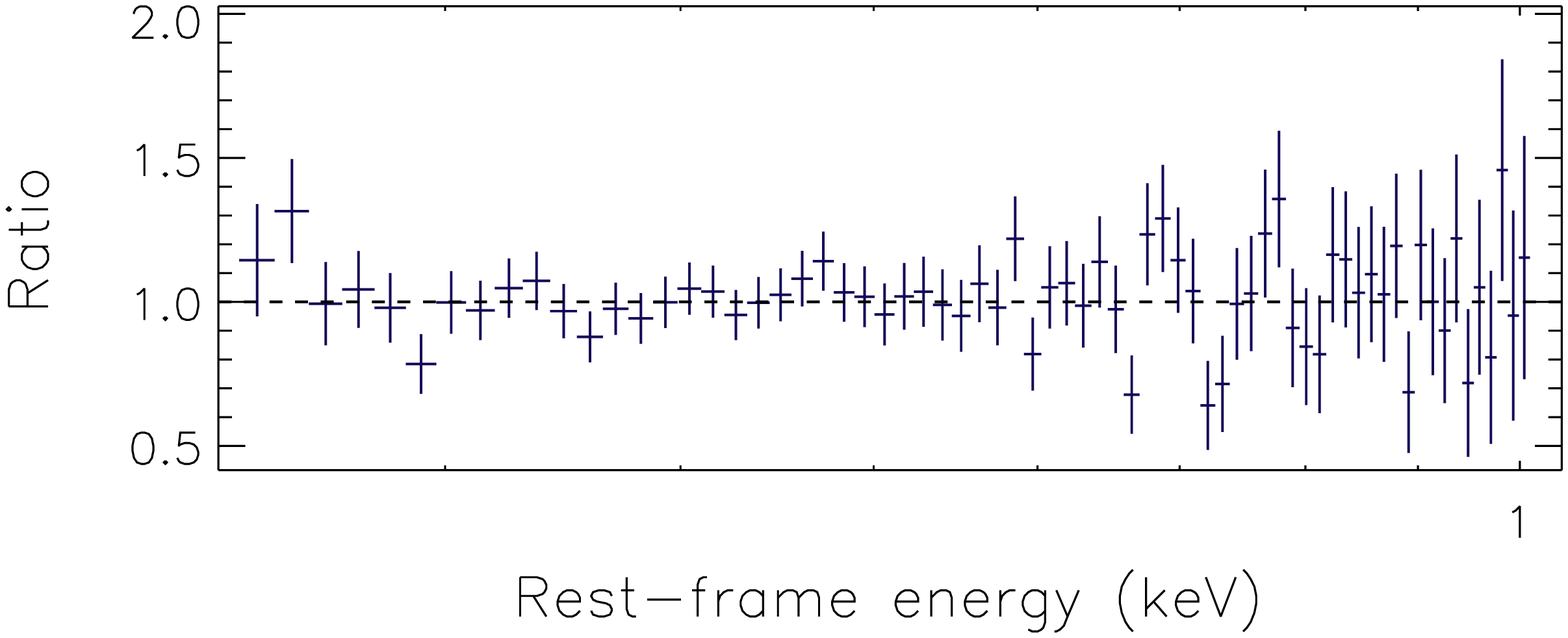}
\includegraphics[width=0.45\textwidth]{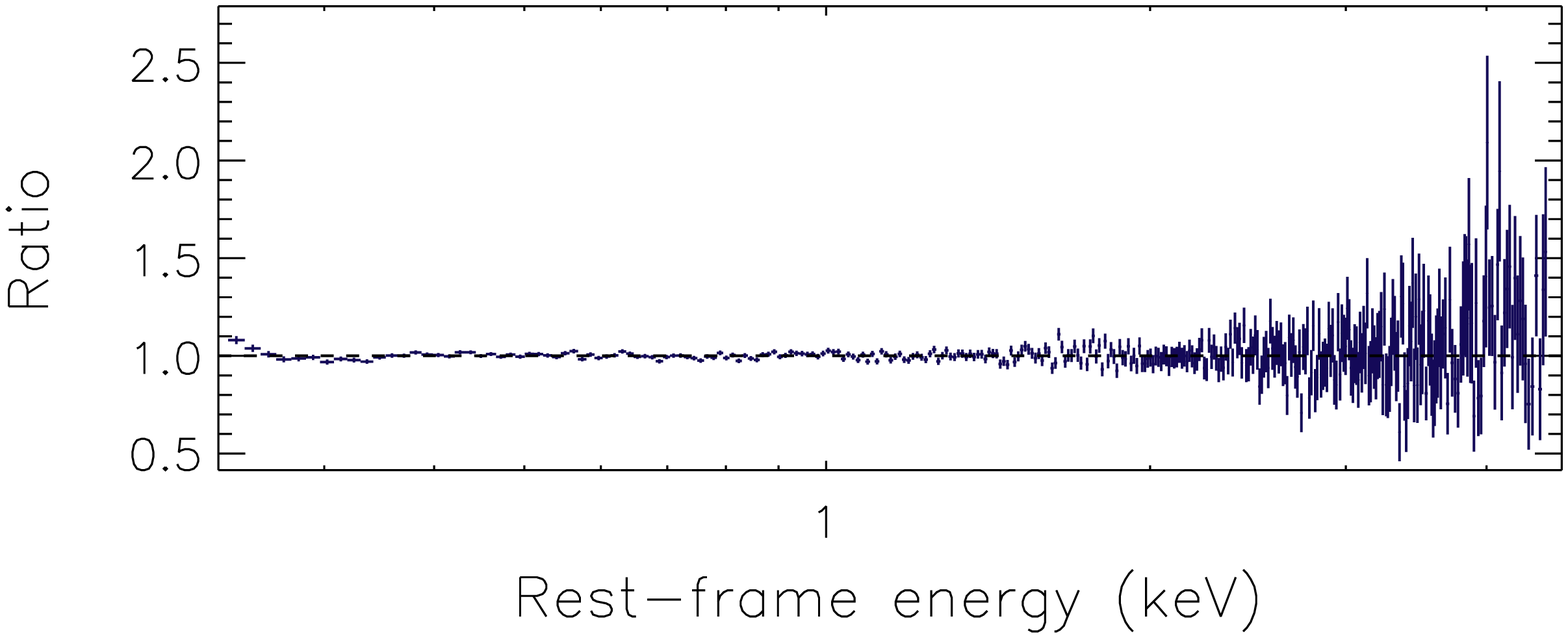}
\caption{Examples of two fits of {\it NICER} spectra obtained during a low ($L_{0.3-2}=9.1\times10^{41}\rm\,erg\,s^{-1}$, left panel) and a high ($L_{0.3-2}=9.7\times10^{43}\rm\,erg\,s^{-1}$, right panel) flux interval. The shaded area in the top panels represent the background, while the bottom panels illustrate the ratio between the best-fitting model (continuous lines in the middle panel) and the data. During the low-luminosity periods no power-law component is needed to reproduce the spectra, which is well fit by a blackbody component (dashed line) and a Gaussian line (dot-dashed line). In the high-luminosity intervals the power-law component (dot-dot dashed line) reappears.}
    \label{fig:NICERspec}
  \end{center}
\end{figure*}

\subsection{X-ray spectroscopy}\label{sect:NICERspectroscopy}

In the following we first analyze the spectra obtained from the individual {\it NICER} exposures (\S\ref{sect:NICERspectroscopy_individual}), and then focus on the spectra obtained by stacking different observations carried out within a few days to a few weeks from each other (\S\ref{sect:NICERspectroscopy_stacked}).

\subsubsection{Individual spectra}\label{sect:NICERspectroscopy_individual}

We fitted a total of 232 {\it NICER} spectra, excluding 33 exposures for which the signal-to-noise ratio was not sufficient to perform spectroscopy. The fits were carried out in an energy range that went from 0.3\,keV to $E_2$, where $E_2$ is the energy where the source and background rates are equal. Given the different signal to noise ratios of the observations, due to the very strong variability of the source, $E_2$ went from 0.5\,keV up to 5\,keV. We therefore used four different models to reproduce the X-ray emission of 1ES\,1927+654.

\begin{itemize}
\item The 11 observations in which the maximum energy was $E_2<1$\,keV were fitted with {\it Model\,1}, which included photo-electric intrinsic and foreground absorption, a blackbody and a power-law component [\textsc{tbabs$\times$ztbabs$\times$(zbb+zpo)}]. As we did for the {\it XMM-Newton} and {\it Swift} observations, foreground absorption was fixed at the Galactic value, and all redshifts were fixed to $z=0.019422$. Below 1\,keV the spectra of these observations were dominated by the blackbody component, therefore the photon index of the power law was fixed to $\Gamma=3$, while the normalization was left free to vary.
\item All the exposures for which $1 \leq E_2/\rm keV < 1.8$ were fitted with {\it Model\,2}, which includes the same component as Model\,1 plus a Gaussian line [(\textsc{tbabs$\times$ztbabs$\times$(zbb+zpo+zgauss)}]. The energy, width and normalization of the line were left free to vary while, as for Model\,1, the photon index of the power law was fixed to $\Gamma=3$. This model was used for 78 observations.
\item The 57 exposures in which $1.8 \leq E_2/\rm keV < 3 $ were fitted with {\it Model\,3}. The only difference between this model and Model\,2 is that the photon index of the power law component was left free to vary. 
\item The remaining 86 exposures were fit with {\it Model\,4}, which considers a cutoff power-law instead of a simple power-law [(\textsc{tbabs$\times$ztbabs$\times$(zbb+zcut+zgauss)}].
\end{itemize}

The models provide typically a good fit, with the median reduced chi-squared being $\chi^2$/DOF=1.03 (see Fig.\,\ref{fig:NICERfit}). The {\it NICER} observations encompass a very broad range of 0.5--10\,keV luminosities, from $\sim 10^{40}\rm\,erg\,s^{-1}$ to $\sim 10^{44}\rm\,erg\,s^{-1}$. In Fig.\,\ref{fig:NICERspec} we illustrate two of the {\it NICER} spectra, during a low (left panel) and a high (right panel) luminosity period.
The harder-when-brighter behaviour discussed in the previous sections is very evident when examining the {\it NICER} spectra. The blackbody temperature varies between $\sim70$\,eV and $\sim 200$\,eV, and is found to correlate with the luminosity of the source (middle panel of Fig.\,\ref{fig:kT_luminosity}). Fitting this trend with the same relation we used for the time-resolved {\it XMM-Newton} observations [$\log (L_{0.3-2}/\rm erg\,s^{-1})=a+b*\log(kT/\rm eV)$, see \S\ref{sec:timeresolved2018}] we find a slope of $b=3.85\pm0.01$. A break in this $kT-L_{0.3-2}$ relation is observed at $\log (L_{0.3-2}/\rm erg\,s^{-1})\simeq 43.7$, with the trend being steeper (flatter) at lower (higher) luminosities. Fitting the two luminosity intervals separately we find $b=7.4\pm0.3$ for $\log (L_{0.3-2}/\rm erg\,s^{-1})< 43.7$ and $b=0.9\pm0.2$ for $\log (L_{0.3-2}/\rm erg\,s^{-1})\geq 43.7$.

The properties of the $\sim 1$\,keV Gaussian line versus the X-ray luminosity are illustrated in the right panel of Fig.\,\ref{fig:line_luminosity}. The energy of the line (top panel) does not vary significantly up to $\log (L_{0.3-2}/\rm erg\,s^{-1})\simeq 43.7$, above which it appears to increase up to $\sim 1.15$\,keV. Similarly to what we have found in the {\it XMM-Newton} observations, we see a clear increase in the flux of the line (middle panel) with the luminosity. Interestingly, above $\log (L_{0.3-2}/\rm erg\,s^{-1})\simeq 43.7$ such a trend disappears, and the flux of the line appears to decrease with the luminosity. A similar behaviour is observed for the width of the line (bottom panel), which increases up to $\simeq 200$\,eV at $\log (L_{0.3-2}/\rm erg\,s^{-1})\simeq 43.7$ for then decreasing.

\subsubsection{Stacked spectra}\label{sect:NICERspectroscopy_stacked}

We also analyzed the spectral variations of 1ES\,1927+654 by stacking {\it NICER} data in sixteen different and successive epochs, spanning intervals between a few days and a few weeks (see Appendix\,\ref{sect:stackednicerspecappendix}). We started by using the same approach to the spectral modelling that we used for the individuals spectra (\S\ref{sect:NICERspectroscopy_individual}), based on the energy range in which the spectra were not dominated by the background. Our spectral models include a blackbody component plus a Gaussian line [{\it bb}; \textsc{tbabs$\times$ztbabs$\times$(zbb+zgauss})], a blackbody, a Gaussian line and either a power law [{\it po}; \textsc{tbabs$\times$ztbabs$\times$(zpo+zbb+zgauss})] or a cutoff power-law [{\it cut}; \textsc{tbabs$\times$ztbabs$\times$(zcut+zbb+zgauss})]. All the spectra for which a power-law or a cutoff power-law model was used were also fitted with a thermally Comptonized plasma [{\it nth}; \textsc{tbabs$\times$ztbabs$\times$(nthcomp+zbb+zgauss})]. In the latter model the temperature of the seed photons was fixed to the temperature of the blackbody component. The results of our analysis are reported in Table\,\ref{tab:fitNICERstackedSpec} in Appendix\,\ref{sect:stackednicerspecappendix}. For four spectra we used the {\it bb} model, while for the remaining twelve spectra we used the {\it po} or {\it cut} models, as well as the {\it nth} models. 
For the first interval the model left strong residuals, suggestive of absorption features, and resulted in a chi-squared of $\chi^2$/DOF=400/98. Adding two ionized absorbers, using a \textsc{xstar} table that assumes $v_{\rm turb}=300\rm\,km\,s^{-1}$ improved the fit significantly ($\chi^2$/DOF=113/92). The other spectra are well fit by the baseline models, and do not require any additional components. 
The temperature of the blackbody component obtained by our fitting changes when considering the {\it nth} model with respect to the {\it po} or {\it cut} models, with $kT$ being typically lower for the former model. However, as shown in the bottom panel of Fig.\,\ref{fig:kT_luminosity}, regardless of the model adopted for the hard X-ray component, we find that the blackbody temperature increases for increasing luminosities.

\section{Discussion}\label{sect:discussion}

\begin{figure*}
\begin{center}
\includegraphics[width=0.48\textwidth]{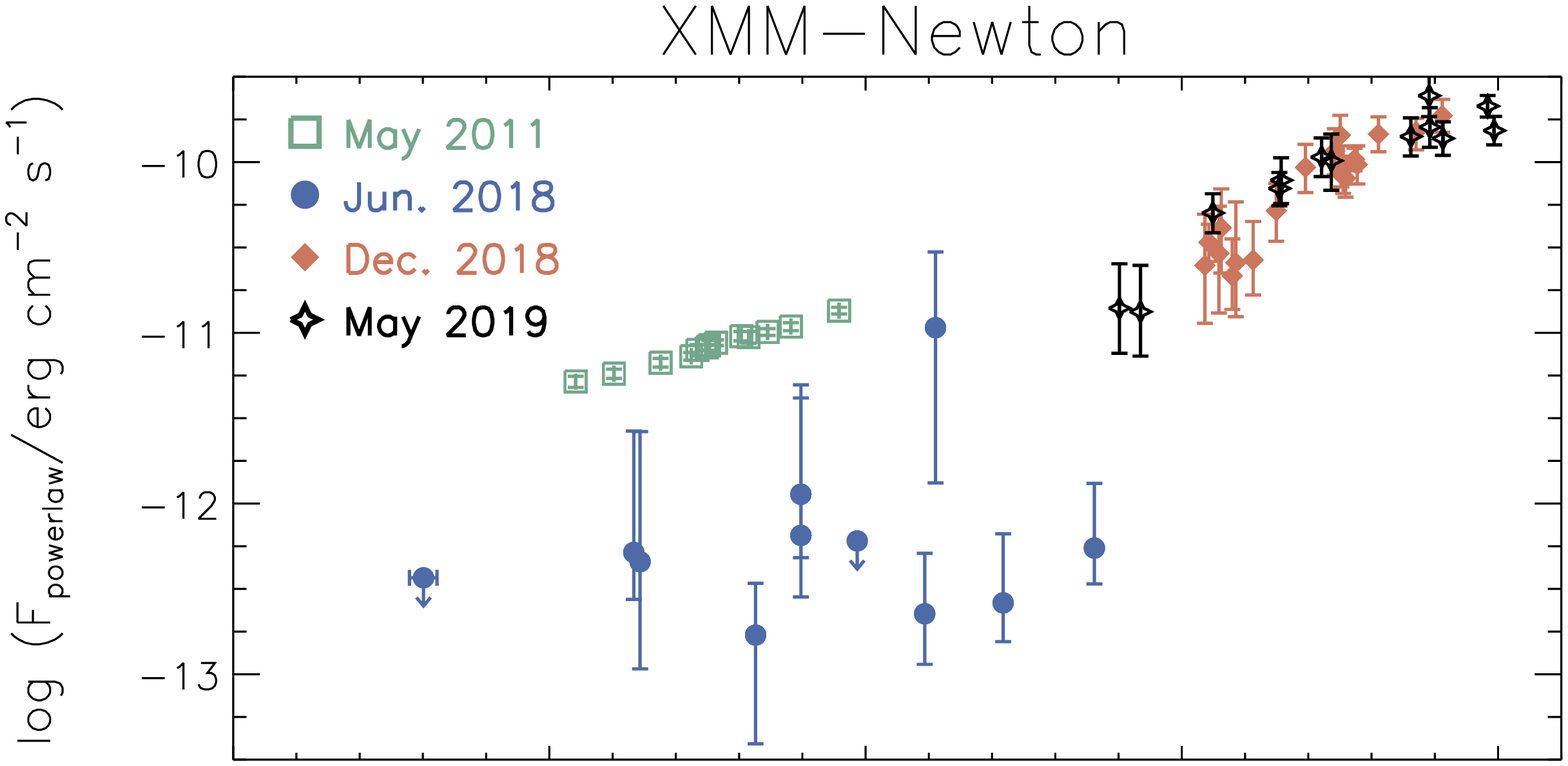}
\includegraphics[width=0.48\textwidth]{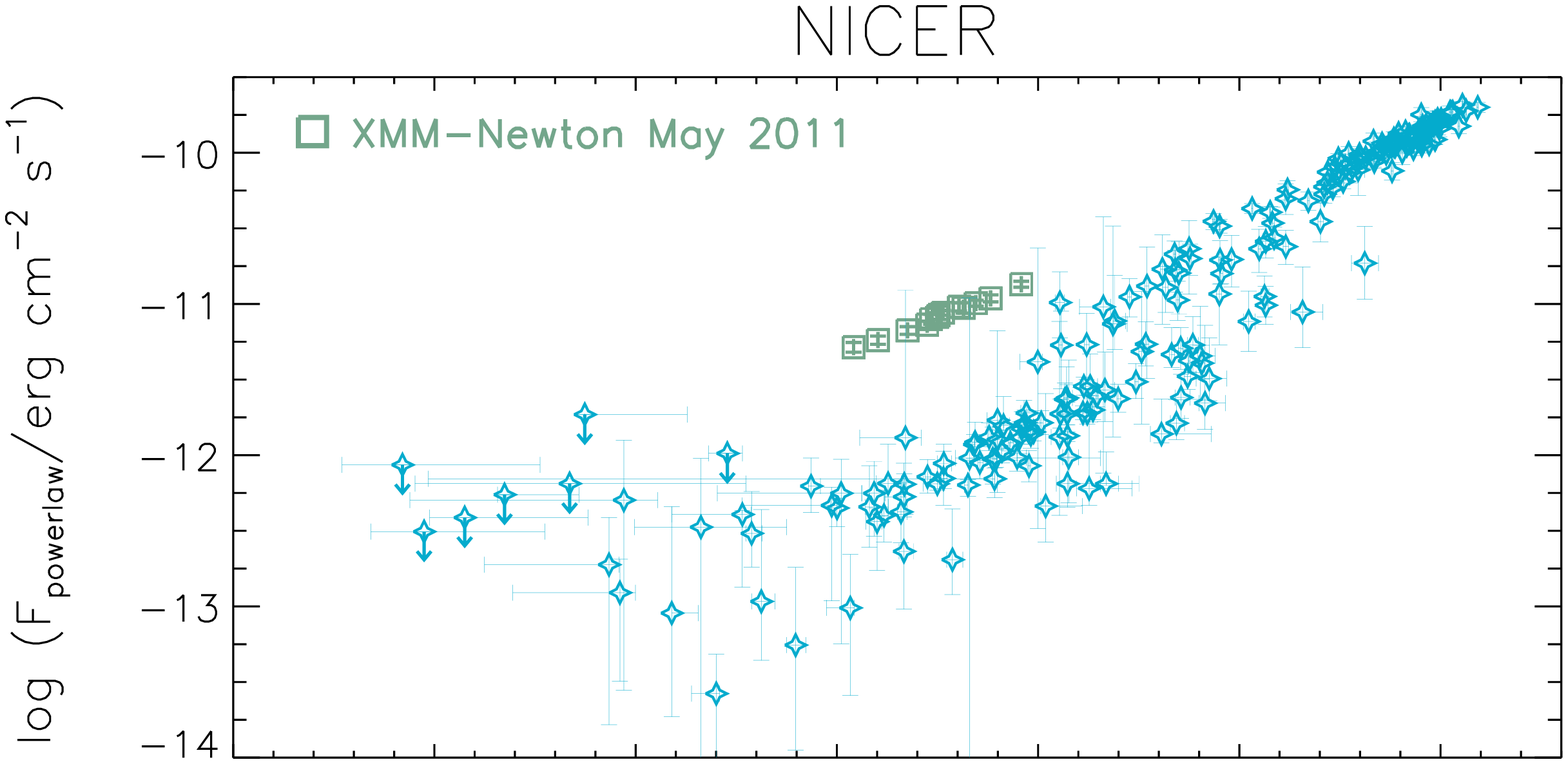}
\includegraphics[width=0.48\textwidth]{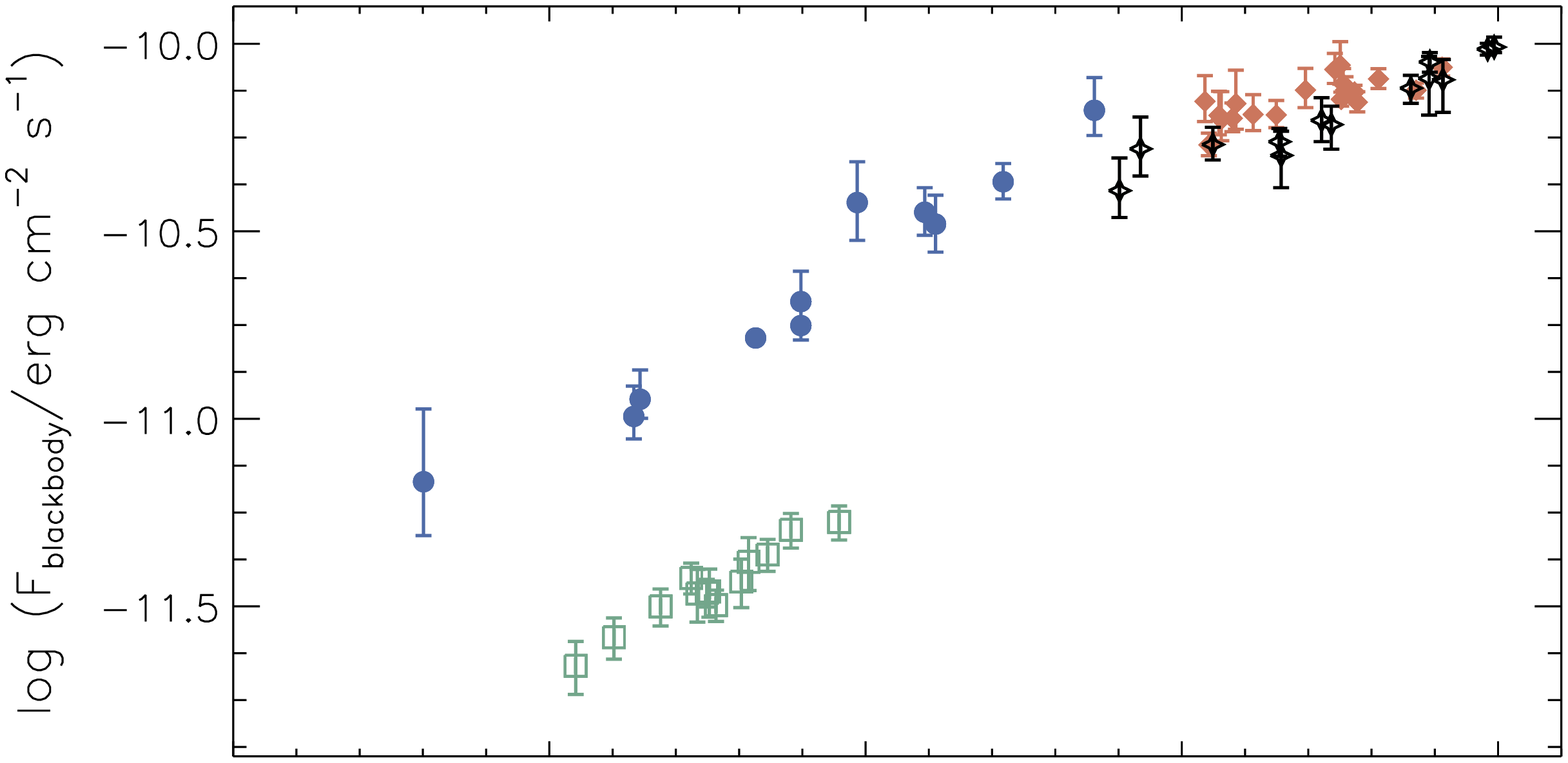}
\includegraphics[width=0.48\textwidth]{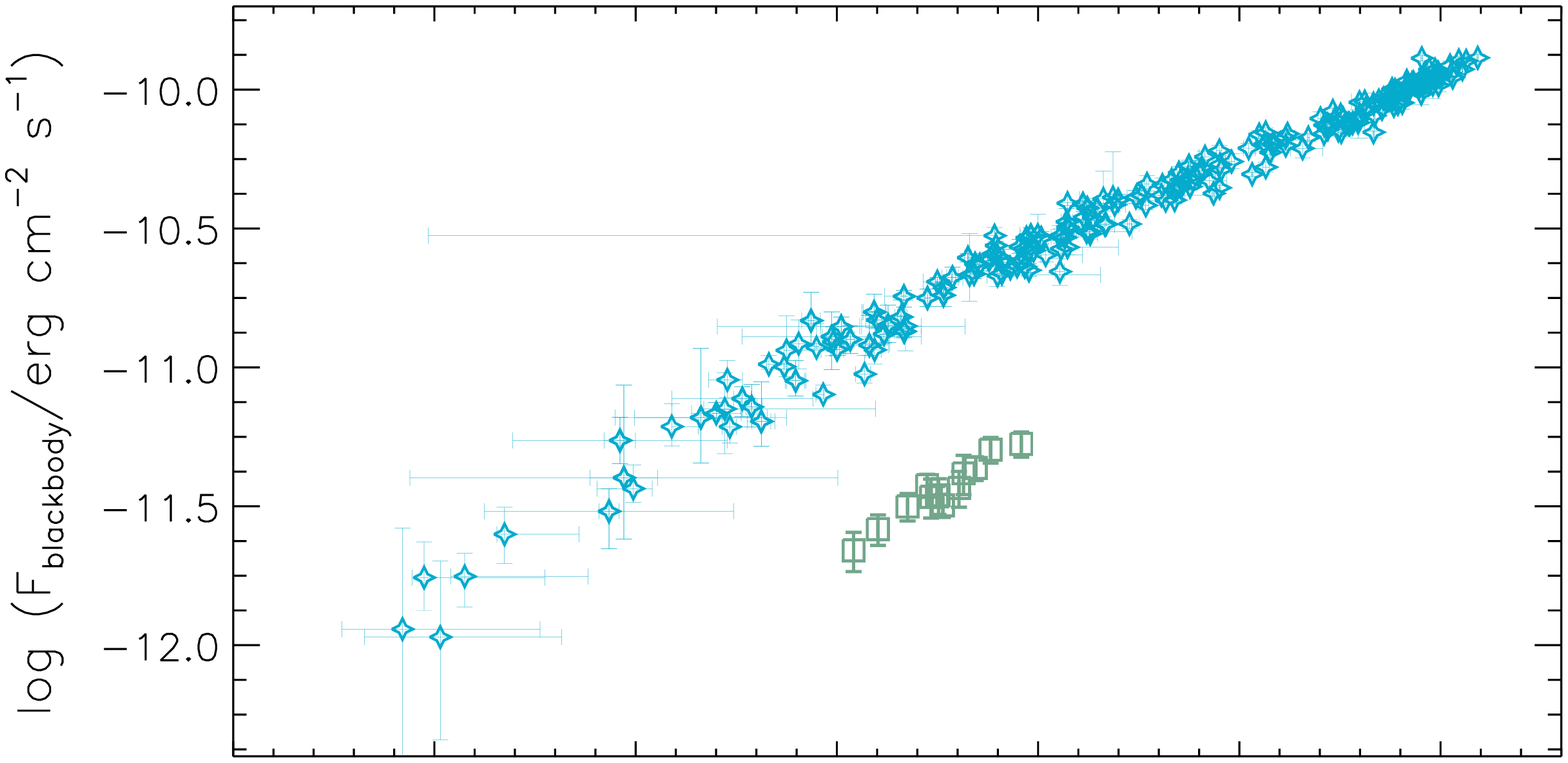}
\includegraphics[width=0.48\textwidth]{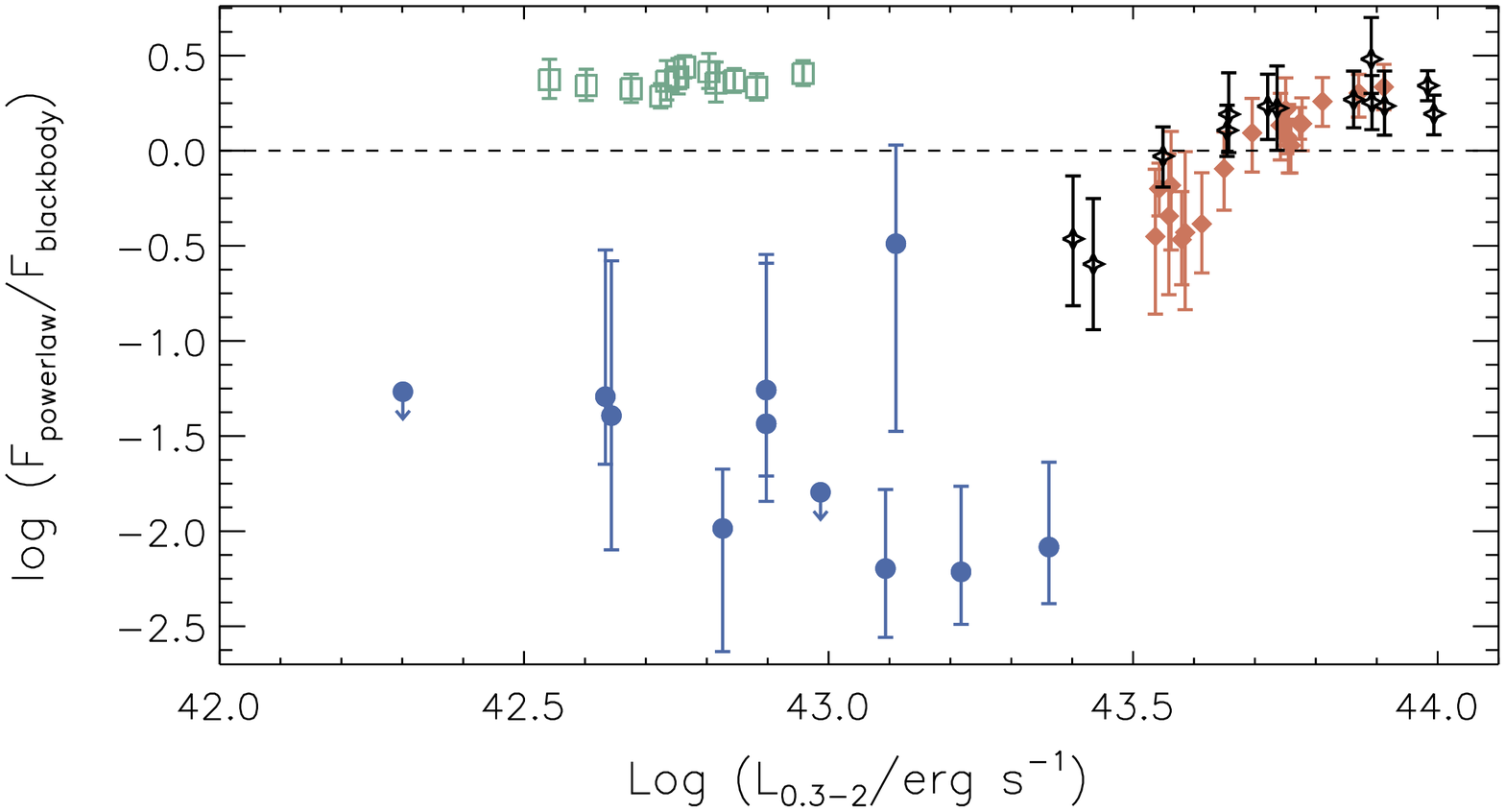}
\includegraphics[width=0.48\textwidth]{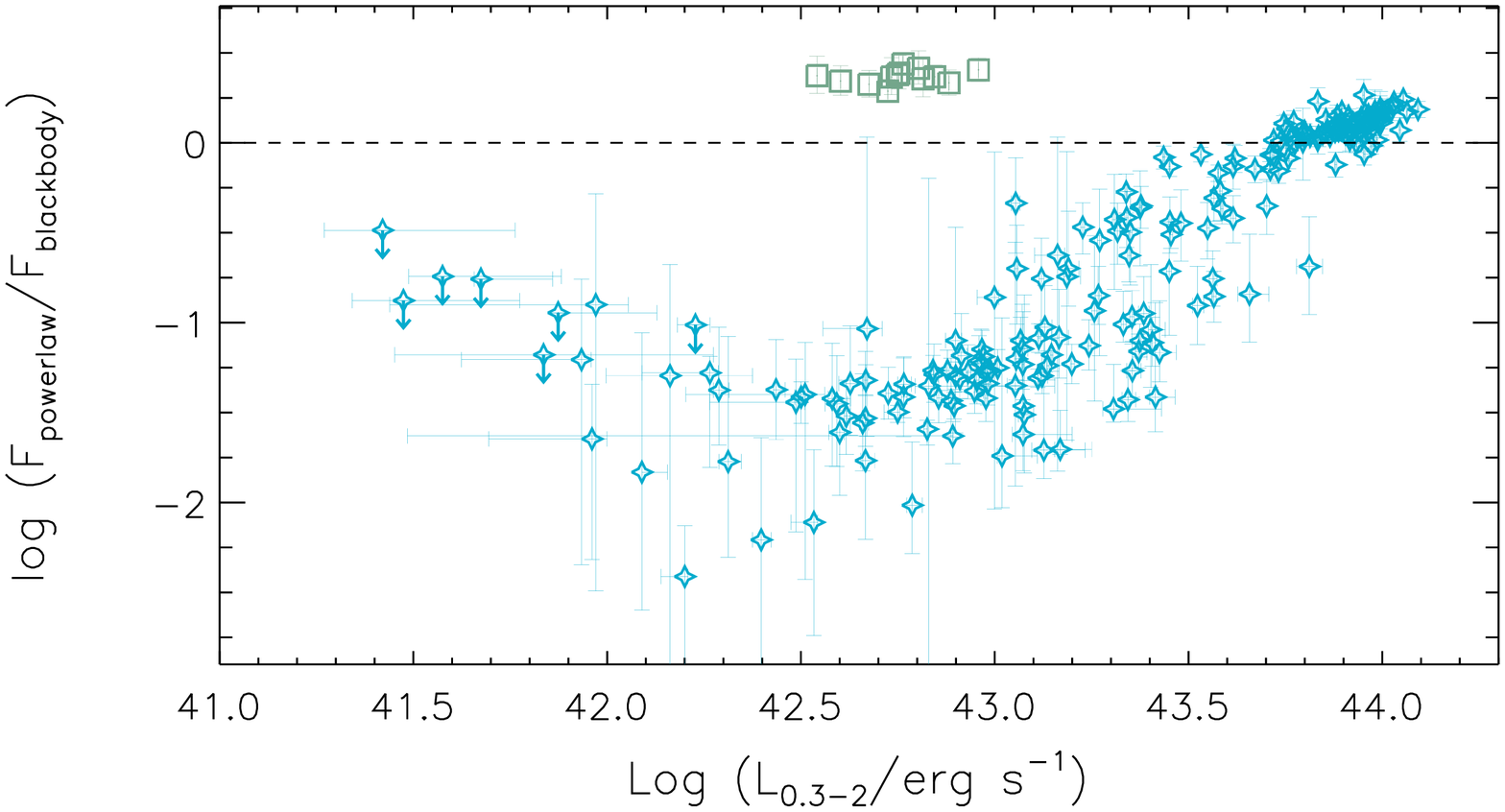}
    \caption{{\it Left panels:} Relation between the contribution of the different spectral components to the 0.3--2\,keV flux and the total luminosity for the June 2018 (filled blue circles), December 2018 (filled red diamonds) and May 2019 (empty black stars) {\it XMM-Newton} observations. 
    The top and middle panel shows the flux of the power-law ($F_{\rm powerlaw}$) and blackbody ($F_{\rm blackbody}$) component, respectively, while the bottom panel shows their ratio. {\it Right panels}: same as left panels for the {\it NICER} observations.}
    \label{fig:bb_po_ratio}
  \end{center}
\end{figure*}

\subsection{X-ray spectral properties 1ES\,1927+654}\label{sect:discXrayprop}
The X-ray spectral shape of 1ES\,1927+654 inferred from our X-ray monitoring campaign is very different from what has been observed in typical AGN; the main peculiarities of this object are the supersoft continuum, dominated by a blackbody component ($kT\sim 80-200$\,eV), the disappearance of the power law component \citep{Ricci:2020}, the broad emission features at $\sim 1$\,keV and $\simeq 1.8$\,keV, and the extremely low cutoff energy of the power-law ($E_{\rm C}\sim 3$\,keV). In the following we discuss first the X-ray characteristics of 1ES\,1927+654 inferred by previous studies (\S\ref{sect:previousXrayobs}), and then the exceptional properties observed over the past $\sim2$\,\,years (\S\ref{sect:Xrayproperties1819}).

\subsubsection{Previous X-ray observations of 1ES\,1927+654}\label{sect:previousXrayobs}
1ES\,1927+654 was first detected in the X-ray band by the {\it Einstein} slew survey, and classified as an AGN (\citealp{Elvis:1992aa}, see also \citealp{Perlman:1996aa} and \citealp{Simcoe:1997zm}). The source was later detected by the {\it ROSAT} mission in the 0.1--2.4\,keV band, and was reported in the all-sky survey bright source catalogue \citep{Voges:1999wq}. {\it ROSAT} observations showed that 1ES\,1927+654 is very variable, and a PSPC pointed observation in December 1998 detected a strong X-ray flare. The 0.1--2.4\,keV luminosities inferred by the all-sky survey and by the pointed observation were $4.6\times 10^{43}\rm\,erg\,s^{-1}$ and $\sim 9\times 10^{43}\rm\,erg\,s^{-1}$, respectively.
\cite{Boller:2003cq} studied the {\it ROSAT} and {\it Chandra} spectra of this AGN, and found a rapid and large-amplitude variation, together with a very steep continuum: $\Gamma=2.6\pm0.3$ in the 0.1--2.4\,keV band and $2.7\pm0.2$ in the 0.3--7\,keV band. During the {\it ROSAT} all-sky survey the source showed significant spectral variability, and the spectral variations did not display any clear correlation with the flux. The pointed PSPC observation showed hardness ratio variability, although also in this case the spectral changes were not correlated with the flux variations.

\cite{Gallo:2013hq} analyzed the 2011 {\it XMM-Newton} observation of 1ES\,1927+654 (28\,ks), together with a $\sim 70$\,ks {\it Suzaku} observation, also carried out in 2011. The luminosity of the source in the 0.3--10\,keV band was $8.63^{+0.05}_{-0.05}\times10^{42}\rm\,erg\,s^{-1}$, very similar to that of the June 2018 {\it XMM-Newton} observation\footnote{$L_{0.3-10}=8.54^{+0.10}_{-0.34}\times10^{42}\rm\,erg\,s^{-1}$}.
\cite{Gallo:2013hq} applied several spectral models to the data, all taking into account a power law component, plus different features, including a blackbody, two neutral partial covering absorbers, two ionized partial covering absorbers and blurred reflection. During these observation the narrow Fe\,K$\alpha$ feature, which is ubiquitously found in AGN (e.g., \citealp{Nandra:1994ly,Shu:2010qv}), was not detected, and only an upper limit on the EW could be obtained ($EW\lesssim 30$\,eV). This is considerably lower than what would be expected for an AGN with a 2--10\,keV luminosity of $3.1\times 10^{42}\rm\,erg\,s^{-1}$: considering the well known anti correlation between the X-ray luminosity and the Fe\,K$\alpha$ EW one would in fact expect $EW\simeq 100$\,eV (e.g., \citealp{Iwasawa:1993ly,Bianchi:2007fr,Ricci:2013cj,Ricci:2013yz,Ricci:2014fr,Boorman:2018ov,Matt:2019ct}).
Using a power-law plus blackbody model, similar to our baseline model, resulted in $\Gamma\simeq 2.3$, a value significantly different from the median value inferred for AGN in the local ($z\lesssim 0.05$) Universe ($\Gamma=1.8-1.9$; e.g., \citealp{Nandra:1994ly,Piconcelli:2005qy,Winter:2009vn,Lubinski:2016ao,Ricci:2011hm,Ricci:2017fj}).
Most unobscured AGN show a strong component below $\simeq 1$\,keV dubbed the soft excess, which can be typically reproduced by a blackbody component with a temperature of $\sim 100$\,eV (e.g., \citealp{Gierlinski:2004dt,Winter:2009vn,Boissay:2016vf,Ricci:2017fj,Garcia:2019cr}), and is believed to be either due to relativistic reflection from the innermost regions of the disk (e.g., \citealp{Crummy:2006cl}) or to emission from a  cooler Comptonizing region, possibly in the form of an atmosphere of the accretion disk (e.g., \citealp{Done:2012uk}). In the case of 1ES\,1927+654 the temperature of the blackbody was higher ($kT\simeq 170$\,eV) than what is typically found in local AGN.

Ionized absorbers have been systematically found in the X-ray spectra of nearby AGN (e.g., \citealp{Costantini:2007js,Pounds:2011ph}), in some cases showing a high degree of complexity (e.g., \citealp{Kaastra:2014zp,Kaastra:2018ua}) and reaching very high velocities (e.g., \citealp{Tombesi:2010zr,Kosec:2018hw,Boissay-Malaquin:2019jh,Walton:2019sw} ). \cite{Gallo:2013hq} found that both a ionized partial covering model and a blurred reflection model are able to well reproduce the data. In the case of the partially covering ionized absorber two outflows with a velocity of $\sim0.3$c and column densities of $3\times10^{22}\rm\,cm^{-2}$ and $6.6\times10^{23}\rm\,cm^{-2}$ were needed, while the blurred model required the system to be observed edge-on. 

\subsubsection{The X-ray spectral shape after the optical/UV outburst}\label{sect:Xrayproperties1819}

After the optical/UV outburst (late December 2017) the X-ray spectral shape of 1ES\,1927+654 drastically changed with respect to the 2011 observation. Our analysis showed that the spectrum is dominated by a blackbody component, and that the power-law was extremely faint in May/June\,\,2018, was undetected in July--August\,\,2018, and re-appeared only a few months later, when the flux of the source increased \citep{Ricci:2020}. In \S\,\ref{sec:alternativemodelling} we discussed how neither partial covering or blurred reflection models can explain the X-ray properties of 1ES\,1927+654 observed after the optical event.

A supersoft X-ray continuum has so far been observed only in handful of objects classified as AGN (e.g., \citealp{Terashima:2012nx,Miniutti:2013ul,Miniutti:2019rg,Sun:2013rr,Lin:2017ec,Shu:2017qy,Shu:2018bh,Giustini:2020qg}), and never before was the power-law observed in the process of disappearing. The {\it XMM-Newton} and the {\it NICER} observations show that the flux of the power-law and that of the blackbody component increases with the 0.3--2\,keV luminosity of the source (top and middle panels of Fig.\,\ref{fig:bb_po_ratio}, see also \citealp{Ricci:2020}). While for the same luminosity interval covered by the 2011 {\it XMM-Newton} observation [$\log (L_{0.3-2}/\rm erg\,s^{-1})\sim 42.5-43$] the flux of the power-law component after the event is considerably lower than in 2011 (green empty squares), that of the blackbody component is $\sim 0.7$\,dex higher. Interestingly, the ratio between the flux of the power law and that of the blackbody shows a clear increase with the luminosity (see Fig.\,3 of \citealp{Ricci:2020}), going from $\sim 10^{-2}$ for $\log (L_{0.3-2}/\rm erg\,s^{-1})\lesssim 43$ to $\sim 10^{0.4}$ for $\log (L_{0.3-2}/\rm erg\,s^{-1})\sim 44$ (Fig.\,\ref{fig:bb_po_ratio}). The transition between having a blackbody-dominated flux to a power-law dominated one is found at $\log (L_{0.3-2}/\rm erg\,s^{-1})\sim 43.7$, and at the highest luminosities the source goes back to having a ratio comparable to that observed in May\,2011 \citep{Ricci:2020}.

The power-law is very steep ($\Gamma \sim 3$), and our joint {\it XMM-Newton}/{\it NuSTAR} observations showed that the addition of a cutoff with a very low energy ($E_{\rm cut}\sim 3$\,keV, see Table\,\ref{tab:fitXMM}) is required (see panels B and C of Fig.\,\ref{fig:XMMspecDec18May19}). 
This value is considerably lower than what is typically observed in nearby accreting SMBHs. From the analysis of the broadband X-ray properties of the $\sim 840$ AGN detected by {\it Swift}/BAT during the first 70\,\,months of operations, \citet{Ricci:2017fj} found that the median cutoff energy is $200\pm29$\,keV ($210\pm36$\,keV for the unobscured AGN). Previous {\it BeppoSAX} observations found consistent results ($230\pm22$\,keV for unobscured AGN, see \citealp{Dadina:2007lc,Dadina:2008sh}). Recent {\it NuSTAR} observations have also confirmed that the cutoff energy is typically found between $\sim40$\,keV \citep{Kara:2017oi} and $\sim 200-300$\,keV (e.g., \citealp{Tortosa:2017ds,Tortosa:2018wj,Kamraj:2018uo,Balokovic:2020ht}), with the highest recorded value thus far being for NGC\,5506 ($E_{\rm cut}=720^{+130}_{-190}$\,keV, \citealp{Matt:2015hp}). A relation between the cutoff energy and the Eddington ratio was recently found by \citet{Ricci:2018mp}, who showed that sources with higher $\lambda_{\rm Edd}$ tend to have cooler coronae. This could be explained by the fact that AGN typically tend to avoid the region in the temperature-compactness parameter space where runaway pair production dominates \citep{Bisnovatyi-Kogan:1971nr,Svensson:1984gs,Fabian:2015ln,Fabian:2017wv}. Extending the relation of \citet{Ricci:2018mp} to higher Eddington ratios, assuming that the trend does not change at $\lambda_{\rm Edd}\gg 1$, one would expect to find such low values of $E_{\rm cut}$ only for $\lambda_{\rm Edd}\sim 30$. Assuming instead the relation between the photon index and the Eddington ratio (e.g., \citealp{Shemmer:2006dk,Brightman:2013bj,Trakhtenbrot:2017pu}), and in particular the one obtained by \citet{Trakhtenbrot:2017pu} considering a cutoff power-law model ($\Gamma=2.00+0.167\log\lambda_{\rm Edd}$), one would need $\lambda_{\rm Edd}\sim 6$ to obtain the photon index found by our observations. It should be however remarked that the physical properties of the corona at these extreme Eddington ratios, where the accretion flow might be very different (e.g., \citealp{Abramowicz:1988sb}), are still unclear.

The black hole mass of 1ES\,1927+654, inferred by \citet{Trakhtenbrot:2019qy} by analyzing broad Balmer lines, is $1.9\times10^{7}\rm\,M_{\odot}$. As discussed in \citet{Ricci:2020}, the black hole mass could however be quite different, due to the fact that the BLR clouds might not have had the time to virialize. This is also shown by the variable FWHM of both H$\alpha$ and H$\beta$ (Li et al. in prep.; see Appendix\,A of \citealp{Ricci:2020}). The stellar mass of the galaxy  ($2.3\times10^{9}M_{\odot}$) suggests $M_{\rm BH}\sim 10^{6}M_{\odot}$ \citep{Kormendy:2013uf}. At $\gtrsim 300$\,days after the optical outburst the optical/UV flux had decreased strongly \citep{Trakhtenbrot:2019qy}, and most of the emission is observed in the X-rays \citep{Ricci:2020}. Assuming that the X-ray bolometric correction  is $\kappa_{\rm X}\sim 1$, and $M_{\rm BH}\sim 10^{6}M_{\odot}$, the maximum observed X-ray luminosity would correspond to $\lambda_{\rm Edd}\sim 1$. Therefore, we can conclude that the maximum Eddington ratio would be $\lambda_{\rm Edd}\simeq \kappa_{\rm X}\gtrsim 1$, and it is likely that the source was super-Eddington for a considerable fraction of its time during our monitoring campaign. In a forthcoming publication (Li et al. in prep.) we will discuss in detail the spectral energy distribution of this source.
It should be noted that some of the properties of 1ES\,1927+654 after the optical outburst, such as its softness, are similar to those of some of the most extreme narrow-line Seyfert\,1s, e.g. 1H\,0707$-$495 and IRAS\,13224$-$3809 (e.g., \citealp{Boller:2003pd,Fabian:2004ff}). However none of these sources shows a power-law component as steep and faint as 1ES\,1927+654. Interestingly, both 1H\,0707$-$495 and IRAS\,13224$-$3809 show an excess due to a relativistically-broadened Fe\,L line at 0.9\,keV \citep{Fabian:2009bz,Fabian:2013ni}. The energy of this line is however lower than that of the feature observed in the X-ray spectrum of 1ES\,1927+654 ($E=1.01\pm0.01$\,keV, see Table\,\ref{tab:fitXMM}).

Similarly to the 2011 observations, we do not see any sign of a Fe\,K$\alpha$ line at 6.4\,keV. This is not surprising, due to the very low flux of the source at 7.1\,keV. However, we could in principle expect to see an Fe K$\alpha$ echo from more distant material (e.g. the BLR, or the torus), which is not found in any of our X-ray observations. This, together with the absence of a Compton hump, suggests that the circumnuclear environment of 1ES\,1927+654 is particularly poor in gas and dust. The two broad emission lines observed at $\sim 1$\,keV  and $\simeq 1.8$\,keV are also extremely puzzling, and they have not been observed in any other AGN thus far. The line at $\sim 1$\,keV could be associated to Ne\,\textsc{x} (1.02\,keV) or to the ionized Fe-L emission from Fe\,\textsc{xx-xiv}. In the latter case the increase in the energy of the line observed at the highest luminosities (top panels of Fig.\,\ref{fig:line_luminosity}) might be related to the increase in the ionization state of the circumnuclear material. The second line is found at $E\sim 1.7$\,keV during the first {\it XMM-Newton} observation and at $\sim 1.9$ during the second and third observations. While the line at $\sim 1.7$\,keV could be Si\,K$\alpha$ (1.74\,keV) or to Al\,\textsc{xiii} ($1.73$\,keV), it is unclear to what transition the line at $\sim 1.9$ could be associated. It should be stressed that we cannot exclude that these two broad features are produced by strong and very turbulent ionized outflows, and we refer to a future publication (Masterson et al. in prep.) for a detailed analysis.

\subsection{The peculiar X-ray spectral variability}\label{sect:peculiarvariability}

Our X-ray monitoring campaign (\S\ref{sect:xmmobs_spec}--\ref{sec:NICERobs}) showed a clear relation between the spectral shape and the flux of the source. This was observed in all X-ray observations, with the spectrum becoming harder as the flux increased. This effect was first illustrated, in a model-independent way, by the fact that: i) the flux ratios between different bands and the softest (0.3--0.5\,keV) band follow the same pattern as the overall variability (see panels C, D and E of Fig.\,\ref{fig:XMMlc_18_bands} and panels D, E, F, G, H of Fig.\,\ref{fig:XMMlc_18_december_bands}); ii) when we plot the flux ratios in each band as a function of the total flux (see Fig.\,\ref{fig:XMMlc_18_ratios} in Appendix\,\ref{appendix:XMM_lightcurves} and Fig.\,\ref{fig:NICER_HR} in Appendix\,\,\ref{appendix:HRnicer} for {\it XMM-Newton} and {\it NICER}, respectively) we see a very clear correlation in all bands, up to the 2--3\,keV bin for {\it XMM-Newton} (i.e., the limit of our EPIC/PN detections); iii) comparing the spectra of the highest and lowest flux intervals (top panels of Fig.\,\ref{fig:XMMspec18_twointervals}), and looking at their ratios (bottom panels of Fig.\,\ref{fig:XMMspec18_twointervals}) we see a clear hardening. This effect is due to both the increase of the temperature of the blackbody with the X-ray luminosity (Fig.\,\ref{fig:kT_luminosity}), and to the increase of the ratio between the flux from the power-law and blackbody component with the luminosity (see Fig.\,\ref{fig:bb_po_ratio} and \citealp{Ricci:2020}). The harder-when-brighter behaviour does not appear to be associated with variations of the outflowing material. Fitting the X-ray spectra with a blackbody component in a range that is not strongly affected by the ionized absorbers ($0.3-0.5$\,keV) we obtain a significant difference in the temperature of the blackbody of low and high-luminosity intervals. In particular for the June\,\,2018 {\it XMM-Newton} observation we obtain $kT=122_{-6}^{+7}\rm\,eV$ and $kT=88_{-6}^{+7}\rm\,eV$ for the brightest (interval\,2) and faintest (interval\,8) periods (see Fig.\,\ref{fig:XMMlc_18_bins}). Moreover, no clear difference is observed in the RGS spectra of two intervals with very different fluxes during the June\,\,2018 observation (Fig.\,\ref{fig:RGS_highlowflux_june18} in Appendix\,\ref{appendix:exposuresintervals}).

Unobscured AGN typically show a softer-when-brighter behaviour (e.g., \citealp{Sobolewska:2009am}) in the X-rays. This is likely related to the steeping of the power-law component for increasing luminosities (e.g., \citealp{Shemmer:2006dk}), which is also found in Galactic black hole binaries (e.g., \citealp{Wu:2008cd}). 
On the other hand, the temperature of the blackbody component, adopted to reproduce the soft excess, does not vary with the source luminosity (e.g., \citealp{Gierlinski:2004dt}).
At low Eddington ratios ($\lambda_{\rm Edd}\lesssim 10^{-2}$) it has been shown that AGN can present a harder-when-brighter behaviour
(e.g., \citealp{Gu:2009rt,Constantin:2009jr,Younes:2011zr,Emmanoulopoulos:2012qy,Connolly:2016kx}), which is however entirely related to the power-law component, and that results in a negative correlation between $\Gamma$ and $\lambda_{\rm Edd}$. Such a trend has also been observed in black hole binaries in the low/hard state (e.g., \citealp{Kalemci:2004qf,Yamaoka:2005pd,Yuan:2007lq}), and it has been argued that such behaviour might be related to the fact that accretion in these objects happens through an advection-dominated accretion flow (ADAF; \citealp{Ichimaru:1977wt,Ho:2008qf,Yuan:2014hb}).
Such a trend has been observed also in blazars (e.g., \citealp{Krawczynski:2004jo}), where shocks and Synchrotron Self-Compton processes in the jet are thought to be responsible for the hardening of the X-ray radiation \citep{Ghisellini:2009uj}.
There are however important differences between the case of 1ES\,1927+654 and that of low Eddington ratio AGN, black hole binaries in the low/hard state and blazars, which suggest that the physical mechanisms responsible for the spectral variability might be very different. The spectral shape of 1ES\,1927+654 is in fact extremely soft, with a very strong blackbody component, while low Eddington ratio AGN and blazars have significantly harder continua. Moreover, the power-law component does not show any spectral variation with luminosity, contrary to what is found in ADAF and strongly jetted AGN.

\subsection{The origin of the blackbody component}

In \citet{Ricci:2020} we discuss how the event that created the broad lines, and drastically transformed the X-ray properties of 1ES\,1927+654, might be related to a TDE in an AGN (e.g., \citealp{Merloni:2015on,Blanchard:2017kb,Liu:2020iz}). The blackbody component, observed in all X-ray spectra of 1ES\,1927+654 after the optical event, could be produced in an accretion disk. However, assuming $M_{\rm BH}\sim 10^6\rm M_{\odot}$, one would expect the maximum temperature to be $\sim 35$\,eV ($\sim 100$\,eV) at the highest luminosity for a non-rotating (maximally rotating) black hole, assuming $\kappa_{\rm X}\simeq 1$ (see Eq.\,\ref{eq:tempmax} in Appendix\,\ref{sect:xmmobs_spec_june_diskbb}). This is considerably lower than that recovered from our fitting ($kT\sim 200$\,eV or $kT\sim 235$\,eV, for a blackbody or a disk blackbody, respectively). As discussed in \S\ref{sect:Xrayproperties1819}, it is likely that $\kappa_{\rm X}\geq 1$, which could lead to a disk temperature consistent with our observations (particularly if $\kappa_{\rm X}\gtrsim 15$). Moreover, the colour temperature correction for  $M_{\rm BH}\sim 10^6\rm M_{\odot}$ could be $\gtrsim 2$ \citep{Done:2012uk}, which would also contribute to increasing the observed blackbody temperature.

The blackbody radius varies between $\sim 4\times 10^{9}\rm\,cm$ (at $\sim 10^{40}\rm\,erg\,s^{-1}$) and $\sim 7\times 10^{10}\rm\,cm$ (at $\sim 10^{44}\rm\,erg\,s^{-1}$), assuming $M_{\rm BH}\sim 10^6\rm M_{\odot}$ and $\kappa_{\rm X}\sim 1$. This is lower than the gravitational radius of the source, $R_{\rm g}=GM_{\rm BH}/c^2\sim1.5\times10^{11}\rm\,cm$ (assuming $M_{\rm BH}\sim 10^6\rm M_{\odot}$), which might suggest that the blackbody emission cannot be easily explained by a standard accretion disk, unless a high inclination angle with respect to the normal to the disk is assumed. Even if the black hole mass was considerably lower (e.g., $\sim 10^5\rm M_{\odot}$), the gravitational radius ($\sim1.5\times10^{10}\rm\,cm$) would still be larger than the blackbody radius at the lowest luminosities observed here. Only a black hole mass of $\lesssim 5\times 10^{4} M_{\odot}$ would be consistent with the observations. In fact, while the blackbody radius would be larger than the gravitational radius at the highest luminosities for $\kappa_{\rm X}\gtrsim 5$, at the lowest luminosities one would need a rather extreme value of the X-ray bolometric correction ($\kappa_{\rm X}\gtrsim 1200$). However, at this time (i.e. $\sim 200$\,days after the event) the optical emission was remarkably stronger than in the X-rays \citep{Trakhtenbrot:2019qy,Ricci:2020}, and the $5100\AA$ luminosity was $\sim 10^{43}\rm\,erg\,s^{-1}$, which could correspond to a bolometric luminosity of $L\sim 10^{44}\rm\,\,erg\,s^{-1}$ (e.g., using the bolometric corrections of \citealp{Kaspi:2000zy}). This would lead to a blackbody radius of $\sim 4\times 10^{11}\rm\,cm>R_{\rm g}$, resolving this discrepancy. Therefore, we conclude that the soft X-ray emission could be associated to an accretion disk, provided that the X-ray bolometric corrections are high enough. 

1ES\,1927+654 shows very strong spectral variability during our campaign, with the temperature of the blackbody component increasing with the luminosity (Fig.\,\ref{fig:kT_luminosity}). This behaviour is very different from what is typically observed in TDEs, which show little evolution in the temperature of the blackbody component. Interestingly, a similar behaviour has been observed also in the narrow-line Seyfert\,1 IRAS\,13224$-$3809 \citep{Chiang:2015mm}, and it was explained as being possibly related to the heating of the inner disc by the X-ray corona and by the effect of light bending, which allows radiation emitted and/or reflected by the disk to illuminate its surface.
If the event were due to a TDE in a pre-existing AGN, part of the X-ray radiation could be produced in the shocks between the debris and the disk, which would produce the observed harder-when-brighter behaviour. The latter scenario will be tested in details in a forthcoming publication (Masterson et al. in prep.). Interestingly, such behaviour has been recently observed also in a candidate TDE with peculiar properties \citep{Lin:2017pi}, as well as in recent observations of supersoft X-ray AGN (e.g., \citealp{Miniutti:2019rg,Shu:2017qy,Giustini:2020qg}), objects which could also be associated to TDEs (e.g., \citealp{Shu:2018bh,King:2020dm}).

\subsection{1ES\,1927+654 and true type-2 AGN}\label{sect:disc_truetype2}

Before becoming a type-1 AGN, 1ES\,1927+654 was considered to be one of the clearest cases of true type-2 AGN (e.g., \citealp{Tran:2011kk}), i.e. unobscured sources lacking the BLR (e.g., \citealp{Tran:2001ox,Panessa:2002cv}). A large fraction of these objects are found to accrete at $\lambda_{\rm Edd}\lesssim 10^{-2}$ (e.g., \citealp{Bianchi:2012bq,Panessa:2009cz}). It has been argued that this would be in agreement with models that predict the disappearance of the BLR at low Eddington ratios (e.g., \citealp{Nicastro:2000fa}), such as the disc-wind scenario (e.g., \citealp{Elitzur:2009pl,Elitzur:2014rp}). In some cases, however, these sources have been found to have high accretion rates (e.g., \citealp{Ho:2012sp}), which led \cite{Elitzur:2016lb} to argue that the dependence on black hole mass might relax the limits on the accretion rate in the disk-wind model. A number of obscured AGN were also thought to lack a BLR, as suggested by the non-detection of polarized broad lines (e.g., \citealp{Tran:2003nj}). \cite{Marinucci:2012ud} found that most these objects have low Eddington ratios ($\lambda_{\rm Edd}<10^{-2}$), possibly also in agreement with the disk-wind scenario. However, thanks to VLT/FORS2 optical spectropolarimetry observations, \citet{Ramos-Almeida:2016kf} were able to find polarized broad lines in several of these objects, questioning whether the lack of detection of broad lines was only related to dilution due to stellar emission in the host galaxy. Recently, using {\it Hubble Space Telescope} ({\it HST}) spectroscopy, \citet{Bianchi:2019uf} found that the archetypal true type-2 AGN, NGC\,3147, actually has very broad lines, which show an asymmetric relativistic profile. The fact that the BLR was not detected before in this objects is due to: i) the very broadened emission profile, caused by the high mass ($2.5\times10^{8}\rm\,M_{\odot}$) of the black hole and by the small distance ($\sim 80\rm\,r_{\rm g}$) between the SMBH and the BLR (due to the low luminosity of the object, e.g. \citealp{Kaspi:2005tu}); ii) the intrinsic weakness of the optical emission with respect to the host galaxy. NGC\,3147 has in fact a 2--10\,keV luminosity of $3.3\pm0.1\times 10^{41}\rm\,erg\,s^{-1}$ ($\lambda_{\rm Edd}\simeq 10^{-4}$, \citealp{Bianchi:2017ya}). 

These results seem to suggest that most true type-2 AGN are just type-1 AGN in which the broad lines are too broad and/or too faint to be detected. This might also be important for changing-look AGN (e.g., \citealp{Dexter:2019yo,Frederick:2019jb,Yan:2019wx}). So far no object has been found to show changes in both the level of obscuration and in the optical type. Therefore, at the beginning/end of the changing-look event, a fraction of sources would be classified as unobscured type-2 AGN. In 2011, before the outburst, 1ES\,1927+654 had a 2--10\,keV luminosity an order of magnitude larger than NGC\,3147 ($3.3\times10^{42}\rm\,erg\,s^{-1}$, \citealp{Gallo:2013hq}). Using the relation between the radius of the BLR and the X-ray luminosity inferred by \citet{Kaspi:2005tu}, we find that the BLR would lie at $\sim 5$\,light days from the SMBH, which corresponds to $\sim 4,500\rm\,r_{\rm g}$ for $M_{\rm BH}=1.9\times10^7 M_{\odot}$ ($\sim 88,000\rm\,r_{\rm g}$ for $M_{\rm BH}=10^6 M_{\odot}$), a factor of $\sim 60$ ($\sim 1200$) larger than NGC\,3147. One would therefore not expect to see extremely broad lines in this object. From the relation between the luminosity of the broad H$\alpha$ line ($L_{\rm bH\alpha}$) and that in the X-ray band \citep{Stern:2012tf}, we would expect $L_{\rm bH\alpha}\simeq 10^{42}\rm\,erg\,s^{-1}$. As shown by \citeauthor{Stern:2012tf} (\citeyear{Stern:2012tf}; see their Fig.\,14) this is considerably higher than the upper limit on the broad H$\alpha$ line ($L_{\rm bH\alpha}\lesssim 3\times10^{39}\rm\,erg\,s^{-1}$) inferred by \citet{Tran:2011kk} for 1ES\,1927+654. This, together with the lack of Fe\,K$\alpha$ emission and of a Compton hump, suggests that before the event 1ES\,1927+654 might have had accretion properties rather different from those of typical AGN.

\subsection{1ES\,1927+654 and changing-look AGN}\label{sect:disc_changinglookAGN}

Over the past years several dozens of changing-look AGN have been discovered at different wavelengths (e.g., \citealp{Yang:2018qt,MacLeod:2019pm,Sheng:2020fc,Hon:2020uq}). A few of these objects have been followed up and studied in detail in the X-ray band. 1ES\,1927+654 shows several extreme characteristics that have not been observed in other changing-look AGN. In particular, 1ES\,1927+654 is the only changing-look AGN that has been found to undergo a complete transformation of its X-ray spectral properties, and to show X-ray variability of over four orders of magnitude on timescales of months. Moreover, another peculiarity of 1ES\,1927+654 is that the broad optical lines were found to disappear a few months after they were created \citep{Trakhtenbrot:2019qy}.

Several changing-look AGN show little X-ray spectral variability, and characteristics consistent with those of typical AGN. NGC\,1566 was found to transition from Sy\,1.8-19 to Sy\,1.2 \citep{Oknyansky:2019qy}, following an increase by a factor of $\sim 40-70$ in the X-rays. After the changing-look event, at the peak of the X-ray luminosity, its overall X-ray spectral properties however are not remarkably different from those of nearby AGN \citep{Parker:2019ua}.
HE\,1136$-$2304 brightened in the X-rays by a factor of $\sim 30$, which was accompanied by an increase in the flux of the Balmer lines, causing a transition from a Sy\,1.9 to a Sy\,1.5 \citep{Parker:2016fr}. The X-ray properties of this source are also not very different from those of typical AGN, with the source showing a well defined power-law component. \cite{Wang:2020pt} found that the X-ray luminosity decreases (increases) as UGC\,3223 transitions from Sy\,1.5 to Sy\,2 (Sy\,2 to Sy\,1.8), but did not find evidence of strong X-ray spectral variations.

On the other hand, \cite{Noda:2018le} showed that the appearance/disappearance of broad optical lines in Mrk\,1018 \citep{Husemann:2016qe,McElroy:2016qy,Krumpe:2017ul} could be directly associated to the presence/absence of a strong soft excess in the X-rays, which extends to the UV and is responsible for the ionizing photons responsible for the emission lines. \citeauthor{Noda:2018le} (\citeyear{Noda:2018le}; see also \citealp{Ruan:2019wc,Ruan:2019rl}) discussed that these variations are similar to the soft-to-hard state transition in black hole binaries, where the inner disc is replaced by an advection dominated accretion flow as the Eddington ratio decreases. This behaviour is remarkably different from what was observed in 1ES\,1927+654, which showed an extremely strong soft X-ray component even after the disappearance of the broad lines.  

The properties of 1ES\,197+654 seem to suggest that the changing-look event is very different from those observed in the aforementioned sources, and that it could be associated to a TDE \citep{Ricci:2020}. An alternative explanation for the origin of the event in 1ES\,1927+654 has been recently proposed by \citet{Scepi:2021lc}, who argued that the inversion of magnetic flux in a magnetically arrested disk could explain its X-ray and optical light-curves, as well as the observed timescales.

\section{Summary and conclusions}\label{sect:conclusion}

We reported here the results obtained by our 2018/2019 X-ray monitoring campaign of the changing-look AGN 1ES\,1927+654. The source, which was previously classified as an unobscured type\,2 AGN ($N_{\rm H}\simeq 10^{20}\rm\,cm^{-2}$), was found to have significantly increased its optical flux in March\,\,2018. Pre-discovery detections from ATLAS showed that the event started on December 23 2017. Our optical spectroscopic follow-up campaign showed, for the first time, evidence of strong broad Balmer emission lines in this source, $\sim 1-3$ months after the optical flux rise \citep{Trakhtenbrot:2019qy}.  Our X-ray monitoring campaign includes 265 {\it NICER} (for a total of 678\,ks) and 14 {\it Swift}/XRT (26\,ks) observations, as well as three simultaneous {\it XMM-Newton}/{\it NuSTAR} (158/169\,ks) exposures.
\smallskip

In the X-rays, 1ES\,1927+654 shows a behaviour unlike any other AGN. The main characteristics of the source are the following.
\begin{itemize}
\item After the optical/UV outburst the source is found to show a super soft X-ray continuum, with the emission being almost entirely due to a blackbody component ($kT\simeq 100$\,eV). The power-law component, ubiquitously found in AGN, and which was present in previous X-ray observations, had almost completely disappeared (top panel of Fig.\,\ref{fig:xmmobseeufs} and left panel of Fig.\,\ref{fig:NICERspec}).
\item When the flux of the source increases a steep power law component re-appears. The ratio between the power-law and blackbody flux increases with the luminosity, with the power-law dominating at $\log (L_{0.3-2}/\rm erg\,s^{-1})\gtrsim 43.7$ (Fig.\,\ref{fig:bb_po_ratio} and \citealp{Ricci:2020}). 
\item The source shows extremely strong X-ray variability, with an amplitude of two orders of magnitude on timescales of a few hours (Fig.\,\ref{fig:XMMlc_18_bins}), and of up to four orders of magnitudes on timescales of a few months (Fig.\,1 of \citealp{Ricci:2020}). 
\item The variability spectrum inferred from {\it XMM-Newton} observations (Fig.\,\ref{fig:XMMspecVarspec18}) shows that the peak of the variability moves to higher energies when the X-ray luminosity increases. The X-ray variability is found to be completely disconnected from the UV variability (\S\ref{sect:UV}).
\item A simple power-law component cannot reproduce the curvature above a few keV, and an extremely low energy cutoff ($E_{\rm cut}\sim 3$\,keV) is needed when {\it NuSTAR} data are included (see panels B and C of Fig.\,\ref{fig:XMMspecDec18May19}).
\item The spectral shape shows a clear relation with the luminosity, becoming harder when brighter. This is due to: i) relative increase of the flux from the power-law component (bottom panels of Fig.\,\ref{fig:bb_po_ratio}); ii) a higher temperature of the blackbody component (see bottom panels of Fig.\,\ref{fig:XMMspec18_twointervals}). We find that $kT$ shows a tight dependence on the luminosity in all our observations (Fig.\,\ref{fig:kT_luminosity}). The harder-when-brighter behaviour is also clearly observed in a model-independent fashion when looking at the hardness ratios (Fig.\,\ref{fig:XMMlc_18_ratios} in Appendix\,\ref{appendix:XMM_lightcurves} and Fig.\,\ref{fig:NICER_HR} in Appendix\,\ref{appendix:HRnicer} for {\it XMM-Newton} and {\it NICER}, respectively).
\item Two prominent features at $\sim 1$\,keV and $\sim 1.8$\,keV were found in the spectra (see panels B of Figs.\,\ref{fig:XMMspecJune18} and \ref{fig:XMMspecDec18May19}), and could be reproduced by using broad Gaussian lines. The broad feature at $\sim 1$\,keV is found in all observations, with the exception of the ones in which the source was most luminous (central panels of Fig.\,\ref{fig:XMMspec18_twointervals}).
\end{itemize}

The unique characteristics of 1ES\,1927+654 in the X-ray band suggests that it belongs to a new type of changing-look AGN, which underwent some catastrophic event that restructured its accretion flow. The lack of X-ray flux above 2\,keV right after the detection of the optical/UV event implies that the event destroyed the X-ray corona, and possibly also the innermost regions of the accretion flow. The corona appears to be in the process of being reformed as the X-ray luminosity of the source increases. In \citet{Ricci:2020} we speculated that the characteristics of 1ES\,1927+654 might match what would be expected from the interaction between a tidally disrupted star with an accretion disk around a SMBH, which would empty the innermost regions of the accretion flow \citep{Chan:2019dx,Chan:2020hu,Chan:2021xa}. Future observational studies and simulations will be able to better understand the link between TDEs and flaring AGN such as 1ES\,1927+654.

\smallskip

In a forthcoming paper (Masterson et al. in prep.) we will test alternative X-ray spectral models, with the goal of understanding the origin of the broad features at $\sim 1$\,keV and $\sim 1.8$\,keV. Soft X-ray surveys, such as the one that is currently being carried out by {\it eROSITA} \citep{Merloni:2012be}, or those that will be performed by the future mission {\it Einstein Probe} \citep{Yuan:2015vg} could potentially detect several objects like 1ES\,1927+654 in the next few years.

\acknowledgments

We acknowledge {\it XMM-Newton}, {\it NuSTAR} and {\it Swift} for the DDT observations they kindly guaranteed us. We thank the referee for their report, which helped us improve the paper. LH acknowledges financial support from the National Key R\&D Program of China grant No. 2016YFA0400702, and the National Science Foundation of China grants No. 11721303 and 11991052. CR acknowledges support from the Fondecyt Iniciacion grant 11190831. BT acknowledges support from the Israel Science Foundation (grant No. 1849/19). IA is a CIFAR Azrieli Global Scholar in the Gravity and the Extreme Universe Program and acknowledges support from that program, from the European Research Council (ERC) under the European Union's Horizon 2020 research and innovation program (grant agreement number 852097), from the Israel Science Foundation (grant number 2752/19), from the United States - Israel Binational Science Foundation (BSF), and from the Israeli Council for Higher Education Alon Fellowship. PG acknowledges support from STFC and a UGC-UKIERI Thematic Partnership. DA acknowledges support from the Royal Society. CHC is partially supported by ERC advanced grant ``TReX". PK acknowledges support from the Science and Technology Facilities Council. CR acknowledges T. Kallman, F. Bauer, C.S. Chang and the Santiago AGN community for useful discussion. Based on observations with the NASA/ESA/CSA Hubble Space Telescope obtained [from the Data Archive] at the Space Telescope Science Institute, which is operated by the Association of Universities for Research in Astronomy, Incorporated, under NASA contract NAS5-26555. Support for Cycle 25 Program GO-15604 was provided through a grant from the STScI under NASA contract NAS5-26555.

{\it Facilities:} \facility{NICER}, \facility{NuSTAR}, \facility{Swift}, \facility{XMM-Newton}

\bibliographystyle{apj} 
\bibliography{1ES1927_xray.bib}

\newpage 

\appendix

\section{Observation log}\label{sect:obslog}

In Table\,\,\ref{tab:obslog} we list all the X-ray observations carried out during our follow-up campaign of 1ES\,1927+654 between 17 May 2018 and 5 August 2019. For completeness we also include the {\it XMM-Newton} observation carried out in May 2011 (PI: L. Gallo).

\tabletypesize{\normalsize}
\LongTables
\begin{deluxetable*}{lcc} 
\tablecaption{X-ray observation log.\label{tab:obslog}}
\tablewidth{0pt}
\tablehead{
\colhead{(1)} & \colhead{(2)} & \colhead{(3)} \\
 \noalign{\smallskip}
\colhead{Date and time} & \colhead{Observation ID} & \colhead{Exposure [s]} 
}
\startdata
\noalign{\bigskip}
\multicolumn3c{{\large \textbf{ \textit{XMM-Newton}}}}\\
\noalign{\smallskip}
\hline \noalign{\smallskip}
2011-05-20 06:08:25	 & 0671860201 & 28600 \\
\noalign{\smallskip}
2018-06-05 17:42:09 & 0830191101	& 46400	\\
\noalign{\smallskip}
2018-12-12 06:10:19 & 0831790301	& 59300	\\
\noalign{\smallskip}
2019-05-06 21:56:01 & 0843270101		& 52000	\\
\noalign{\bigskip}
\hline \noalign{\bigskip}
\multicolumn3c{{\large \textbf{ \textit{NuSTAR}}}}\\
\noalign{\smallskip}
\hline \noalign{\smallskip}
2018-06-05 13:06:09	& 90401625002 &	45877 \\
\noalign{\smallskip}
2018-12-12 00:01:09	& 90401641002 &	64748 \\
\noalign{\smallskip}
2019-05-06 20:06:09	& 90501618002 &	58249 \\
\noalign{\bigskip}
\hline 
\noalign{\bigskip}
\multicolumn3c{{\large \textbf{ \textit{Swift/XRT}}}}\\
\noalign{\smallskip}
\hline 
\noalign{\smallskip}
	2018-05-17 13:50:57&	00010682001	& 2179	\\
\noalign{\smallskip}
	2018-05-31 06:09:31&	00010682002	& 1770	\\
\noalign{\smallskip}
	2018-06-14 09:10:57&	00010682003	& 2114	\\
\noalign{\smallskip}
	2018-07-10 15:06:57&	00010682004	& 1592	\\
\noalign{\smallskip}
	2018-07-24 00:52:57	&	 00010682005 &	2288	\\
\noalign{\smallskip}
	2018-08-07 20:12:57	&	00010682006	& 2159	\\
\noalign{\smallskip}
	2018-08-23 11:19:57&	00010682007	& 1963	\\
\noalign{\smallskip}
	2018-10-03 08:46:57&	00010682008	& 1240	\\
\noalign{\smallskip}
	2018-10-19 14:09:57&	00010682009	& 960	\\
\noalign{\smallskip}
	2018-10-23 13:29:05&	00010682010	& 1585	\\
\noalign{\smallskip}
	2018-11-21 00:48:35&	00010682011	& 2161	\\
\noalign{\smallskip}
	2018-12-06 11:07:35&	00010682012	& 1554	\\
\noalign{\smallskip}
	2018-12-12 02:30:36&	00010682013	& 1976	\\
\noalign{\smallskip}
	2019-03-28 06:54:36	 &	00010682014	& 2113	\\
\noalign{\bigskip}
\hline \noalign{\bigskip}
\multicolumn3c{{\large \textbf{ \textit{NICER}}}}\\
\noalign{\smallskip}
\hline \noalign{\smallskip}
2018-05-22 19:21:20     &   	1200190101	&  2156  	\\
\noalign{\smallskip}
2018-05-23 14:07:53     &   	1200190102	&  8833  	\\
\noalign{\smallskip}
2018-05-24 00:41:20     &   	1200190103	&  14577  	\\
\noalign{\smallskip}
2018-05-24 23:48:00     &   	1200190104	&  5846  	\\
\noalign{\smallskip}
2018-05-26 14:23:40     &   	1200190105	&  2236  	\\
\noalign{\smallskip}
2018-05-26 23:41:35     &   	1200190106	&  5233	\\
\noalign{\smallskip}
2018-05-28 00:23:35     &   	1200190107	&  11270	\\	
\noalign{\smallskip}
2018-05-29 01:05:34     &   	1200190108	&  7694	\\
\noalign{\smallskip}
2018-05-31 08:38:00     &   	1200190109	&  1445  	\\
\noalign{\smallskip}
2018-06-01 04:42:00     &   	1200190110	&  604  	\\
\noalign{\smallskip}
2018-06-02 13:07:00     &   	1200190111	&  1838  	\\
\noalign{\smallskip}
2018-06-03 07:38:00     &   	1200190112	&  2101  	\\
\noalign{\smallskip}
2018-06-04 03:42:00     &   	1200190113	&  1058  	\\
\noalign{\smallskip}
2018-06-05 04:22:18     &   	1200190114	&  1249  	\\
\noalign{\smallskip}
2018-06-06 06:49:20     &   	1200190115	&  515  	\\
\noalign{\smallskip}
2018-06-07 01:31:52     &   	1200190116	&  1021  	\\
\noalign{\smallskip}
2018-06-09 21:11:00     &   	1200190117	&   9  	\\	
\noalign{\smallskip}
2018-06-10 04:52:49     &   	1200190118	&  62  	\\
\noalign{\smallskip}
2018-06-14 10:31:40     &   	1200190119	&  871  	\\
\noalign{\smallskip}
2018-06-15 12:46:21     &   	1200190120	&  440  	\\
\noalign{\smallskip}
2018-06-16 05:44:03     &   	1200190121	&  768  	\\
\noalign{\smallskip}
2018-06-17 00:14:25     &   	1200190122	&  1063  	\\
\noalign{\smallskip}
2018-06-18 08:55:20     &   	1200190123	&  24  	\\
\noalign{\smallskip}
2018-06-19 00:19:20     &   	1200190124	&  2703	 	\\
\noalign{\smallskip}
2018-06-20 05:25:40     &   	1200190125	&  1101	\\
\noalign{\smallskip}
2018-06-21 01:26:40     &   	1200190126	&  1105	\\
\noalign{\smallskip}
2018-06-22 02:00:40     &   	1200190127	&  2489	\\
\noalign{\smallskip}
2018-06-22 23:38:40     &   	1200190128	&  6883  	\\
\noalign{\smallskip}
2018-06-24 03:27:20     &   	1200190129	&  5778	\\
\noalign{\smallskip}
2018-06-25 10:21:20     &   	1200190130	&  2436	\\
\noalign{\smallskip}
2018-06-26 00:18:40     &   	1200190131	&  4766	\\
\noalign{\smallskip}
2018-06-27 01:04:16     &   	1200190132	&  4872	\\
\noalign{\smallskip}
2018-06-28 00:17:20     &   	1200190133	&  5037	\\
\noalign{\smallskip}
2018-06-29 02:32:00     &   	1200190134	&  1697  	\\
\noalign{\smallskip}
2018-06-30 00:16:00     &   	1200190135	&  736	\\
\noalign{\smallskip}
2018-07-01 21:03:20     &   	1200190136	&  434	\\
\noalign{\smallskip}
2018-07-04 21:27:40     &   	1200190137	&  974	\\
\noalign{\smallskip}
2018-07-05 17:40:41     &   	1200190138	&  324  	\\
\noalign{\smallskip}
2018-07-06 16:42:01     &   	1200190139	&  854  	\\
\noalign{\smallskip}
2018-07-07 12:51:20     &   	1200190140	&  1220	\\
\noalign{\smallskip}
2018-07-08 10:21:00     &   	1200190141	&  2550  	\\
\noalign{\smallskip}
2018-07-09 01:47:20     &   	1200190142	&  984  	\\
\noalign{\smallskip}
2018-07-11 04:33:00     &   	1200190143	&   1024  	\\
\noalign{\smallskip}
2018-07-13 18:18:00     &   	1200190144	&  772  	\\
\noalign{\smallskip}
2018-07-14 14:30:53     &   	1200190145	&  787  	 	\\
\noalign{\smallskip}
2018-07-16 15:55:00     &   	1200190146	&  1454  	\\
\noalign{\smallskip}
2018-07-17 04:11:40     &   	1200190147	&   2448  	\\
\noalign{\smallskip}
2018-07-18 00:20:00     &   	1200190148	&  294  	\\
\noalign{\smallskip}
2018-07-19 10:16:00     &   	1200190149	&  289  	\\
\noalign{\smallskip}
2018-07-21 03:55:40     &   	1200190150	&  966  	\\
\noalign{\smallskip}
2018-07-24 02:41:20     &   	1200190151	&  3466  	\\
\noalign{\smallskip}
2018-07-25 03:26:00     &   	1200190152	&  295  	\\
\noalign{\smallskip}
2018-07-26 07:12:00     &   	1200190153	&  358  	\\
\noalign{\smallskip}
2018-07-27 03:18:40     &   	1200190154	&  304  	\\
\noalign{\smallskip}
2018-08-03 16:14:40     &   	1200190155	&  573  	\\
\noalign{\smallskip}
2018-08-06 04:28:40     &   	1200190156	&  2050	\\
\noalign{\smallskip}
2018-08-14 22:09:00     &   	1200190157	&  166  	\\
\noalign{\smallskip}
2018-08-15 15:04:20     &   	1200190158	&  2372  	\\
\noalign{\smallskip}
2018-08-17 07:24:59     &   	1200190159	&  5573  	\\
\noalign{\smallskip}
2018-08-18 00:17:40     &   	1200190160	&  1800  	\\
\noalign{\smallskip}
2018-08-19 08:47:19     &   	1200190161	&  1442  	\\
\noalign{\smallskip}
2018-08-20 00:13:19     &   	1200190162	&  3314  	\\
\noalign{\smallskip}
2018-09-07 14:15:40     &   	1200190163	&  731  	\\
\noalign{\smallskip}
2018-09-08 04:10:11     &   	1200190164	&   1753  	\\
\noalign{\smallskip}
2018-09-09 15:40:51     &   	1200190165	&  191  	\\
\noalign{\smallskip}
2018-09-12 12:48:01     &   	1200190166	&  1858  	 	\\
\noalign{\smallskip}
2018-09-14 09:33:59     &   	1200190167	&  8696  \\
\noalign{\smallskip}
2018-09-18 09:13:59     &   	1200190168	&  414  	\\
\noalign{\smallskip}
2018-09-23 00:26:40     &   	1200190169	&  1879  	\\
\noalign{\smallskip}
2018-09-24 02:41:20     &   	1200190170	&  3574  	\\
\noalign{\smallskip}
2018-09-26 19:31:38     &   	1200190171	&  667  	\\
\noalign{\smallskip}
2018-09-28 19:26:00     &   	1200190172	&  152  	\\
\noalign{\smallskip}
2018-09-30 02:20:20     &   	1200190173	&  1256  	\\
\noalign{\smallskip}
2018-09-30 23:54:40     &   	1200190174	&  1733  	\\
\noalign{\smallskip}
2018-10-02 06:51:00     &   	1200190175	&  2948  	\\
\noalign{\smallskip}
2018-10-03 07:32:40     &   	1200190176	&  479  	\\
\noalign{\smallskip}
2018-10-04 05:06:57     &   	1200190177	&  399  	\\
\noalign{\smallskip}
2018-10-05 12:01:58     &   	1200190178	&  455  	\\
\noalign{\smallskip}
2018-10-06 17:18:20     &   	1200190179	&  922  	\\
\noalign{\smallskip}
2018-10-08 00:09:58     &   	1200190180	&  1560  	\\
\noalign{\smallskip}
2018-10-10 23:11:00     &   	1200190181	&   173  	\\
\noalign{\smallskip}
2018-10-11 16:09:40     &   	1200190182	&   5  	\\	
\noalign{\smallskip}
2018-10-12 23:03:25     &   	1200190183	&  744	\\
\noalign{\smallskip}
2018-10-13 00:35:32     &   	1200190184	&  8265  	\\
\noalign{\smallskip}
2018-10-15 00:26:32     &   	1200190185	&  3948  	\\
\noalign{\smallskip}
2018-10-16 01:08:51     &   	1200190186	&  3578	\\
\noalign{\smallskip}
2018-10-17 01:50:51     &   	1200190187	&  2142  	 	\\
\noalign{\smallskip}
2018-10-18 21:18:21     &   	1200190188	&  1362  	\\
\noalign{\smallskip}
2018-10-19 18:42:42     &   	1200190189	&  2079	\\
\noalign{\smallskip}
2018-10-20 14:47:00     &   	1200190190	&  1701	\\
\noalign{\smallskip}
2018-10-21 12:22:26     &   	1200190191	&  3956	\\
\noalign{\smallskip}
2018-10-22 19:16:16     &   	1200190192	&  1889  	\\
\noalign{\smallskip}
2018-10-23 18:25:35     &   	1200190193	&  1948  	\\
\noalign{\smallskip}
2018-10-24 00:35:54     &   	1200190194	&  1644  	\\
\noalign{\smallskip}
2018-10-27 10:27:20     &   	1200190195	&  2356  	\\
\noalign{\smallskip}
2018-10-28 00:23:40     &   	1200190196	&  3993  	\\
\noalign{\smallskip}
2018-10-29 02:28:00     &   	1200190197	&  4834  	\\
\noalign{\smallskip}
2018-10-30 00:05:40     &   	1200190198	&  3064  	\\
\noalign{\smallskip}
2018-10-31 08:42:59     &   	1200190199	&  2243  	\\
\noalign{\smallskip}
2018-11-01 03:04:59     &   	1200190201	&   699  	\\
\noalign{\smallskip}
2018-11-02 00:55:40     &   	1200190202	&   38  	\\
\noalign{\smallskip}
2018-11-06 15:38:37     &   	1200190203	&   438  	\\
\noalign{\smallskip}
2018-11-07 07:02:47     &   	1200190204	&   1503  	\\
\noalign{\smallskip}
2018-11-08 10:48:11	& 1200190205	& 2349 \\
\noalign{\smallskip}
2018-11-09 09:57:17	& 1200190206	& 2460 \\
\noalign{\smallskip}
2018-11-10 20:06:00	& 1200190207	& 1324 \\
\noalign{\smallskip}
2018-11-11 11:19:05	& 1200190208	& 1154 \\
\noalign{\smallskip}
2018-11-12 01:12:08	& 1200190209	& 9771 \\
\noalign{\smallskip}
2018-11-13 08:04:14	& 1200190210	& 3233 \\
\noalign{\smallskip}
2018-11-14 10:20:59	& 1200190211	& 4223 \\
\noalign{\smallskip}
2018-11-15 03:19:25	& 1200190212	& 8357 \\
\noalign{\smallskip}
2018-11-16 04:00:25	& 1200190213	& 4411 \\
\noalign{\smallskip}
2018-11-17 12:21:00	& 1200190214	& 2128 \\
\noalign{\smallskip}
2018-11-18 03:49:20	& 1200190215	& 3487 \\
\noalign{\smallskip}
2018-11-19 12:19:40	& 1200190216	& 1902 \\
\noalign{\smallskip}
2018-11-20 03:46:00	& 1200190217	& 3378 \\
\noalign{\smallskip}
2018-11-21 12:10:00	& 1200190218	& 1947 \\
\noalign{\smallskip}
2018-11-22 00:32:20	& 1200190219	& 3311 \\
\noalign{\smallskip}
2018-11-23 01:06:59	& 1200190220	& 4171 \\
\noalign{\smallskip}
2018-11-24 01:52:20	& 1200190221	& 3984 \\
\noalign{\smallskip}
2018-11-25 14:51:16	& 1200190222	& 2143 \\
\noalign{\smallskip}
2018-11-26 06:17:16	& 1200190223	& 2162 \\
\noalign{\smallskip}
2018-11-27 06:58:57	& 1200190224	& 2180 \\
\noalign{\smallskip}
2018-11-28 01:31:00	& 1200190225	& 4432 \\
\noalign{\smallskip}
2018-11-29 08:30:40	& 1200190226	& 1920 \\
\noalign{\smallskip}
2018-11-29 23:49:22	& 1200190227	& 4153 \\
\noalign{\smallskip}
2018-12-01 15:57:14	& 1200190228	& 1009 \\
\noalign{\smallskip}
2018-12-02 18:08:40	& 1200190229	& 1290 \\
\noalign{\smallskip}
2018-12-04 04:33:40	& 1200190230	& 7625 \\
\noalign{\smallskip}
2018-12-05 11:01:31	& 1200190231	& 2274 \\
\noalign{\smallskip}
2018-12-06 02:28:19	& 1200190232	& 4559 \\
\noalign{\smallskip}
2018-12-07 09:20:40	& 1200190233	& 1136 \\
\noalign{\smallskip}
2018-12-08 03:52:20	& 1200190234	& 1145 \\
\noalign{\smallskip}
2018-12-09 04:31:20	& 1200190235	& 2288 \\
\noalign{\smallskip}
2018-12-10 06:46:40	& 1200190236	& 2289 \\
\noalign{\smallskip}
2018-12-11 04:23:40	& 1200190237	& 2903 \\
\noalign{\smallskip}
2018-12-12 17:50:00	& 1200190238	& 7212 \\
\noalign{\smallskip}
2018-12-18 19:00:00	& 1200190239	& 3502 \\
\noalign{\smallskip}
2018-12-21 19:15:40	& 1200190240	& 3751		\\
\noalign{\smallskip}
2018-12-22 19:51:59	& 1200190241	& 3090			\\
\noalign{\smallskip}
2018-12-23 19:16:40	& 1200190242	& 2144			\\
\noalign{\smallskip}
2018-12-23 23:55:00	& 1200190243	& 1168		\\	
\noalign{\smallskip}
2018-12-25 00:17:00	& 1200190244	& 6414		\\
\noalign{\smallskip}
2018-12-27 00:18:40	& 1200190245	& 3965			\\
\noalign{\smallskip}
2018-12-28 01:02:00	& 1200190246	& 1275			\\
\noalign{\smallskip}
2018-12-29 06:47:15	& 1200190247	& 2606		\\	
\noalign{\smallskip}
2018-12-31 00:11:40	& 1200190248	& 3491		\\
\noalign{\smallskip}
2019-01-04 16:36:10	& 1200190249	& 1513			\\
\noalign{\smallskip}
2019-01-05 15:43:12	& 1200190250	& 2100			\\
\noalign{\smallskip}
2019-01-06 07:11:31	& 1200190251	& 441 		\\	
\noalign{\smallskip}
2019-01-08 17:48:38	& 1200190252	& 1963		\\
\noalign{\smallskip}
2019-01-09 20:02:59	& 1200190253	& 1056			\\
\noalign{\smallskip}
2019-01-10 14:39:20	& 1200190254	& 2058			\\
\noalign{\smallskip}
2019-01-12 08:15:31	& 1200190255	& 1265		\\	
\noalign{\smallskip}
2019-01-13 01:13:10	& 1200190256	& 1337		\\
\noalign{\smallskip}
2019-01-14 03:26:53	& 1200190257	& 1488			\\
\noalign{\smallskip}
2019-01-15 08:46:08	& 1200190258	& 1849			\\
\noalign{\smallskip}
2019-01-16 17:11:08	& 1200190259	& 1987		\\	
\noalign{\smallskip}
2019-01-17 08:38:25	& 1200190260	& 1836		\\
\noalign{\smallskip}
2019-01-18 10:54:21	& 1200190261	& 2039		\\	
\noalign{\smallskip}
2019-01-19 05:25:02	& 1200190262	& 2049		\\	
\noalign{\smallskip}
2019-01-20 07:38:47	& 1200190263	& 1932	\\		
\noalign{\smallskip}
2019-01-21 23:43:37	& 1200190264	& 1757	\\	
\noalign{\smallskip}
2019-01-23 12:50:06	& 1200190265	& 1268		\\	
\noalign{\smallskip}
2019-01-24 19:57:21	& 1200190266	& 1115		\\	
\noalign{\smallskip}
2019-01-30 20:53:40	& 1200190267	& 299 	\\	 	
\noalign{\smallskip}
2019-01-30 23:59:00	& 1200190268	& 945 	\\ 	
\noalign{\smallskip}
2019-02-01 00:32:20	& 1200190269	& 2858		\\
\noalign{\smallskip}
2019-02-02 04:39:42	& 1200190270	& 1139		\\
\noalign{\smallskip}
2019-02-03 12:54:40	& 1200190271	& 3201	\\	
\noalign{\smallskip}
2019-02-04 01:14:20	& 1200190272	& 6509	\\
\noalign{\smallskip}
2019-02-05 00:21:00& 1200190273 	& 5881	\\
\noalign{\smallskip}
2019-02-06 00:57:56	& 1200190274	& 5198	\\	
\noalign{\smallskip}
2019-02-07 10:55:00	& 1200190275	& 6386 		\\
\noalign{\smallskip}
2019-02-08 17:50:00	& 1200190276	& 3867 		\\
\noalign{\smallskip}
2019-02-09 03:15:59	& 1200190277	& 19523 		\\
\noalign{\smallskip}
2019-02-10 00:49:00	& 1200190278	& 7586 		\\
\noalign{\smallskip}
2019-02-15 05:53:40	& 1200190279	& 1911 		\\
\noalign{\smallskip}
2019-02-16 03:28:22	& 1200190280	& 1749		\\
\noalign{\smallskip}
2019-02-19 22:45:00	& 1200190281	& 478 		\\
\noalign{\smallskip}
2019-02-20 03:19:00& 1200190282		& 2197 		\\
\noalign{\smallskip}
2019-02-22 17:14:20& 1200190283		& 4012 		\\
\noalign{\smallskip}
2019-02-23 10:14:00& 1200190284		& 6828		\\
\noalign{\smallskip}
2019-02-24 18:47:20	& 1200190285	& 1880 		\\
\noalign{\smallskip}
2019-02-25 00:57:40	& 1200190286	& 3092 		\\
\noalign{\smallskip}
2019-02-26 09:09:20	& 1200190287	& 890 		\\
\noalign{\smallskip}
2019-03-05 20:15:43	& 2200190201	& 4534 		\\
\noalign{\smallskip}
2019-03-06 19:25:00	& 2200190202	& 1715 		\\
\noalign{\smallskip}
2019-03-07 09:20:00	 & 2200190203	& 1538 		\\
\noalign{\smallskip}
2019-03-08 05:24:48 & 2200190204	& 2663 		\\
\noalign{\smallskip}
2019-03-10 00:41:40 & 2200190205	& 2168 		\\
\noalign{\smallskip}
2019-03-14 02:09:00 &	2200190206	& 1156 		\\
\noalign{\smallskip}
2019-03-16 21:58:46 &	2200190207	& 683 		\\
\noalign{\smallskip}
2019-03-17 02:37:06 &	2200190208	& 2860 		\\
\noalign{\smallskip}
2019-03-22 23:14:40 &	2200190209	& 850 		\\
\noalign{\smallskip}
2019-03-23 00:54:20 &	2200190210	& 573 		\\
\noalign{\smallskip}
2019-03-24 14:14:40 &	2200190211	& 436 		\\
\noalign{\smallskip}
2019-03-27 00:29:20 &	2200190212	& 1511 		\\
\noalign{\smallskip}
2019-03-28 18:22:40 &	2200190213	& 1663 		\\
\noalign{\smallskip}
2019-03-29 00:45:40 &	2200190214	& 945 		\\
\noalign{\smallskip}
2019-03-30 10:36:48 &	2200190215	& 1191 		\\
\noalign{\smallskip}
2019-04-03 05:27:27 &	2200190216	& 1960 		\\
\noalign{\smallskip}
2019-04-05 17:39:20 &	2200190217	& 373 		\\
\noalign{\smallskip}
2019-04-06 07:32:14 &	2200190218	& 27 		\\
\noalign{\smallskip}
2019-04-10 15:03:20 &	2200190219	& 326 		\\
\noalign{\smallskip}
2019-04-13 00:20:40 &	2200190220	& 3822 		\\
\noalign{\smallskip}
2019-04-19 05:58:25 &	2200190221	& 4177 		\\
\noalign{\smallskip}
2019-04-28 09:30:40 & 	2200190222	& 818 		\\
\noalign{\smallskip}
2019-04-29 22:44:00 &  	2200190223	& 1098 		\\
\noalign{\smallskip}
2019-04-30 00:17:40 &  	2200190224	& 1032 		\\
\noalign{\smallskip}
2019-05-02 08:01:00 &  	2200190225	& 809 		\\
\noalign{\smallskip}
2019-05-04 03:09:40 &	2200190226	& 1988 		\\
\noalign{\smallskip}
2019-05-06 19:56:40	& 2200190227	& 5238 \\  	
\noalign{\smallskip}
2019-05-07 00:34:40	& 2200190228	& 20505 \\  	
\noalign{\smallskip}
2019-05-10 18:18:40	& 2200190229	& 567 \\  	
\noalign{\smallskip}
2019-05-14 10:18:00	& 2200190230	& 8005 \\  	
\noalign{\smallskip}
2019-05-16 03:55:18	& 2200190231	& 2490 \\  	
\noalign{\smallskip}
2019-05-19 15:21:57	& 2200190232	& 1800 \\  	
\noalign{\smallskip}
2019-05-20 00:38:38	& 2200190233	& 2650 \\  	
\noalign{\smallskip}
2019-05-22 14:24:05	& 2200190234	& 1101 \\  	
\noalign{\smallskip}
2019-05-23 16:39:46	& 2200190235	& 800 \\  	
\noalign{\smallskip}
2019-05-24 06:32:36	& 2200190236	& 1904 \\  	
\noalign{\smallskip}
2019-05-25 02:41:03	& 2200190237	& 2052 \\  	
\noalign{\smallskip}
2019-05-27 04:08:21	& 2200190238	& 767 \\  	
\noalign{\smallskip}
2019-05-28 01:52:00	& 2200190239	& 4145 \\  	
\noalign{\smallskip}
2019-05-30 12:35:57	& 2200190240	& 1018 \\  	
\noalign{\smallskip}
2019-05-31 07:07:57	& 2200190241	& 434 \\  	
\noalign{\smallskip}
2019-06-01 21:47:38	& 2200190242	& 79 \\  	
\noalign{\smallskip}
2019-06-02 00:53:18	& 2200190243	& 1834 \\  	
\noalign{\smallskip}
2019-06-03 03:16:06	& 2200190244	& 56 \\  	
\noalign{\smallskip}
2019-06-04 08:39:00	& 2200190245	& 684 \\  	
\noalign{\smallskip}
2019-06-14 12:42:15	& 2200190246	& 1564 \\  	
\noalign{\smallskip}
2019-06-15 01:07:59	& 2200190247	& 7162 \\  	
\noalign{\smallskip}
2019-06-16 00:20:20	& 2200190248	& 7154 \\  	
\noalign{\smallskip}
2019-06-19 08:40:00	& 2200190249	& 2012 \\  	
\noalign{\smallskip}
2019-06-21 19:33:00	& 2200190250	& 1316 \\  	
\noalign{\smallskip}
2019-06-24 03:15:20	& 2200190251	& 736 \\  	
\noalign{\smallskip}
2019-06-25 07:17:40	& 2200190252	& 3811 \\  	
\noalign{\smallskip}
2019-06-28 21:52:23	& 2200190253	& 875 \\  	
\noalign{\smallskip}
2019-06-29 02:18:00	& 2200190254	& 4093 \\  	
\noalign{\smallskip}
2019-06-29 23:59:55	& 2200190255	& 3576 \\  	
\noalign{\smallskip}
2019-07-01 00:51:00	& 2200190256	& 983 \\  	
\noalign{\smallskip}
2019-07-01 23:50:00	& 2200190257	& 1172 \\  	
\noalign{\smallskip}
2019-07-03 06:50:20	& 2200190258	& 3098 \\  	
\noalign{\smallskip}
2019-07-03 23:50:20	& 2200190259	& 1798 \\  	
\noalign{\smallskip}
2019-07-05 08:29:40	& 2200190260	& 4144 \\  	
\noalign{\smallskip}
2019-07-05 23:51:20	& 2200190261	& 9888 \\  	
\noalign{\smallskip}
2019-07-07 12:57:40	& 2200190262	& 5823 \\  	
\noalign{\smallskip}
2019-07-08 07:32:00	& 2200190263	& 797 \\  	
\noalign{\smallskip}
2019-07-11 05:09:20	& 2200190264	& 2519 \\  	
\noalign{\smallskip}
2019-07-12 04:17:40	& 2200190265	& 3674 \\  	
\noalign{\smallskip}
2019-07-15 06:37:00	& 2200190266	& 913 \\  	
\noalign{\smallskip}
2019-07-19 13:49:04	& 2200190267	& 1052 \\  	
\noalign{\smallskip}
2019-07-20 05:44:15	& 2200190268	& 3049 \\  	
\noalign{\smallskip}
2019-07-21 04:55:14	& 2200190269	& 1642 \\    	
\noalign{\smallskip}
2019-07-22 03:43:25	&	2200190270		&1095  \\  
\noalign{\smallskip}
2019-07-22 23:49:26	&	2200190271		&1239  \\  	
\noalign{\smallskip}
2019-07-24 06:46:28		&2200190272		&1510  \\  
\noalign{\smallskip}
2019-07-26 20:38:59	&	2200190273		&4292  \\  	
\noalign{\smallskip}
2019-07-29 19:54:40	&	2200190274		&530  \\  
\noalign{\smallskip}
2019-08-01 12:45:48	&	2200190275		&2053  \\  	
\noalign{\smallskip}
2019-08-02 05:48:11	&	2200190276		&4681  \\  	
\noalign{\smallskip}
2019-08-03 00:19:20	&	2200190277		&2300  \\  
\noalign{\smallskip}
2019-08-04 22:46:06 &	2200190278		&674 	 \\  
\noalign{\smallskip}
2019-08-05 15:54:38	&	2200190279		&424 	
\tablecomments{Table \ref{tab:obslog} is published in its entirety in the electronic edition of the {\it Astrophysical Journal}.}
\end{deluxetable*}

\clearpage

\section{The {\it Swift}/XRT observations }\label{appendix:XRT_spectra}

The results of our spectral analysis of the 14 {\it Swift}/XRT observations of 1ES\,1927+654 are reported in Table\,\ref{tab:fitXRT}, while the spectra are illustrated in Figure\,\ref{fig:XRTobs}. In the following we describe the procedure adopted for the data reduction (\S\ref{sect:swift_datared}) and our spectral analysis (\S\ref{sec:xrtobs_spec}).

\subsection{XRT and UVOT data analysis}\label{sect:swift_datared}

A total of fourteen observations of 1ES\,1927+654 were performed in 2018 by the X-ray Telescope (XRT, \citealp{Burrows:2005vn}) and the Ultraviolet Optical Telescope (UVOT, \citealp{Roming:2005rt}) on board the {\it Neil Gehrels Swift Observatory} \citep{Gehrels:2004dq}. The typical exposures were of $\sim 2$\,ks. The data were retrieved from the High Energy Astrophysics Science Archive Research Center (HEASARC) {\it Swift} archive\footnote{https://heasarc.gsfc.nasa.gov/cgi-bin/W3Browse/swift.pl}.

For all the observations the XRT data were reduced using the \textsc{xrtpipeline v0.13.4}, which is part of the XRT Data Analysis Software within Heasoft\,v6.24, following the standard guidelines \citep{Evans:2009fk}. The UVOT data were instead reduced using the \textsc{UVOTimsum} and \textsc{UVOTsource} routines which are a part of the \textsc{FTools} package provided by HEASARC\footnote{http://heasarc.gsfc.nasa.gov/ftools/} \citep{Blackburn:1995mz}. The scripts were run via the Python \textsc{readswift} pipeline written by S. Valenti on a {\it Swift}-reduction Docker container created by C. McCully. We used an aperture radius of $5\arcsec$ both at the position of the event and at a nearby position with no obvious sources for sky subtraction. We did not remove host contamination, which will be part of a dedicated forthcoming publication (Li et al. in prep.).

\subsection{{\it Swift}/XRT spectral analysis}\label{sec:xrtobs_spec}

The fourteen {\it Swift}/XRT observations were carried out several weeks apart, and show very different spectral characteristics, depending on the flux level of the source. The spectra obtained from the fourteen XRT observations are shown in Fig.\,\ref{fig:XRTobs} in Appendix\,\ref{appendix:XRT_spectra}. During the first observation, in mid-May 2018, the source was found to have a 0.3--2\,keV luminosity of $L_{0.3-2}\simeq 10^{43}$, i.e. about a factor of two higher than the previous 2011 {\it XMM-Newton} observation in the same energy band ($5.4\times 10^{42}\rm\,erg\,s^{-1}$), while the overall 0.5--10\,keV flux was the same. A decline in flux was observed over the following observations, and the source was very weak in July and during the first week of August\,\,2018, with its 0.3--2\,keV luminosity decreasing below $10^{41}\rm\,erg\,s^{-1}$. An increase in flux was then observed in the following observations, with the AGN going back to an X-ray luminosity of $10^{43}\rm\,erg\,s^{-1}$ in October\,\,2018, and showing a similar flux in March\,\,2019.

Due to the lower signal-to-noise ratio with respect to the {\it XMM-Newton} observations, the spectra were fitted using a model that includes a blackbody and, if it significantly improved the fit, a power-law component [\textsc{tbabs$\times$(zbb+zpo)}]. The column density of the absorber was fixed to the Galactic value. We fitted the model using Cash statistics, similarly to what was done for the time-resolved {\it XMM-Newton} spectra (\S\ref{sec:timeresolved2018}). The results of the spectral fitting are reported in Table\,\,\ref{tab:fitXRT} in Appendix\,\ref{appendix:XRT_spectra}. Similarly to what was observed in the three {\it XMM-Newton}/{\it NuSTAR} observations, the temperature of the blackbody component increases with the X-ray luminosity of the source, reaching $kT=205^{+26}_{-24}$\,eV for $L_{0.3-2}=3.6\times 10^{43}\rm\,erg\,s^{-1}$. The powerlaw component was observed only in the observations in which $L_{0.3-2}\gtrsim 7\times10^{42}\rm\,erg\,s^{-1}$, and the photon index varied in the range $\Gamma=3-4.5$, consistent with what we found for the {\it XMM-Newton} observations.

\tabletypesize{\normalsize}
\begin{deluxetable*}{ccccc} 
\tablecaption{Spectral parameters obtained for the fourteen {\it Swift}/XRT observations (see \S\ref{sec:xrtobs_spec} and Fig.\,\ref{fig:XRTobs}). \label{tab:fitXRT}}
\tablehead{
\colhead{(1)} & \colhead{(2)} & \colhead{(3)} & \colhead{(4)}  & \colhead{(5)}  \\
\noalign{\smallskip}
 \colhead{Obs. ID } & \colhead{$kT$ } & \colhead{$\Gamma$ }  & \colhead{C-stat/DOF } & \colhead{$L_{\rm 0.3-2}$  }  \\
\noalign{\smallskip}
  \colhead{ } & \colhead{[eV]} & \colhead{}  & \colhead{ } & \colhead{ [$\rm erg\,s^{-1}$] }  
}
\startdata
\noalign{\smallskip}
00010682001	& $149\pm10$	& $4.52^{+0.56}_{-0.41}$ & 161/129 & $1.26_{-0.06}^{+0.03}\times10^{43}$  \\
\noalign{\smallskip}
00010682002	& $101^{+7}_{-6}$	& \nodata & 54/68 & $2.60_{-0.19}^{+0.15}\times10^{42}$  \\
\noalign{\smallskip}
00010682003	& $105^{+10}_{-9}$	& \nodata & 47/58 & $1.47_{-0.15}^{+0.12}\times10^{42}$  \\
\noalign{\smallskip}
00010682004	& $56^{+41}_{-23}$	& \nodata & 8/5 & $1.40_{-1.40}^{+0.12}\times10^{41}$  \\
\noalign{\smallskip}
00010682005	& $\geq 71$	& \nodata & 13/3 & $3.10_{-3.10}^{+2.68}\times10^{41}$  \\
\noalign{\smallskip}
00010682006	& $85^{+19}_{-16}$	& \nodata & 10/18 & $3.31_{-0.90}^{+0.49}\times10^{41}$  \\
\noalign{\smallskip}
00010682007	& $94\pm8$	& \nodata & 52/59 & $1.62_{-0.15}^{+0.12}\times10^{42}$  \\
\noalign{\smallskip}
00010682008	& $141^{+13}_{-10}$	& $2.95^{+0.51}_{-0.70}$ & 146/128 & $1.42_{-0.05}^{+0.12}\times10^{43}$  \\
\noalign{\smallskip}
00010682009	& $205^{+26}_{-23}$	& $3.48^{+0.37}_{-0.32}$ & 151/165 & $3.60_{-0.15}^{+0.13}\times10^{43}$  \\
\noalign{\smallskip}
00010682010	& $159^{+13}_{-12}$	& $3.96^{+0.80}_{-0.70}$ & 126/124 & $2.09_{-0.09}^{+0.08}\times10^{43}$  \\
\noalign{\smallskip}
00010682011	& $148^{+12}_{-8}$	& $3.11^{+0.77}_{-1.63}$ & 127/123 & $1.68_{-0.05}^{+0.13}\times10^{43}$  \\
\noalign{\smallskip}
00010682012	& $132^{+11}_{-10}$	& $4.36^{+2.60}_{-4.36}$ &  101/104 & $ 7.33_{-0.65}^{+0.92}\times10^{42}$  \\
\noalign{\smallskip}
00010682013	& $ 153^{+11}_{-7}$	& $3.16^{+0.46}_{-0.78}$ &  122/149 & $  2.68_{-0.14}^{+0.18}\times10^{43}$  \\
\noalign{\smallskip}
00010682014	& $  182^{+10}_{-8}$	& $ 2.84^{+0.22}_{-0.27}$ &   214/218  & $ 4.54\pm0.10\times10^{43}$  
\enddata
\tablecomments{The columns report (1) the observation ID, (2) the temperature of the blackbody component, (3) the photon index of the power law component, (4) the C-stat and the number of DOF, (5) the 0.3--2\,keV luminosity.}
\end{deluxetable*}

\begin{figure*}[h!]
  \begin{center}
  \center{\Large{Swift/XRT observations}}\par\medskip
\includegraphics[width=0.48\textwidth]{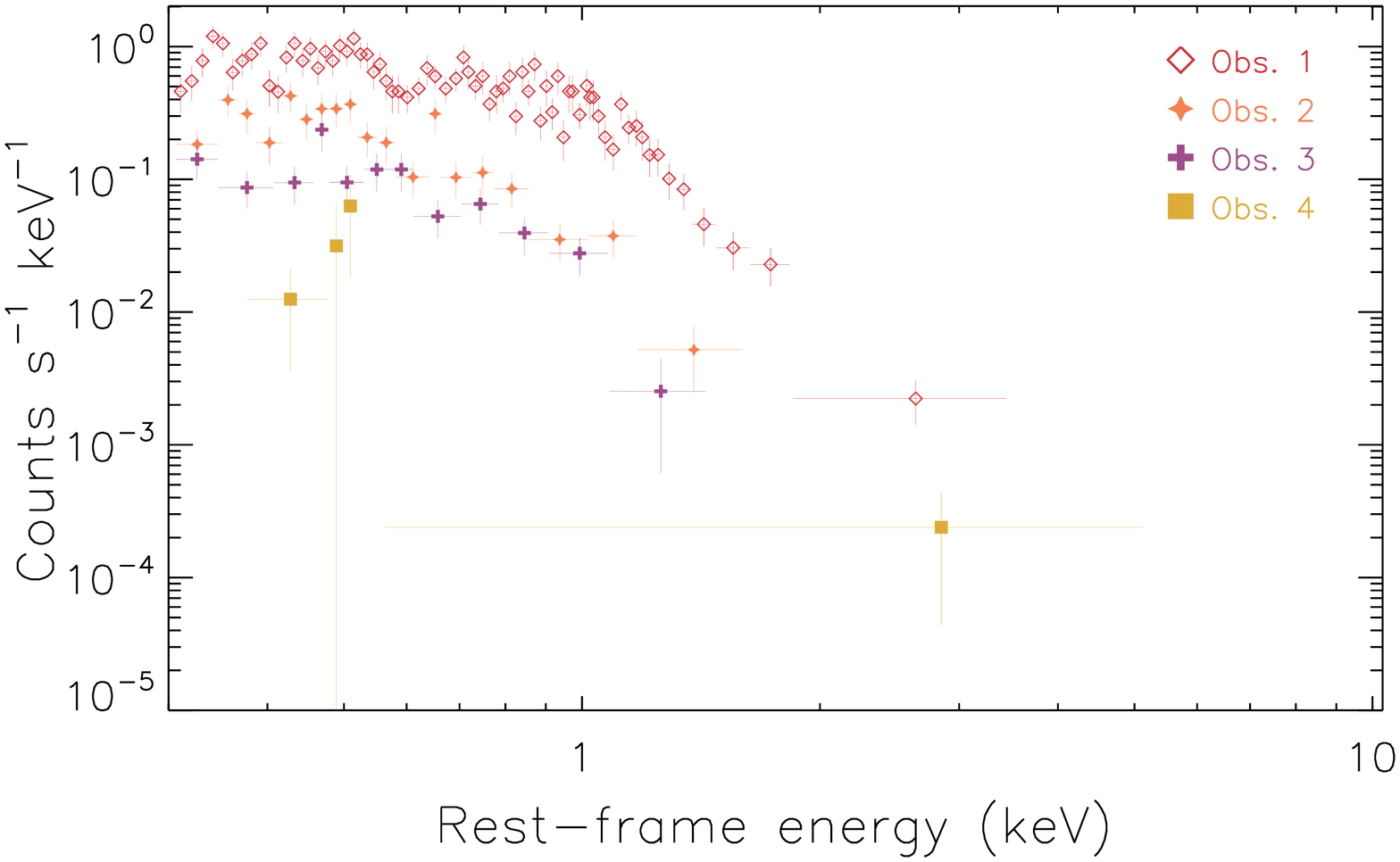}
\includegraphics[width=0.48\textwidth]{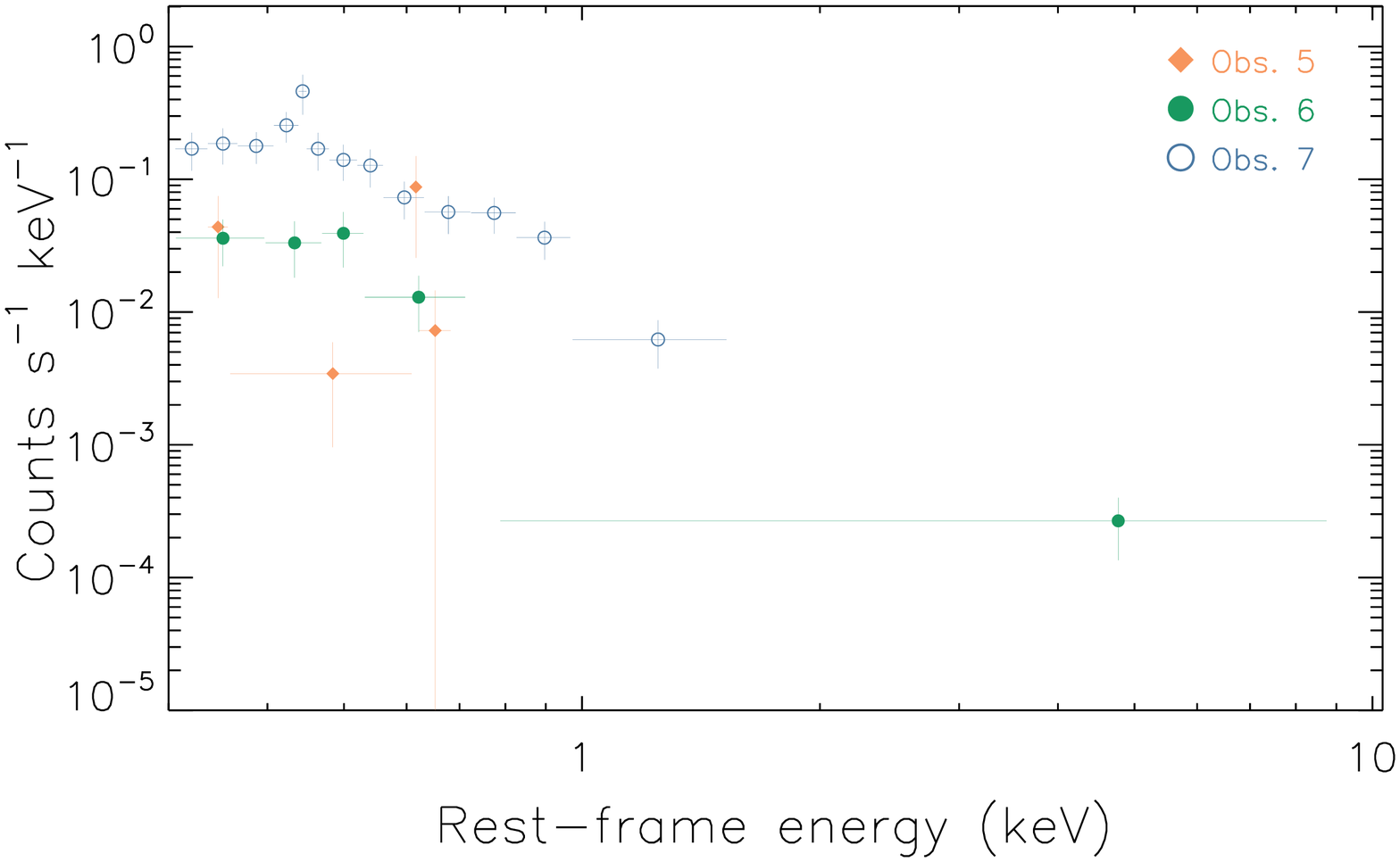}
\par\medskip
\includegraphics[width=0.48\textwidth]{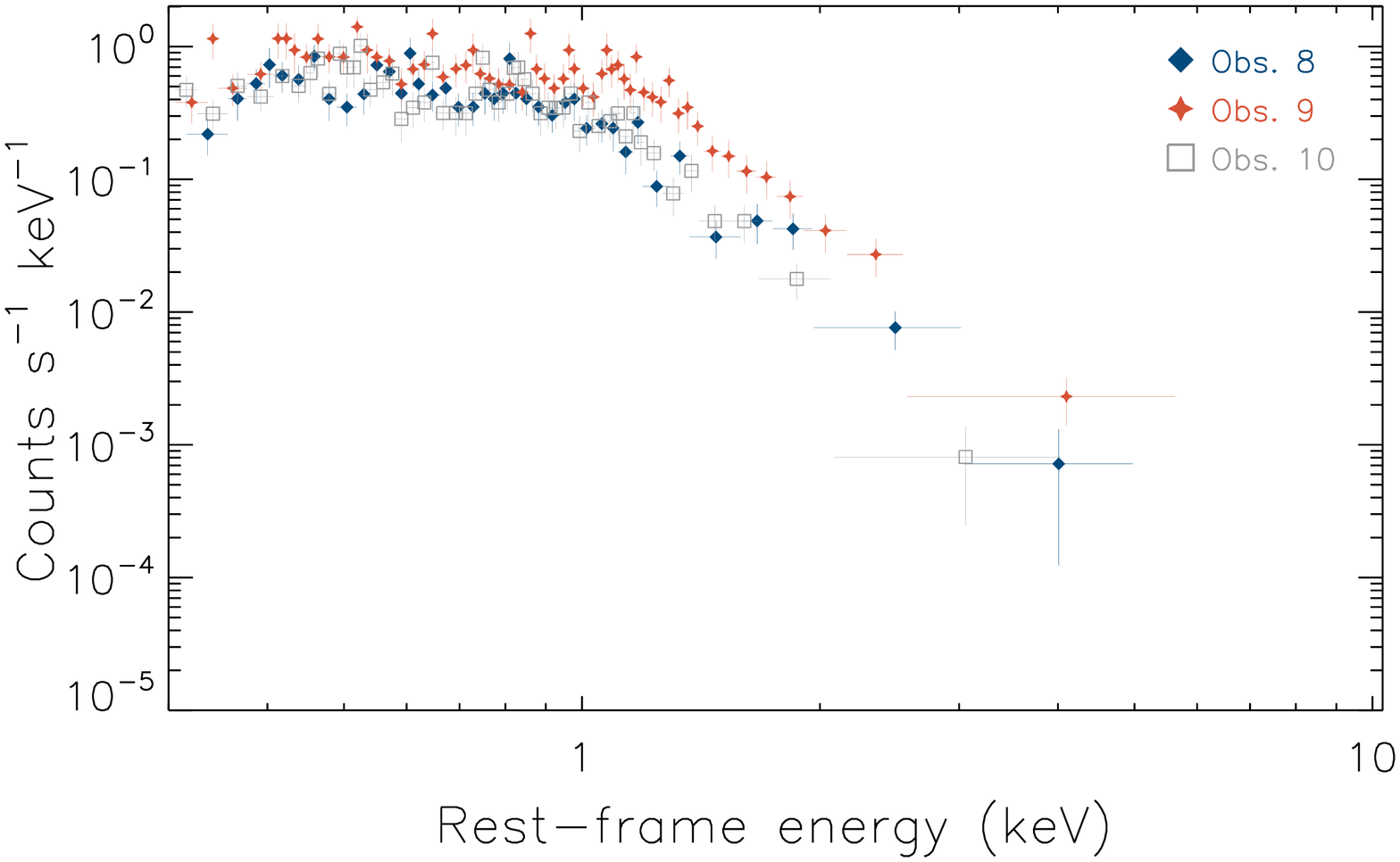}
\includegraphics[width=0.48\textwidth]{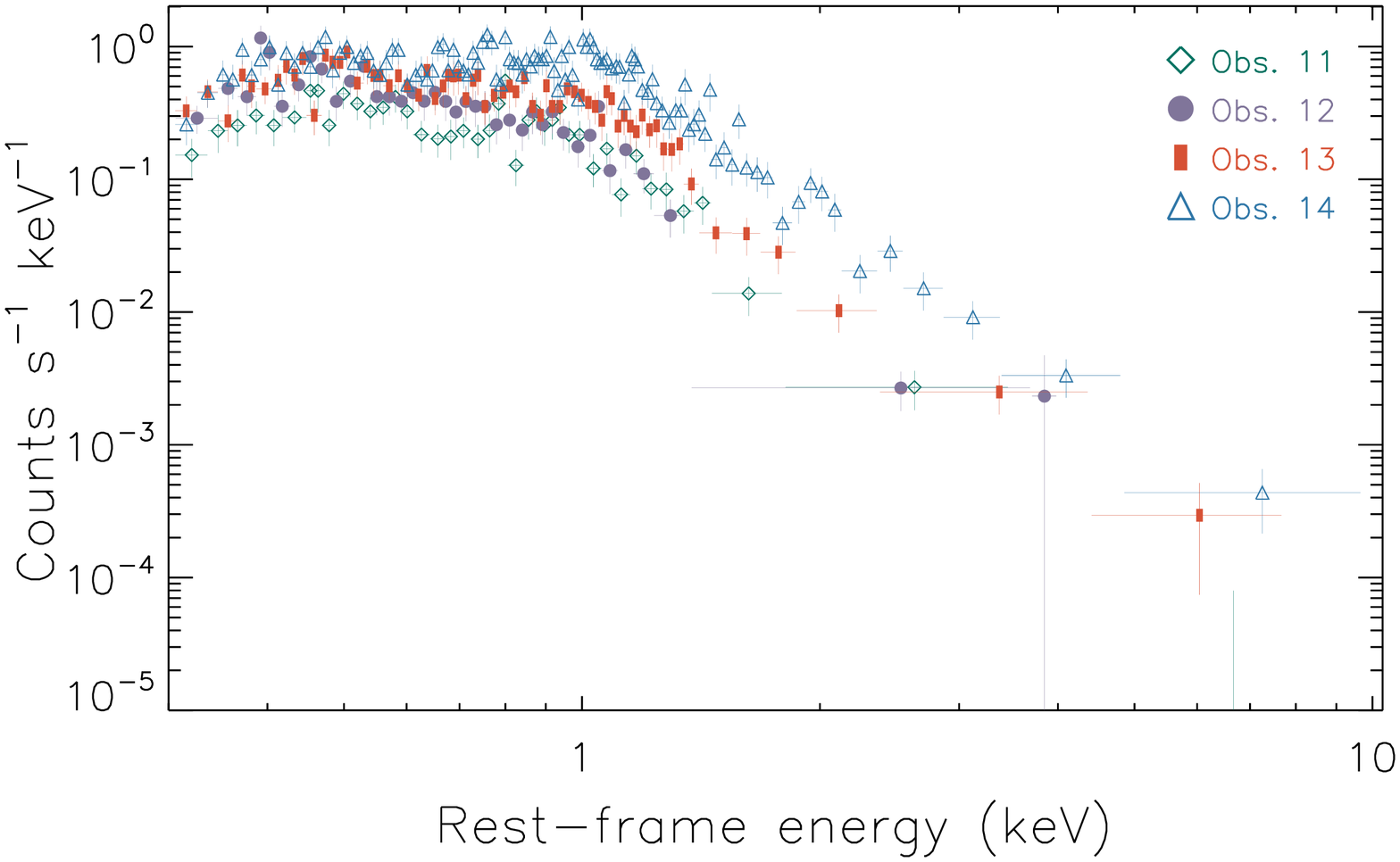}
    \caption{{\it Swift}/XRT observations of 1ES\,1927+654 carried out in 2018, illustrating the dramatic variability of the source. All the observations showed a very soft spectral shape, with very little flux above 2\,keV. The source was in a very low-flux state during observations four, five and six. Further details about the spectral analysis can be found in \S\ref{sec:xrtobs_spec} and Table\,\,\ref{tab:fitXRT}. }
    \label{fig:XRTobs}
  \end{center}
\end{figure*}

\clearpage

\section{The simultaneous {\it XMM-Newton} and {\it NICER} observation in December 2018}\label{appendix:simultaneousNICER}

\begin{figure*}[h!]    
 \begin{center}
    \includegraphics[width=0.65\textwidth]{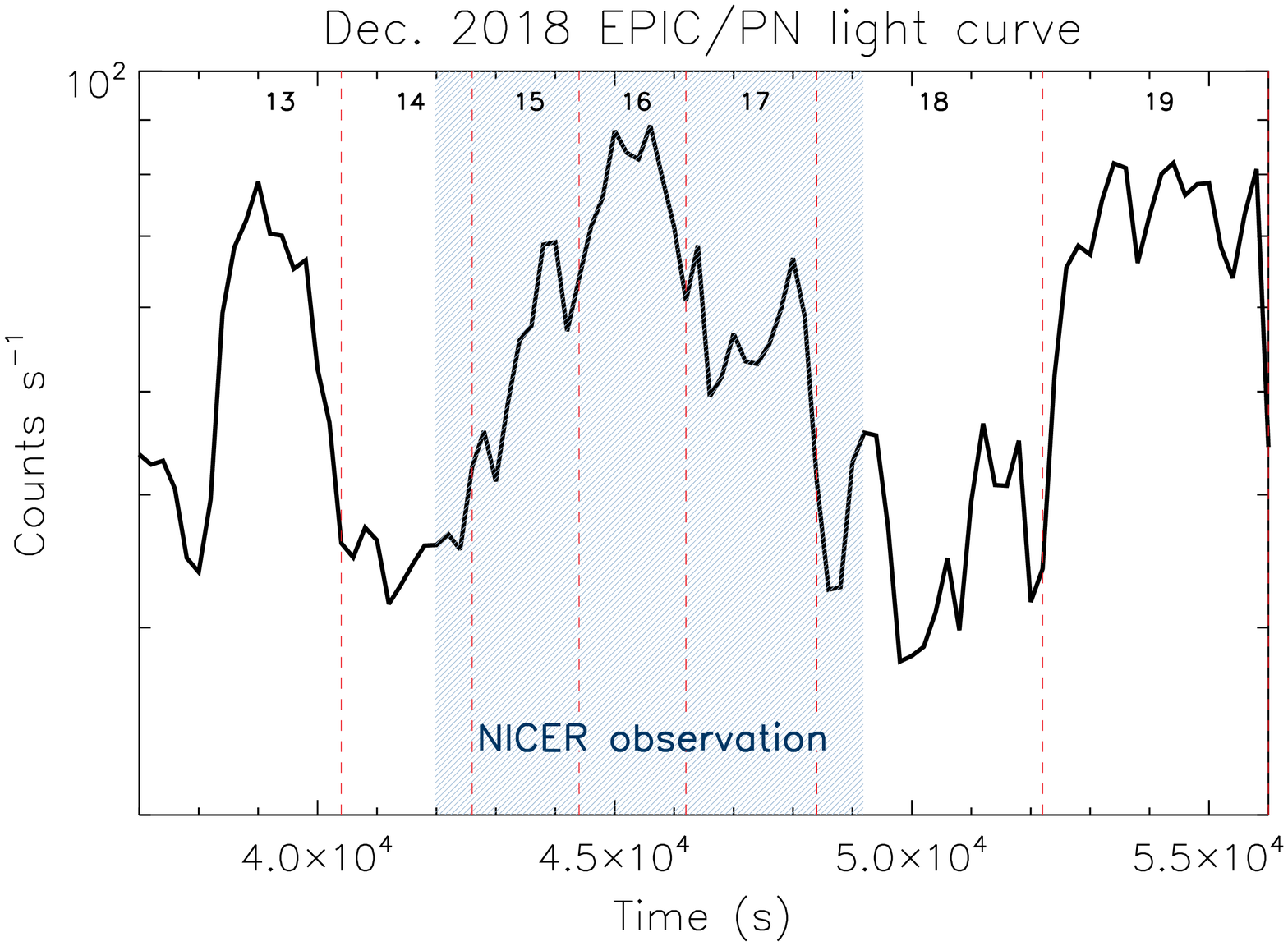}
 \end{center}
    \caption{Extract of the December 2018 {\it XMM-Newton} EPIC/PN light curve (top right panel of Fig.\,\ref{fig:XMMlc_18_bins}) around the 7.2\,ks {\it NICER} observation (blue shaded area). The EPIC/PN spectrum used for the comparison was extracted from that same interval. }
    \label{fig:xmmobs2018lcnicer}
\end{figure*}

{\it NICER} observed 1ES\,1927+654 on December 12 2018 (Observation ID 1200190238, see Table\,\,\ref{tab:obslog}), at the same time of the second {\it XMM-Newton} observation of our campaign. This observation lasted 7.2\,ks, and started $\sim 42$\,ks after the beginning of the {\it XMM-Newton} observation. Given the extremely strong spectral and flux variability of the source on timescales of a few kiloseconds (see Fig.\,\ref{fig:XMMlc_18_bins}), we could not use the {\it NICER} spectrum for our combined spectral analysis with {\it XMM-Newton} and {\it NuSTAR}. In Fig.\,\ref{fig:xmmobs2018lcnicer} we show the EPIC/PN light curve of the December 2018 observation around the {\it NICER} observation. {\it NICER} covers interval 15, 16 and 17, and part of intervals 14 and 18, all of which show significantly different spectra (see Table\,\,\ref{tab:fitXMM18epochs_Dec}), with temperatures of the blackbody component varying between $kT\simeq 120$\,eV and $kT\simeq 180$\,eV and 0.3--2\,keV luminosities ranging between $\simeq 3.6\times10^{43}\rm\,erg\,s^{-1}$ and $\simeq 8.2\times10^{43}\rm\,erg\,s^{-1}$. 

\begin{figure*}[h!]    
 \begin{center}
    \includegraphics[width=0.5\textwidth]{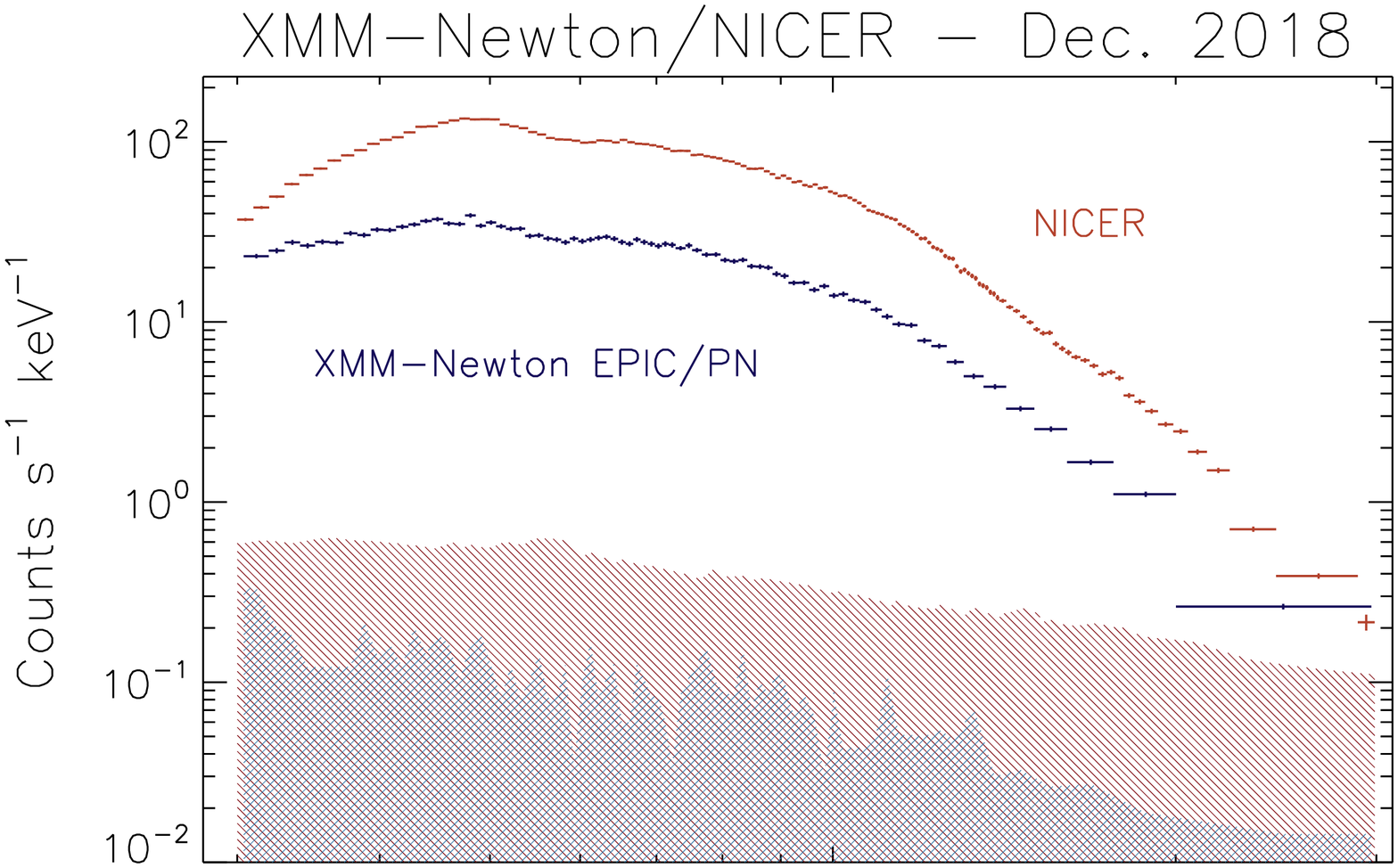}
    \includegraphics[width=0.5\textwidth]{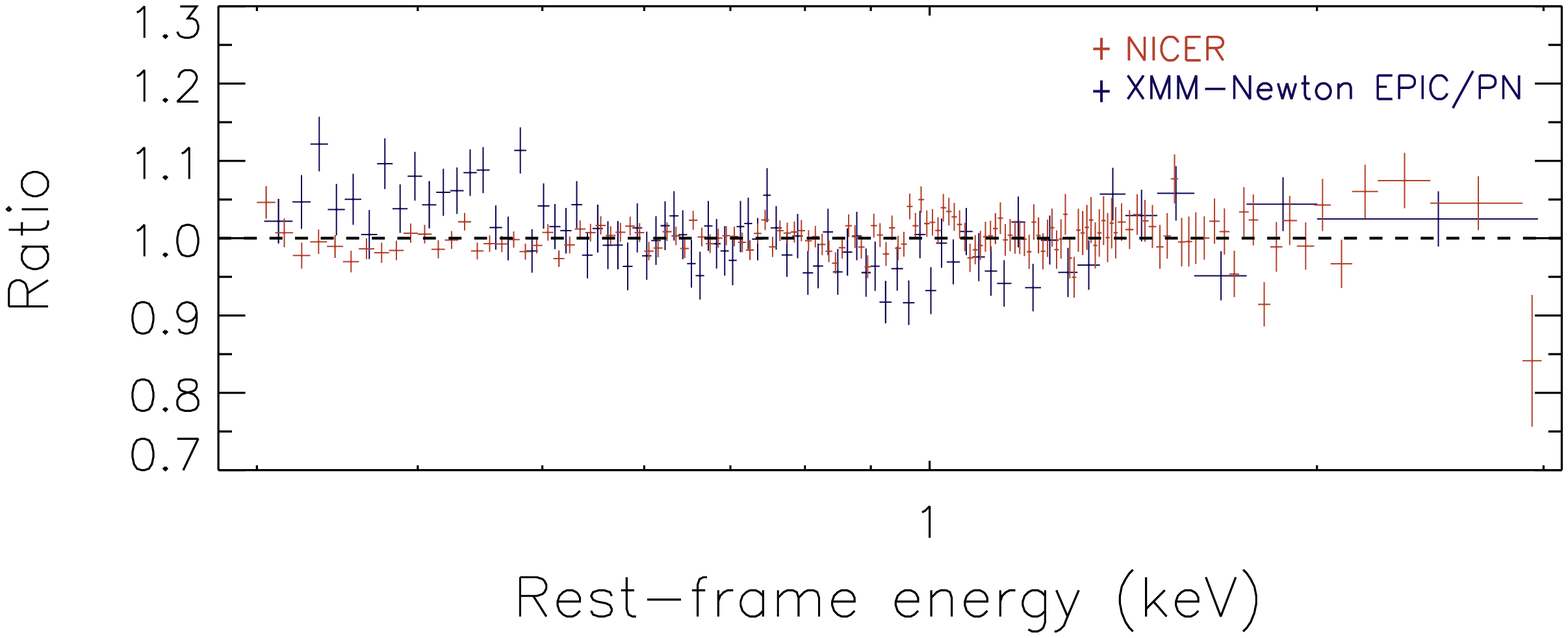}
 \end{center}
    \caption{ {\it Top panel:} December 2018 simultaneous {\it XMM-Newton} EPIC/PN (blue points) and {\it NICER} (red points) 0.3--3\,keV spectra, rebinned to 30$\sigma$. The EPIC/PN was extracted from the blue shaded region of Fig.\,\ref{fig:xmmobs2018lcnicer}, in order to be simultaneous with {\it NICER}. The blue and red shaded area are the background spectra of EPIC/PN and {\it NICER}, respectively. {\it Bottom panel:} ratio between the model and the data. We fitted both data sets simultaneously, fixing all parameters to have the same values, in order to highlight possible differences. }
    \label{fig:xmmobs2018specnicer}
\end{figure*}

We used this {\it NICER} observation to confirm that there were no issues related to pile-up or X-ray loading in the EPIC/PN spectrum, since the observation was carried out when the source was extremely bright. We extracted the EPIC/PN spectrum in the same interval covered by the {\it NICER} observation (blue shaded area in Fig.\,\ref{fig:xmmobs2018lcnicer}), and compared the spectra, and the spectral parameters obtained by fitting the two spectra with the same model. The two spectra are shown in the top panel of Fig.\,\ref{fig:xmmobs2018specnicer}.
The spectral analysis was carried out in the same band (0.3--3\,keV) for the two spectra, and we used our best model, which includes neutral and ionized absorption, a blackbody, a power law and two broad Gaussian lines [\textsc{cons$\times$tbabs$\times$ztbabs$\times$mtable\{xstar\}} \textsc{$\times$(zbb+zpo+zgauss+zgauss)}]. 
We started fitting the two datasets together, fixing the parameters of the two spectra to have the same values, and found that the model can reproduce well the spectra ($\chi^2=751$ for 622\,DOF); the residuals to this joint fit are illustrated in the bottom panel of Fig.\,\ref{fig:xmmobs2018specnicer}.

Next we fitted the two spectra separately using the same model. We fixed the parameters of the second Gaussian line ($E\sim 1.92$\,keV) to those obtained by the fit to the whole December 2018 {\it XMM-Newton} observation (see \S\ref{sect:xmmobs_spec_december_may} and Table\,\,\ref{tab:fitXMMdec18}). The results of the fits to the {\it NICER} and {\it XMM-Newton} spectra are reported in column\,2 and 3 of Table\,\,\ref{tab:fitXMMnicer_dec18}, respectively. Overall EPIC/PN data are not able to constrain the parameters as well as {\it NICER}, due to the lower count rate ($23.1\rm\,ct\,s^{-1}$ versus $78.6\rm\,ct\,s^{-1}$). All parameters are consistent within their uncertainties, with the exception of the column density of the neutral absorber, which is slightly higher in the best-fit model to the {\it XMM-Newton} spectrum. The photon index obtained by fitting the {\it XMM-Newton} observation is consistent ($3.7^{+0.1}_{-0.2}$) within the uncertainties with that inferred from the {\it NICER} observation ($3.44^{+0.06}_{-0.07}$). The small difference is likely related to the $\sim 5-10\%$ excess observed in the $0.3-0.5$\,keV band in the EPIC/PN spectrum (see bottom panel of Fig.\,\ref{fig:xmmobs2018specnicer}). The blackbody temperature is on the other hand very consistent between the two fits: $160\pm1$\,eV and $158^{+4}_{-5}$\,eV for the {\it NICER} and {\it XMM-Newton} observation, respectively.

\clearpage

\tabletypesize{\normalsize}
\begin{deluxetable*}{llcc} 
\tablecaption{Spectral parameters obtained fitting individually the {\it NICER} (column 2) and {\it XMM-Newton} EPIC/PN (column 3) spectra obtained during the 7.2\,ks simultaneous December 2018 observation. The model used consists of a blackbody component, a power law, two Gaussian absorption lines and two Gaussian emission lines, a neutral and an ionized absorber.\label{tab:fitXMMnicer_dec18}}
\tablehead{
 \colhead{ } & \colhead{(1) } & \colhead{(2)} &  \colhead{(3)}   \\
\noalign{\smallskip}
 \colhead{ } & \colhead{ } & \colhead{{\it NICER}} & \colhead{{\it XMM-Newton}} 
}
\startdata
\noalign{\smallskip}
a) & $N_{\rm H}$ ($10^{20}\rm\,cm^{-2}$)	& $3.5_{-0.7}^{+0.5}$& $  6.2_{-1.8}^{+1.5 }$	 \\
\noalign{\smallskip}
b) & $N_{\rm H}^W$ ($10^{20}\rm\,cm^{-2}$)		& $2.0^{+0.3}_{-0.2}$& $\leq 2.1$	  \\
\noalign{\smallskip}
c) &	$\log \xi$ ($\rm\,erg\,cm\,s^{-1}$) & $  \leq 1.3$& $   \leq 3.7$	  \\
\noalign{\smallskip}
d) & $z$	& $ -0.26\pm0.01$	& $-0.29_{-NC}^{+0.07}$    \\
\noalign{\smallskip}
e) & $\Gamma$	& $  3.44_{-0.07}^{+0.06 }$& $  3.7_{-0.2}^{+0.1 }$  \\
\noalign{\smallskip}
f) & $E_1$ (keV)	& $  1.12\pm0.01$& $   1.13\pm 0.05$	  \\
\noalign{\smallskip}
g) & $\sigma_1$ (eV) & $ 113\pm15$& $   84 _{-44}^{+302}$	  \\
\noalign{\smallskip}
h) & $EW_1$ (eV)	& $  32\pm3$& $  21\pm8$	  \\
\noalign{\smallskip}
i) & $kT$ (eV)	& $  160\pm1$& $158^{+4}_{-5}$	  \\
\noalign{\smallskip}
j) & $\chi^2$/DOF	&  339/257 &  341/356	
\enddata
\tablecomments{The lines report: the column density of the neutral absorber (a); the column density (b), ionization parameter (c) and redshift (d) of the ionized absorber; the photon index of the power-law component (e); the energy (f), width (g) and equivalent width (h) of the Gaussian emission line; the temperature of the blackbody (i), and the value of the chi-squared and the number of degrees of freedom (j).}
\end{deluxetable*}

\clearpage
\section{The December 2018 {\it Nustar} observations of 1ES\,1927+654}\label{sect:XMMvsNustar}
\begin{figure*}[h!]    
 \begin{center}
    \includegraphics[width=0.75\textwidth]{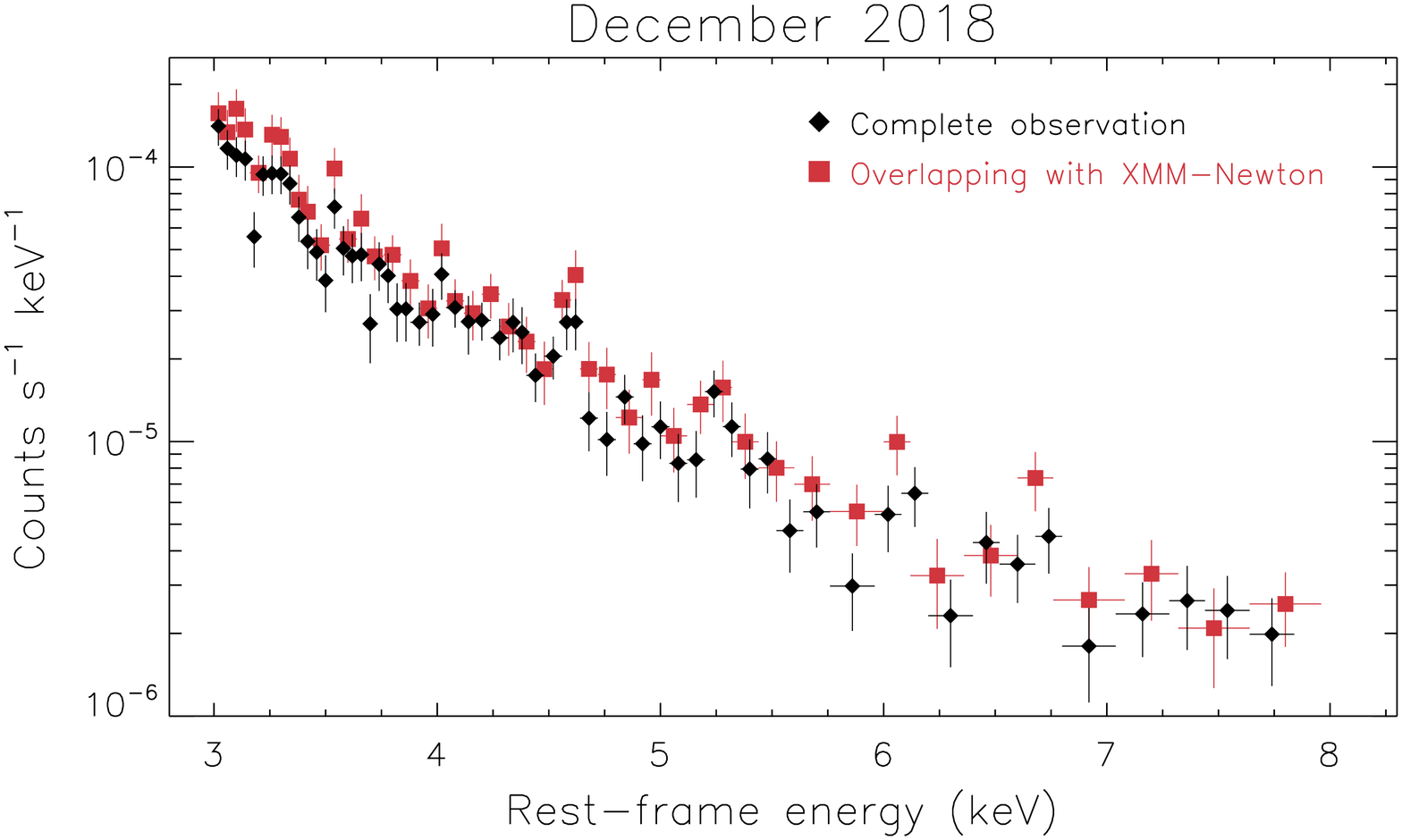}
 \end{center}
    \caption{{\it NuSTAR} spectrum from the whole observation (black diamonds) and for the part of the observation overlapping with {\it XMM-Newton} (red squares).}
    \label{fig:nustarobservations_gti}
\end{figure*}

Given the strong variability of 1ES\,1927+654 on ks timescales we looked for potential differences between the {\it NuSTAR} spectrum obtained from the whole observations and that obtained by considering only the interval of time in which the observations was simultaneous with {\it XMM-Newton}. The spectra are illustrated Fig.\,\ref{fig:nustarobservations_gti}.

We fitted both spectra (Fig.\,\ref{fig:nustarobservations_gti}) in the 3--8\,keV range using a simple power law mode, to test the differences obtained in the spectral parameters. For the complete observation (black diamonds) we obtained $\Gamma=4.79^{+0.22}_{-0.21}$ and a normalization of $n_{\rm po}=2.8^{+1.0}_{-0.7}\times 10^{-2}\rm\,photons\,keV^{-1}\,cm^{-2}\,s^{-1}$. Fitting the spectrum obtained from the interval consistent with the {\it XMM-Newton} observation (red squares) we found a very similar spectral shape ($\Gamma=4.78^{+0.26}_{-0.25}$) and normalization ($n_{\rm po}=3.5^{+1.5}_{-1.0}\times 10^{-2}\rm\,photons\,keV^{-1}\,cm^{-2}\,s^{-1}$). We concluded that, given the negligible spectral differences, it is safe to fit the {\it XMM-Newton} spectra with the {\it NuSTAR} spectrum extracted from the whole observation. We nevertheless included a constant when fitting {\it NuSTAR} and {\it XMM-Newton} data to take into account possible differences in flux and calibrations.

\clearpage

\section{The {\it XMM-Newton}/RGS spectra }\label{appendix:RGS_spectra}

\begin{figure*}[h!]    
\begin{center}
    \includegraphics[width=0.85\textwidth]{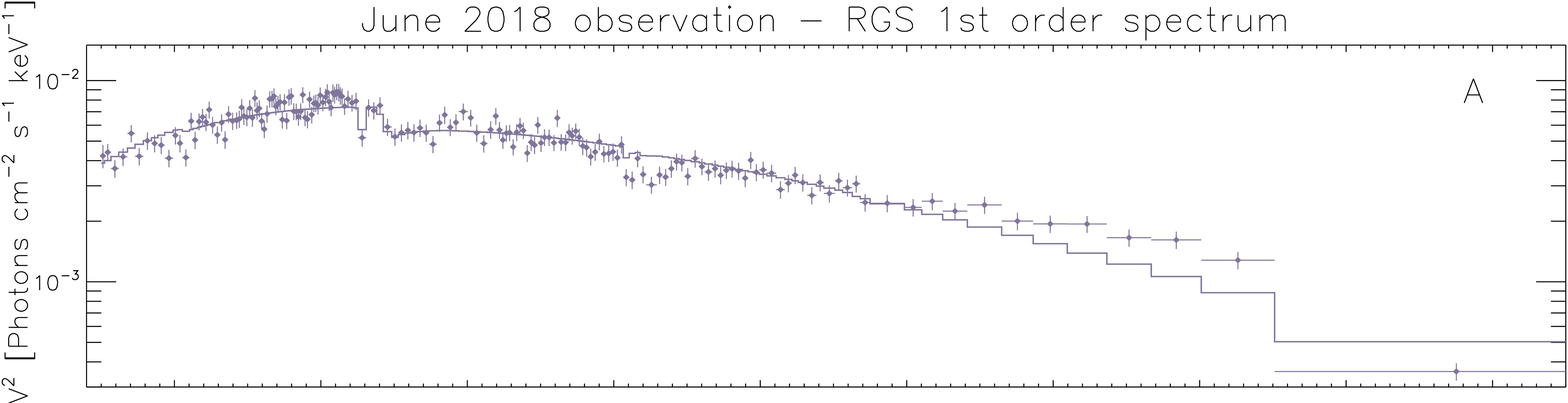}
    \includegraphics[width=0.85\textwidth]{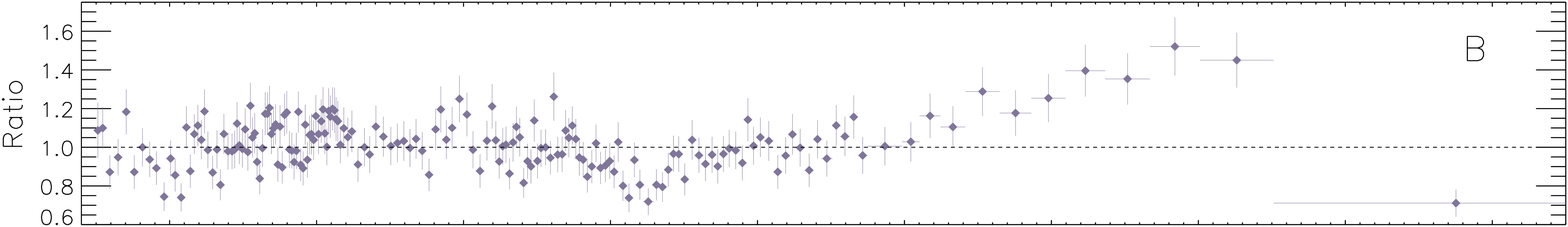}
    \includegraphics[width=0.85\textwidth]{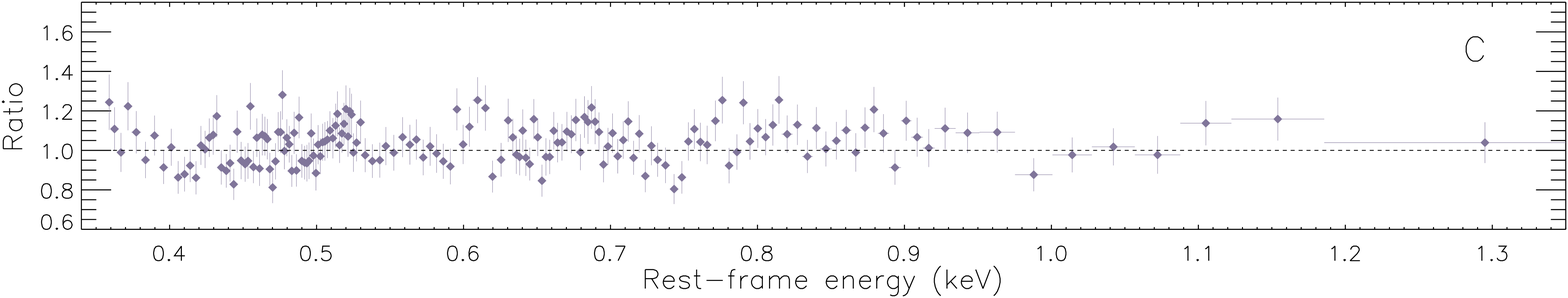}
    \par\bigskip
    \par\medskip
     \includegraphics[width=0.85\textwidth]{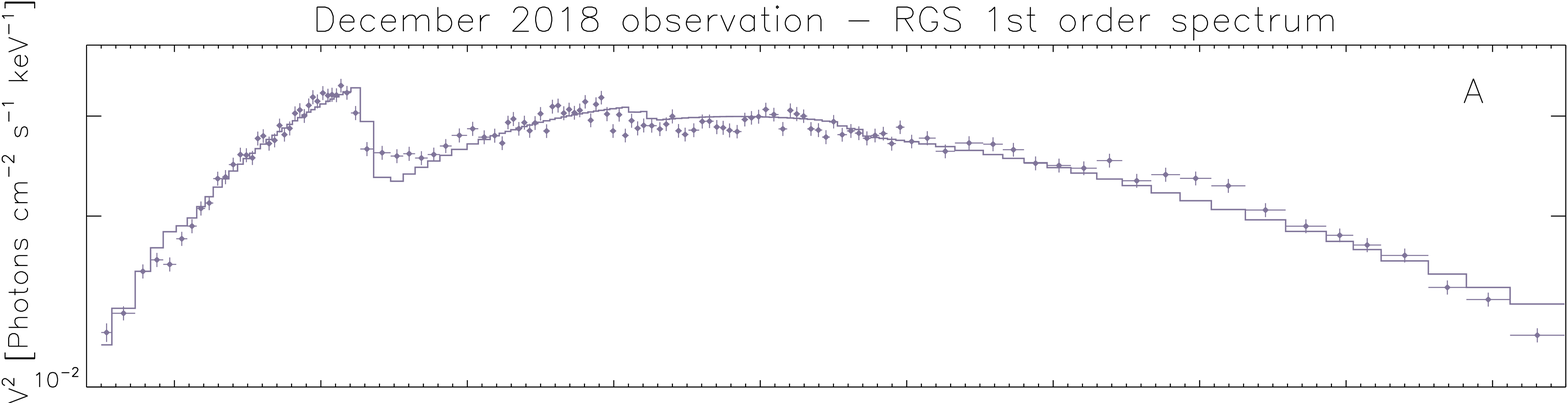}
    \includegraphics[width=0.85\textwidth]{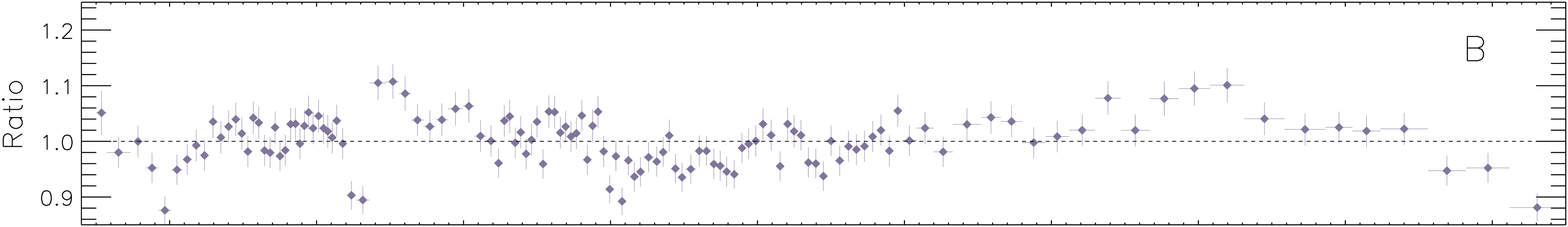}
    \includegraphics[width=0.85\textwidth]{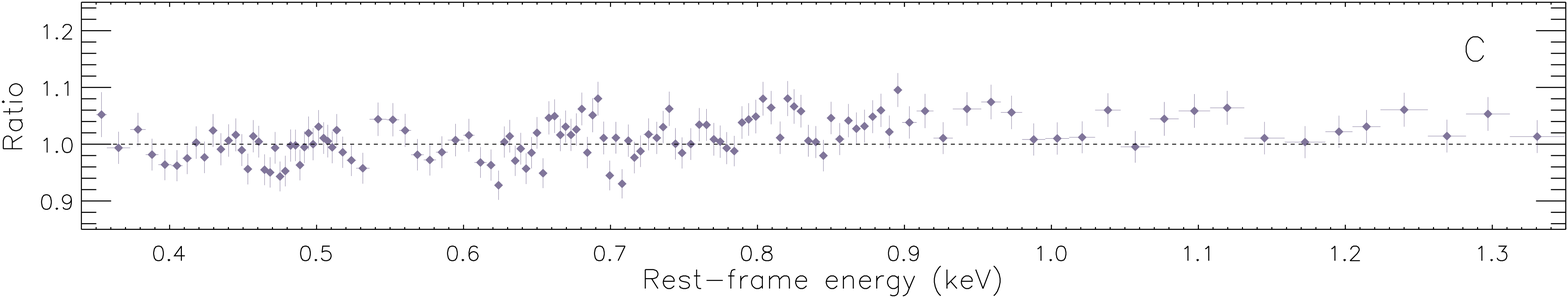}
       
    \caption{{\it XMM-Newton}/RGS 2018 observations. {\it Panels A:} first order spectra obtained from the June (top figure) and December 2018 (bottom figure) {\it XMM-Newton} observations. For visual clarity the spectra were rebinned to 10$\sigma$ and 35$\sigma$, respectively. The model for the continuum (continuous lines) includes a blackbody and a powerlaw [\textsc{ztbabs*(zbb+zpow)} in \textsc{XSPEC}]. {\it Panels B:} ratios between the model illustrated in Panels A and the data. {\it Panels C:} ratios obtained when considering the best-fitting model [\textsc{tbabs$\times$ztbabs$\times$mtable\{xstar\}} \textsc{(zbb+zpo+zgauss+zgauss)}].}
    \label{fig:RGS_spec_18}
\end{center}
\end{figure*}

\begin{figure*}[h!]    
\begin{center}
     \includegraphics[width=0.85\textwidth]{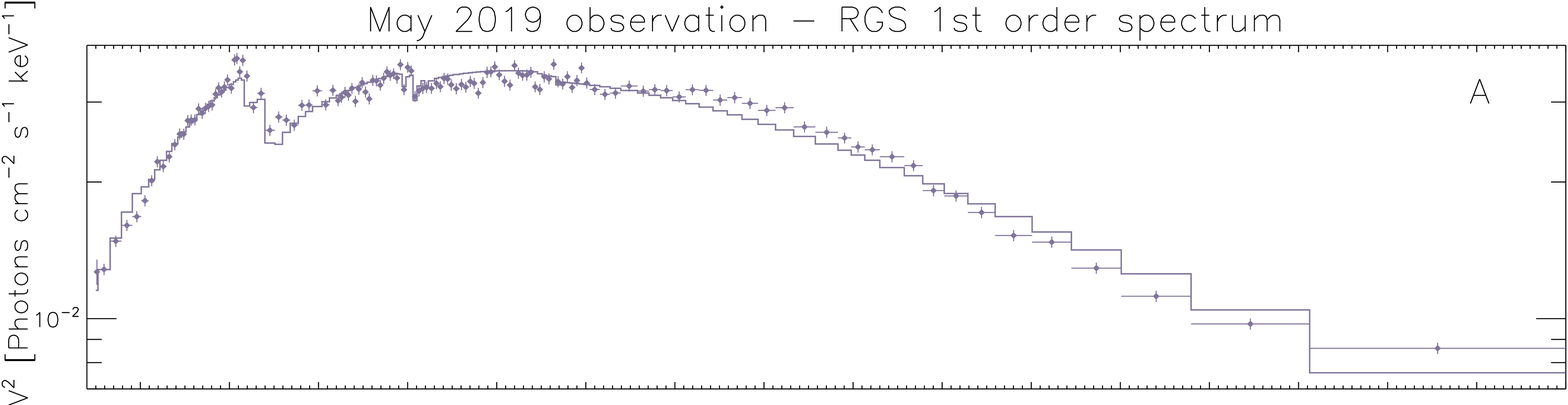}
    \includegraphics[width=0.85\textwidth]{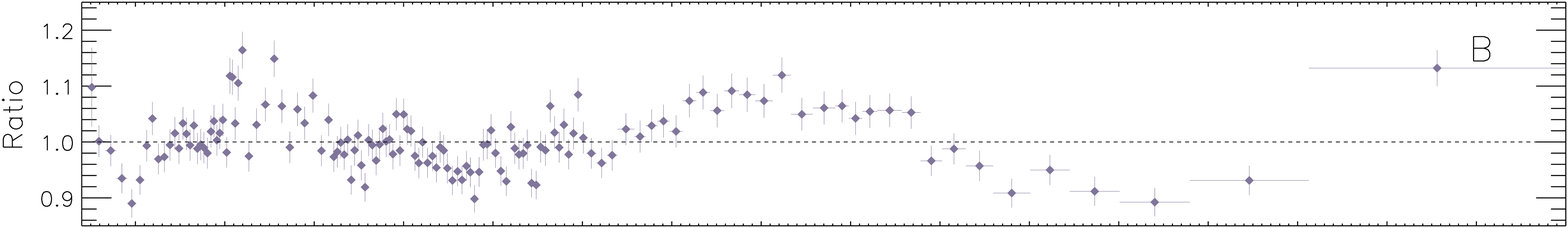}
    \includegraphics[width=0.85\textwidth]{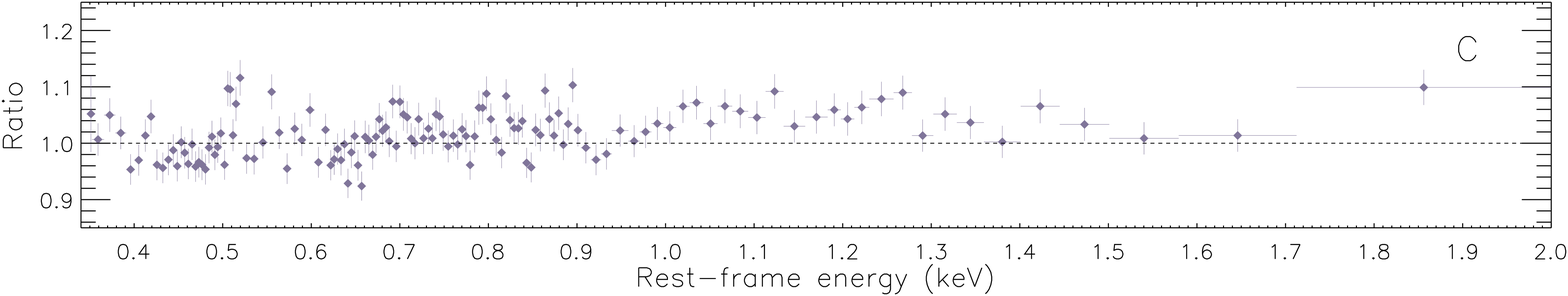}
           \caption{{\it XMM-Newton}/RGS May 2019 observations. {\it Panels A:} first order spectrum of the May 2019 observation. For visual clarity the spectra was rebinned to 35$\sigma$. The model for the continuum (continuous line) includes a blackbody and a powerlaw [\textsc{ztbabs*(zbb+zpow)} in \textsc{XSPEC}]. {\it Panels B:} ratios between the model illustrated in Panels A and the data. {\it Panels C:} ratios obtained when considering the best-fitting emission lines model [\textsc{tbabs$\times$ztbabs$\times$mtable\{xstar\}$\times$(zbb+zpo+zgauss+zgauss)}]. }
    \label{fig:RGS_spec_19}
\end{center}
\end{figure*}

\begin{figure*}[h!]    
\begin{center}
    \includegraphics[width=0.7\textwidth]{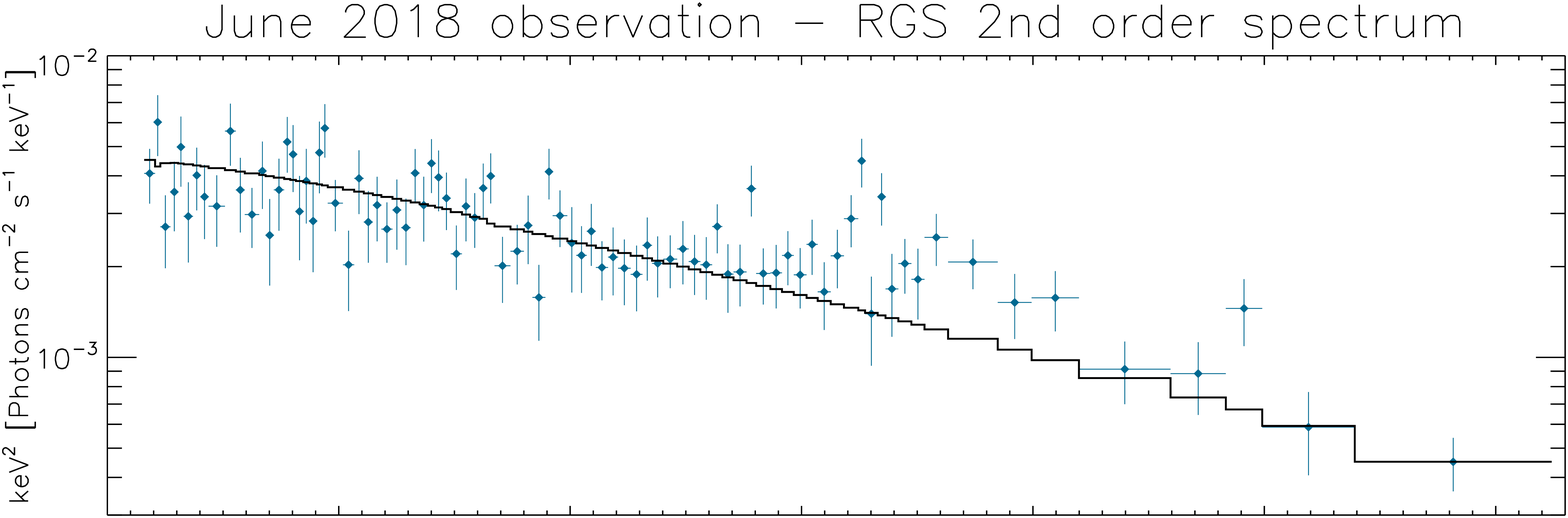}
    \includegraphics[width=0.7\textwidth]{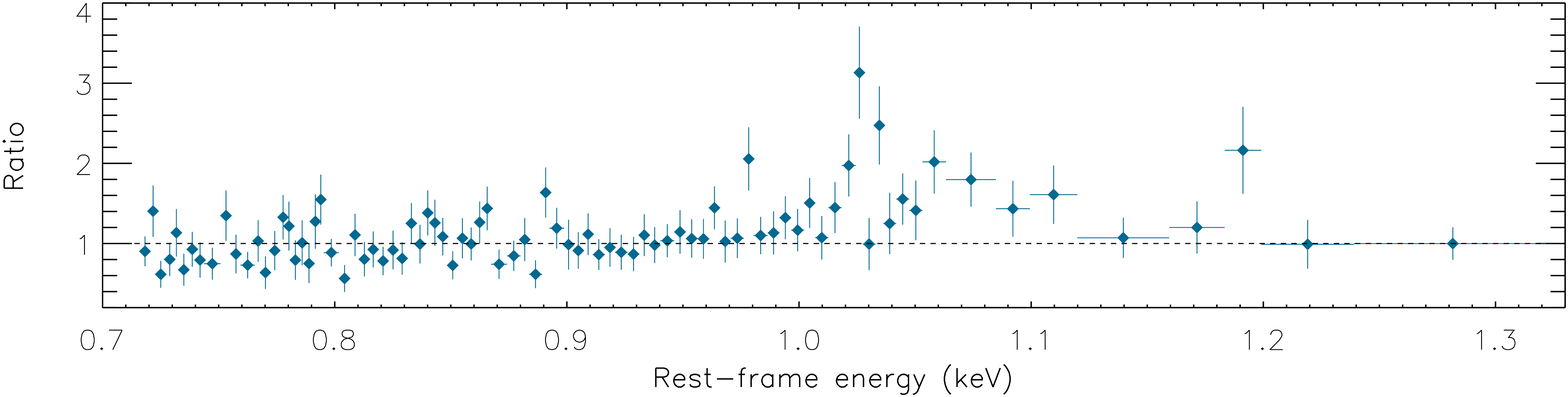}
\par\bigskip
    \includegraphics[width=0.7\textwidth]{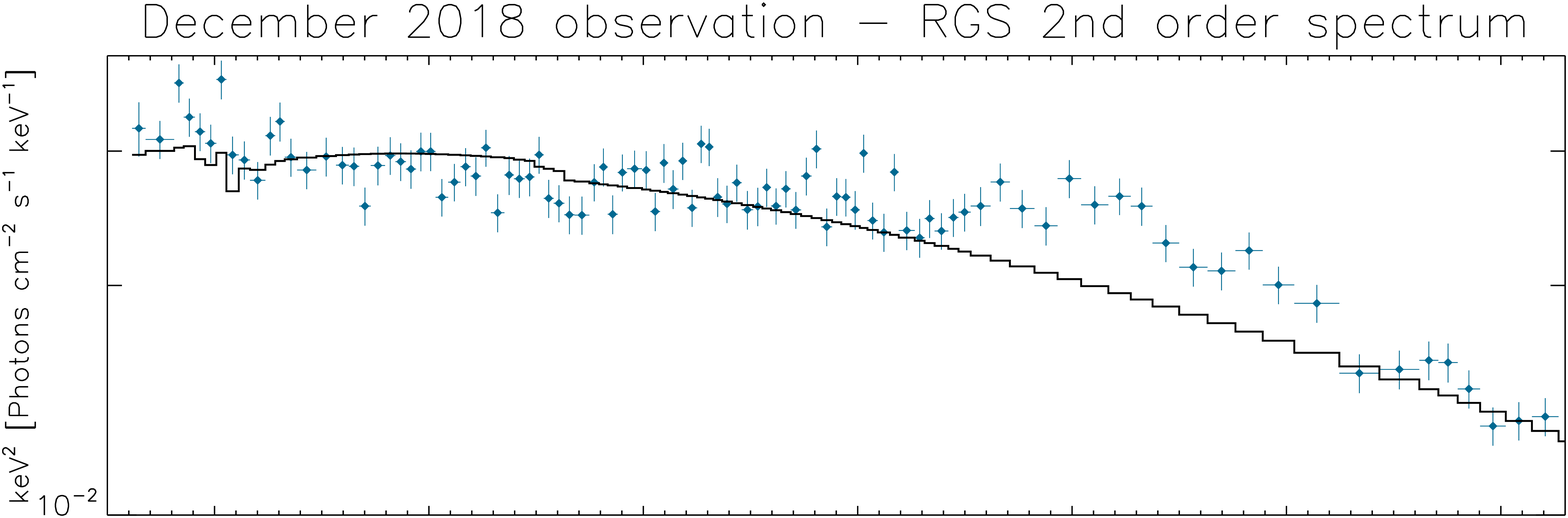}
    \includegraphics[width=0.7\textwidth]{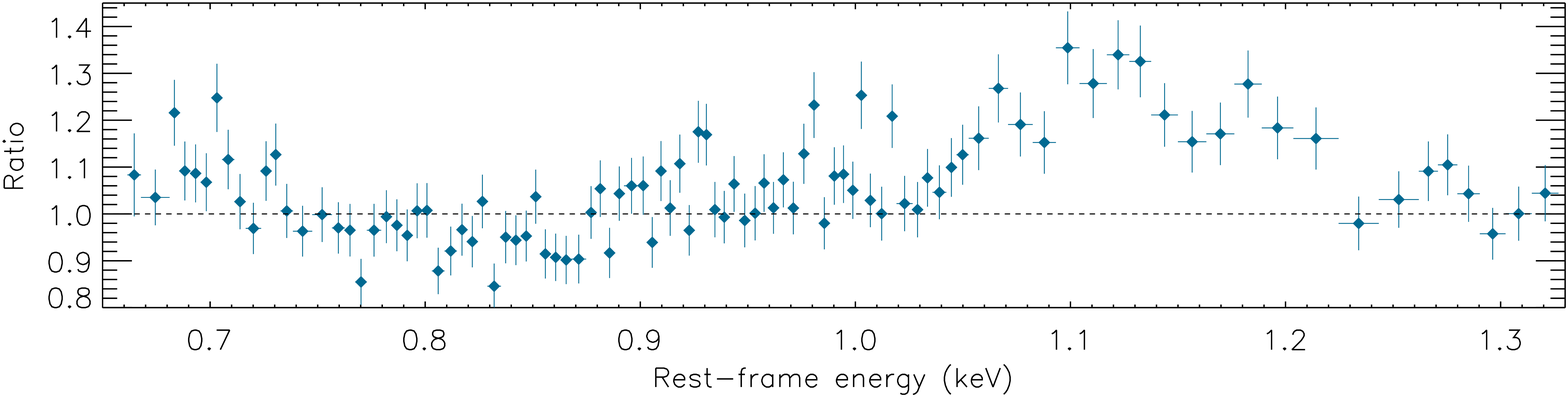}
\par\bigskip
    \includegraphics[width=0.7\textwidth]{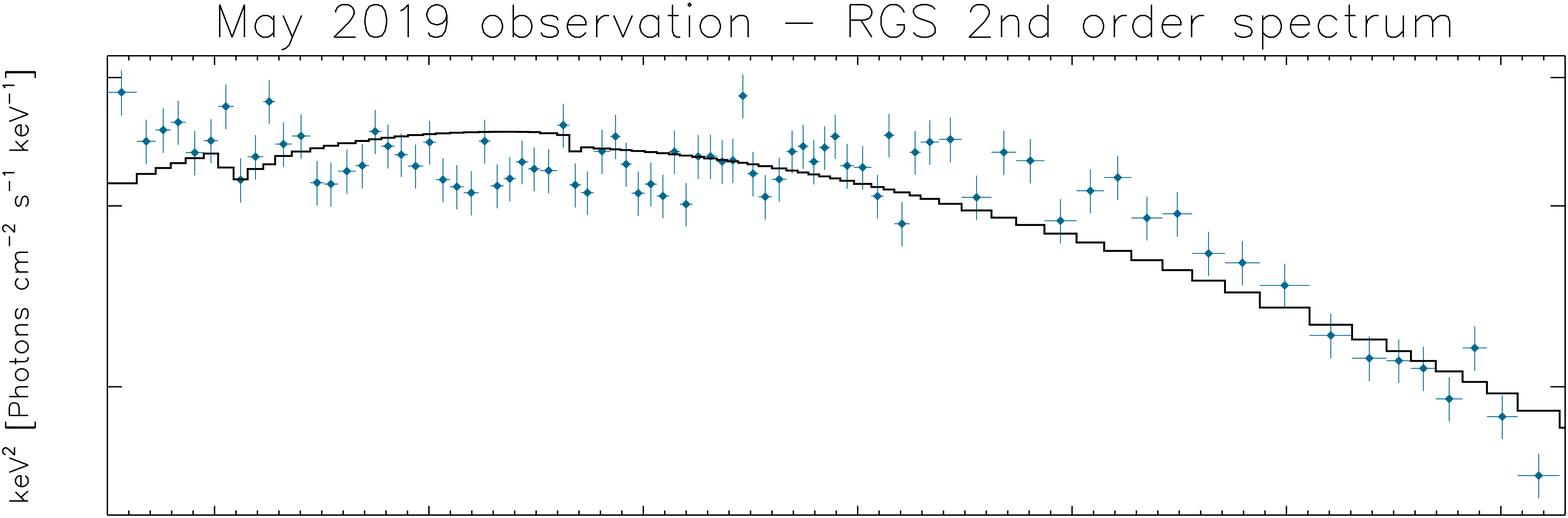}
    \includegraphics[width=0.7\textwidth]{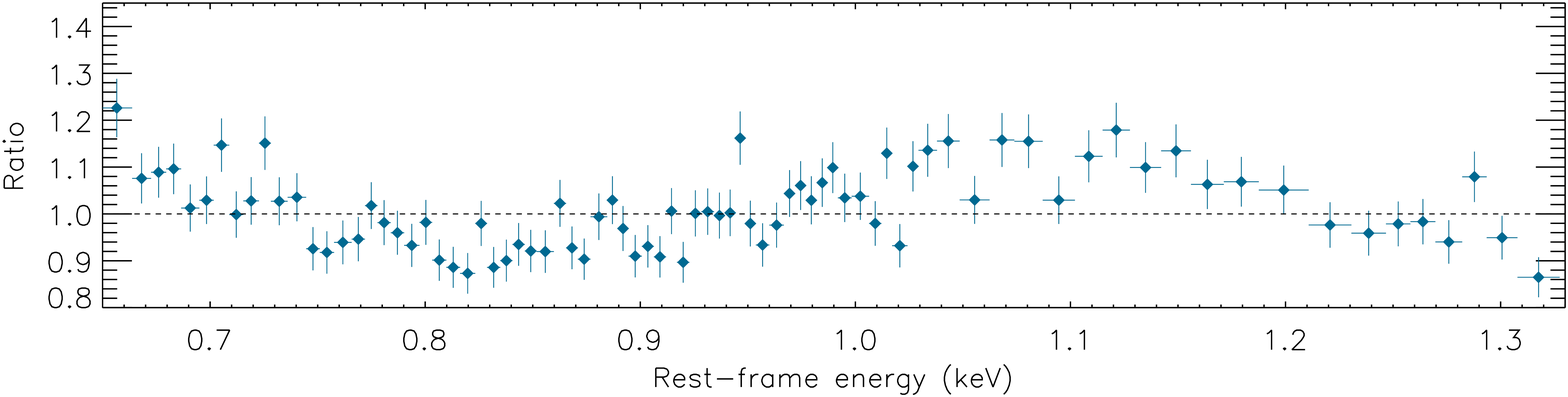}   \caption{RGS second order spectra obtained from the June\,\,2018 (top panel), December\,\,2018 (middle panel) and May\,\,2019 (bottom panel) {\it XMM-Newton} observation. For visual clarity the June\,\,2018, December\,\,2018 and May\,\,2019 spectra were rebinned to 3$\sigma$, 17$\sigma$ and 17$\sigma$, respectively. The model for the continuum (black lines) includes a blackbody and a powerlaw [\textsc{ztbabs*(zbb+zpow)} in \textsc{XSPEC}]. The bottom panels show the ratio between the model and the data. }
    \label{fig:RGS_spec_order2}
\end{center}
\end{figure*}

In this section we show the {\it XMM-Newton}/RGS spectra of 1ES\,1927+654 obtained by our campaign. In Fig.\,\ref{fig:RGS_spec_18} we illustrate the two 2018 first order spectra, while in Fig.\,\ref{fig:RGS_spec_19} we show the May\,\,2019 one. The second-order RGS spectra are illustrated in Fig.\,\ref{fig:RGS_spec_order2}.

\clearpage

\section{Alternative continuum models }\label{appendix:XMM_alternativecontinuummodels}

For the two {\it XMM-Newton} observations of 1ES\,1927+654 carried out in 2018 we explored two different models of the continuum, the first including a Multicolour disk model (\S\ref{sect:xmmobs_spec_june_diskbb} and \ref{sect:xmmobs_spec_dec_diskbb}), while the second considered a Comptonized disk model ($\S$\ref{sect:xmmobs_spec_june_diskbb} and \ref{sect:xmmobs_spec_dec_diskbb}).

\subsection{The {\it XMM-Newton} June 2018 observation}\label{appendix:june2018}

\subsubsection{Multicolour disk model}\label{sect:xmmobs_spec_june_diskbb}

As a first test we adopted, instead of the blackbody component, a multicolor disk (MCD) blackbody (\textsc{diskbb} in \textsc{XSPEC}, \citealp{Mitsuda:1984os,Makishima:1986qo}). Since this model does not consider redshift we used it in combination with the \textsc{zashift} model. The parameters of \textsc{diskbb} are the same of the classical blackbody used in \S\ref{sect:xmmobs_spec_june}, and we left free to vary both the temperature ($kT_{\rm MCD}$) and the normalization.

We started from a model that includes neutral absorption and a power-law component [\textsc{cons$\times$tbabs$\times$ztbabs$\times$(zashift$\times$diskbb+zpo)}], and found that, similarly to what was found in the case of the blackbody component, this model cannot reproduce well the X-ray spectrum ($\chi^{2}=1755$ for 317 DOF), leaving strong residuals in the 1--2\,keV range. The addition of a Gaussian line improves significantly the fit ($\chi^{2}=538$ for 314 DOF), however still leaving residuals between 1.5 and 2\,keV. Adding a second Gaussian line improves the fit, and yields $\chi^{2}=480$ for 311 DOF, while adding a warm absorber to this model improved only marginally the fit ($\chi^{2}=477$ for 308 DOF). Including the RGS data to the fit we obtained Stat$=3218$ for 2847 DOF, and a temperature of the disk of $kT_{\rm MD}=121\pm1$\,eV (see column\,2 of Table\,\,\ref{tab:fitXMM1819}). Overall the multicolour disk model does not reproduce the X-ray spectrum of 1ES\,1927+654 as well as the blackbody model.

\subsubsection{Comptonized disk model}\label{sect:xmmobs_spec_june_diskbb}

As a further test we used a Comptonized disk model (\textsc{compTT}, \citealp{Titarchuk:1994zt}) to reproduce the strong soft component observed in the spectrum of 1ES\,1927+654. The geometry parameter of the Comptonizing region, which was set to 1 (disk geometry), while the temperature of the seed photons ($T_{0}$) was assumed to follow Wien's law. 
The maximum effective temperature of a classic Shakura-Sunyaev accretion disk \citep{Shakura:1973qy,Bonning:2007yq} is:
\begin{equation}\label{eq:tempmax}
kT_{\rm max}=C\times L_{46}^{0.25}/M_8^{0.5}\rm\,K=C_1\times L_{46}^{0.25}/M_8^{0.5}\rm\,eV,
\end{equation}
where $L_{46}$ and $M_8$ are the bolometric luminosity and the black hole mass in units of $10^{46}\rm\,erg\,s^{-1}$ and $10^{8}\,M_{\odot}$, while the values of the constants are $C=10^{5.08}$ ($C_1=10.4$) for a non-rotating Schwarzschild black hole and $C=10^{5.54}$ ($C_1=29.9$) for a maximally rotating one. We assumed a black hole mass of $M_{\rm BH}=1.9\times10^7 M_{\odot}$ \citep{Trakhtenbrot:2019qy}, while the bolometric luminosity was calculated from the 0.3--10\,keV luminosity, assuming a bolometric correction of $\kappa_{\rm X}=20$ ($L_{46}=1.69\times 10^{-2}$). It should be noted that this is a rather arbitrary value, in fact since most of the X-ray radiation is emitted below 2\,keV we cannot use the typical 2--10\,keV bolometric corrections (e.g., \citealp{Vasudevan:2007qt,Vasudevan:2009ng}). From this luminosity we obtained $kT_{\rm max}\simeq 9$\,eV for a Schwarzschild black hole and $kT_{\rm max}\simeq25$\,eV for a maximally rotating one. We started assuming a maximally rotating black hole, and set $kT_{0}=25$\,eV. Interestingly, this value is considerably lower than that inferred for 1ES\,1927+654 by using a simple blackbody model. As we discussed in \S\ref{sect:broadline}, fitting with a relativistic line model the broad feature at $\sim 1\rm\,keV$ suggests an intermediate spin, which would imply an even lower temperature. 

In the fit both the temperature ($kT_{\rm Comp.}$) and optical depth ($\tau$) of the Comptonizing plasma were left free to vary. A model including only the Comptonized disk fails to reproduce the shape of the continuum ($\chi^2$/DOF=11334/318). The addition of a power-law component [\textsc{cons$\times$tbabs$\times$ztbabs$\times$(compTT+zpo)}] improves the fit, but still results in a very large chi-squared ($\chi^2=11316$ for 316 DOF).
Adding a Gaussian line improves the fit ($\chi^2=3467$ for 313 DOF), and the line energy is $0.561\pm0.002$\,keV, while the width exceeds 200\,eV, contributing to most of the X-ray emission between 0.5 and 1\,keV, and improves the fit only because the \textsc{compTT} fails to reproduce the shape of the continuum above $\simeq 0.5$\,keV. A second Gaussian line improves significantly the fit ($\chi^2=410$ for 310 DOF), and the energy (width) of the line is $0.927^{+0.002}_{-0.012}$\,keV ($141\pm5$\,eV), similar to one of the two broad lines needed when applying the blackbody model.
The addition of a warm absorber, similarly to what was done using the blackbody model (\S\ref{sect:xmmobs_spec_june}) further improves the fit ($\chi^2=340$ for 307 DOF), to a level comparable to what we obtained for the blackbody model. It should be however noted that, while this model can reproduce rather well the data, most of the continuum below 1\,keV comes from an unrealistically large line ($\sigma=271^{+46}_{-84}$\,eV) at $E\simeq 0.2$. The results obtained applying this last model to the EPIC and RGS data are reported in column\,3 of Table\,\,\ref{tab:fitXMM1819}. In this last model we left the energy of the lines completely free to vary.
Considering the case of a non-rotating Schwarzschild SMBH ($kT_{\rm max}\simeq 9$\,eV) we obtained very similar results, with the model failing to reproduce the continuum and needing an unrealistically broad ($\sigma \simeq 300$\,eV) Gaussian line at $E<200$\,eV.

\tabletypesize{\normalsize}
\begin{deluxetable*}{llccc} 
\tablecaption{Spectral parameters obtained for the {\it XMM-Newton} June 2018 observation (EPIC+RGS) using two models: i) a multi-colour disk component, a power law, two Gaussian lines, a neutral and an ionized absorber (column\,2); ii) a Comptonized disk component, a power law, two Gaussian lines, a neutral and an ionized absorber (column\,3).\label{tab:fitXMM1819}}
\tablehead{
 \colhead{ } & \colhead{(1) } & \colhead{(2)} &  \colhead{(3)}  \\
\noalign{\smallskip}
 \colhead{ } & \colhead{ } & \colhead{Multi-colour disk} &  \colhead{Comptonized disk}  
}
\startdata
\noalign{\smallskip}
a) & $N_{\rm H}$ ($10^{20}\rm\,cm^{-2}$) & $3.3\pm0.5$	&  $ 1.5_{-0.9 }^{+1.1 }$ \\
\noalign{\smallskip}
b) & $N_{\rm H}^W$ ($10^{20}\rm\,cm^{-2}$)	 & $\leq 1.7$	&  $3.6\pm0.7$ \\
\noalign{\smallskip}
c) &	$\log \xi$ ($\rm\,erg\,cm\,s^{-1}$)  & $ 3.0_{-0.9}^{+0.2 }$	&  $2.7_{-0.8 }^{+0.2}$ \\
\noalign{\smallskip}
d) & $z$	 & $-0.216 _{-0.018}^{+0.017 }$   &  $-0.191_{-0.006 }^{+0.007 }$ \\
\noalign{\smallskip}
e) & $\Gamma$	 & $ 0.7_{-0.7}^{+0.6 }$	&  $ -0.2_{-2.7 }^{+1.9 }$ \\
\noalign{\smallskip}
f) & $E_1$ (keV)	 & $ 1.00\pm 0.01$	&  $ 1.00_{-0.01 }^{+0.02 }$ \\
\noalign{\smallskip}
g) & $\sigma_1$ (eV)  & $ 92\pm6$	&  $109_{-8 }^{+6 }$ \\
\noalign{\smallskip}
h) & $EW_1$ (eV)	 & $ 142_{-9}^{+6 }$	&  $268_{-206}^{+404 }$ \\
\noalign{\smallskip}
i) & $E_2$ (keV)	 & $ 1.81_{-0.04}^{+0.05 }$	&  $ 0.02_{-0.02 }^{+0.16}$ \\
\noalign{\smallskip}
j) & $\sigma_2$ (eV)  & $ 162_{-44}^{+56}$	&  $ 312_{-16 }^{+10 }$ \\
\noalign{\smallskip}
k) & $EW_2$ (eV)	 & $ 767_{-163}^{+ 381}$	&  $ 1076_{-184 }^{+9436 }$ \\
\noalign{\smallskip}
l) & $C_{\rm MOS}$ 	 & $ 1.04\pm 0.01$	&  $1.05\pm0.01$ \\
\noalign{\smallskip}
m) & $C_{\rm RGS}$	 & $ 0.92\pm 0.01$	&  $ 0.93\pm0.01$ \\
\noalign{\smallskip}
n) & $kT_{\rm MD}$ (eV)	 	& $ 122\pm 1$  & \nodata \\
\noalign{\smallskip}
o) & $kT_{\rm Comp.}$	 & \nodata	&  $ 5.7_{-5.7 }^{+3.1 }$ \\
\noalign{\smallskip}
p) & $\tau$	  & \nodata	&  $ 0.6_{-0.3 }^{+0.6 }$ \\
\noalign{\smallskip}
q) & Stat/DOF	  &  3218/2847	&  2940/2822  
\enddata
\tablecomments{The table reports: the column density of the neutral absorber (a); the column density (b), ionization parameter (c) and redshift (d) of the ionized absorber; the photon index of the power-law component (e); the energy (f), width (g) and equivalent width (h) of the first Gaussian line; the energy (i), width (j) and equivalent width (k) of the second Gaussian line; the cross-calibration constant of the MOS (l) and RGS (m) spectra; the temperature of the multi-colour disk component (n); the temperature (o) and optical depth (p) of the Comptonized disk; the value of the statistic and the number of degrees of freedom (q).}
\end{deluxetable*}

\subsection{The {\it XMM-Newton}  December 2018 observation}\label{appendix:dec2018}

\subsubsection{Multicolour disk model}\label{sect:xmmobs_spec_dec_diskbb}

Following what was done for the June 2018 observation (\S\ref{sect:xmmobs_spec_june_diskbb}) we applied a multicolour disk model instead of a blackbody model [\textsc{cons$\times$tbabs$\times$ztbabs$\times$(zashift$\times$diskbb+zpo)}]. This resulted in a poor fit ($\chi^{2}$/DOF=1822/567), and was unable to reproduce correctly the {\it NuSTAR} spectrum, besides leaving strong residuals in the 1--3\,keV range. This model also resulted in a low cross-calibration constant ($C_{\rm Nu}=0.47\pm 0.04$).
Adding a Gaussian line improved the fit ($\chi^{2}$/DOF=888/564), still leaving however strong residuals above 1\,keV and low cross-calibration constant ($C_{\rm Nu}=0.34\pm 0.03$). A cutoff powerlaw component instead of the simple powerlaw [\textsc{cons$\times$tbabs$\times$ztbabs$\times$(zashift$\times$diskbb+zcut)}] allowed to reproduce better the $E\gtrsim 3$\,keV part of the X-ray spectrum ($\chi^{2}$/DOF=679/563).
We then added a warm absorber and a second Gaussian emission line [\textsc{cons$\times$tbabs$\times$ztbabs$\times$mtable\{xstar\}$\times$} \textsc{(zashift$\times$diskbb+zcut+zgauss+zgauss)}], and similarly to what found with the other models, this yields a better fit ($\chi^{2}$/DOF=634/557). We then included the RGS spectrum and added two Gaussian absorption lines. Overall the model can reproduce the X-ray spectrum of 1ES\,1927+654, but not as well as the blackbody model (Stat/DOF=4481/3712). The results of this fit are reported in column\,2 of Table\,\,\ref{tab:fitXMMdec18}.

\subsubsection{Comptonized disk model}\label{sect:xmmobs_spec_dec_diskbb}

The Comptonized disk model was also applied to the {\it XMM-Newton} spectrum of the December\,\,2018 observation. Considering the higher luminosity of the source ($L_{46}=8.8\times10^{-2}$), using Eq.\,\ref{eq:tempmax}, we find that the temperature of the seed photons would be $kT_0=37$\,eV and $kT_0=13$\,eV for a maximally-rotating and Schwarzschild black hole, respectively. As done for the June 2018 observation, we started using the temperature for a maximally rotating SMBH.

Similarly to what we found for the previous {\it XMM-Newton} observation (\S\ref{sect:xmmobs_spec_june_diskbb}), this model cannot reproduce the spectral shape of the source. Considering a simple model including a Comptonized disk and a power-law component [\textsc{cons$\times$tbabs$\times$ztbabs$\times$(compTT+zpo)}] yields a chi-squared of $\chi^{2}=9048$ for 566 DOF. Strong residuals are still observed when adding a Gaussian line and, as observed for the other models, the power-law component does not reproduce correctly the curvature of the spectrum above $\sim 3$\,keV. We therefore used a cutoff power-law, which improves the fit, but still not to an acceptable level ($\chi^{2}$/DOF= 1859/562). We included a warm absorber and a second Gaussian line, as done for the previous {\it XMM-Newton} observation. This improves the fit ($\chi^{2}$/DOF=624/556) but most of the emission in the $E\gtrsim 0.5-1.5$\,keV interval comes from two extremely large Gaussian lines, suggesting that the model cannot really reproduce the X-ray spectrum. Considering a temperature of the seed photons of $kT_0=13$\,eV (for a Schwarzschild black hole) does not improve the fit significantly. We used a model including two emission lines, two absorption lines, a neutral and a warm absorber, a Comptonized disk (with $kT_0=37$\,eV), and a cutoff power law to fit all data, including the RGS spectrum. The results of this fit reported in column\,3 of Table\,\,\ref{tab:fitXMMdec18}.

\tabletypesize{\normalsize}
\begin{deluxetable*}{llcc} 
\tablecaption{Spectral parameters obtained for the {\it XMM-Newton} December 2018 observation (EPIC+RGS) using two models: i) a multi-colour disk component, a power law, two Gaussian absorption lines and two emission lines, a neutral and an ionized absorber (column\,2); ii) a Comptonized disk component, a power law, two Gaussian absorption lines and two emission lines, a neutral and an ionized absorber (column\,3).\label{tab:fitXMMdec18}}
\tablehead{
 \colhead{ } & \colhead{(1) } & \colhead{(2)} &  \colhead{(3)}  \\
\noalign{\smallskip}
 \colhead{ } & \colhead{ } & \colhead{Multi-colour disk} &  \colhead{Comptonized disk}  
}
\startdata
\noalign{\smallskip}
a) & $N_{\rm H}$ ($10^{20}\rm\,cm^{-2}$)	& $ 2.1_{-0.4}^{+0.7 }$	&  $  3.9\pm0.1$ \\
\noalign{\smallskip}
b) & $N_{\rm H}^W$ ($10^{20}\rm\,cm^{-2}$)		& $\leq 1.1$	& $\leq 1.1$ \\
\noalign{\smallskip}
c) &	$\log \xi$ ($\rm\,erg\,cm\,s^{-1}$) & $  2.91_{-0.4}^{+0.05 }$	&   $\leq 1.2$ \\
\noalign{\smallskip}
d) & $z$	&  $-0.434_{-0.005}^{+0.007 }$   &  $-0.168_{-0.004 }^{+0.003 }$ \\
\noalign{\smallskip}
e) & $\Gamma$	& $ 2.7\pm0.2$	&  $ 0.63_{-0.21 }^{+0.03 }$ \\
\noalign{\smallskip}
f) & $E_{\rm cut}$	& $  2.30_{-0.3}^{+0.03 }$	&  $ 0.57_{-0.09 }^{+0.11 }$ \\
\noalign{\smallskip}
g) & $E_1$ (keV)	& $  1.022\pm 0.004$	&  $  0.92_{-0.04 }^{+0.01 }$ \\
\noalign{\smallskip}
h) & $\sigma_1$ (eV) &  $  175\pm4$	&  $ 222_{-2 }^{+13 }$ \\
\noalign{\smallskip}
i) & $EW_1$ (eV)	&  $  130_{-4}^{+5 }$	&  $ 416_{-206}^{+404 }$ \\
\noalign{\smallskip}
j) & $E_2$ (keV)	& $  1.95\pm0.05$	&  $  0.455\pm0.002$ \\
\noalign{\smallskip}
k) & $\sigma_2$ (eV) &  $  \geq 219$	&  $ \geq 200$ \\
\noalign{\smallskip}
l) & $EW_2$ (eV)	& $  88_{-8 }^{+10  }$	&  $  274_{-121}^{+NC}$ \\
\noalign{\smallskip}
m) & $C_{\rm Nu}$ 	& $  0.76\pm 0.04$	&  $ 0.97\pm0.05$ \\
\noalign{\smallskip}
n) & $C_{\rm RGS}$	& $ 0.917 \pm 0.004$	&  $  0.92\pm0.01$ \\
\noalign{\smallskip}
o) & $E_{\rm abs1}$ (keV)	&  $  0.383\pm0.004$	 & $  0.396\pm0.001$ \\
\noalign{\smallskip}
p) & $\sigma_{\rm abs1}$ (eV) & $ 56\pm 4$	&  $ 0.2_{-0.1 }^{+0.5 }$ \\
\noalign{\smallskip}
q) & Strength ($10^{-3}$)   & $  25\pm1$	&  $ 2.7_{-1.8 }^{+11.4 }$ \\
\noalign{\smallskip}
r) & $E_{\rm abs2}$ (keV)	& $  0.527\pm0.001$	 & $  0.527\pm0.001$ \\
\noalign{\smallskip}
s) & $\sigma_{\rm abs2}$ (eV) & $ 0.5\pm0.1$	&  $ 0.8_{-0.5 }^{+1.0 }$ \\
\noalign{\smallskip}
t) & Strength ($10^{-3}$)   & $ 3.0_{-0.8}^{+0.9}$	&  $ 1.0\pm 0.5$ \\
\noalign{\smallskip}
u) & $kT_{\rm MD}$ (eV)	& $ 181 \pm 1$  & \nodata \\
\noalign{\smallskip}
v) & $kT_{\rm Comp.}$	& \nodata	&  $  31.5\pm0.2$ \\
\noalign{\smallskip}
y) & $\tau$	& \nodata	&  $  0.97_{-0.02 }^{+0.01 }$ \\
\noalign{\smallskip}
x) & Stat/DOF	 &  4481/3712	&  4152/3711
\enddata
\tablecomments{The lines report: the column density of the neutral absorber (a); the column density (b), ionization parameter (c) and redshift (d) of the ionized absorber; the photon index (e) and cutoff (f) of the power-law component; the energy (g), width (h) and equivalent width (i) of the first Gaussian emission line; the energy (j), width (k) and equivalent width (l) of the second Gaussian line; the cross-calibration constant of the {\it NuSTAR}/FPM (m) and {\it XMM-Newton} RGS (n) spectra; the energy (o), width (p) and strength (q) of the first Gaussian absorption line; the energy (r), width (s) and strength (t) of the second Gaussian absorption line; the temperature of the multi-colour disk component (u); the temperature (v) and optical depth (y) of the Comptonized disk; the value of the statistic and the number of degrees of freedom (x).}
\end{deluxetable*}

\clearpage

\section{The {\it XMM-Newton} light curves }\label{appendix:XMM_lightcurves}

\begin{figure}[h!]
  \begin{center}
  \par\medskip
\includegraphics[width=0.51\textwidth]{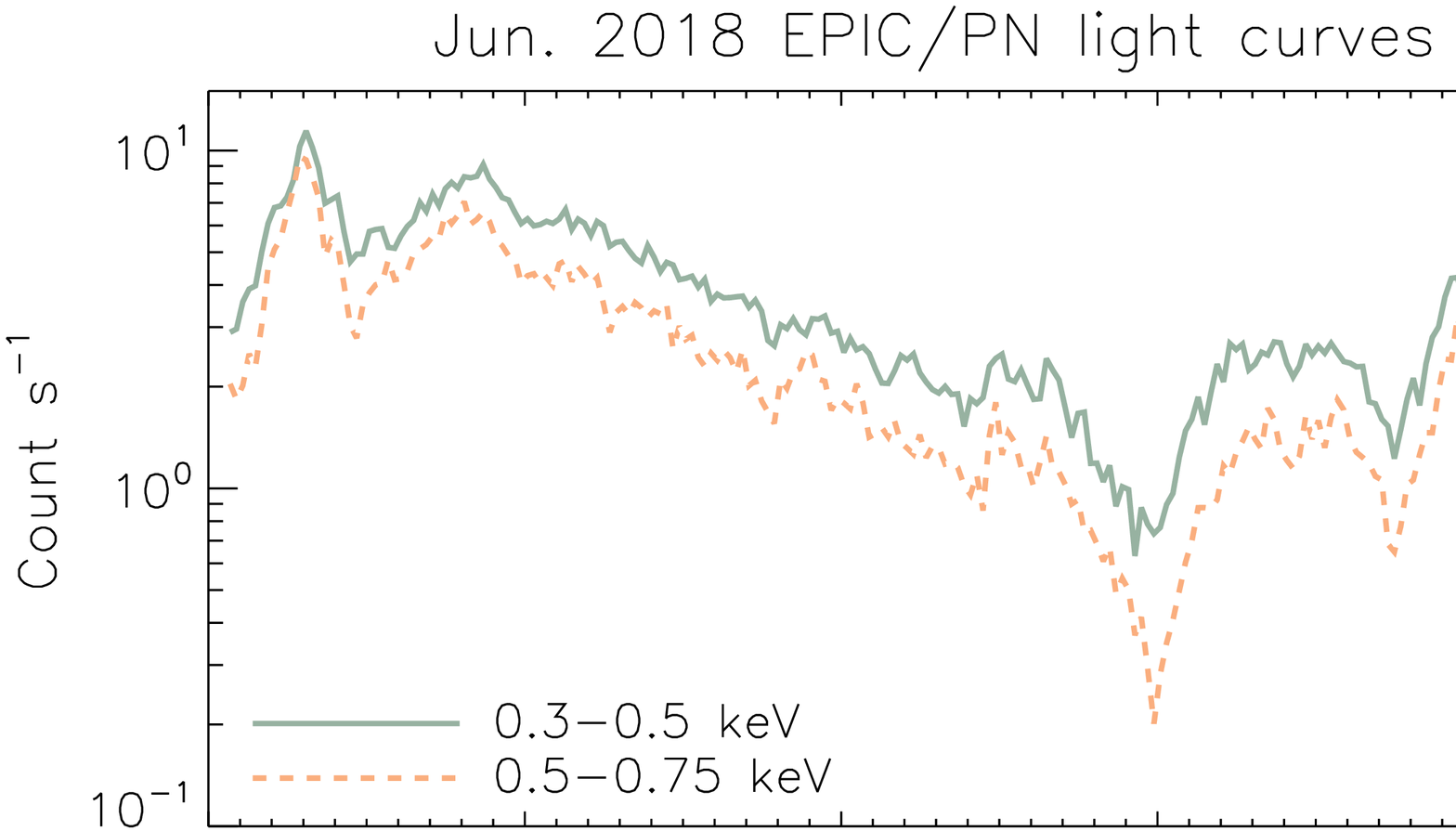}
\includegraphics[width=0.51\textwidth]{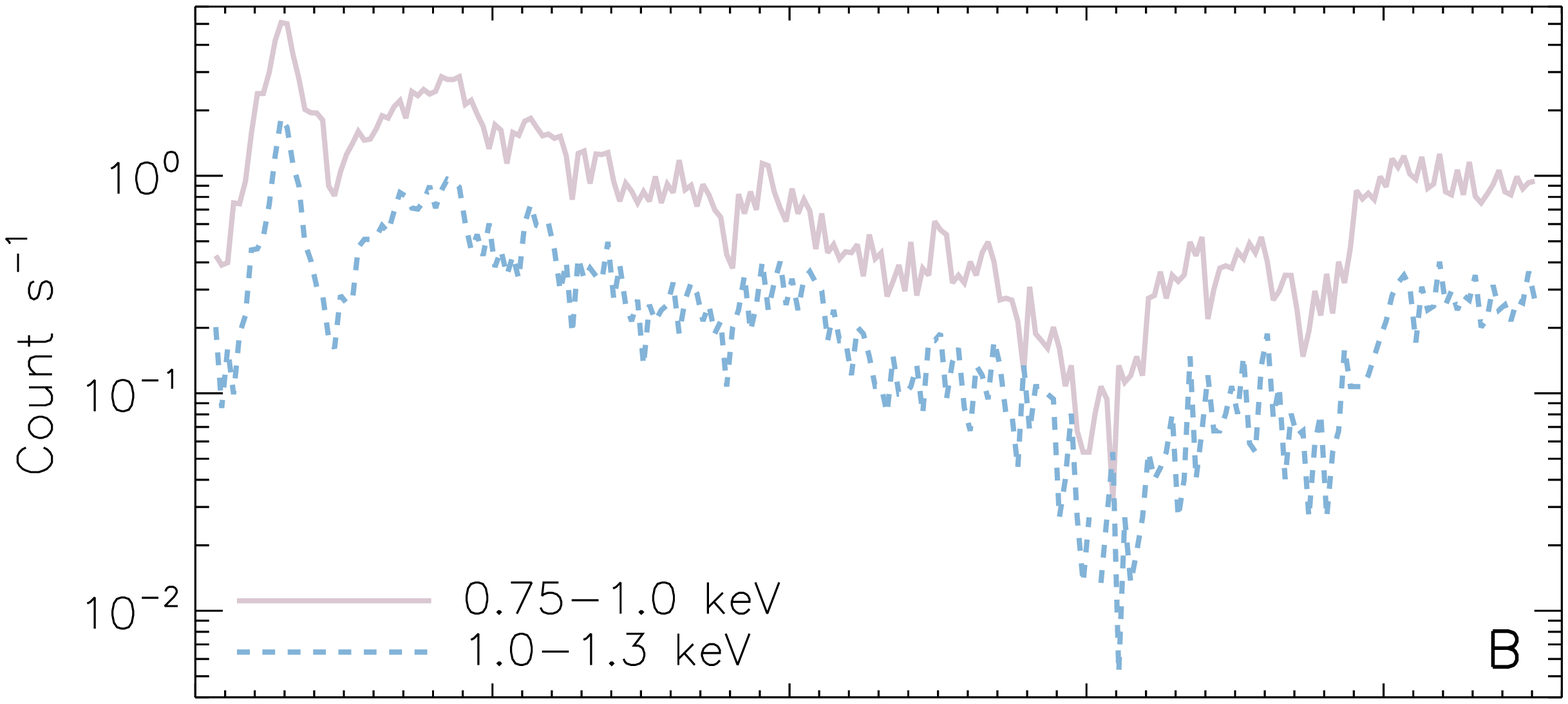}
  \includegraphics[width=0.51\textwidth]{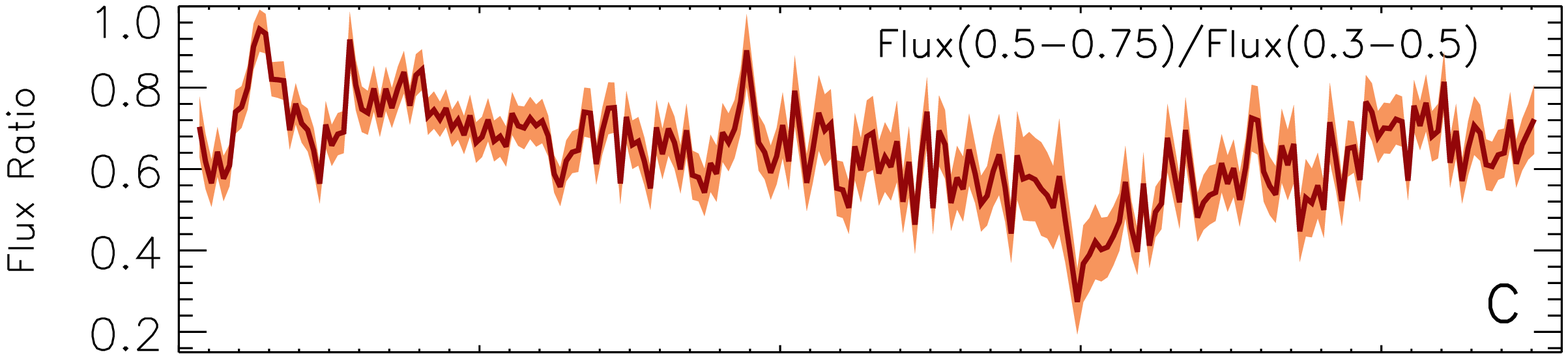}
  \includegraphics[width=0.51\textwidth]{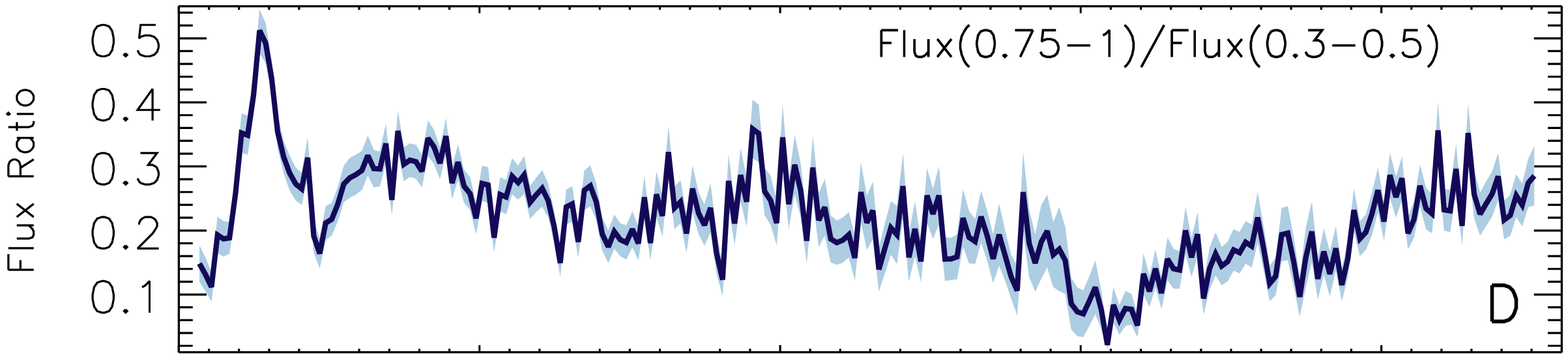}
  \includegraphics[width=0.51\textwidth]{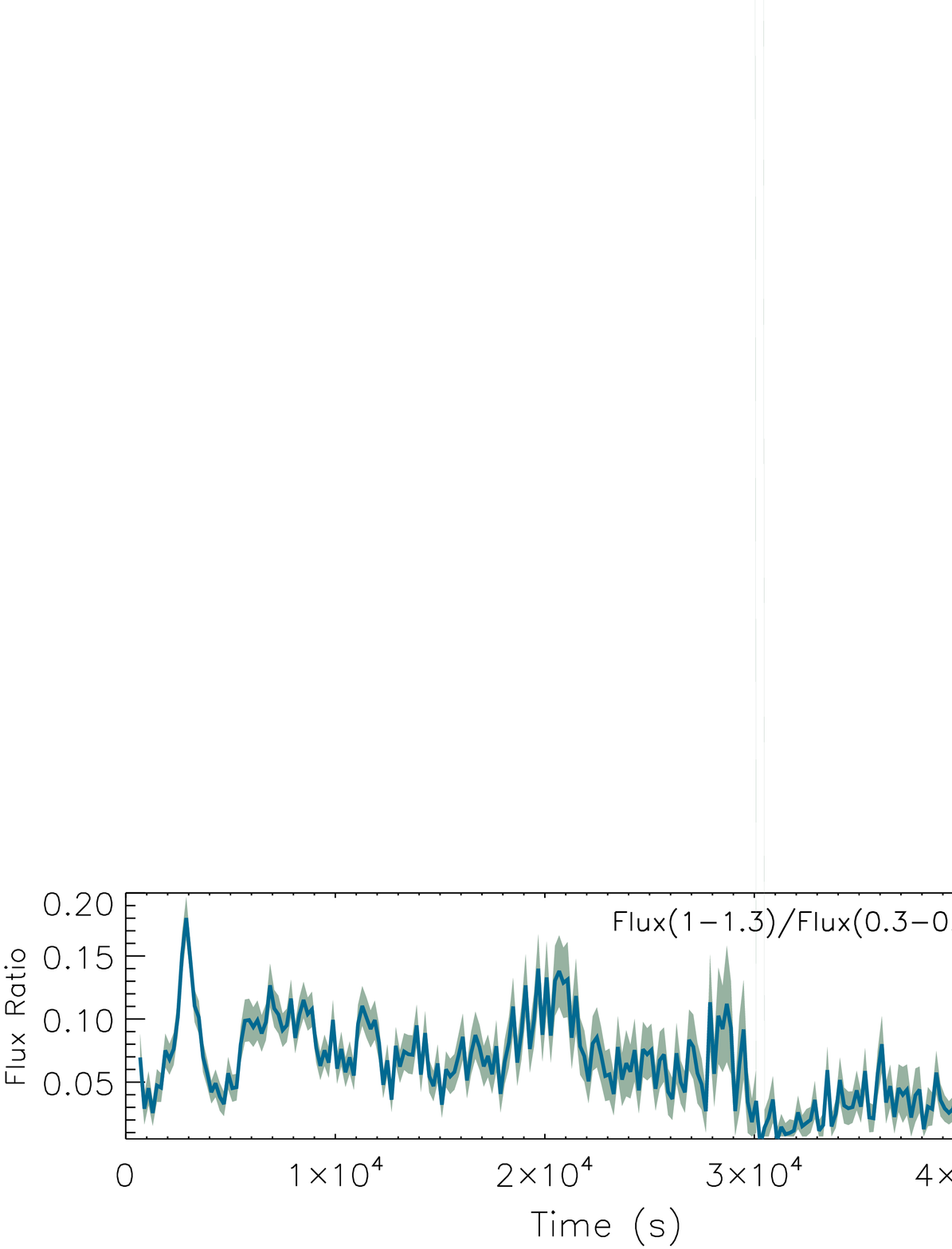}
    \caption{{\it XMM-Newton} EPIC/PN light curves and flux ratios of 1ES\,1927+654 during the June\,\,2018 observation. Panels A and B: light curves in four different energy bands, showing that the strongest variability is observed in the 0.75--1\,keV (continuous pink line) and 1.0--1.3\,keV (dashed cyan line) bands. Panels C--E: ratios between the flux in three different bands and that in the 0.3--0.5\,keV range.}
    \label{fig:XMMlc_18_bands}
  \end{center}
\end{figure}

\begin{figure*}[b!]
  \begin{center}
  \par\medskip
\includegraphics[width=0.4\textwidth]{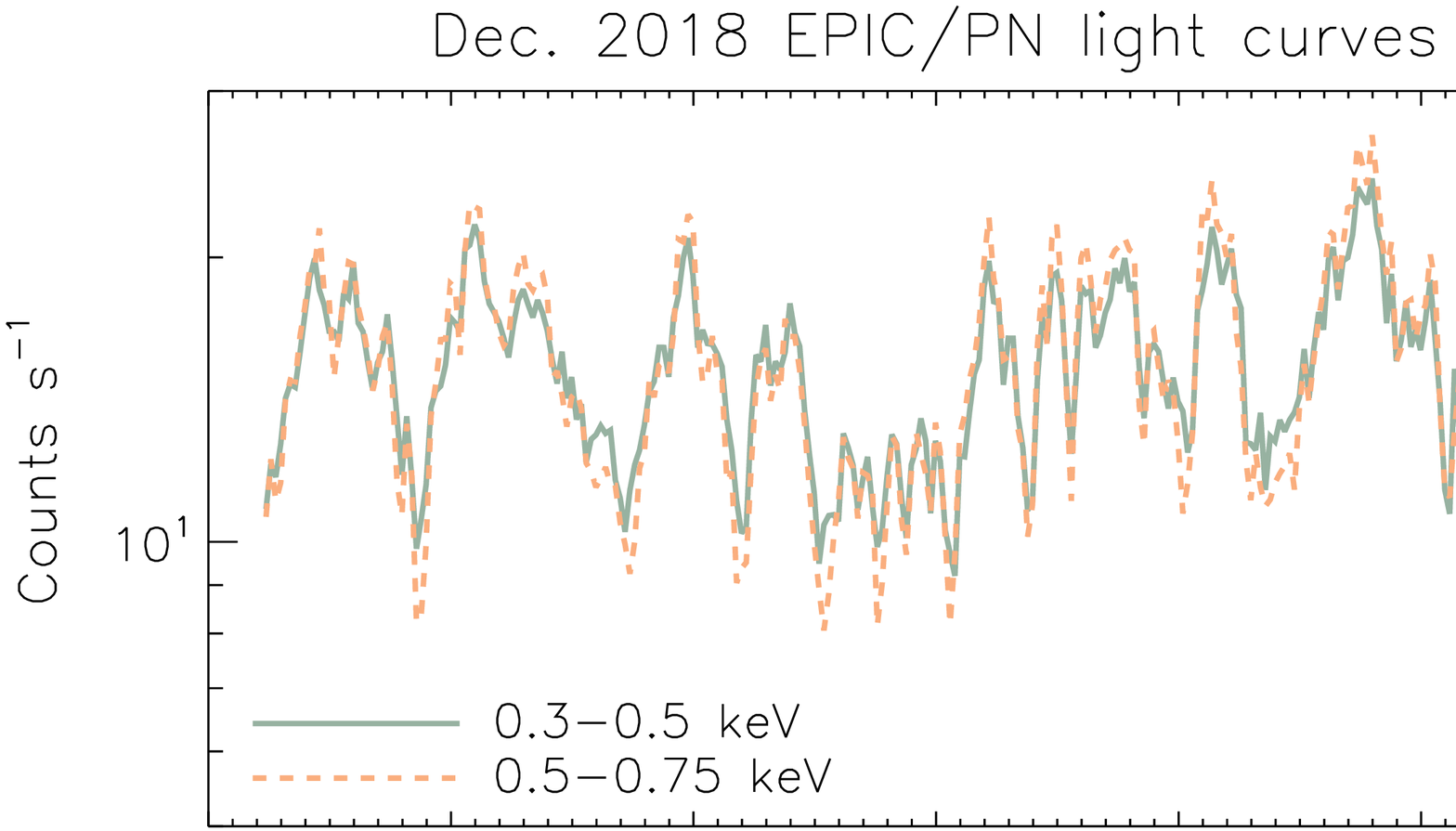}
\includegraphics[width=0.4\textwidth]{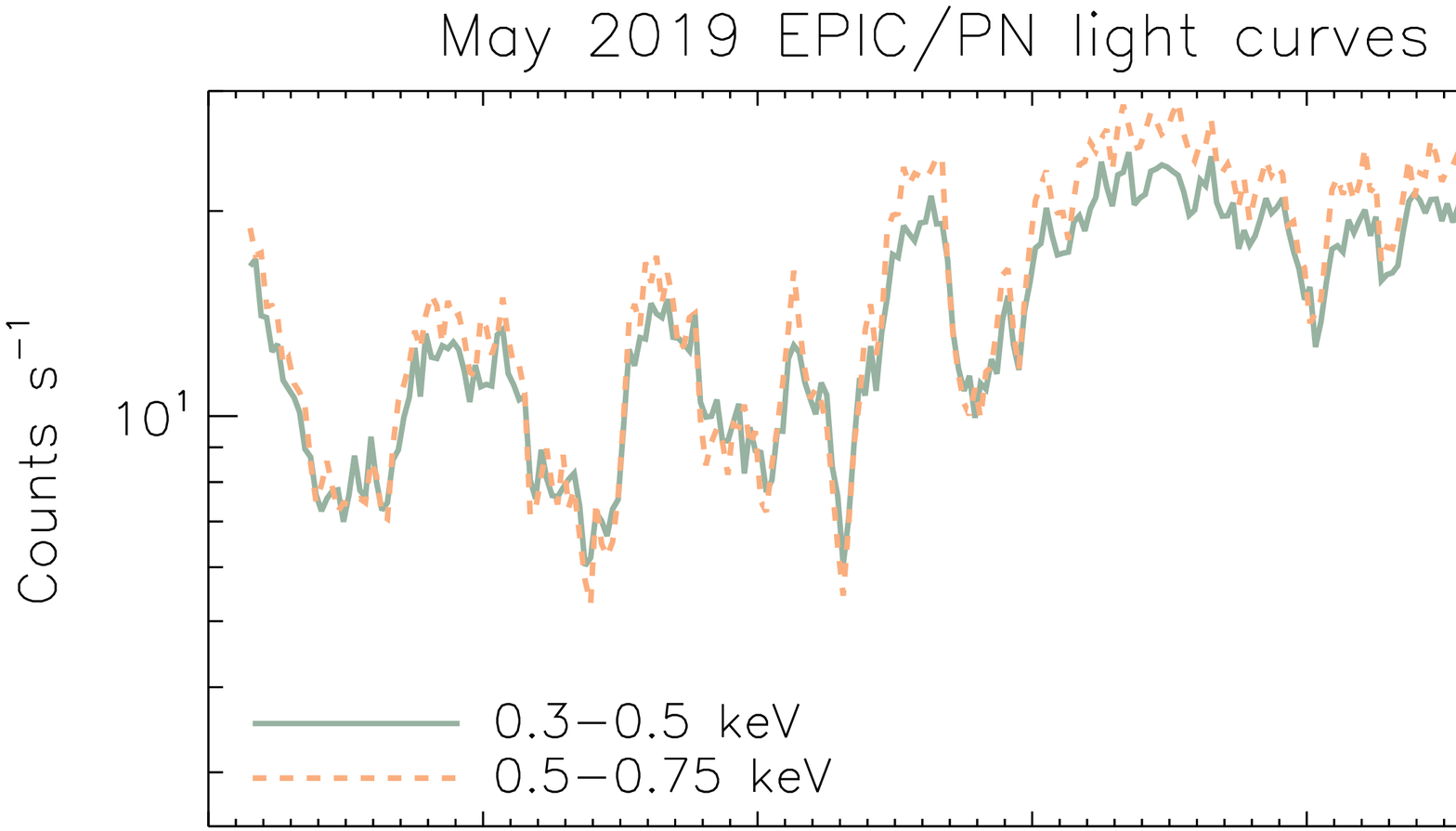}
\includegraphics[width=0.4\textwidth]{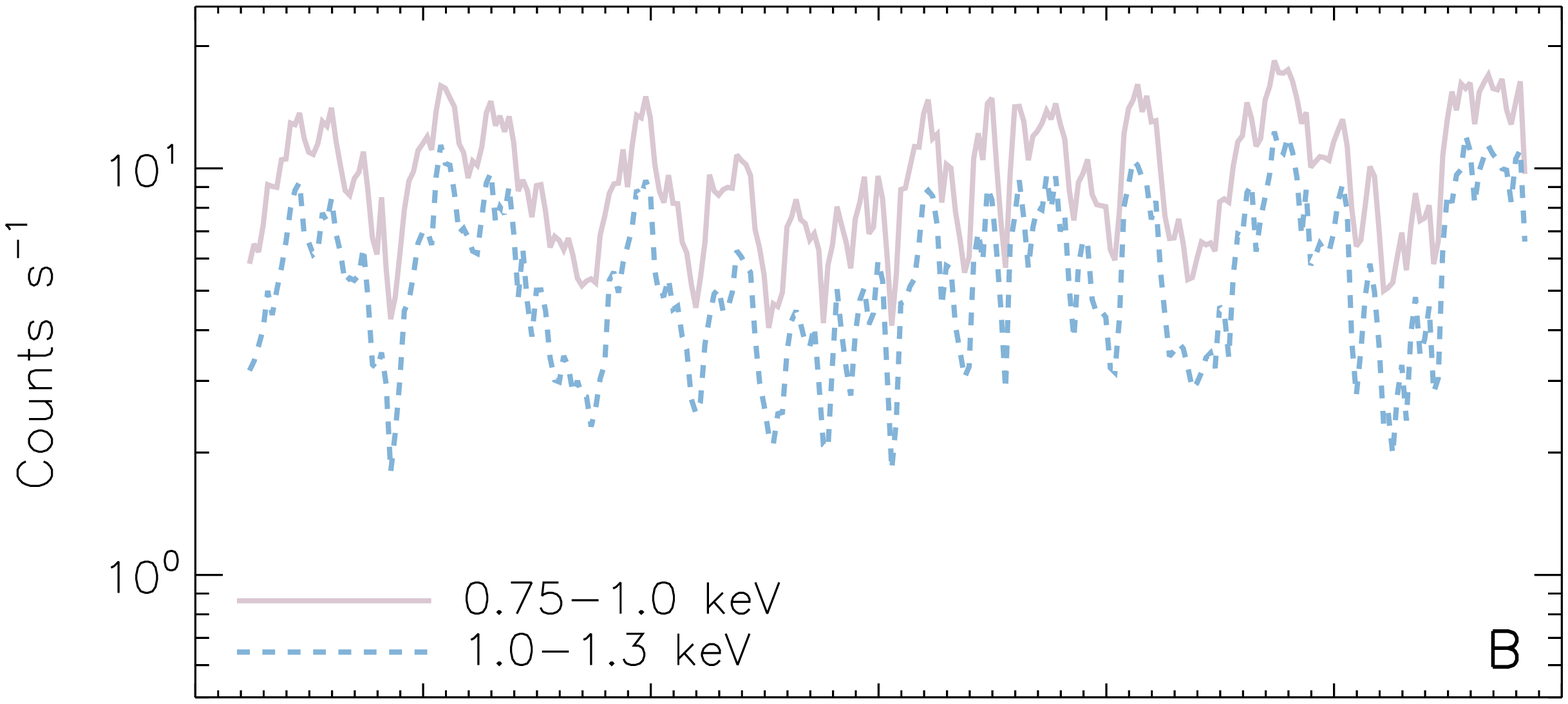}
\includegraphics[width=0.4\textwidth]{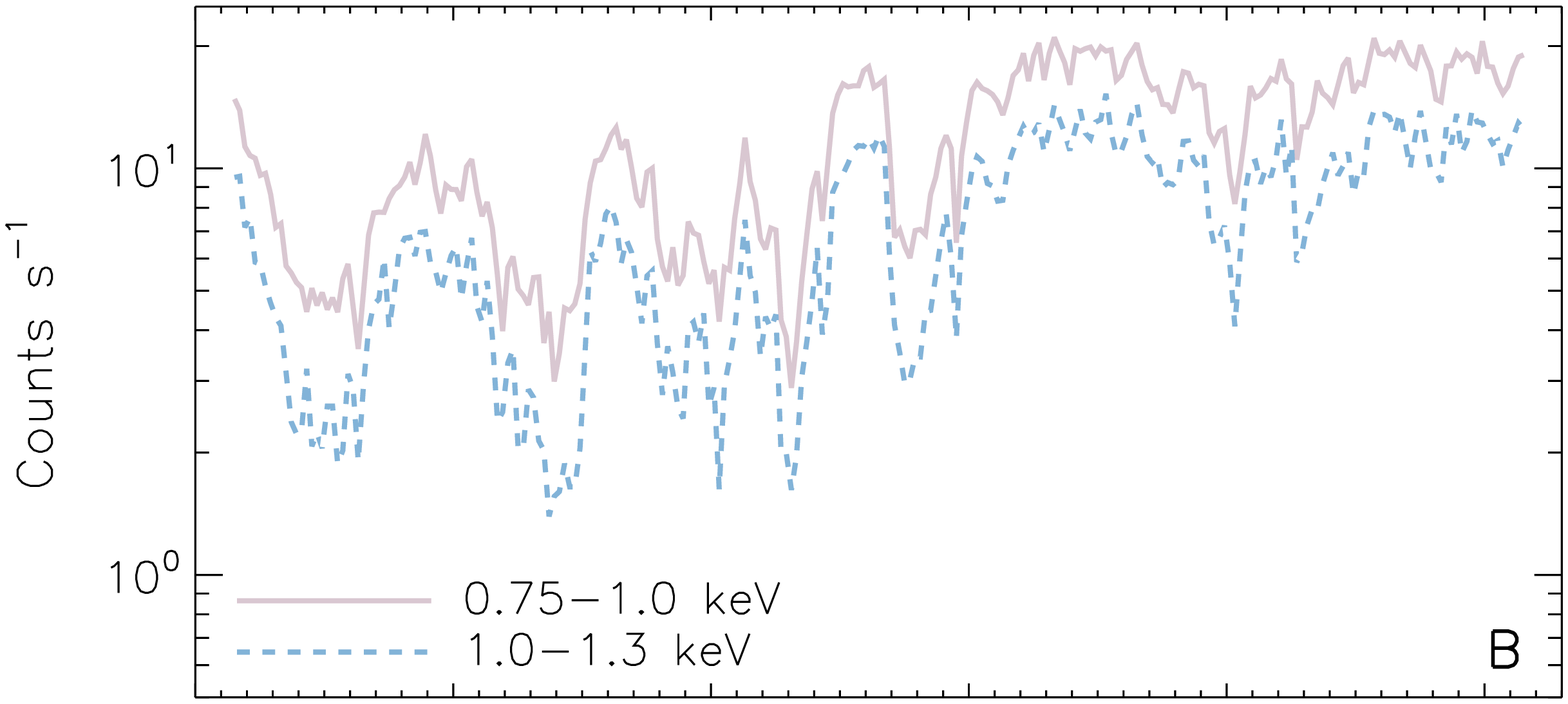}
  \includegraphics[width=0.4\textwidth]{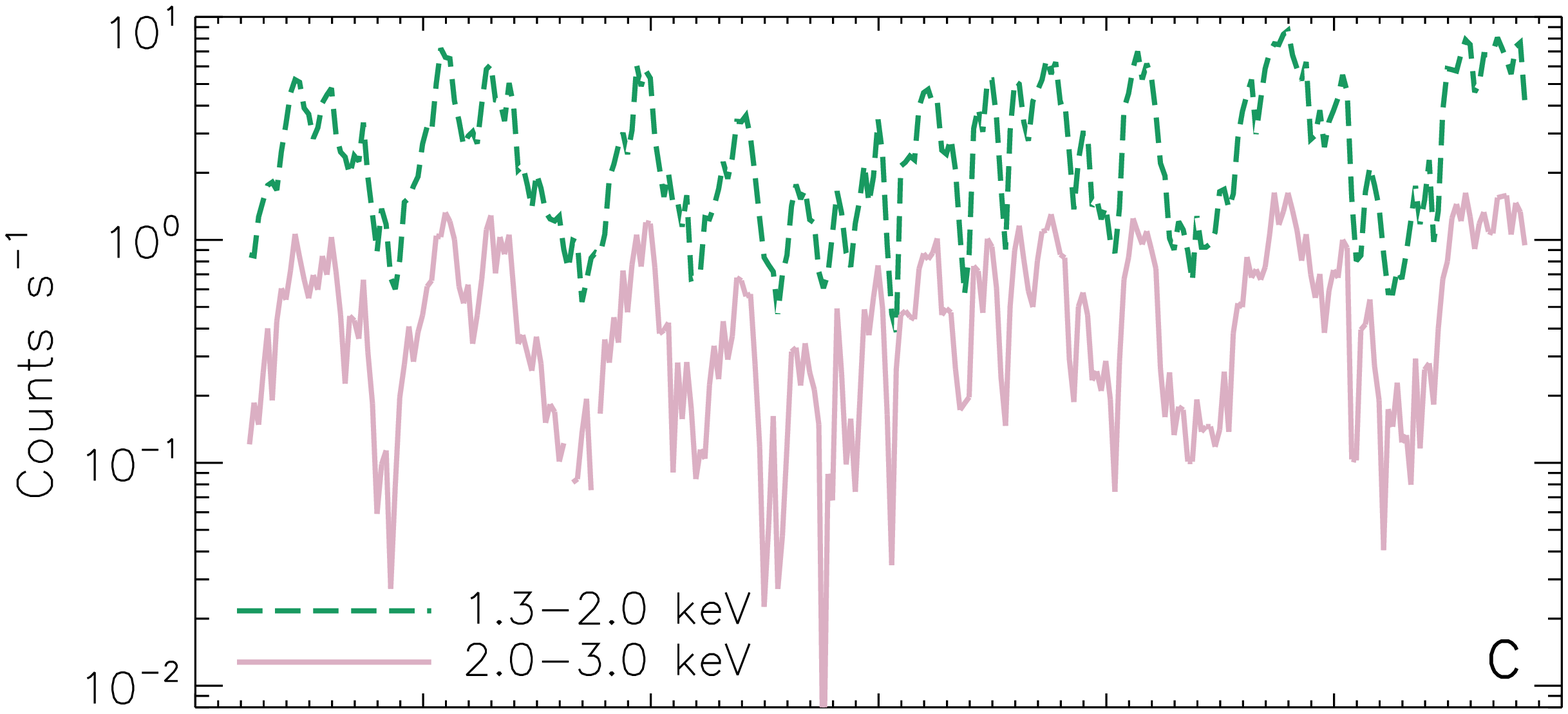}
  \includegraphics[width=0.4\textwidth]{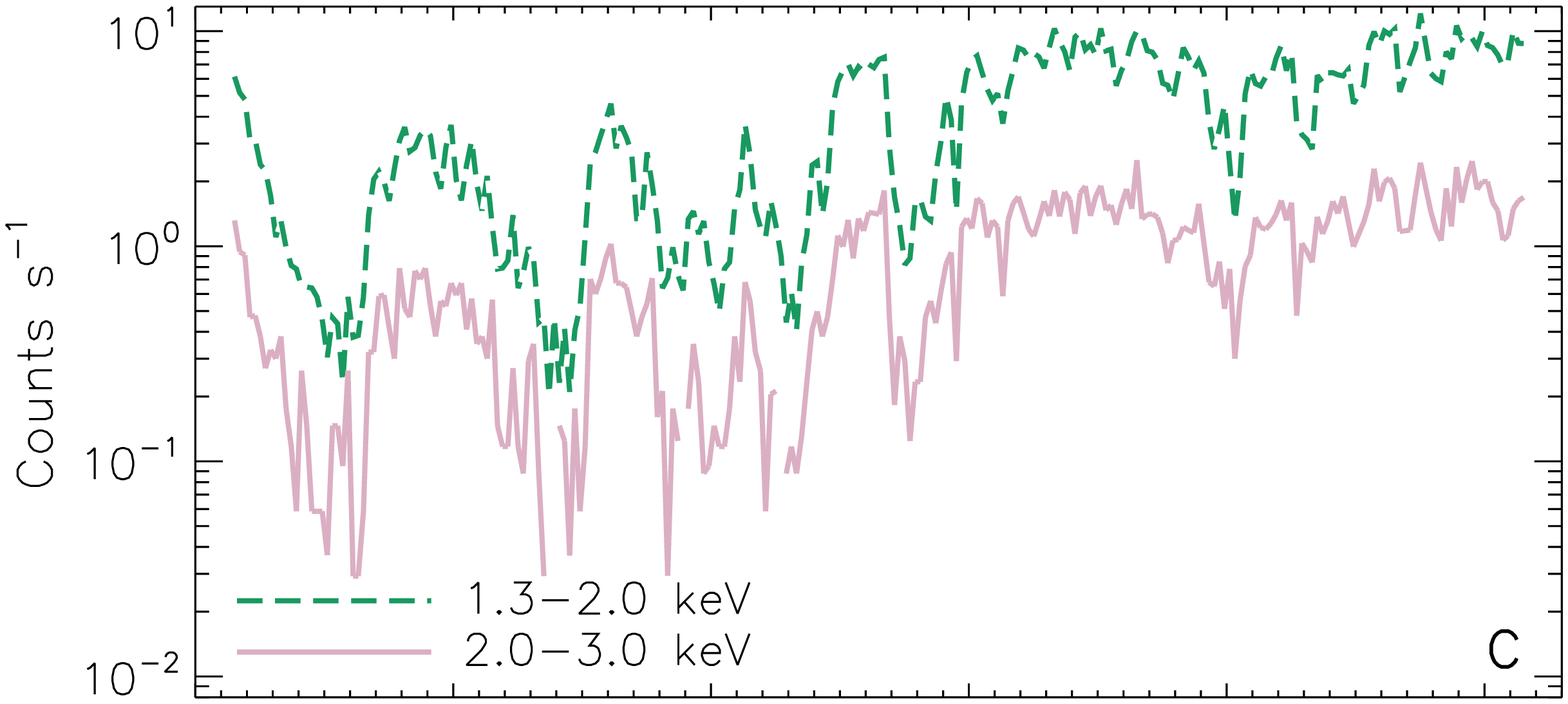}
  \includegraphics[width=0.4\textwidth]{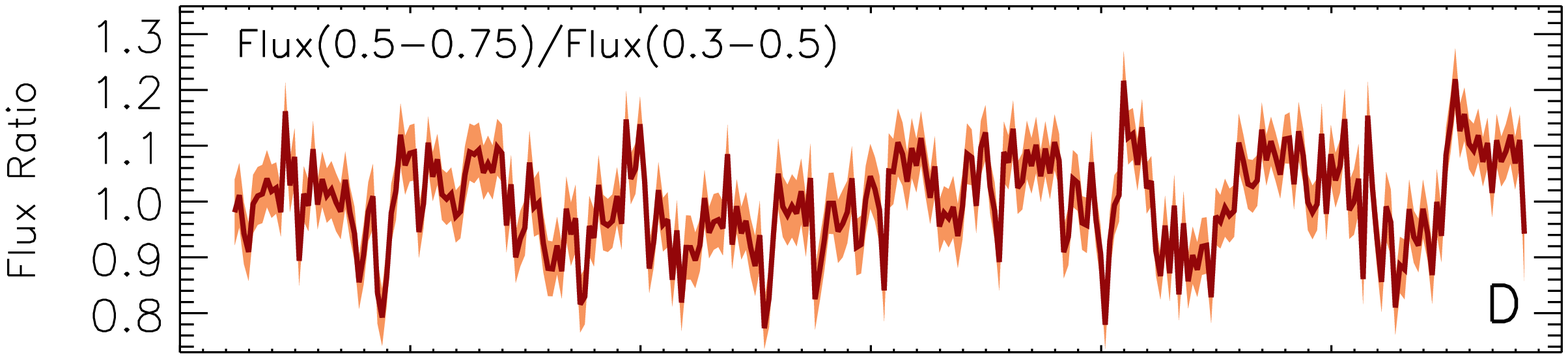}
  \includegraphics[width=0.4\textwidth]{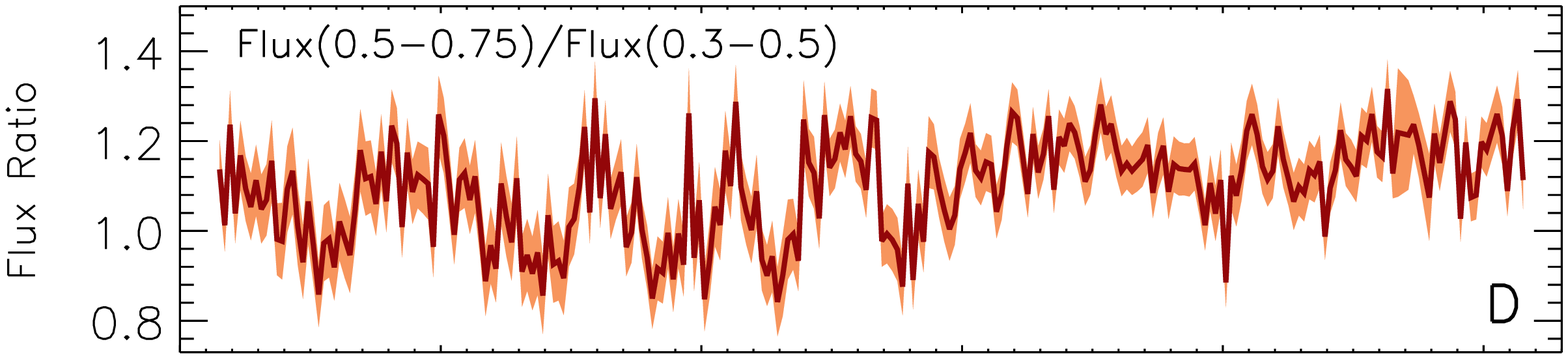}
    \includegraphics[width=0.4\textwidth]{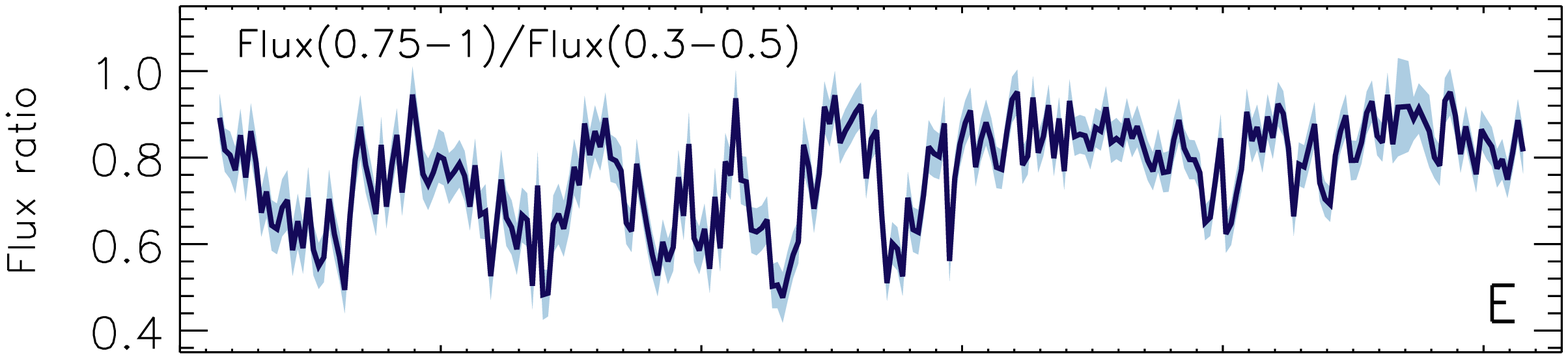}
  \includegraphics[width=0.4\textwidth]{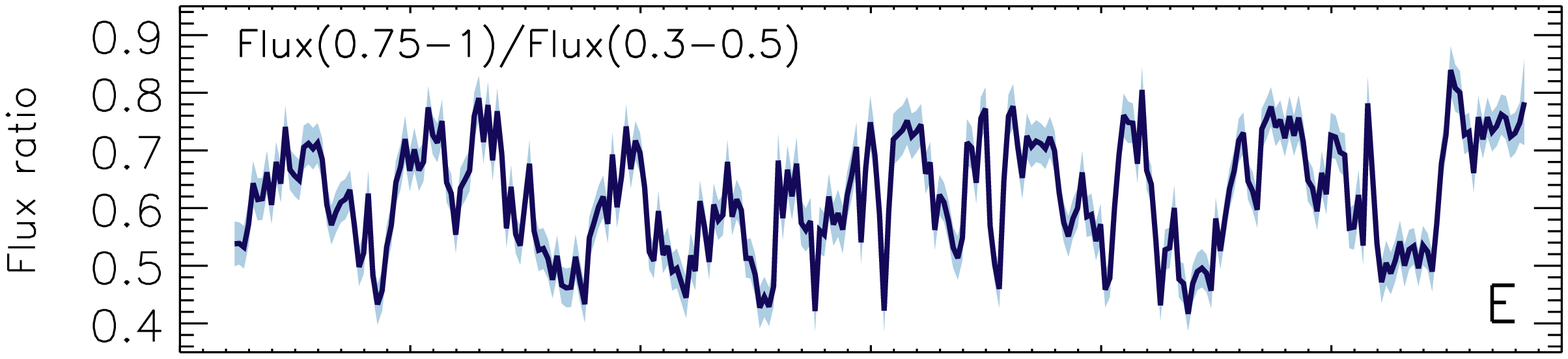}
  \includegraphics[width=0.4\textwidth]{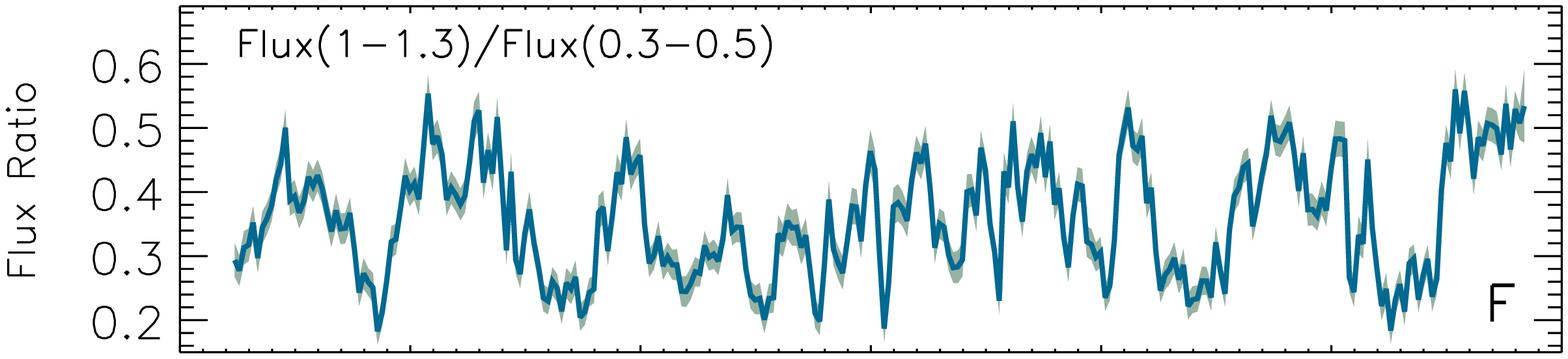}
  \includegraphics[width=0.4\textwidth]{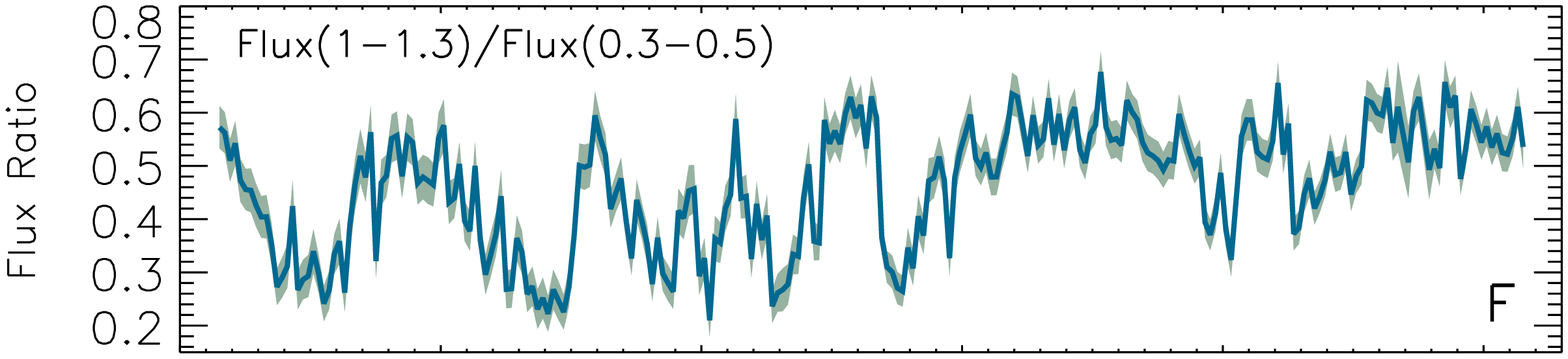}
  \includegraphics[width=0.4\textwidth]{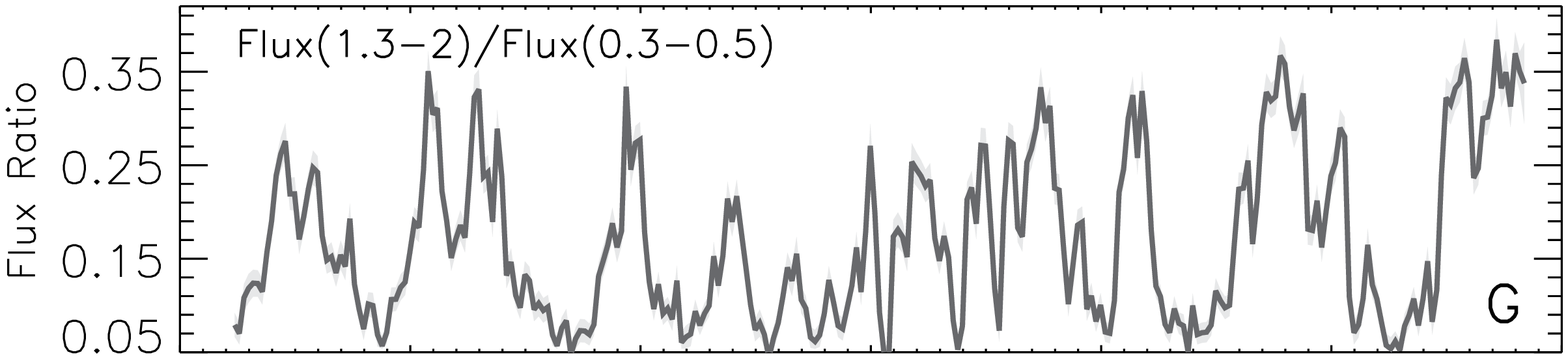}
  \includegraphics[width=0.4\textwidth]{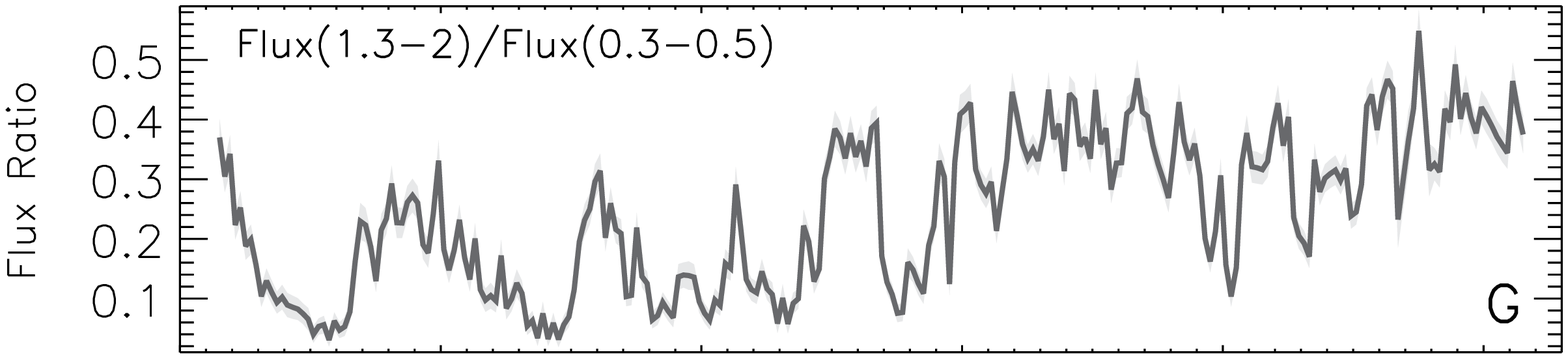}
  \includegraphics[width=0.4\textwidth]{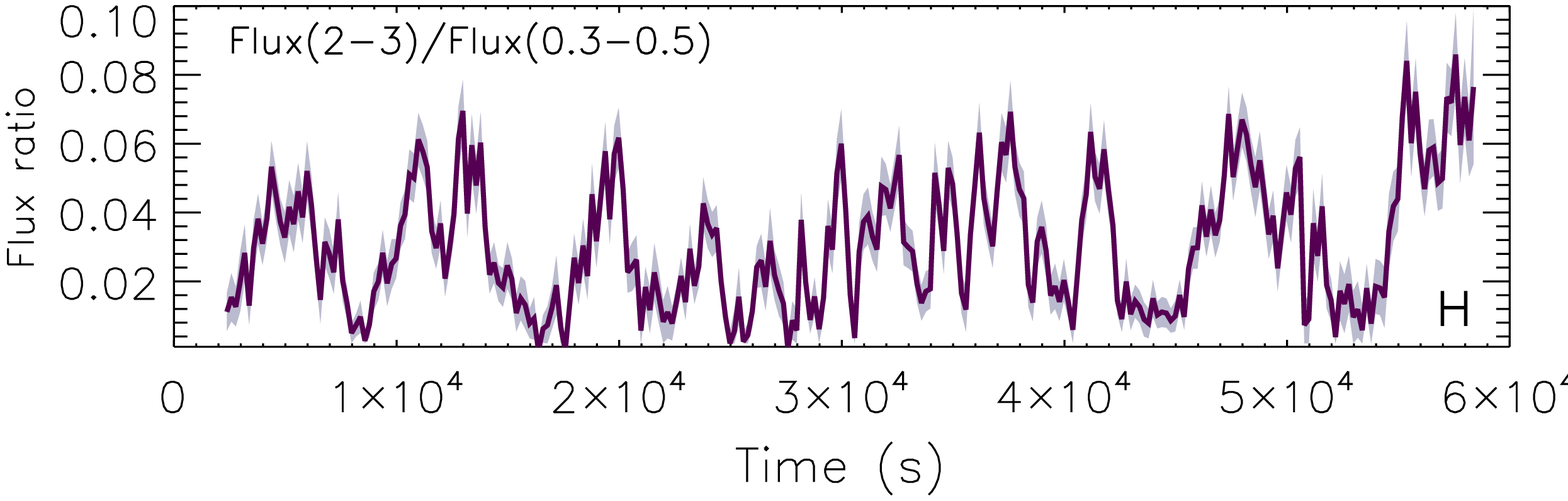}
  \includegraphics[width=0.4\textwidth]{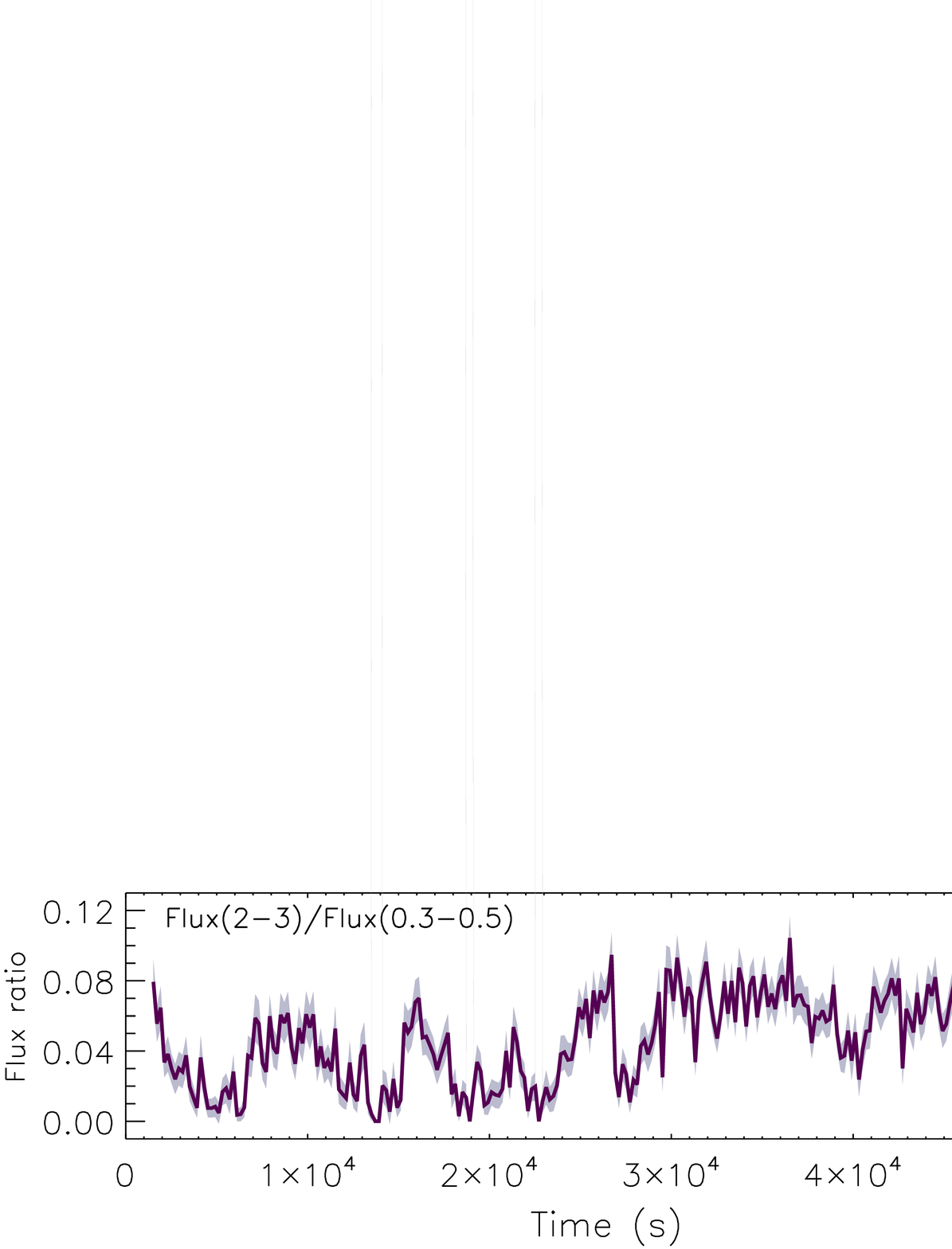}
    \caption{{\it XMM-Newton} EPIC/PN light curves and flux ratios of 1ES\,1927+654 during the December\,\,2018 ({\it left panels}) and May\,\,2019 ({\it right panels}) observations. Panels A--C: light curves in six different energy bands, showing that the strongest variability is observed in the 1.3--2\,keV (dashed green line) and 2--3\,keV (continuous pink line) bands. Panels D--H: ratios between the flux in five different bands and that in the 0.3--0.5\,keV range.}
    \label{fig:XMMlc_18_december_bands}
  \end{center}
\end{figure*}

\begin{figure*}
  \begin{center}
\includegraphics[width=0.48\textwidth]{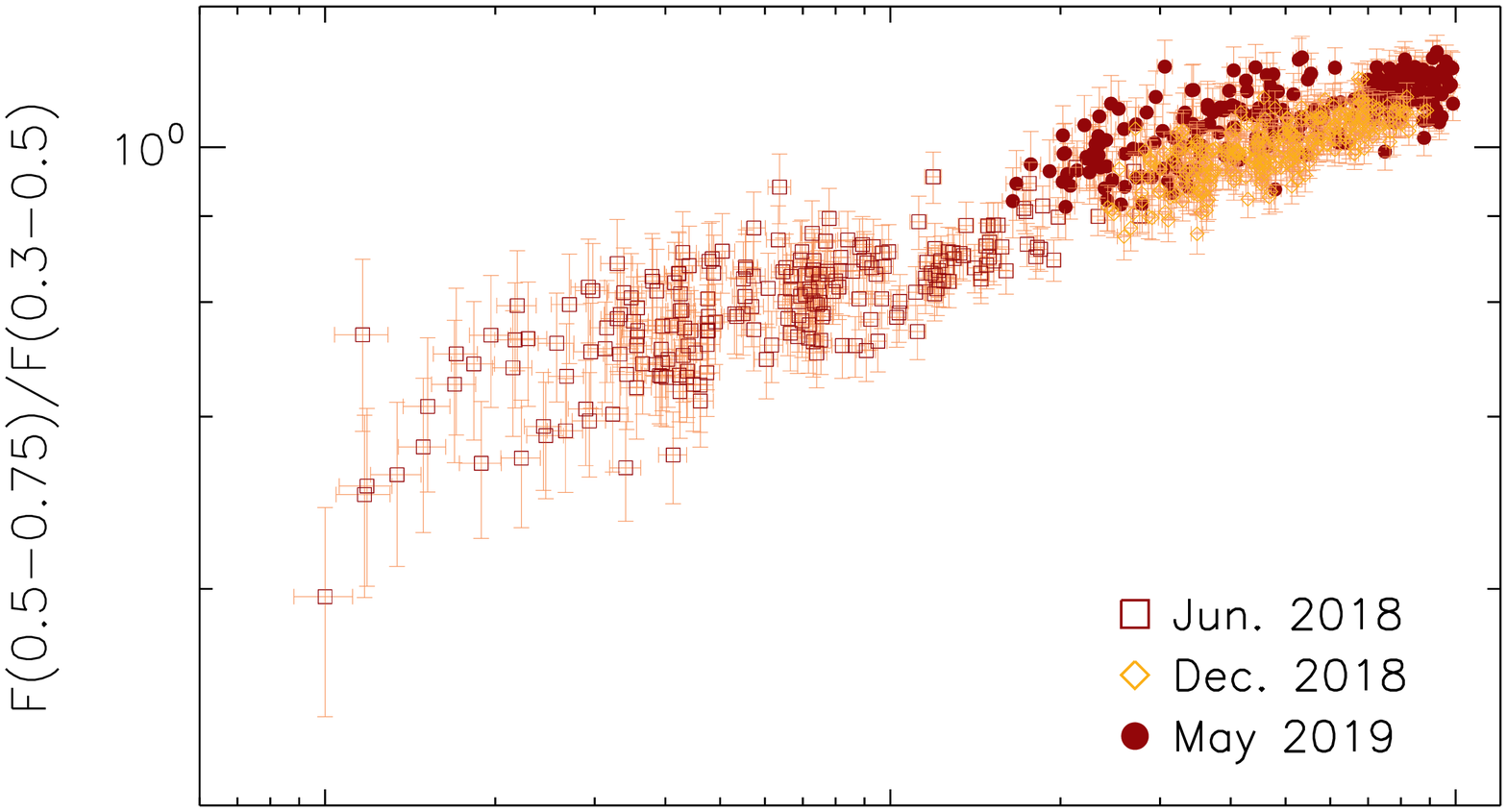}
\includegraphics[width=0.48\textwidth]{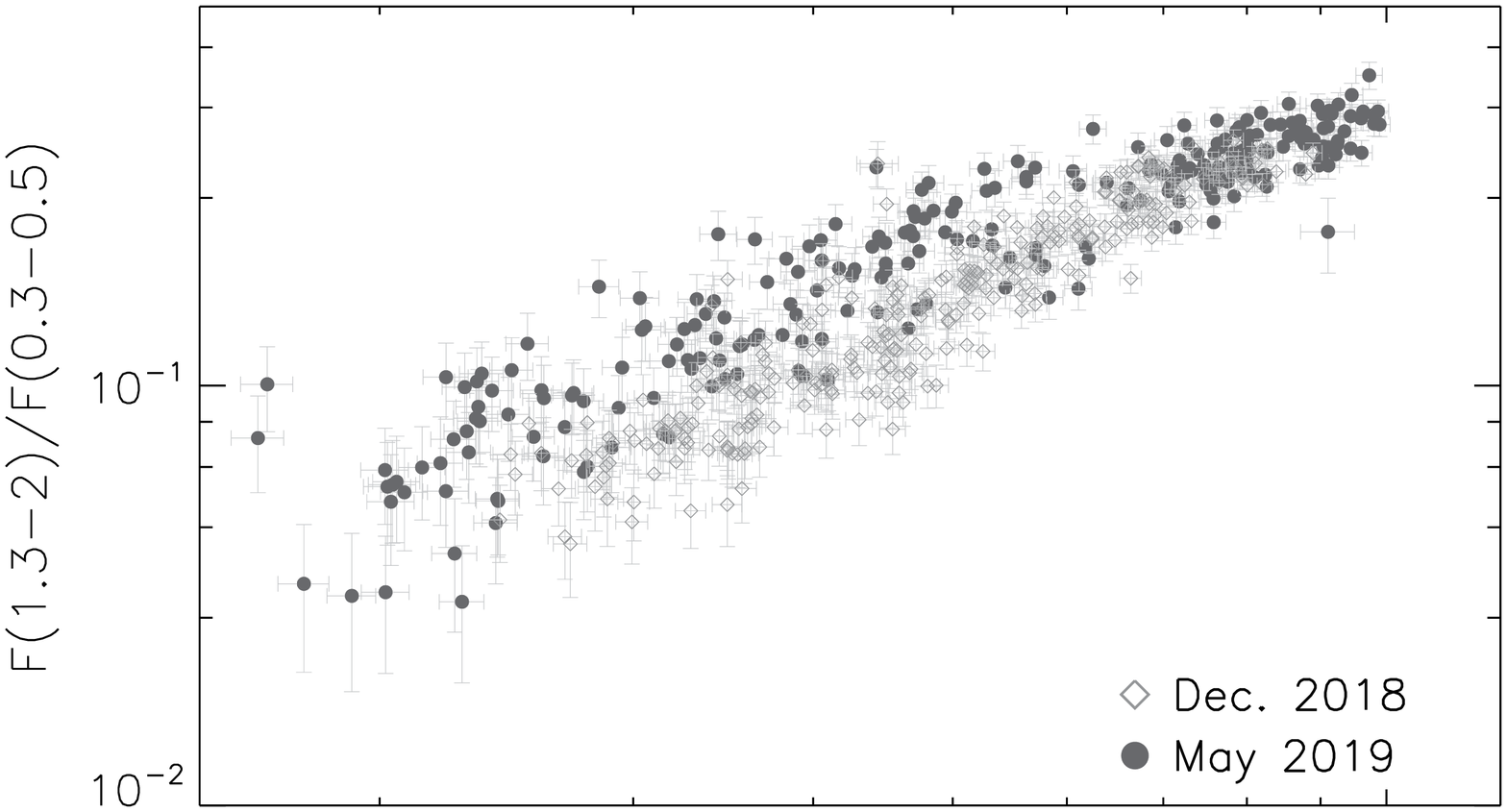}
\includegraphics[width=0.48\textwidth]{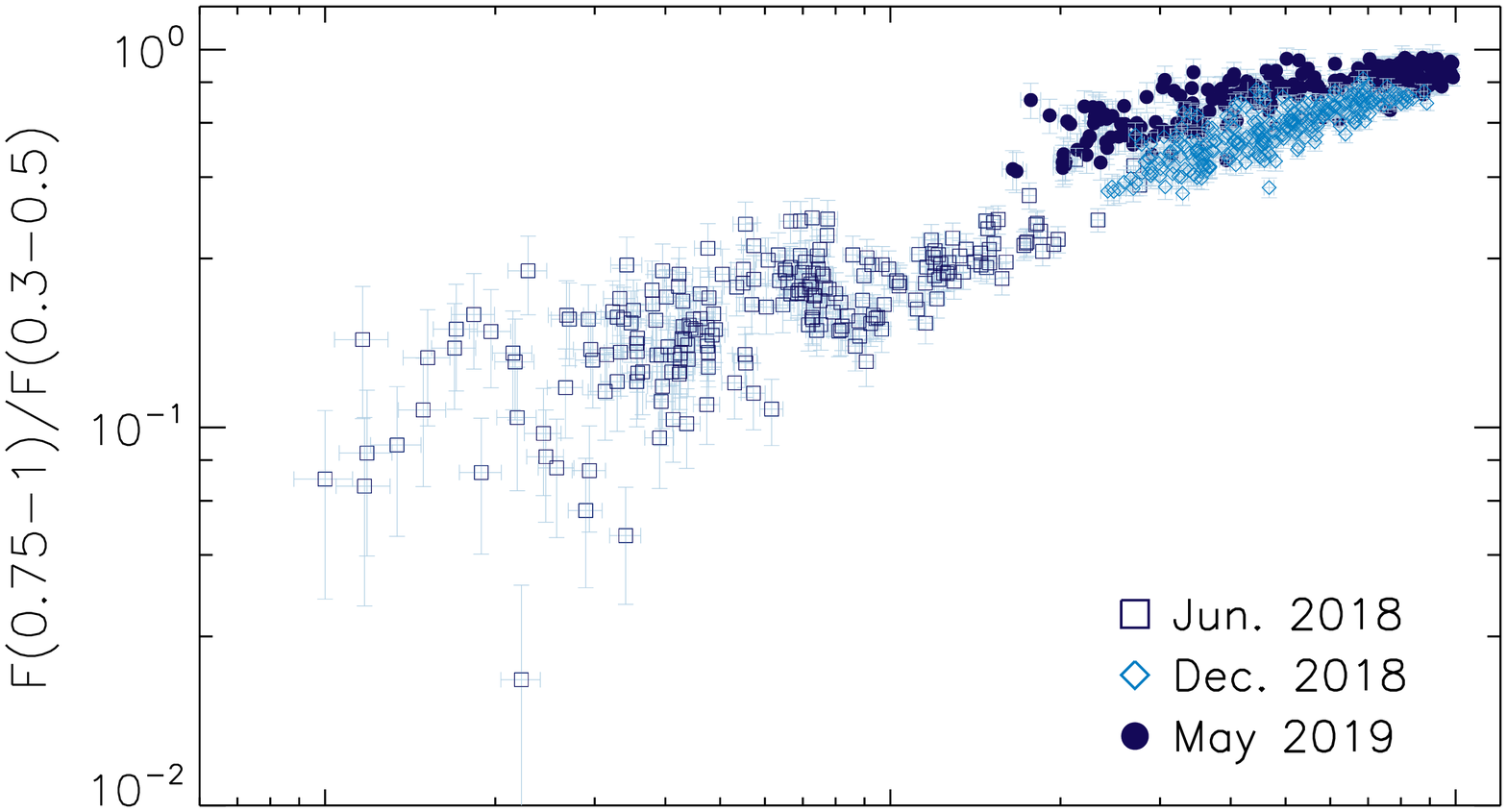}
\includegraphics[width=0.48\textwidth]{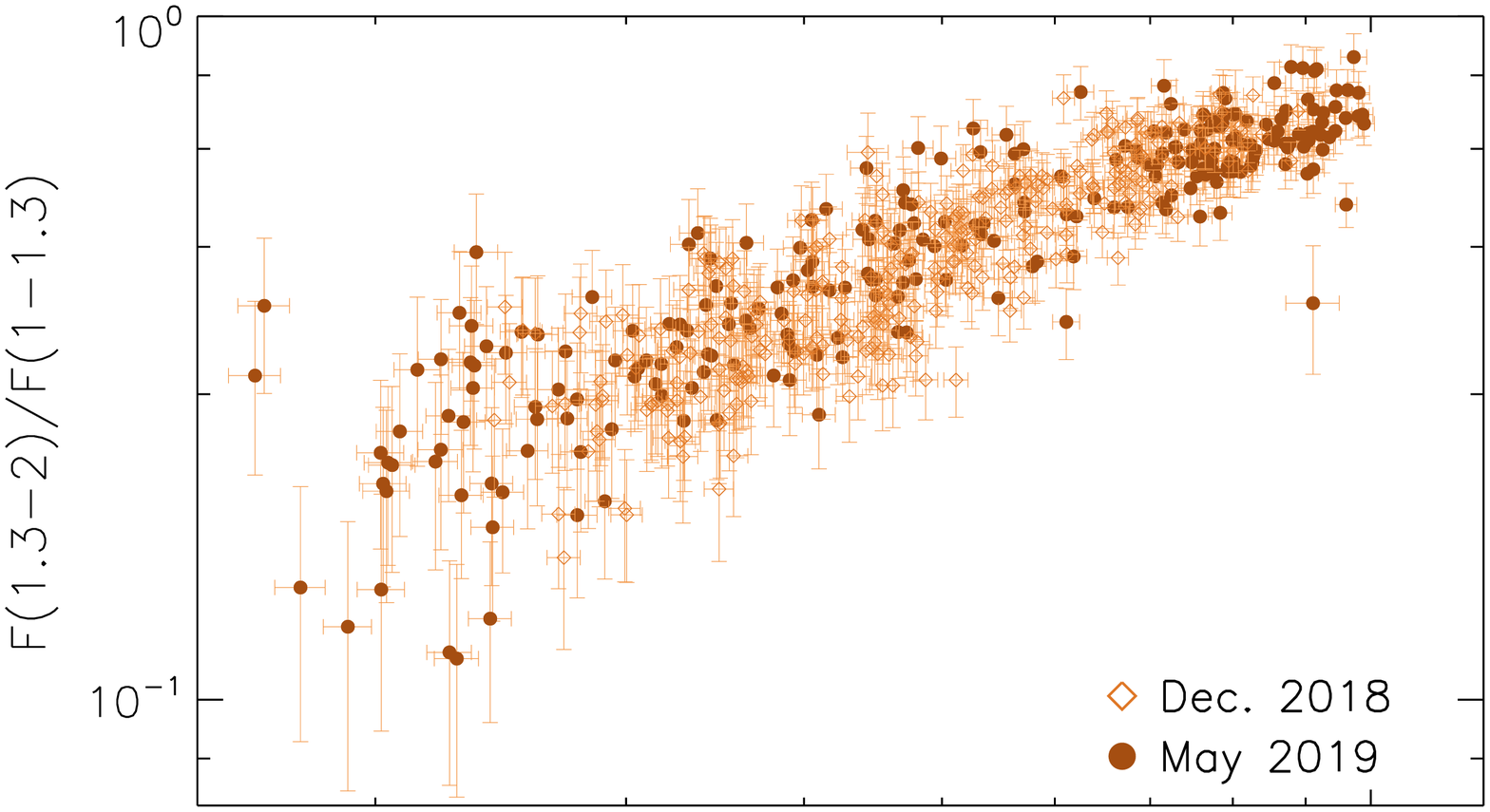}
\includegraphics[width=0.48\textwidth]{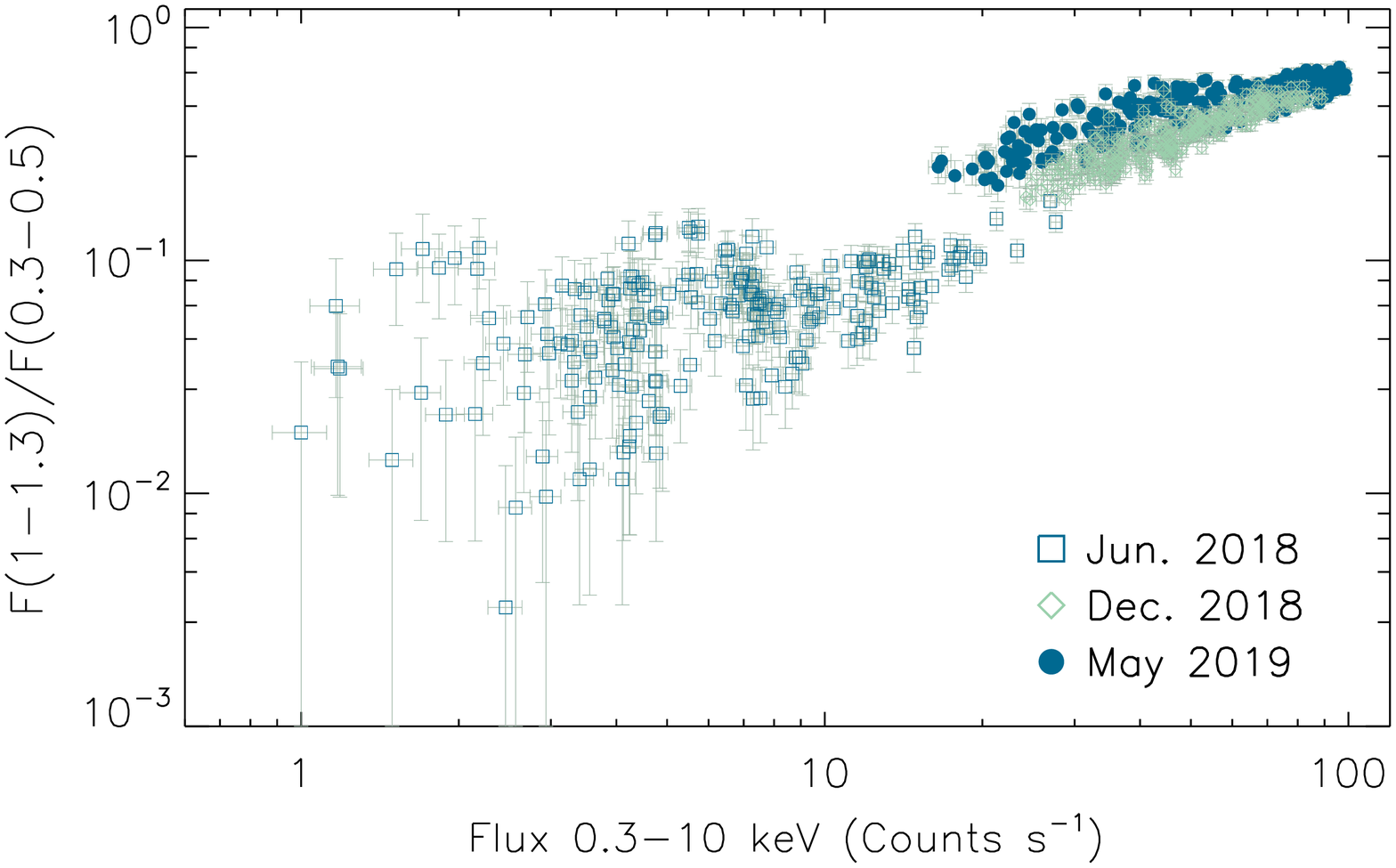}
\includegraphics[width=0.48\textwidth]{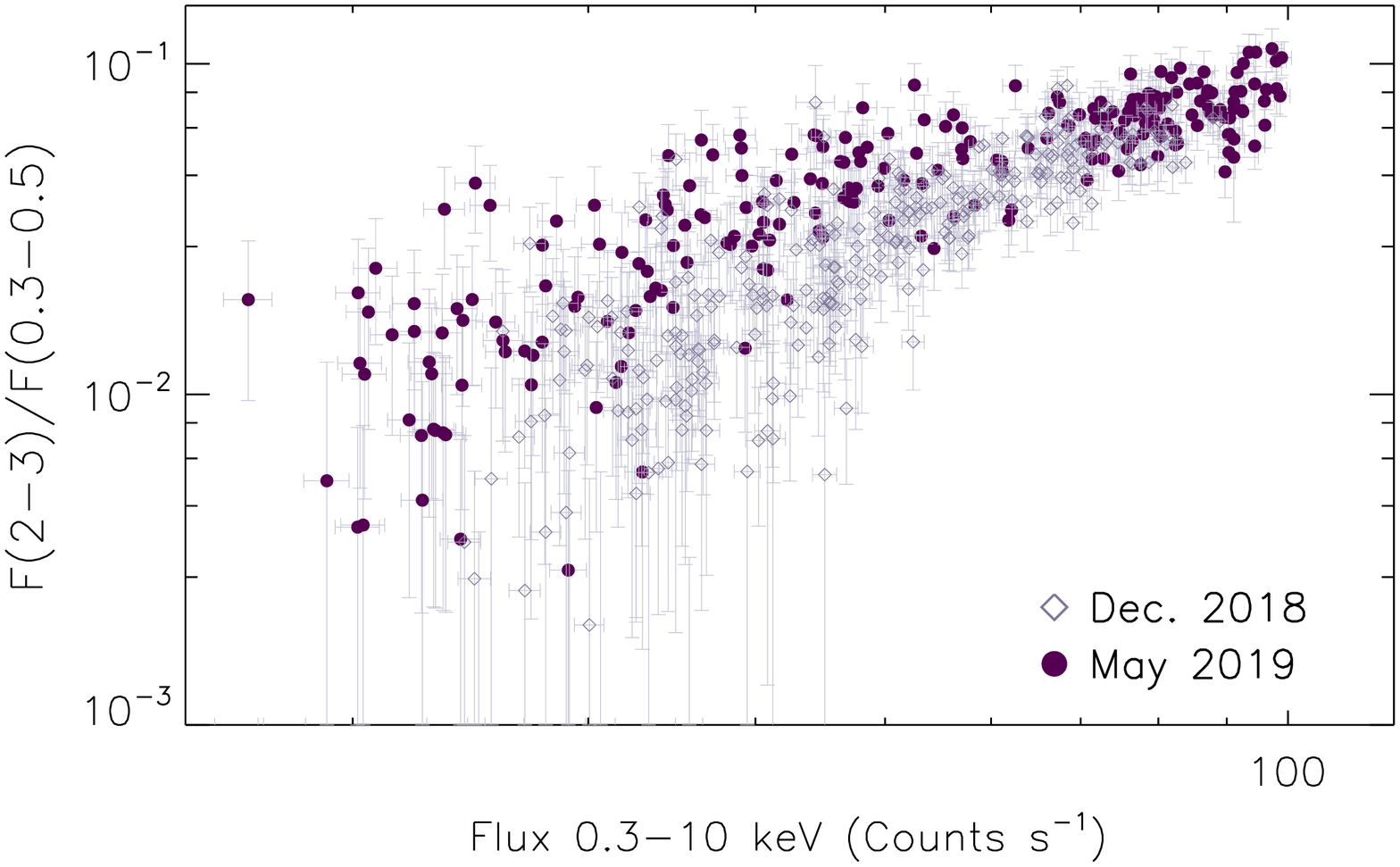}
    \caption{ {\it Left panel:} {\it XMM-Newton} EPIC/PN flux ratios in different bands (0.5--0.75/0.3--0.5\,keV, 0.75--1/0.3--0.5\,keV and 1--1.3/0.3--0.5\,keV, for the top, middle and bottom panel, respectively) versus the 0.3--10\,keV count rate for the June\,\,2018 (empty squares), December\,\,2018 (empty diamonds) and May\,\,2019 (filled circles) observations. The fluxes were obtained integrating over intervals of 200\,s. {\it Right panel:} same as left panel using the 1.3--2/0.3--0.5\,keV (top panel), 1.3--2/1--1.3\,keV (middle panel) and 2--3/0.3--0.5\,keV flux ratios for the December\,\,2018 (empty diamonds) and May\,\,2019 (filled circles) observations.}
    \label{fig:XMMlc_18_ratios}
  \end{center}
\end{figure*}

In this section we show the {\it XMM-Newton} EPIC/PN light curves of 1ES\,1927+654 in different energy bands for the June\,\,2018 (Fig.\,\ref{fig:XMMlc_18_bands}), December\,\,2018 and May\,\,2019 (Fig.\,\ref{fig:XMMlc_18_december_bands}) observations. The figures also show the ratio between the flux in different bands and that in the 0.3--0.5\,keV interval. This illustrates how the hardness of the sources follows the same pattern as the total flux. The flux ratios in different bands versus the 0.3--10\,keV count rate are illustrated for all observations in Fig.\,\ref{fig:XMMlc_18_ratios}, and clearly show that the source becomes harder when brighter.

\clearpage

\section{Time-resolved spectroscopy of the 2018/2019 {\it XMM-Newton} observations}\label{appendix:exposuresintervals}
\begin{figure*}[h!]    
 \begin{center}
    \includegraphics[width=0.75\textwidth]{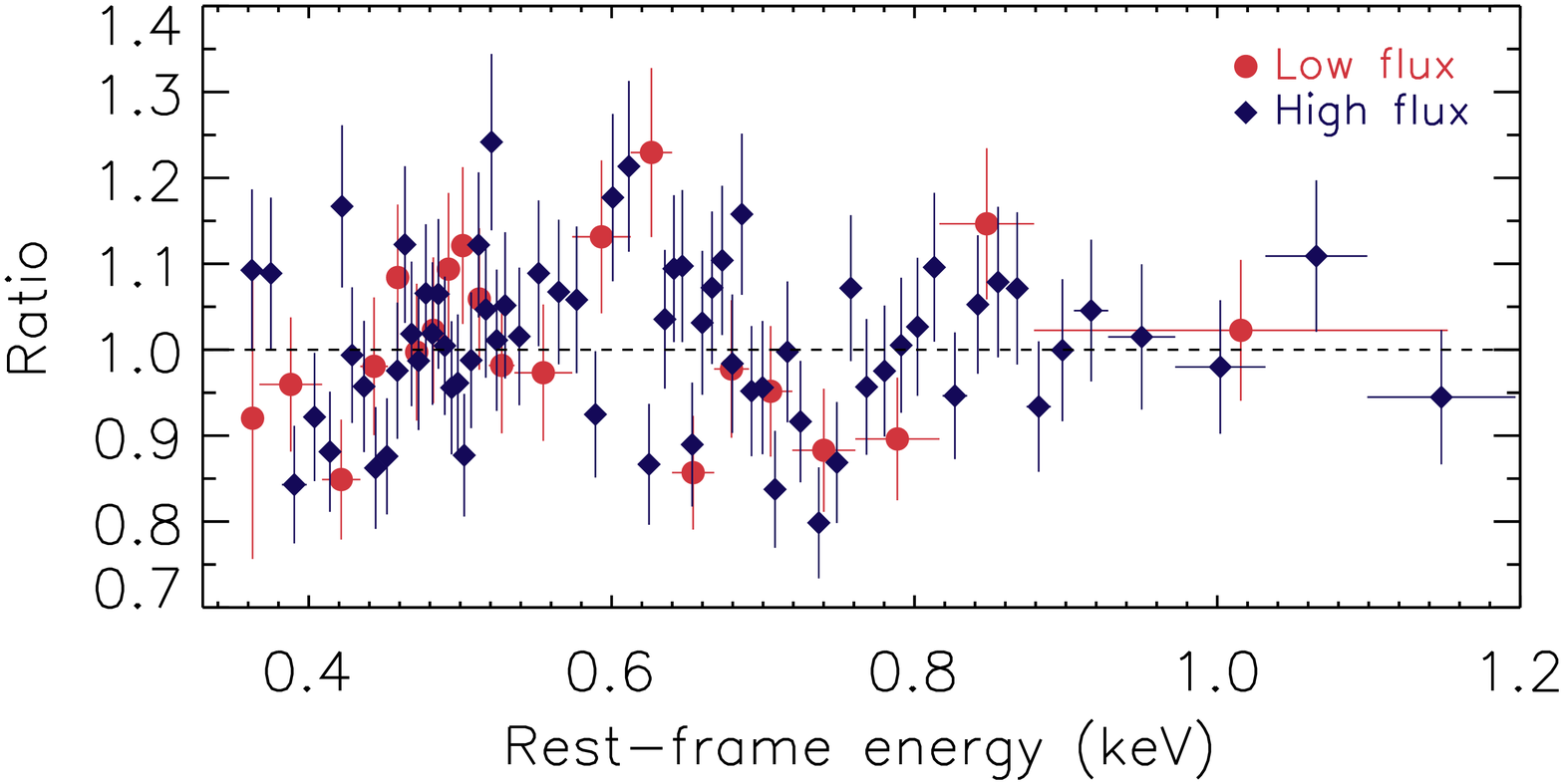}
 \end{center}
    \caption{Ratio between the {\it XMM-Newton}/RGS spectra for the low (red circles) and high (black diamonds) flux intervals of the June\,\,2018 observation and a model that considers a blackbody and power law (see \S\ref{sec:timeresolved2018_2019}). For visual clarity both spectra have been rebinned to 12$\sigma$.}
    \label{fig:RGS_highlowflux_june18}
\end{figure*}

The spectral parameters obtained by fitting the different intervals of the three {\it XMM-Newton} observations are reported in Tables\,\,\ref{tab:fitXMM18epochs_Jun}--\ref{tab:fitXMM19epochs_May}. The ratio between the {\it XMM-Newton}/RGS spectra for the low and high flux intervals of the June\,\,2018 observation and a model including a blackbody and power law are illustrated in Fig.\,\ref{fig:RGS_highlowflux_june18}. No clear difference is found in the residuals, suggesting that the ionized outflows do not change strongly between the two intervals.

\tabletypesize{\normalsize}
\begin{deluxetable*}{cccccccccc}[t!]    
\tablecaption{Spectral parameters obtained for the 11 time intervals of the {\it XMM-Newton} June 2018 observation (see top panel of Fig.\,\ref{fig:XMMlc_18_bins}). The spectral model includes a blackbody component, a power law and a Gaussian line [\textsc{tbabs$\times$ztbabs$\times$(zpo+zbb+zgauss})].\label{tab:fitXMM18epochs_Jun}}
\tablehead{
 \colhead{(1) } & \colhead{(2)} &  \colhead{(3)} &\colhead{(4)} & \colhead{(5) }  & \colhead{(6) }  & \colhead{(7)} & \colhead{(8)} & \colhead{(9)} & \colhead{(10)} \\
\noalign{\smallskip}
 \colhead{Interval } & \colhead{$L_{\rm 0.3-2}$} &  \colhead{$N_{\rm H}$} &\colhead{$kT$} & \colhead{$\Gamma$ }  & \colhead{C-stat/DOF }  & \colhead{Energy} & \colhead{$\sigma$} & \colhead{EW} & \colhead{Flux line} \\
\noalign{\smallskip}
 \colhead{ } & \colhead{[$\rm 10^{43}\,erg\,s^{-1}$]} & \colhead{[$\rm 10^{20}\,cm^{-2}$]}& \colhead{[eV]} & \colhead{ }  & \colhead{ }  & \colhead{[keV]} & \colhead{[eV]} & \colhead{[eV]} & \colhead{[$10^{-4}\rm\,ph\,cm^{-2}\,s^{-1}$]}    
}
\startdata
\noalign{\smallskip}
1	& $0.97\pm0.01$	&  $4.7^{+2.7}_{-2.4}$     &  $94\pm5$	& \nodata & 193/193 & $0.99^{+0.03}_{-0.04} $ & $84^{+29}_{-27}$ & $202_{-44}^{+50}$ &  $4.3^{+2.4}_{-1.6}$ \\
\noalign{\smallskip}
2	& $2.30\pm0.02$	&   $3.8^{+2.2}_{-1.9}$    &   $105\pm5$	&$1.92^{+1.15}_{-1.09}$ &231/250  & $0.94^{+0.04}_{-0.05}$& $125^{+28}_{-26}$ & $288^{+56}_{-53}$ & $28^{+15}_{-9}$ \\
\noalign{\smallskip}
3	& $1.24_{-0.02}^{+0.01}$	&   $2.2^{+1.7}_{-1.6}$    &   $100\pm4$	& $0.75^{+1.23}_{-1.50}$ & 242/236 & $0.97^{+0.04}_{-0.05}$ &$100^{+31}_{-28}$ & $172^{+35}_{-30}$ &  $5.6_{-1.9}^{+3.2}$  \\
\noalign{\smallskip}
4	&  $1.65_{-0.02}^{+0.01}$	&   $1.6^{+1.3}_{-1.2}$     &   $104\pm3$	& $2.06_{-1.08}^{+1.11}$ & 279/284 & $0.99^{+0.03}_{-0.02}$& $117^{+19}_{-16}$ & $222_{-28}^{+30}$ & $9.5^{+2.1}_{-2.8}$  \\
\noalign{\smallskip}
5	& $1.29_{-0.03}^{+0.01}$	&    $\leq 7.1$    &   $100^{+4}_{-5}$	& $5.51_{-1.80}^{+0.86}$ & 229/231 & $1.03_{-0.06}^{+0.03}$ & $91_{-40}^{+43}$ & $186_{-42}^{+41}$ & $4.2^{+3.1}_{-1.7}$   \\
\noalign{\smallskip}
6	&$0.79\pm0.01$	&   $\leq 1.8$    &    $99\pm3$	& $3.13_{-3.00}^{+1.59}$ & 259/265 & $1.01^{+0.02}_{-0.03}$ &$117\pm25$ & $253_{-48}^{+47}$ & $3.4^{+1.0}_{-0.9}$   \\
\noalign{\smallskip}
7	& $0.43\pm0.01$ 	&    $\leq 2.3$   &    $93^{+3}_{-4}$	& $2.84_{-2.94}^{+1.55}$ & 248/248 & $1.00^{+0.03}_{-0.04}$& $127^{+35}_{-31}$ & $316^{+85}_{-93}$ & $2.0^{+0.9}_{-0.6}$  \\
\noalign{\smallskip}
8	& $0.20\pm0.01$	&    $\leq 7.3$   &   $79\pm7$	& \nodata & 214/181 & $1.00^{+0.07}_{-0.13}$ & $143^{+69}_{-41}$ & $966^{+334}_{-178}$ & $1.3^{+1.2}_{-0.5}$   \\
\noalign{\smallskip}
9	&  $0.44\pm0.01$ 	&   $\leq 2.4$    &   $89\pm2$	& $3.58_{-3.25}^{+1.45}$ & 239/217 & $0.98^{+0.04}_{-0.06}$ & $94^{+55}_{-42}$ & $136_{-53}^{+38}$ & $0.9_{-0.4}^{+0.7}$  \\
\noalign{\smallskip}
10	& $0.67\pm0.01$	&   $\leq 2.2$     &   $99^{+2}_{-4}$	& \nodata & 225/215  & $0.99\pm0.04$ & $80_{-32}^{+36}$ & $111_{-28}^{+29}$ & $1.6_{-0.7}^{+0.9}$ \\
\noalign{\smallskip}
11	& $0.79\pm0.01$	&     $\leq 3.9$   &   $97\pm5$	& $3.18^{+1.25}_{-1.35}$ & 245/266  & $0.98_{-0.04}^{+0.03}$ & $109_{-24}^{+28}$ & $216_{-43}^{+44}$ & $2.2_{-1.3}^{+1.9}$ 
\enddata
\tablecomments{The columns report (1) the interval used for the spectral analysis, (2) the 0.3--2\,keV luminosity of the source, (3) the column density of the cold absorber, (4) the temperature of the blackbody component, (5) the photon index of the power-law component, (6) the value of C-stat and the number of DOF, the (7) energy, (8) width, (9) equivalent width and (10) flux  of the Gaussian line.}
\end{deluxetable*}

\tabletypesize{\normalsize}
\begin{deluxetable*}{cccccccccc} 
\tablecaption{Spectral parameters obtained for the 19 time intervals of the {\it XMM-Newton} December 2018 observation (see middle panel of Fig.\,\ref{fig:XMMlc_18_bins}). The spectral model includes a blackbody component, a power law and a Gaussian line [\textsc{tbabs$\times$ztbabs$\times$(zpo+zbb+zgauss})].   \label{tab:fitXMM18epochs_Dec}}
\tablehead{
 \colhead{(1) } & \colhead{(2)} &  \colhead{(3)} &\colhead{(4)} & \colhead{(5) }  & \colhead{(6) }  & \colhead{(7)} & \colhead{(8)} & \colhead{(9)} & \colhead{(10)} \\
\noalign{\smallskip}
 \colhead{Interval } & \colhead{$L_{\rm 0.3-2}$} &  \colhead{$N_{\rm H}$} & \colhead{$kT$} & \colhead{$\Gamma$ }  & \colhead{C-stat/DOF }  & \colhead{Energy} & \colhead{$\sigma$} & \colhead{EW} & \colhead{Flux line} \\
\noalign{\smallskip}
 \colhead{ } & \colhead{[$\rm 10^{43}\,erg\,s^{-1}$]} & \colhead{[$\rm 10^{20}\,cm^{-2}$]} & \colhead{[eV]} & \colhead{ }  & \colhead{ }  & \colhead{[keV]} & \colhead{[eV]} & \colhead{[eV]} & \colhead{[$10^{-4}\rm\,ph\,cm^{-2}\,s^{-1}$]}    
}
\startdata
\noalign{\smallskip}
1	& $ 3.85^{+0.09}_{-0.10}$& $ \leq 6$ & $ 125^{+12 }_{-13 }$	& $ 3.15^{+0.71 }_{-1.03 }$ &  337/361  & $ 1.03^{+0.05 }_{-0.10 } $ & $158^{+63}_{-64}$ & $ 221_{-90}^{+89}$ &  $33^{+42}_{-18}$ \\
\noalign{\smallskip}
2	& $ 5.74^{+0.02}_{-0.07}$ & $ 3.8\pm1.5$& $ 138^{+9}_{-14}$	& $ 3.36^{+0.19}_{-0.22}$ &  571/523  & $1.03^{+0.06}_{-0.13} $ & $ 216^{+65}_{-48}$ & $165_{-77}^{+61}$ &  $46^{+63}_{-22}$ \\
\noalign{\smallskip}
3	& $ 3.65^{+0.03 }_{-0.07}$ & $3.8^{+2.6 }_{-2.5 }$& $121^{+11 }_{- 13}$	& $ 3.74^{+0.41}_{-0.50}$ &  404/397  & $ 0.99^{+0.05 }_{-0.10} $ & $172^{+56 }_{-49 }$ & $225_{-85}^{+80}$ &  $37^{+41}_{-17}$ \\
\noalign{\smallskip}
4	& $ 5.93^{+0.02 }_{-0.03 }$ & $4.8\pm1.2$& $ 155^{+4 }_{-5 }$	& $ 3.58_{-0.10}^{+0.11}$ &  738/688  & $ 1.11^{+0.02 }_{-0.03 } $ & $ 126^{+27 }_{-25 }$ & $69_{-11}^{+12}$ &  $18^{+7}_{-5}$ \\
\noalign{\smallskip}
5	& $ 3.44^{+0.12 }_{-0.03}$ & $ 3.5^{+2.5 }_{-2.3 }$& $ 117\pm7$	& $ 3.52^{+0.55 }_{-0.83 }$ &  336/351  & $ 1.02^{+0.04 }_{-0.03 } $ & $169^{+41 }_{-32 }$ & $247_{-79}^{+81}$ &  $31^{+20}_{-11}$ \\
\noalign{\smallskip}
6	& $ 5.63^{+0.03}_{-0.07}$ & $ 8.8\pm2.1$& $132^{+10}_{-9}$	& $ 3.70^{+0.20}_{-0.22}$ &  459/509  & $1.10^{+0.05}_{-0.07} $ & $ 145^{+49}_{-38}$ & $ 102_{-34}^{+33}$ &  $26^{+22}_{-11}$ \\
\noalign{\smallskip}
7	& $ 4.10^{+0.03 }_{-0.13 }$ & $ \leq 3.1$& $129^{+7}_{-8 }$	& $ 3.06^{+0.47 }_{-0.59 }$ &  363/437  & $ 1.07^{+0.03 }_{-0.05 } $ & $ 142^{+43 }_{-33 }$ & $ 181_{-28}^{+57}$ &  $25^{+16}_{-8}$ \\
\noalign{\smallskip}
8	& $4.96^{+0.03 }_{-0.08}$ & $ 5.9^{+2.2 }_{-2.1 }$& $130^{+10 }_{-13 }$	& $3.73^{+0.23 }_{-0.26 }$ &  413/465  & $1.03^{+0.07 }_{-0.12 } $ & $180^{+64 }_{-51 }$ & $133_{-77}^{+58}$ &  $33^{+45}_{-17}$ \\
\noalign{\smallskip}
9	& $ 3.49\pm0.02$ & $ 2.5^{+1.4}_{-1.3 }$& $ 129^{+6 }_{-7 }$	& $ 3.46^{+0.19 }_{-0.21 }$ &  505/541  & $ 1.00^{+0.03 }_{-0.05 } $ & $175^{+28 }_{-24 }$ & $ 201_{-41}^{+39}$ &  $32^{+15}_{-9}$ \\
\noalign{\smallskip}
10	& $ 5.69^{+0.04 }_{-0.06 }$ & $ 4.8^{+2.6 }_{-2.5 }$& $ 149^{+10 }_{-8 }$	& $ 3.42^{+0.22 }_{-0.24 }$ &  553/504  & $1.06^{+0.04 }_{-0.06 } $ & $113^{+45 }_{-32 }$ & $67_{-22}^{+23}$ &  $20^{+15}_{-8}$ \\
\noalign{\smallskip}
11	& $ 5.65\pm0.02$ & $ 3.5\pm1.3 $& $ 152^{+4 }_{-5 }$	& $ 3.34\pm0.13 $ &  566/526  & $ 1.11\pm0.03  $ & $ 119^{+34 }_{-32 }$ & $ 46_{-12}^{+11}$ &  $11^{+5}_{-4}$ \\
\noalign{\smallskip}
12	& $ 4.46\pm0.04$ & $ 2.6\pm2.3 $& $ 140\pm7 $	& $ 3.62^{+0.28 }_{-0.34 }$ &  425/447  & $1.10^{+0.03 }_{-0.04 } $ & $115^{+39}_{-32}$ & $116\pm27$ &  $18^{+10}_{-6}$ \\
\noalign{\smallskip}
13	& $ 6.47^{+0.03}_{-0.05}$ & $ 6.6^{+2.0}_{-1.9}$& $ 157^{+6}_{-7}$	& $3.70^{+0.15}_{-0.16}$ &  536/562  & $ 1.16^{+0.03 }_{-0.04 } $ & $ 72^{+47 }_{-37 }$ & $34_{-11}^{+10}$ &  $9^{+6}_{-4}$ \\
\noalign{\smallskip}
14	& $3.62^{+0.05}_{-0.06}$ & $2.6^{+2.6 }_{-2.4 }$& $ 120^{+9 }_{-11}$	& $3.66^{+0.49}_{-0.71}$ &  342/378  & $1.01^{+0.06 }_{-0.10 } $ & $ 184^{+55 }_{-43 }$ & $257_{-99}^{+85}$ &  $35^{+37}_{-15}$ \\
\noalign{\smallskip}
15	& $ 5.52^{+0.03 }_{-0.06 }$ & $ 7.0^{+2.4}_{-2.3}$& $ 142^{+7}_{-9}$	& $ 3.81^{+0.22 }_{-0.21 }$ &  417/489  & $ 1.09^{+0.05 }_{-0.07 } $ & $ 131^{+55 }_{-39}$ & $61_{-27}^{+24}$ &  $16^{+16}_{-8}$ \\
\noalign{\smallskip}
16	& $ 8.16\pm0.05 $ & $ 6.3^{+2.0 }_{-1.9}$& $ 177\pm5 $	& $3.57\pm0.15 $ &  506/557  & \nodata &  \nodata &  \nodata &   \nodata \\
\noalign{\smallskip}
17	& $ 5.99^{+0.03}_{-0.04}$ & $ 3.7\pm2.1 $& $ 159^{+7 }_{-8 }$	& $ 3.37^{+0.18 }_{-0.20 }$ &  553/554  & $ 1.12\pm0.04 $ & $ 88^{+61 }_{-50 }$ & $ 36_{-13}^{+11}$ &  $9^{+8}_{-5}$ \\
\noalign{\smallskip}
18	& $ 3.80^{+0.03 }_{-0.07 }$ & $ \leq 2.9$ & $ 128^{+6 }_{-7 }$	& $ 3.05^{+0.48 }_{-0.63 }$ &  363/386  & $ 1.02^{+0.03 }_{-0.05 } $ & $ 151^{+33 }_{-29 }$ & $181_{-39}^{+38}$ &  $27^{+15}_{-9}$ \\
\noalign{\smallskip}
19	& $ 7.42\pm0.04 $ & $ 4.0\pm1.5 $& $ 175\pm4 $	& $ 3.28^{+0.13 }_{-0.14 }$ &  465/477  & \nodata & \nodata & \nodata&  \nodata
\enddata
\tablecomments{The columns report (1) the interval used for the spectral analysis, (2) the 0.3--2\,keV luminosity of the source, (3) the column density of the cold absorber, (4) the temperature of the blackbody component, (5) the photon index of the power-law component, (6) the value of C-stat and the number of DOF, the (7) energy, (8) width, (9) equivalent width and (10) flux  of the Gaussian line.}
\end{deluxetable*}

\tabletypesize{\normalsize}
\begin{deluxetable*}{cccccccccc} 
\tablecaption{Spectral parameters obtained for the 13 time intervals of the {\it XMM-Newton} May 2019 observation (see bottom panel of Fig.\,\ref{fig:XMMlc_18_bins}). The spectral model includes a blackbody component, a power law and a Gaussian line [\textsc{tbabs$\times$ztbabs$\times$(zpo+zbb+zgauss})].   \label{tab:fitXMM19epochs_May}}
\tablehead{
 \colhead{(1) } & \colhead{(2)} &  \colhead{(3)} &\colhead{(4)} & \colhead{(5) }  & \colhead{(6) }  & \colhead{(7)} & \colhead{(8)} & \colhead{(9)} & \colhead{(10)} \\
\noalign{\smallskip}
 \colhead{Interval } & \colhead{$L_{\rm 0.3-2}$} &  \colhead{$N_{\rm H}$} & \colhead{$kT$} & \colhead{$\Gamma$ }  & \colhead{C-stat/DOF }  & \colhead{Energy} & \colhead{$\sigma$} & \colhead{EW} & \colhead{Flux line} \\
\noalign{\smallskip}
 \colhead{ } & \colhead{[$\rm 10^{43}\,erg\,s^{-1}$]} & \colhead{[$\rm 10^{20}\,cm^{-2}$]} & \colhead{[eV]} & \colhead{ }  & \colhead{ }  & \colhead{[keV]} & \colhead{[eV]} & \colhead{[eV]} & \colhead{[$10^{-4}\rm\,ph\,cm^{-2}\,s^{-1}$]}    
}
\startdata
\noalign{\smallskip}
1	& $ 5.45^{+0.04}_{-0.07}$& $5.7^{+3.2}_{-3.1}$ & $ 158^{+10}_{-14}$	& $ 3.65^{+0.23}_{-0.25}$ &   423/495   & $ 1.04^{+0.09}_{-0.06}$  & $149^{+60}_{-51}$ & $ 83^{+36}_{-40}$ &  $25^{+29}_{-13}$ \\
\noalign{\smallskip}
2	& $ 2.72^{+0.01}_{-0.08}$& $3.1^{+2.8}_{-2.6}$ & $ 118\pm8$	& $ 3.32^{+0.49}_{-0.51}$ &   350/384   & $ 1.00\pm0.03$  & $137^{+24}_{-21}$ & $ 332^{+70}_{-59}$ &  $34^{+13}_{-9}$ \\
\noalign{\smallskip}
3	& $ 4.51^{+0.02}_{-0.04}$& $4.8\pm1.8$ & $ 148\pm8$	& $ 3.58\pm0.15$ &   674/605   & $ 1.04^{+0.04}_{-0.05}$  & $154^{+33}_{-29}$ & $ 126\pm30$ &  $28^{+15}_{-9}$ \\
\noalign{\smallskip}
4	& $ 2.52^{+0.01}_{-0.11}$& $\leq 4.7$ & $ 121^{+10}_{-14}$	& $ 3.35^{+0.46}_{-0.25}$ &   427/386   & $ 0.96^{+0.05}_{-0.08}$  & $164^{+47}_{-35}$ & $ 285^{+85}_{-89}$ &  $32^{+30}_{-12}$ \\
\noalign{\smallskip}
5	& $ 5.26^{+0.02}_{-0.07}$& $6.8\pm2.2$ & $ 136^{+11}_{-13}$	& $ 3.67^{+0.17}_{-0.18}$ &   537/548   & $ 1.01^{+0.03}_{-0.05}$  & $151^{+27}_{-32}$ & $ 161^{+44}_{-42}$ &  $47^{+26}_{-15}$ \\
\noalign{\smallskip}
6	& $  3.54^{+0.02}_{-0.03}$& $ 5.3\pm1.7$ & $  127^{+7}_{-8}$	& $  3.77^{+0.18}_{-0.19}$ &    533/538    & $  1.01^{+0.03}_{-0.04}$  & $ 152^{+25}_{-22}$ & $  213^{+25}_{-35}$ &  $ 35^{+14}_{-9}$ \\
\noalign{\smallskip}
7	& $  7.80^{+0.09}_{-0.06}$& $ 6.8\pm2.3$ & $ 173^{+7}_{-8}$	& $  3.58^{+0.16}_{-0.17}$ &    543/603    & $  1.17^{+0.07}_{-0.11}$  & $ \leq 143$ & $  15^{+7}_{-6}$ &  $ 5^{+8}_{-3}$ \\
\noalign{\smallskip}
8	& $  4.54^{+0.03}_{-0.08}$& $ 4.6^{+2.4}_{-2.3}$ & $  134^{+13}_{-22}$	& $  3.67^{+0.21}_{-0.22}$ &    482/490    & $  0.96^{+0.06}_{-0.14}$  & $185^{+72}_{-49}$ & $  159^{+58}_{-86}$ &  $ 42^{+76}_{-21}$ \\
\noalign{\smallskip}
9	& $  7.78^{+0.04}_{-0.07}$& $ 10.0^{+2.7}_{-2.6}$ & $ 150^{+11}_{-18}$	& $  3.68^{+0.17}_{-0.18}$ &    524/573    & $  1.00^{+0.07}_{-0.13}$  & $ 143^{+84}_{-55}$ & $  51^{+26}_{-30}$ &  $ 30^{+59}_{-17}$ \\
\noalign{\smallskip}
10	& $  9.63^{+0.04}_{-0.05}$& $ 6.8^{+1.3}_{-1.2}$ & $  183^{+4}_{-3}$	& $  3.56\pm0.09$ &    826/784    & $  \leq 1.29$  & $ NC$ & $ \leq 9$ &  $\leq 6.3$ \\
\noalign{\smallskip}
11	& $  7.28^{+0.03}_{-0.07}$& $ 5.6^{+2.1}_{-2.2}$ & $  158^{+8}_{-11}$	& $ 3.48^{+0.18}_{-0.19}$ &    493/502    & $  1.09^{+0.06}_{-0.10}$  & $ 143^{+69}_{-57}$ & $  54^{+27}_{-28}$ &  $ 18^{+25}_{-10}$ \\
\noalign{\smallskip}
12	& $  8.18\pm0.08$& $ 4.5\pm1.8$ & $  155^{+10}_{-18}$	& $ 3.27^{+0.17}_{-0.16}$ &    543/524    & $  \leq 1.05$  & $ \geq 142$ & $  \leq 130$ &  $ 55^{+70}_{-41}$ \\
\noalign{\smallskip}
13	& $  9.86^{+0.14}_{-0.27}$& $ 4.0\pm1.6$ & $  167^{+13}_{-11}$	& $  3.10^{+0.15}_{-0.16}$ &    555/534    & $  \geq 1.00$  & $ \geq 124$ & $  \leq 111$ &  $31^{+48}_{-26}$

\enddata
\tablecomments{The columns report (1) the interval used for the spectral analysis, (2) the 0.3--2\,keV luminosity of the source, (3) the column density of the cold absorber, (4) the temperature of the blackbody component, (5) the photon index of the power-law component, (6) the value of C-stat and the number of DOF, the (7) energy, (8) width, (9) equivalent width and (10) flux  of the Gaussian line.}
\end{deluxetable*}

\clearpage

\section{Hardness ratio of {\it NICER} observations}\label{appendix:HRnicer}

\begin{figure*}[h!]    
 \begin{center}
    \includegraphics[width=0.75\textwidth]{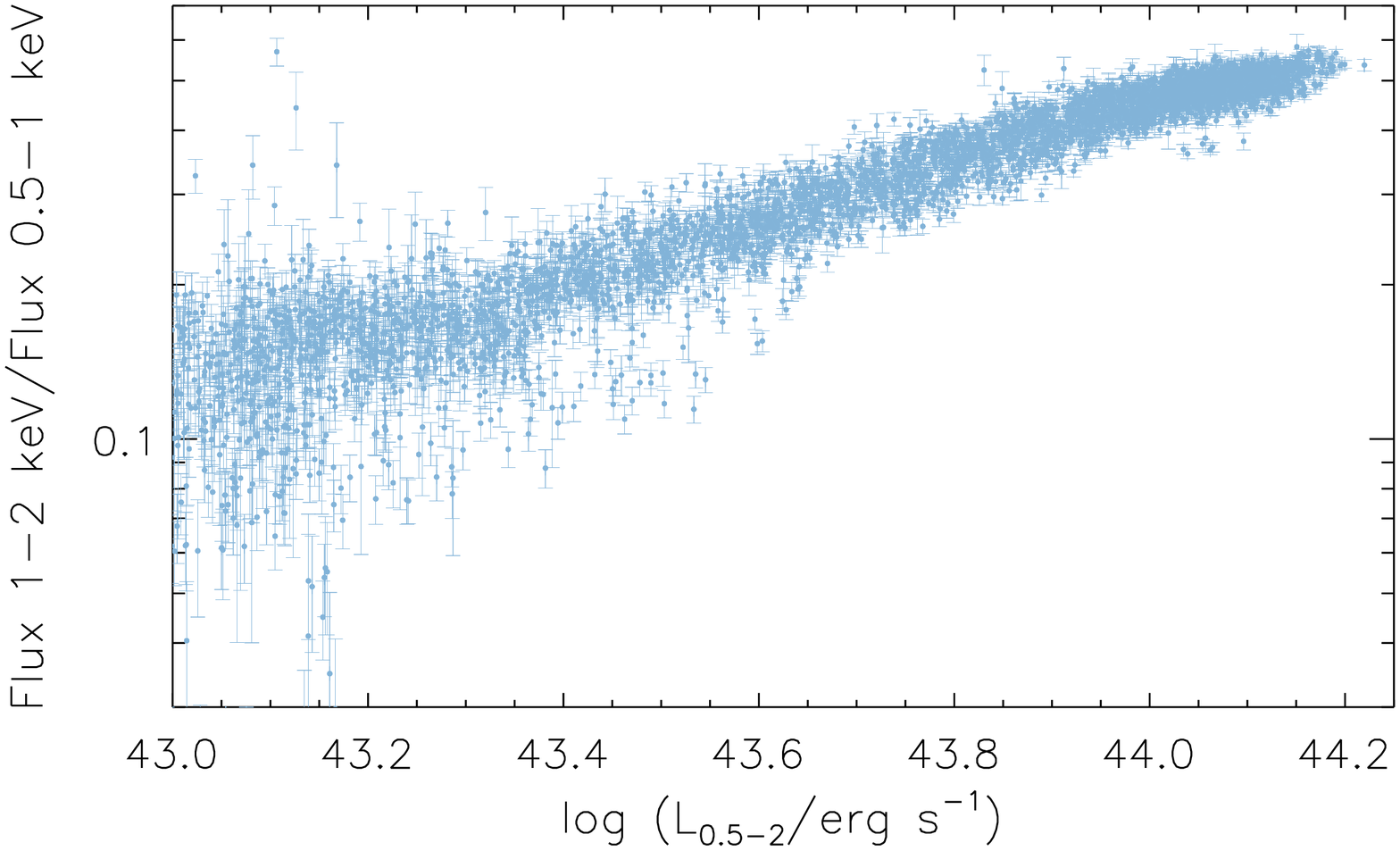}
 \end{center}
    \caption{Hardness ratio of {\it NICER} observations, showing the same harder-when-brighter behaviour observed in the {\it XMM-Newton} observations.}
    \label{fig:NICER_HR}
\end{figure*}

In Fig.\,\ref{fig:NICER_HR} we illustrate the ratio between the 0.5--1\,keV and the 1--2\,keV flux versus the 0.5--2\,keV luminosity. The figure shows a well defined harder-when-brighter behaviour.

\clearpage
\section{Stacked {\it NICER} spectra}\label{sect:stackednicerspecappendix}

The spectral parameters obtained by fitting the 16 stacked {\it NICER} spectra are reported in Table\,\ref{tab:fitNICERstackedSpec}. The fits include either only a blackbody ({\it bb}), a blackbody plus a cutoff power law component ({\it cut} in the table) or a blackbody plus a thermally Comptonized continuum ({\it nth}). For the latter model we fixed the temperature of the seed photons to that of the blackbody component.

\tabletypesize{\scriptsize}
\begin{deluxetable*}{cccccccccccc} 
\tablecaption{Spectral parameters obtained by fitting the stacked {\it NICER} spectra.  \label{tab:fitNICERstackedSpec}}
\tablehead{
 \colhead{(1) } & \colhead{(2)} &  \colhead{(3)} &\colhead{(4)} & \colhead{(5) }  & \colhead{(6) }  & \colhead{(7)} & \colhead{(8)} & \colhead{(9)} & \colhead{(10)}& \colhead{(11)}& \colhead{(12)} \\
\noalign{\smallskip}
 \colhead{Interval } & \colhead{Dates} &  \colhead{Model } &  \colhead{$N_{\rm H}$} & \colhead{$kT$} & \colhead{$\Gamma$ }  & \colhead{$E_{\rm cut}$ }&\colhead{$kT_{\rm e}$ }  & \colhead{Energy} & \colhead{$\sigma$}  & \colhead{$L_{0.3-2}$ }  & \colhead{$\chi^2$/DOF }   \\
\noalign{\smallskip}
 \colhead{ }  & \colhead{ [YY/MM/DD]}  & \colhead{ }  & \colhead{[$\rm 10^{20}\,cm^{-2}$]} & \colhead{[eV]} & \colhead{ }  & \colhead{ keV} & \colhead{ keV}  & \colhead{[keV]} & \colhead{[eV]}  & \colhead{[$\rm 10^{43}\,erg\,s^{-1}$]}   & \colhead{ }  
}
\startdata
\noalign{\smallskip}
1	& 18/5/22$-$18/6/7 	& bb	 &$0.6^{+0.2}_{-0.3}$	 &	$99^{+3}_{-2}$ &	\nodata &\nodata	 & 	\nodata& $0.95^{+0.01}_{-0.02}$ 	& $157^{+8}_{-6}$	&   0.9	& 113/92	  \\
\noalign{\medskip}
2	&  18/6/9$-$18/7/1	& bb	 &$0.8^{+0.6}_{-0.5}$	 &	$97\pm2$ &	\nodata & \nodata	 & 	\nodata & $0.99\pm0.03$ 	& $83^{+36}_{-38}$	&  0.11 	& 97/82	  \\
\noalign{\medskip}

3	&  18/7/4$-$18/7/2	& bb	 &$11.1^{+9.3}_{-7.4}$	 &	$64^{+12}_{-11}$ &\nodata	 & 	\nodata 	& \nodata & \nodata & \nodata	&  0.05 	& 127/116	  \\
\noalign{\medskip}
4	&  18/8/3$-$18/8/20	& bb	 &$1.0^{+0.8}_{-0.9}$	 &	$91^{+3}_{-3}$ &	\nodata	 & 	\nodata	& \nodata &  $0.98\pm0.15$ & NC	&  0.12 	& 118/104	  \\
\noalign{\medskip}
5	&  18/9/7$-$18/10/8	& po	 &$\leq 2.3$	 &	$111\pm3$ &	$2.1\pm0.7$ & \nodata & 	\nodata   & $0.97\pm0.01$ 	& $148^{+9}_{-8}$	&  1.1 	& 180/141	  \\
\noalign{\smallskip}
	&  	& nth	 &$\leq 2.8$	 &	$111\pm3$ &	$2^{+0.8}_{-0.4}$ &\nodata	 & 	NC  & $0.97\pm0.01$ 	& $148^{+9}_{-8}$	&  1.1 	& 180/140	  \\
\noalign{\medskip}
6	&  18/10/10$-$18/10/31	& po	 &$0.8\pm0.7$	 &	$134\pm2$ &	$3.25\pm0.07$ & \nodata	 & 	\nodata & $1.01\pm0.01$ 	& $155\pm6$	& 2.7  	& 287/191	  \\
\noalign{\smallskip}
	&  	& nth	 	 &	$1.3\pm0.7$ &	$119\pm2$ &$3.17^{+0.10}_{-0.08}$	 & 	\nodata & NC & $0.98\pm0.01$ 	& $172\pm6$	&   2.7	& 239/190	  \\
\noalign{\medskip}
7	&  18/11/1$-$18/11/29	& po	 &$\leq 1.3$	 &	$116\pm2$ &	$2.50\pm0.12$ &\nodata	 & 	\nodata   & $0.97\pm0.01$ 	& $175\pm5$	&  1.7  	& 225/191	  \\
\noalign{\smallskip}
	&  	& nth	 &$0.6^{+0.8}_{-0.4}$	 &	$115\pm2$ &	$2.54^{+0.11}_{-0.19}$ &\nodata	 & 	 NC & $0.97\pm0.1$ 	& $176^{+3}_{-4}$	&   1.7	& 225/190	  \\
\noalign{\medskip}
8	&  18/12/1$-$18/12/31	& po	 &$0.7^{+0.7}_{-0.6}$	 &	$135\pm2$ &	$3.22\pm0.05$ & \nodata	 & 	\nodata   & $1.00\pm0.01$ 	& $158^{+6}_{-5}$	&  2.7 	& 226/161	  \\
\noalign{\smallskip}
	&  	& nth	 &$0.8^{+0.7}_{-0.5}$	 &	$118\pm2$ &	$3.51^{+0.30}_{-0.04}$ & \nodata & 	NC 	& $0.98\pm0.01$   & $173\pm5$	&  2.7 	& 214/160	  \\
\noalign{\medskip}
9	&  19/1/4$-$19/1/30	& cut	 &$\leq 1.6$	 &	$122^{+7}_{-5}$ &	$2.4^{+0.6}_{-0.7}$ &$1.8^{+2.5}_{-0.7}$	 & 	\nodata	& $0.97\pm0.02$ 	& $183^{+12}_{-13}$	&   2.5	& 262/240	  \\
\noalign{\smallskip}
	&  	& nth	 &$1.0^{+1.0}_{-0.9}$	 &	$114^{+1}_{-3}$ &	$3.22^{+0.4}_{-0.03}$ &\nodata	 & 	NC & $0.96^{+0.01}_{-0.01}$ 	& $187^{+8}_{-9}$	&  2.5 	& 260/240	  \\
\noalign{\medskip}
10	&  19/2/1$-$19/2/26	& cut	 &$0.8\pm0.4$	 &	$136^{+2}_{-2}$ &	$1.54\pm0.11$ & $0.88\pm0.04$	 & 	\nodata  & $1.00\pm0.01$ 	& $206^{+9}_{-6}$	&  6.2 	& 391/340	  \\
\noalign{\smallskip}
	&  	& nth	 &	$1.4\pm0.4$	 &	$121\pm1$ &	$3.44^{+0.2}_{-0.08}$ &\nodata	 & 	NC	& $0.99\pm0.01$	&  $199^{+45}_{-4}$	&  6.2 	& 486/340	  \\
\noalign{\medskip}
11	&  19/3/5$-$19/3/30	& cut	 &$\leq 1.5$	 &	$126\pm2$ &	$0.8_{-0.3}^{+0.5}$ &$0.72_{-0.07}^{+0.13}$	 & 	\nodata  & $0.95\pm0.02$ 	& $227^{+9}_{-11}$	&   5.3	& 455/340	  \\
\noalign{\smallskip}
	&  	& nth	 &$1.2\pm0.8$	 &	$121^{+3}_{-2}$ &	$3.26^{+0.03}_{-0.04}$ & 	\nodata  & NC	 & 	$0.96\pm0.01$& $209\pm7$ 		&  5.3 	& 513/340	  \\
\noalign{\medskip}
12	&  19/4/3$-$19/4/30	& cut	 &$0.9^{+0.9}_{-0.8}$	 &	$131^{+7}_{-4}$ &	$1.1^{+0.4}_{-0.3}$ &$0.68^{+0.09}_{-0.06}$	 & 	\nodata & $0.98\pm0.02$ 	& $216^{+13}_{-16}$	&  6.9 	& 352/340	  \\
\noalign{\smallskip}
	&  	& nth	 &$1.3^{+0.9}_{-0.8}$	 &	$117\pm4$ &	$3.67^{+0.04}_{-0.11}$ & \nodata	 & 	 NC & $1.00\pm0.01$ 	& $190^{+10}_{-9}$	&  6.9 	& 405/340	  \\
\noalign{\medskip}
13	&  19/5/2$-$19/5/31	& cut	 &$1.5\pm0.5$	 &	$132^{+2}_{-4}$ &	$1.27^{+0.21}_{-0.14}$ &$0.79^{+0.07}_{-0.04}$	 & 	\nodata& $0.98\pm0.01$ 	& $221^{+7}_{-9}$	&  6.8 	& 500/440	  \\
\noalign{\smallskip}
	&  	& nth	 &$2.1\pm0.5$	 &	$119\pm2$ &	$3.47^{+0.12}_{-0.09}$ & \nodata	 & 	 NC & $0.99\pm0.01$ 	& $202\pm5$	&   6.8	& 669/440	  \\
\noalign{\medskip}
14	&  19/6/01$-$2019/6/29	& cut	 &$\leq 0.8$	 &	$156^{+4}_{-5}$ &	$1.52^{+0.09}_{-0.10}$ &$0.88\pm0.04$	 & 	\nodata & $1.00\pm0.02$ 	& $206^{+13}_{-12}$	&   9.1	&  446/440	  \\
\noalign{\smallskip}
	&  	& nth	 &$0.9^{+0.5}_{-0.4}$	 &	$120^{+2}_{-3}$ &	$3.56^{+0.12}_{-0.2}$ & \nodata	 & 	NC  & $0.99\pm0.01$ 	& $200\pm7$	&   9.1	& 514/440	  \\
\noalign{\medskip}
15	&  19/7/1$-$19/7/29	& cut	 &$\leq 0.8$	 &	$167\pm3$ &	$1.58\pm0.07$ &$0.91\pm0.03$	 & 	\nodata   & $1.03\pm0.01$ 	& $184^{+12}_{-11}$	&  9.8 	& 500/440	  \\
\noalign{\smallskip}
	&  	& nth	 &$1.0\pm0.5$	 &	$126^{+2}_{-2}$ &	$3.56\pm0.01$ & \nodata	 & 	 NC & $1.02\pm0.01$ 	& $185^{+6}_{-7}$	&  9.8 	& 601/440	  \\
\noalign{\medskip}
16	&  19/8/1$-$19/8/5	& cut	 &$\leq 1.0$	 &	$182^{+4}_{-5}$ &	$1.71^{+0.12}_{-0.06}$ &$1.02^{+0.07}_{-0.05}$	 & 	\nodata & $1.06\pm0.03$ 	& $156^{+30}_{-24}$	&  11.6 	& 452/440	  \\
\noalign{\smallskip}
	&  	& nth	 &$\leq 1.0$	 &	$133^{+2}_{-5}$ &	$3.50\pm0.03$ & \nodata & 	 NC & $1.05^{+0.01}_{-0.02}$ 	& $186^{+19}_{-16}$	&   11.6	& 501/440

\enddata
\tablecomments{The columns report (1) the interval number; (2) the range of dates for which the {\it NICER} spectra were stacked; (3) the model applied, which include only a blackbody component plus a Gaussian line ({\it bb}), a blackbody, a Gaussian line and either a power law [{\it po}; \textsc{tbabs$\times$ztbabs$\times$(zpo+zbb+zgauss})], a cutoff power-law [{\it cut}; \textsc{tbabs$\times$ztbabs$\times$(zcut+zbb+zgauss})] or a thermally Comptonized plasma [{\it nth}; \textsc{tbabs$\times$ztbabs$\times$(nthcomp+zbb+zgauss})]. In the latter model the temperature of the seed photons was fixed to the temperature of the blackbody component; (4) the column density of the cold absorber; (5) the temperature of the blackbody component; (6) the photon index of the power-law or thermally Comptonized continuum; (7) the cutoff energy; (8) the temperature of the electrons; the (9) energy and (10) width of the Gaussian line; (11) the 0.3--2\,keV luminosity of the source; (12) the Chi-squared and the number of degrees of freedom. NC: not constrained.}
\end{deluxetable*}

\clearpage
\section{The May 2011 {\it XMM-Newton} observation of 1ES\,1927+654}\label{sect:2011xmmobs}

\begin{figure*}[h!]    
 \begin{center}
    \includegraphics[width=0.7\textwidth]{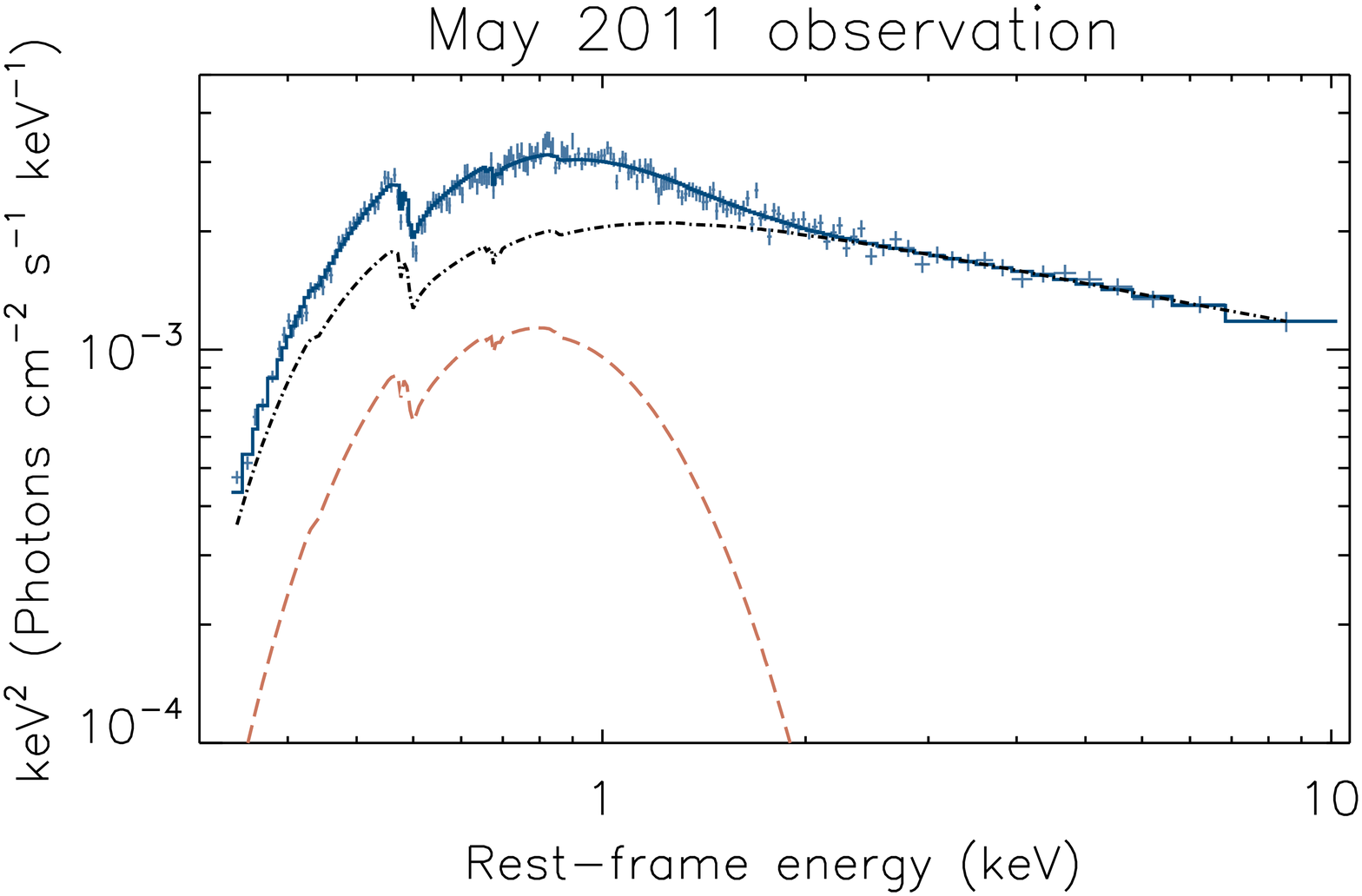}
    \includegraphics[width=0.7\textwidth]{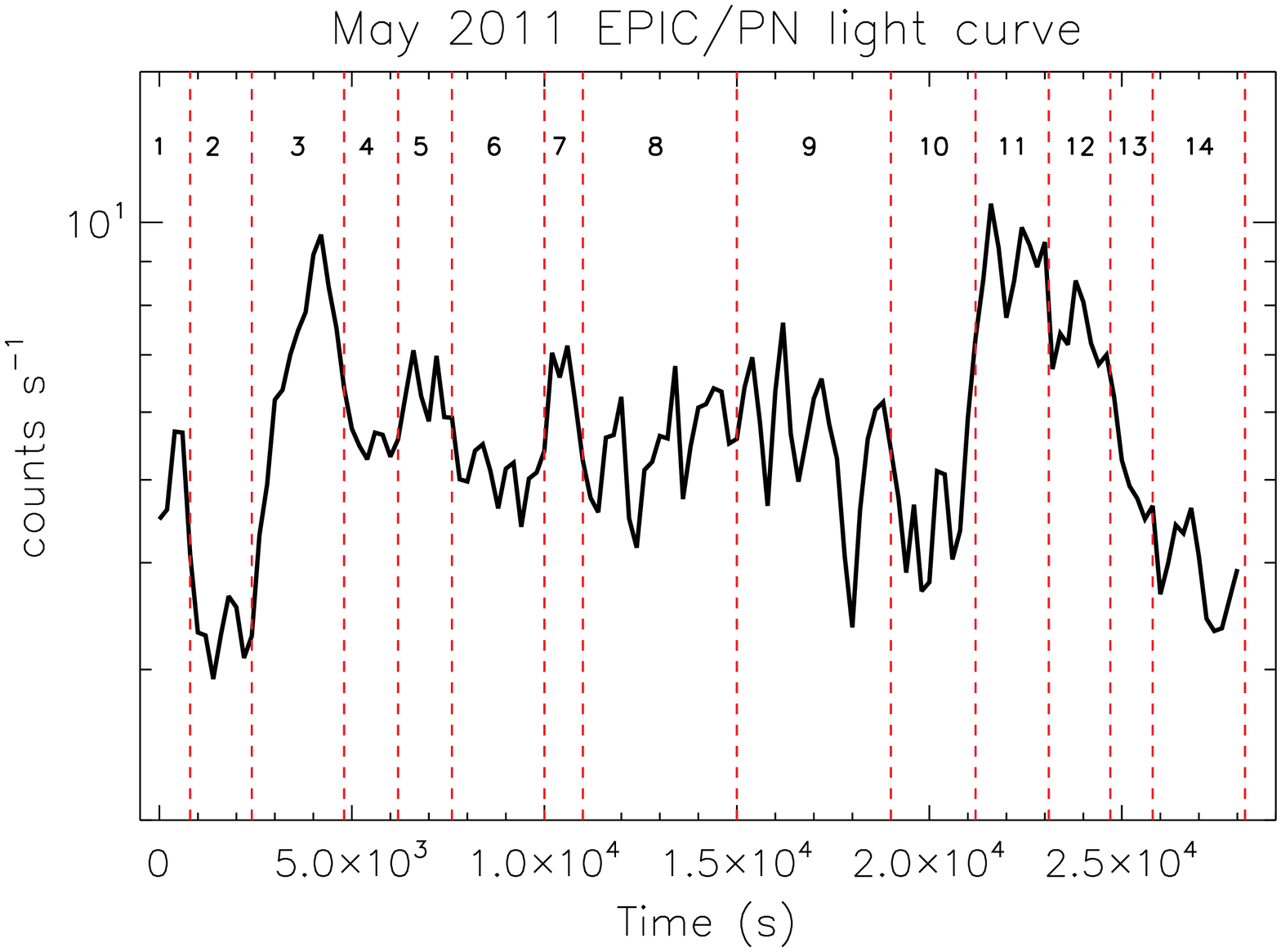}
 \end{center}
    \caption{{\it Top panel:} May 2011 {\it XMM-Newton} EPIC/PN spectrum of 1ES\,1927+654. The continuous lines show the best-fitting model, which includes a blackbody (dashed red line), and a power-law (black dot-dashed line), absorbed by low-column density neutral gas. {\it Bottom panel:} May 2011 {\it XMM-Newton} EPIC/PN light curve of 1ES\,1927+654. The figure also shows the intervals used for the time-resolved spectroscopy, which are denoted by the vertical red dashed lines. The results of the spectral analysis of these intervals are reported in Table\,\,\ref{tab:fitXMM11epochs}.}
    \label{fig:xmmobs2011eeufs}
\end{figure*}

\begin{figure*}[t!]
  \begin{center}
  \center{\Large{2011 {\it XMM-Newton} observation}}\par\medskip
\includegraphics[width=0.48\textwidth]{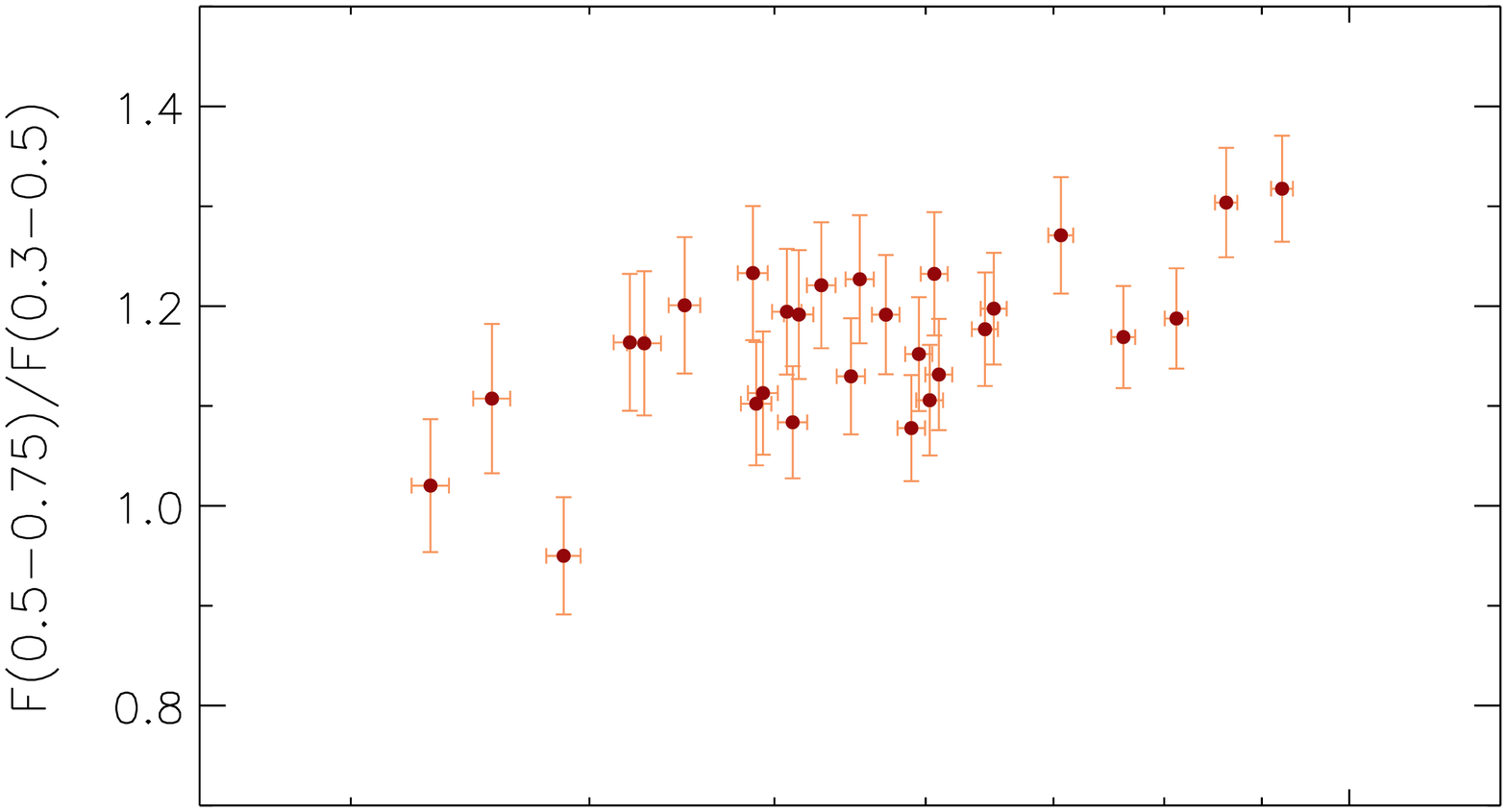}
\includegraphics[width=0.48\textwidth]{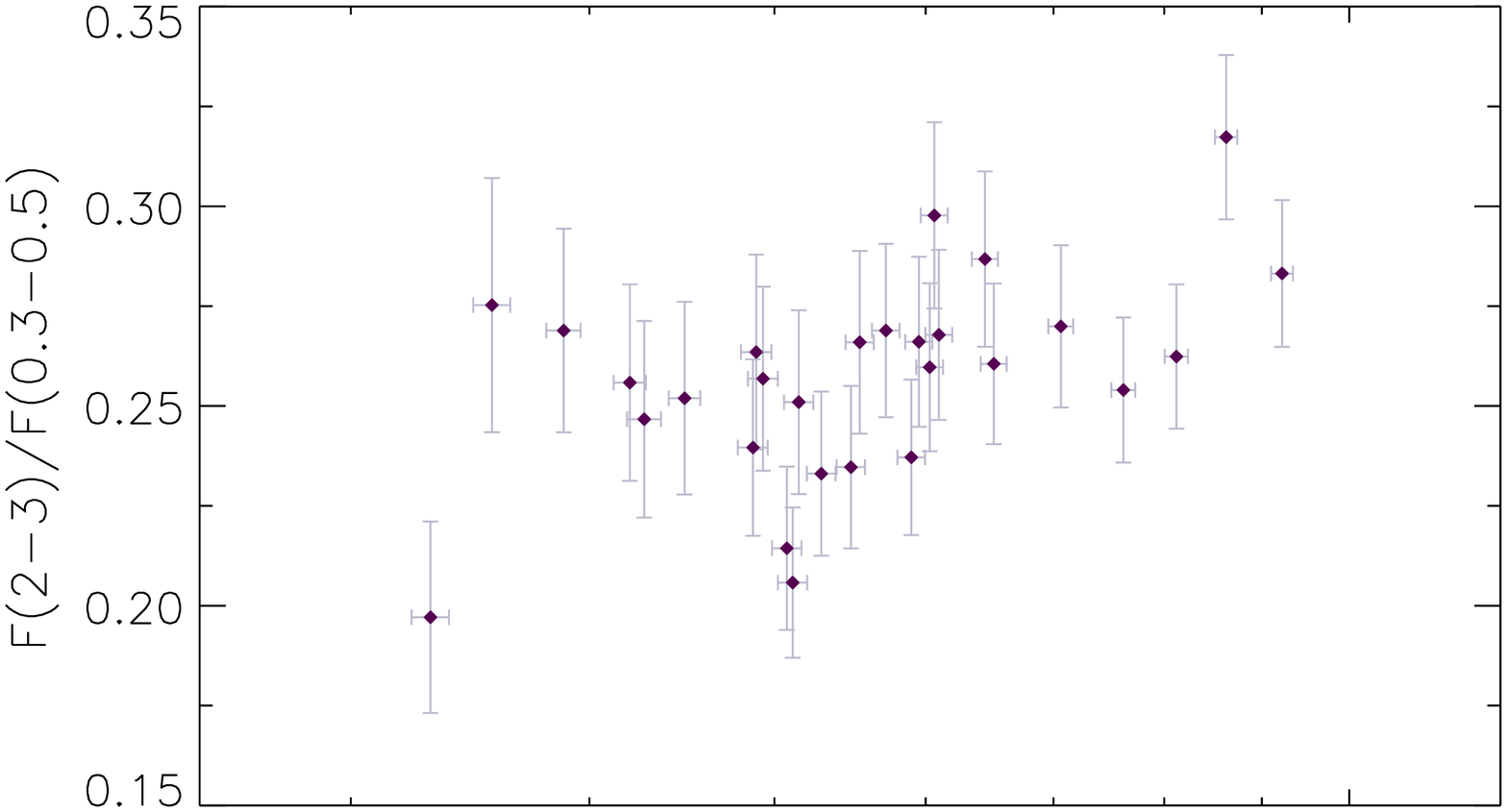}
\includegraphics[width=0.48\textwidth]{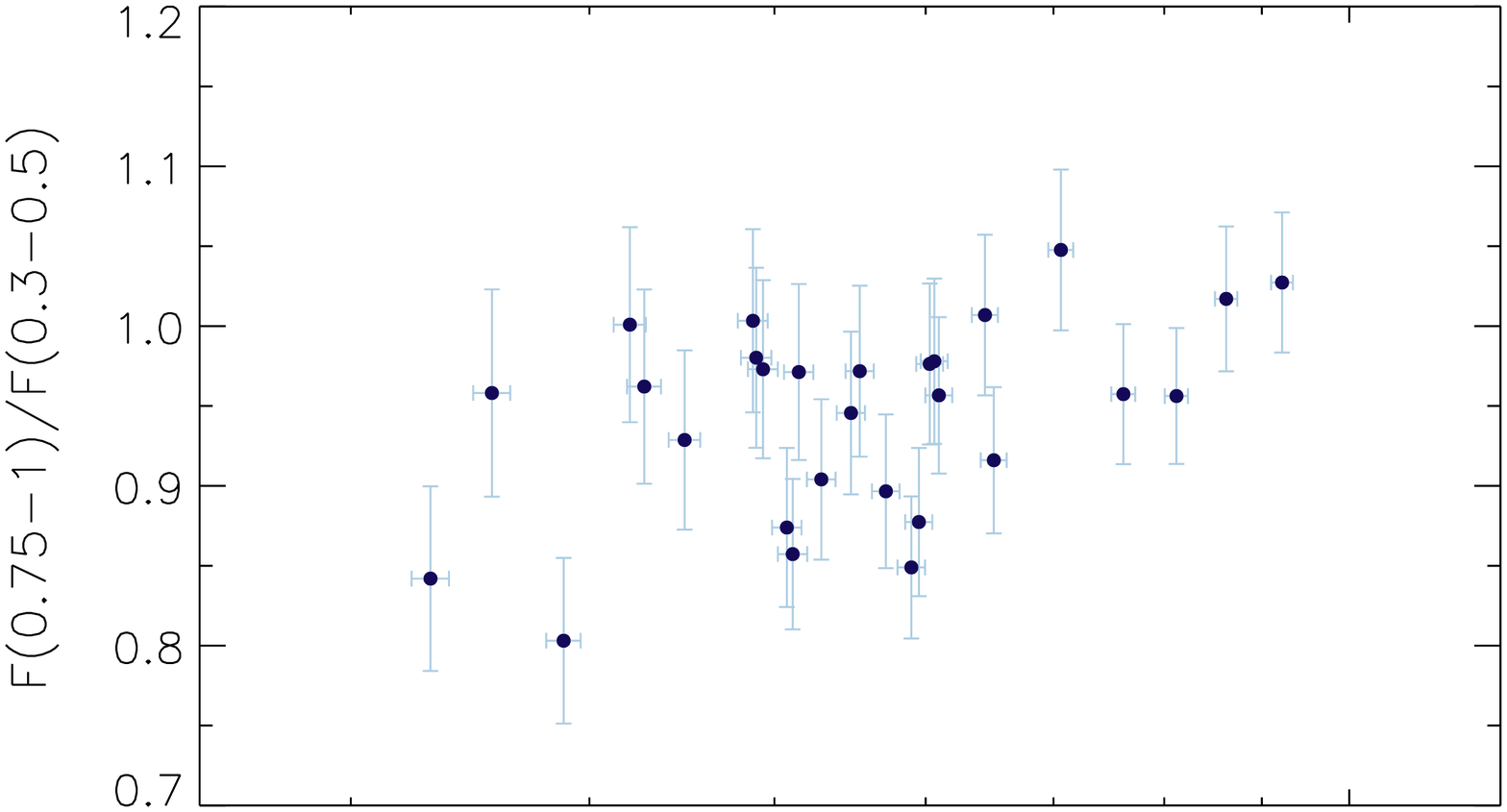}
\includegraphics[width=0.48\textwidth]{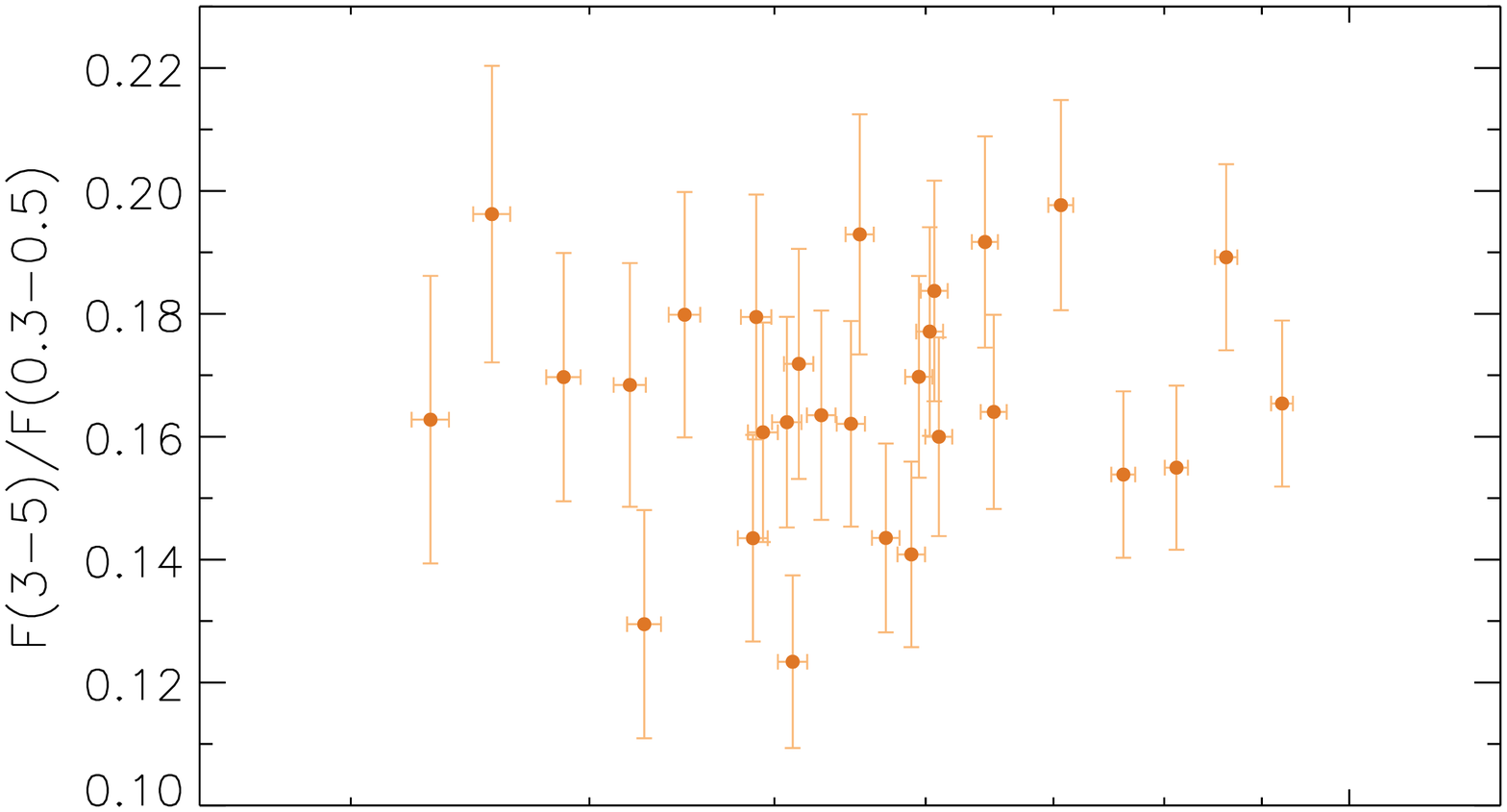}
\includegraphics[width=0.48\textwidth]{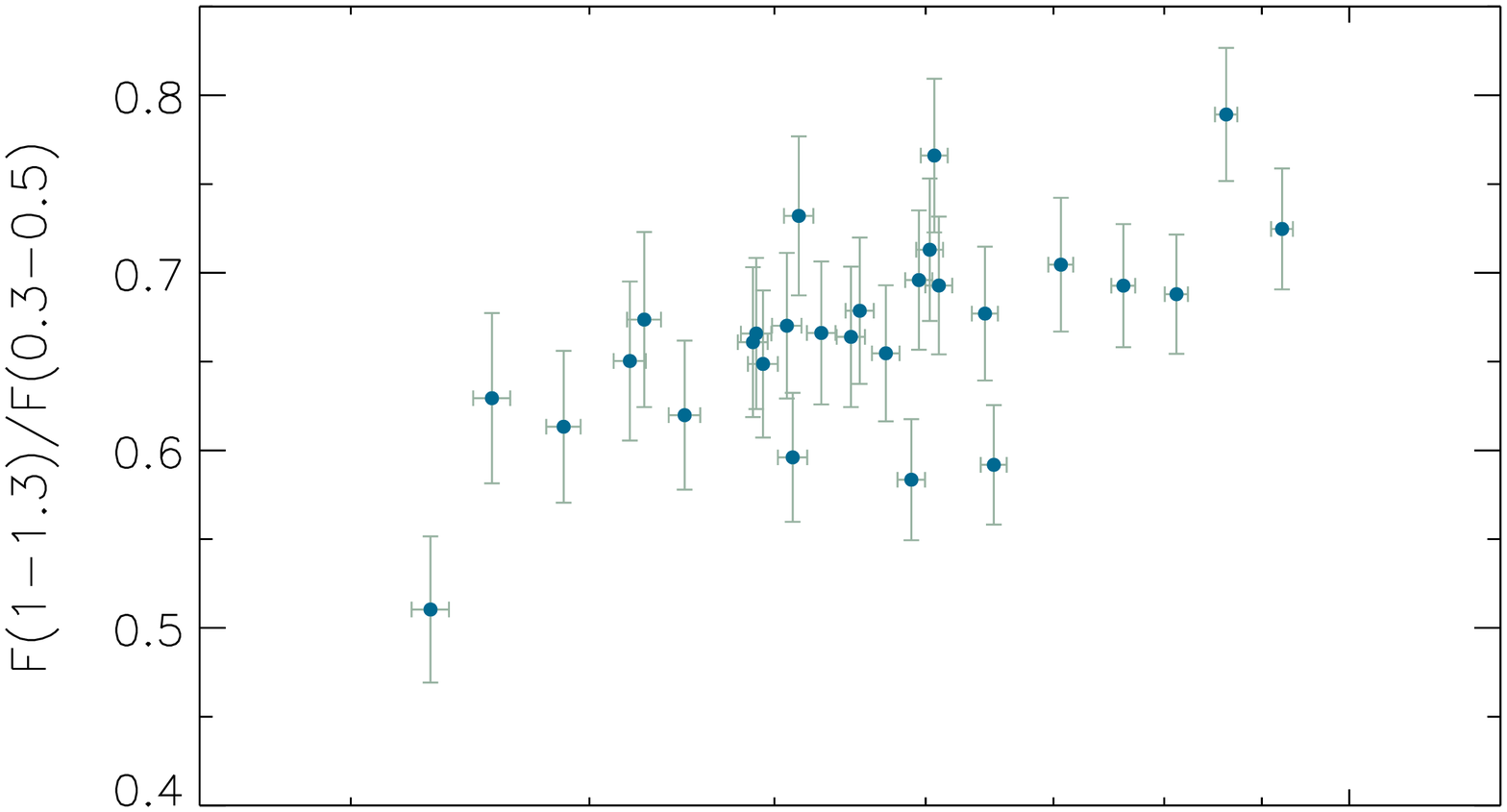}
\includegraphics[width=0.48\textwidth]{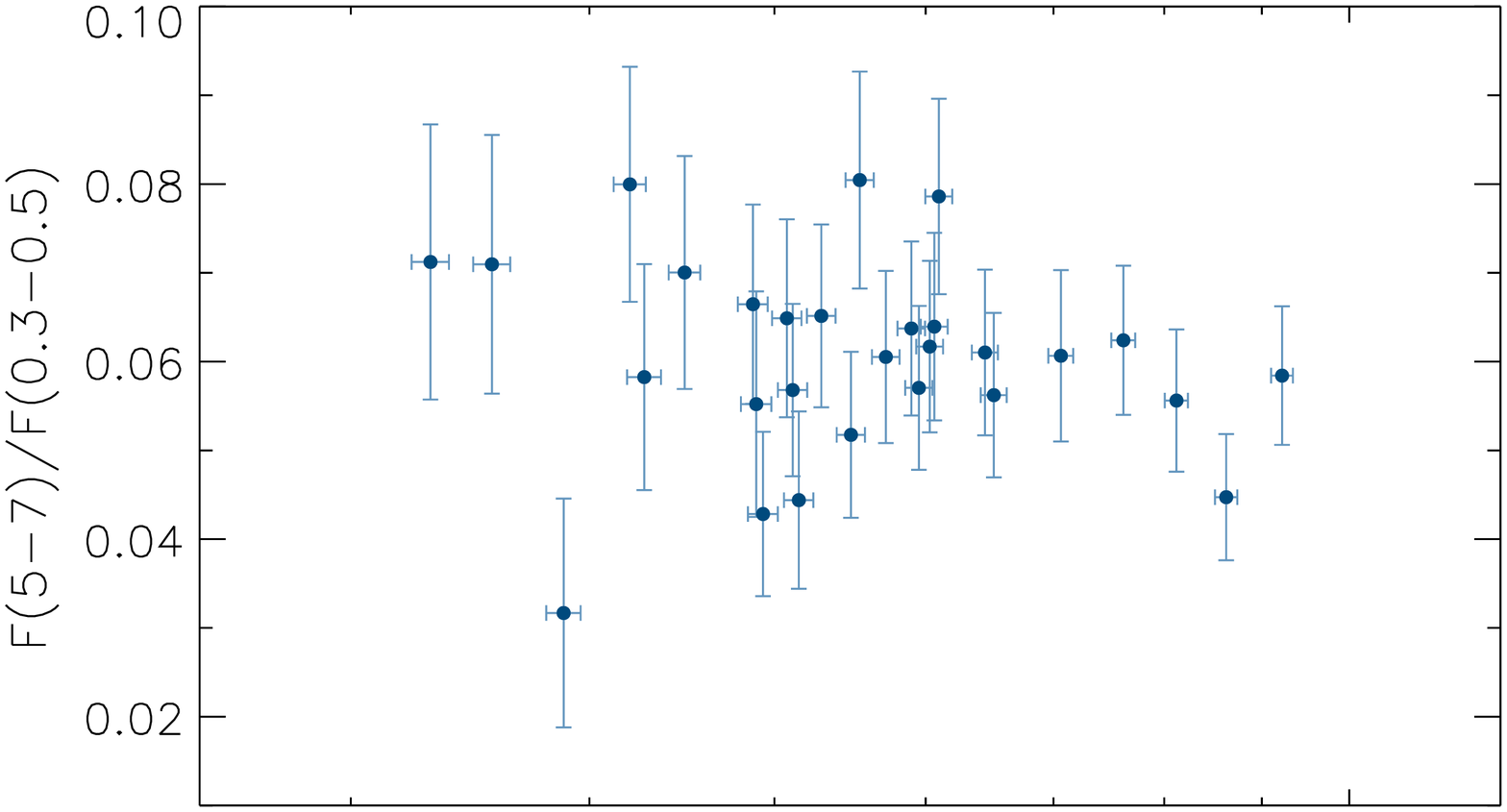}
\includegraphics[width=0.48\textwidth]{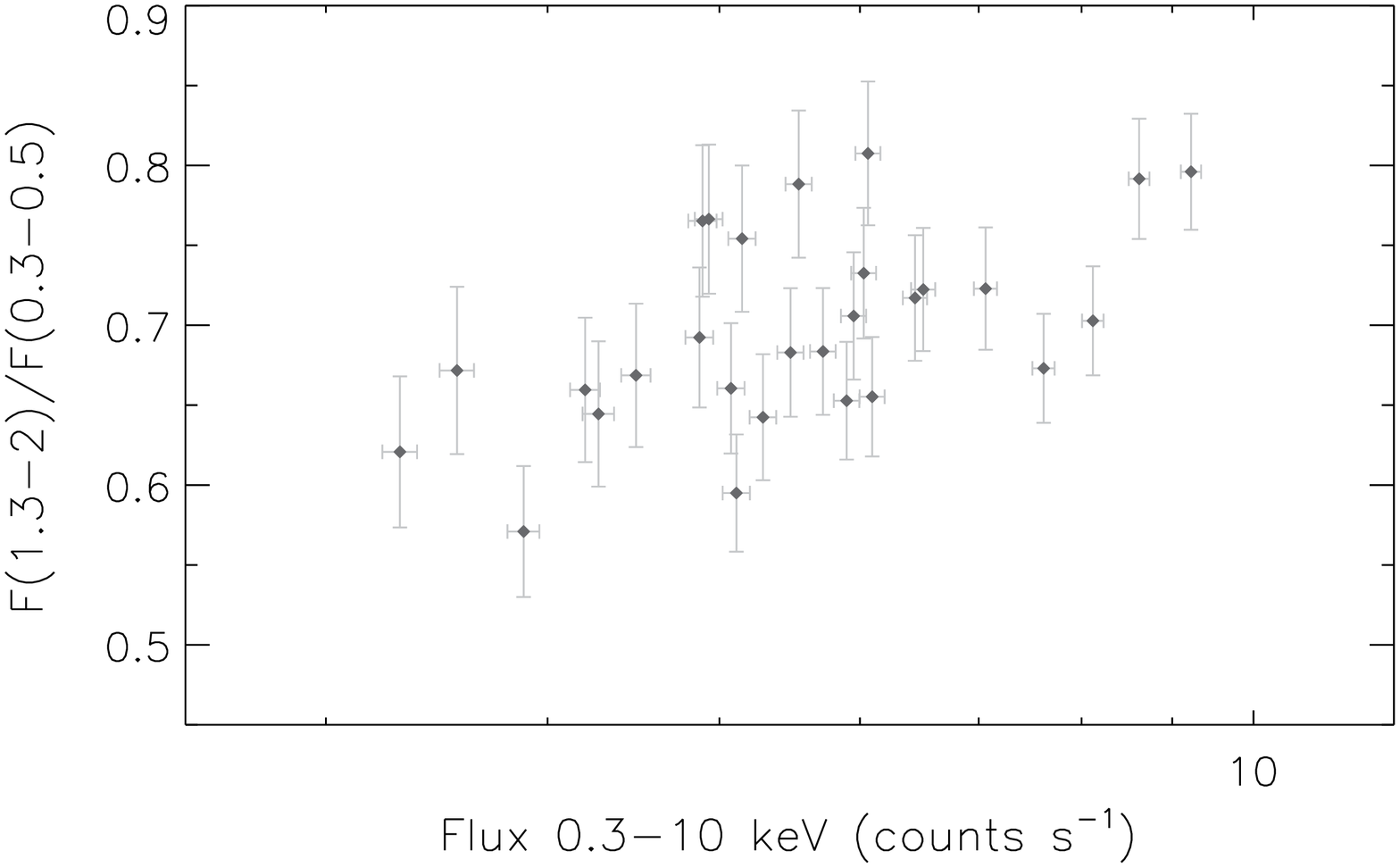}
\includegraphics[width=0.48\textwidth]{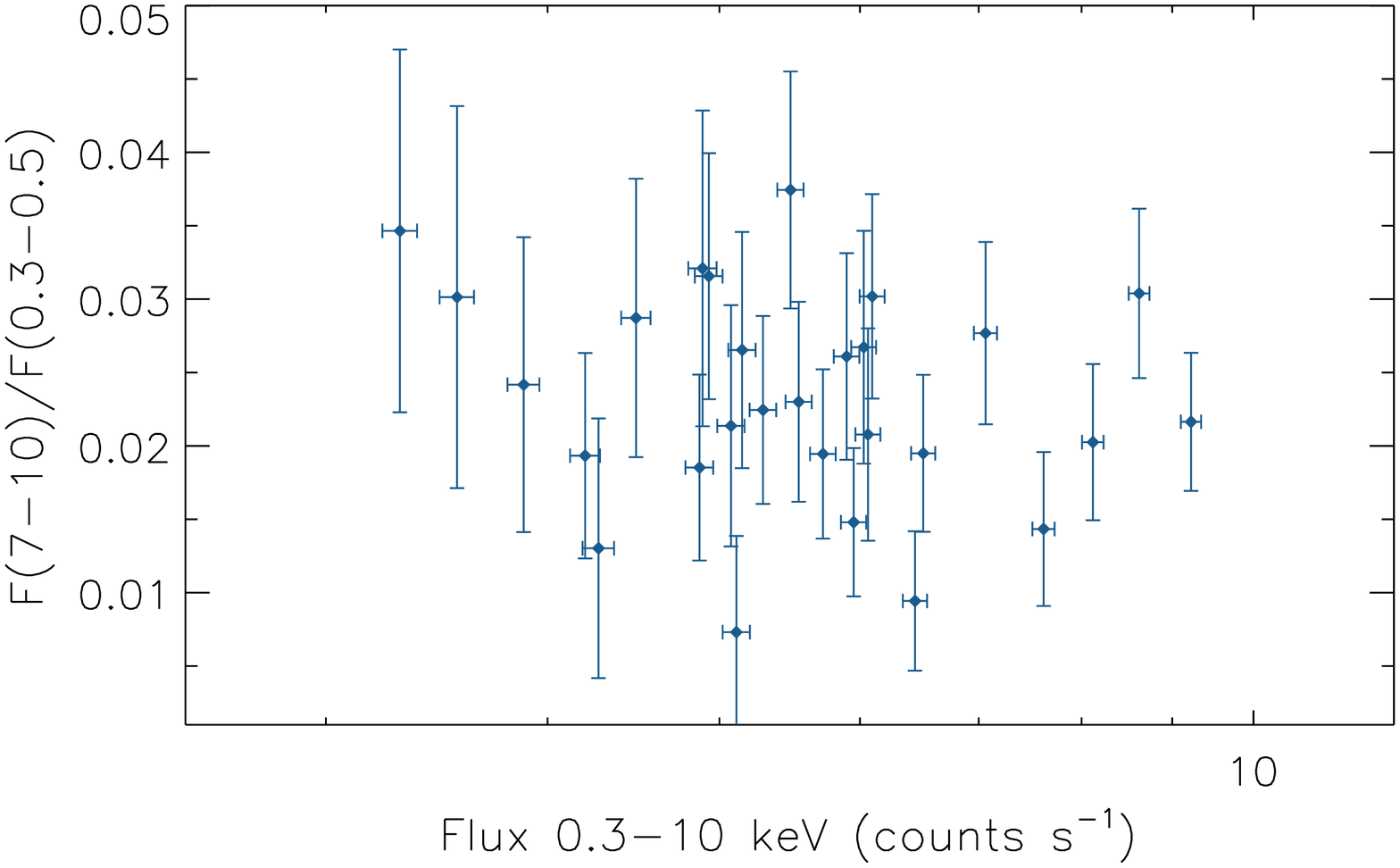}
    \caption{EPIC/PN flux ratios in different energy bands versus the total 0.3--10\,keV flux for the 2011 {\it XMM-Newton} observation of 1ES\,1927+654. The fluxes were obtained integrating over intervals of 1\,ks. 
    }
    \label{fig:XMMlc_11_bands_ratios}
  \end{center}
\end{figure*}

The {\it XMM-Newton} EPIC/PN data of the May 2011 observation were reduced following the same guidelines outlined in \S\ref{sect:xmm_datared}. The spectrum was fitted with a model that included a power-law, a blackbody and neutral absorption, which yields a good fit ($\chi^{2}$/DOF=1727/1674, see top panel of Fig.\,\ref{fig:xmmobs2011eeufs}). The source is highly variable (see bottom panel of Fig.\,\ref{fig:xmmobs2011eeufs}), and we divided the observation into 14 intervals, with exposures between $\sim 1$\,ks and $4$\,ks, similarly to what we did for the 2018 and 2019 observations. We fitted all the exposures using the same model adopted for the complete X-ray observation, and report the results in Table\,\,\ref{tab:fitXMM11epochs}. No clear relation was found between the temperature of the blackbody and the luminosity of the source. The relation between the hardness ratios and the total flux is shown in Fig.\,\ref{fig:XMMlc_11_bands_ratios}.

\tabletypesize{\normalsize}
\begin{deluxetable*}{ccccccccc} 
\tablecaption{Spectral parameters obtained for the fourteen time intervals of the {\it XMM-Newton} May 2011 observation (see bottom panel of Fig.\,\ref{fig:xmmobs2011eeufs}). The spectral model includes a blackbody component and a power law, both absorbed by neutral gas [\textsc{tbabs$\times$ztbabs$\times$(zpo+zbb})].\label{tab:fitXMM11epochs}
}
\tablehead{
 \colhead{(1) } & \colhead{(2)} &  \colhead{(3)} &\colhead{(4)} & \colhead{(5) } & \colhead{(6) }  \\
\noalign{\smallskip}
 \colhead{Interval }& \colhead{$L_{\rm 0.3-2}$}&   \colhead{$N_{\rm H}$} &\colhead{$kT$} & \colhead{$\Gamma$ }  & \colhead{C-stat/DOF }   \\
\noalign{\smallskip}
 \colhead{ }  & \colhead{[$\rm 10^{42}\,erg\,s^{-1}$]}  & \colhead{[$\rm 10^{20}\,cm^{-2}$]} & \colhead{[eV]} & \colhead{ }   
}
\startdata
\noalign{\smallskip}
1	&   $5.25^{+0.08}_{-0.35}$     &   $\leq 6.7$  &  $ 208^{+43}_{-55}$	& $ 2.43^{+0.30}_{-0.16}$ &  535/582   \\
\noalign{\smallskip}
2	&   $3.48^{+0.06}_{-0.08}$     &   $\leq 1.8$  &  $ 171^{+26}_{-17}$	& $ 2.35^{+0.13}_{-0.15}$ &  573/659   \\
\noalign{\smallskip}
3	&   $7.00^{+0.06}_{-0.11}$     &   $2.6\pm2.2$  &  $ 182\pm20$	& $ 2.42\pm0.14$ &  808/924   \\
\noalign{\smallskip}
4	&   $5.66^{+0.05}_{-0.27}$     &   $7.1^{+4.4}_{-3.6}$  &  $ 141^{+27}_{-22}$	& $ 2.52\pm0.17$ &  629/716   \\
\noalign{\smallskip}
5	&   $6.36^{+0.06}_{-0.17}$     &   $\leq 6.3$  &  $ 171^{+32}_{-31}$	& $2.43^{+0.19}_{-0.21}$ &  667/750   \\
\noalign{\smallskip}
6	&   $5.30^{+0.05}_{-0.14}$     &   $\leq 3.9$   &  $ 180^{+22}_{-25}$	& $ 2.46^{+0.17}_{-0.12}$ &  688/797   \\
\noalign{\smallskip}
7	&   $6.53^{+0.06}_{-0.22}$     &   $7.5^{+4.3}_{-3.7}$  &  $ 147^{+33}_{-28}$	& $2.65^{+0.20}_{-0.21}$ &  511/631   \\
\noalign{\smallskip}
8	&   $5.61^{+0.04}_{-0.07}$     &   $3.7^{+2.0}_{-1.9}$  &  $170\pm19$	& $2.49^{+0.11}_{-0.12}$ &  831/998   \\
\noalign{\smallskip}
9	&   $5.80^{+0.04}_{-0.07}$     &   $3.9\pm2.0$  &  $ 180\pm22$	& $2.51\pm0.12$ &  864/987   \\
\noalign{\smallskip}
10	&   $4.74^{+0.04}_{-0.13}$     &   $\leq 4.4$   &  $181^{+23}_{-24}$	& $ 2.40^{+0.17}_{-0.14}$ &  645/764   \\
\noalign{\smallskip}
11	&   $9.08^{+0.07}_{-0.17}$     &   $6.3\pm2.3$  &  $156^{+23}_{-20}$	& $2.58\pm0.12$ &  725/897   \\
\noalign{\smallskip}
12	&   $7.62^{+0.06}_{-0.20}$     &   $7.1^{+3.0}_{-2.7}$  &  $144^{+23}_{-20}$	& $2.62\pm0.14$ &  600/785   \\
\noalign{\smallskip}
13	&   $5.42^{+0.02}_{-0.56}$     &   $\leq 3.5$  &  $182^{+19}_{-31}$	& $2.30^{+0.22}_{-0.08}$ &  601/650   \\
\noalign{\smallskip}
14	&   $4.00^{+0.04}_{-0.11}$     &    $\leq 5.5$  &  $ 170^{+22}_{-23}$	& $ 2.36\pm0.18$ &  704/782   
\enddata
\tablecomments{The columns report (1) the interval used for the spectral analysis, (2) the 0.3--2\,keV luminosity of the source, (3) the column density of the cold absorber, (4) the temperature of the blackbody component, (5) the photon index of the power-law component, and (6) the value of C-stat and the number of DOF.}
\end{deluxetable*}

\clearpage

\section{{\it NICER} spectra}\label{appendix:NICERspectra}

All the {\it NICER} spectra used in our analysis are reported in Fig.\,\ref{fig:NICERplots0}. The shaded area in the top panels represent the background, while the bottom panels illustrate the ratio between the best-fitting model (continuous lines in the middle panel) and the data. During the low-luminosity periods no power-law component is needed to reproduce the spectra, which is well fit by a blackbody component (dashed line) and a Gaussian line (dot-dashed line). In the high-luminosity intervals the power-law component (dot-dot dashed line) can be observed again.

\renewcommand\thefigure{\thesection\arabic{figure}}

\begin{figure*}
\begin{center}
\includegraphics[width=0.42\textwidth]{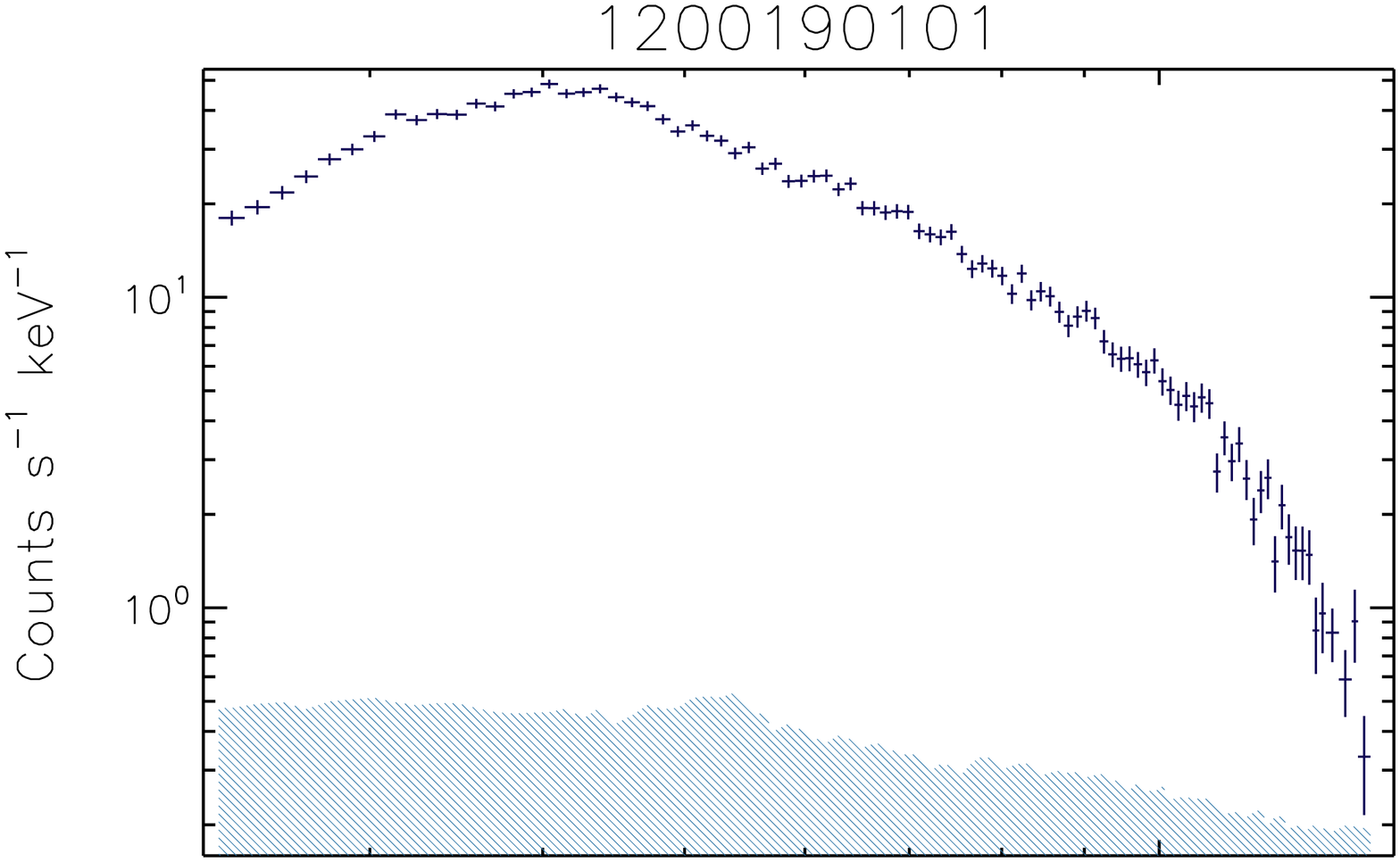}
\includegraphics[width=0.42\textwidth]{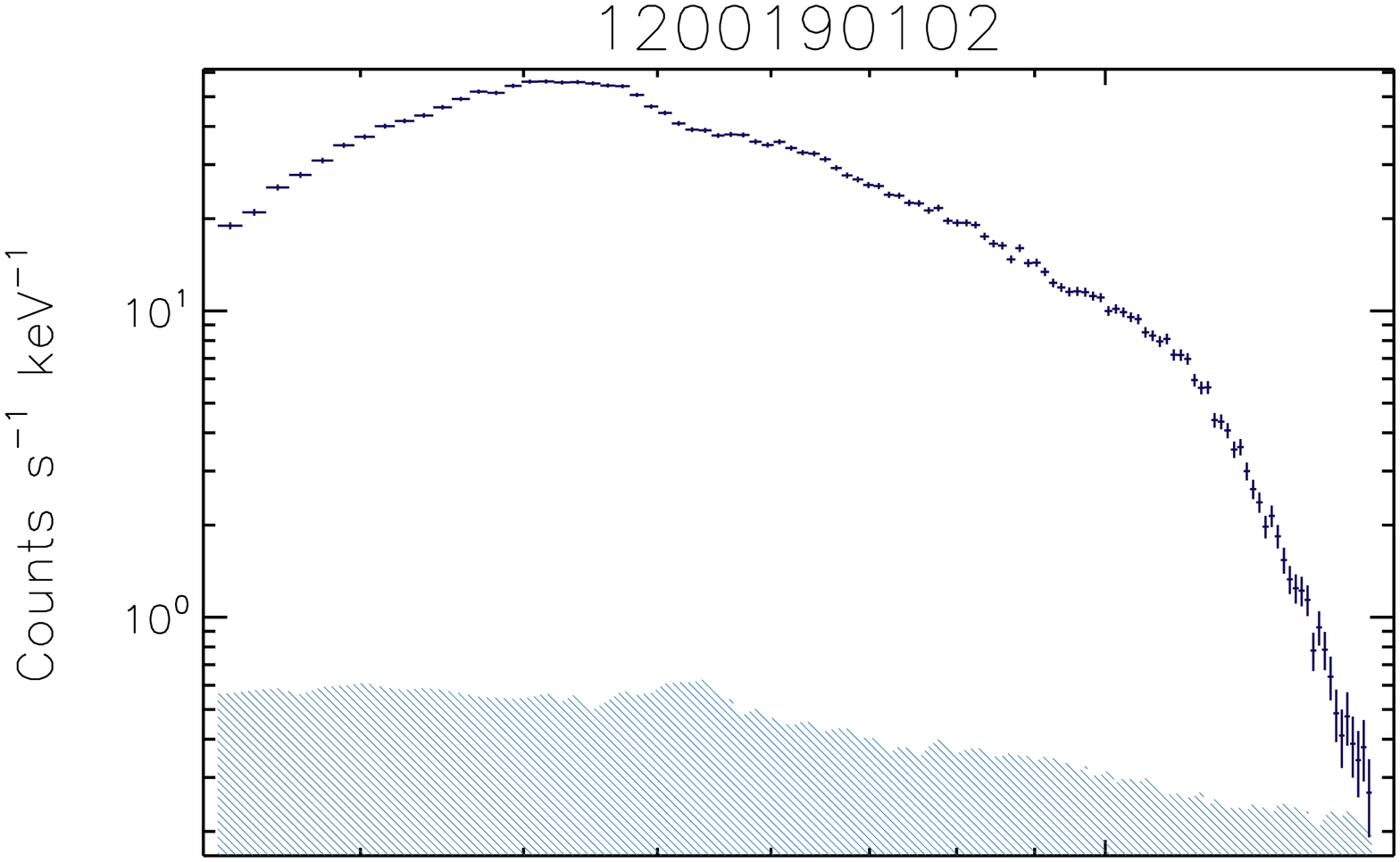}
\includegraphics[width=0.42\textwidth]{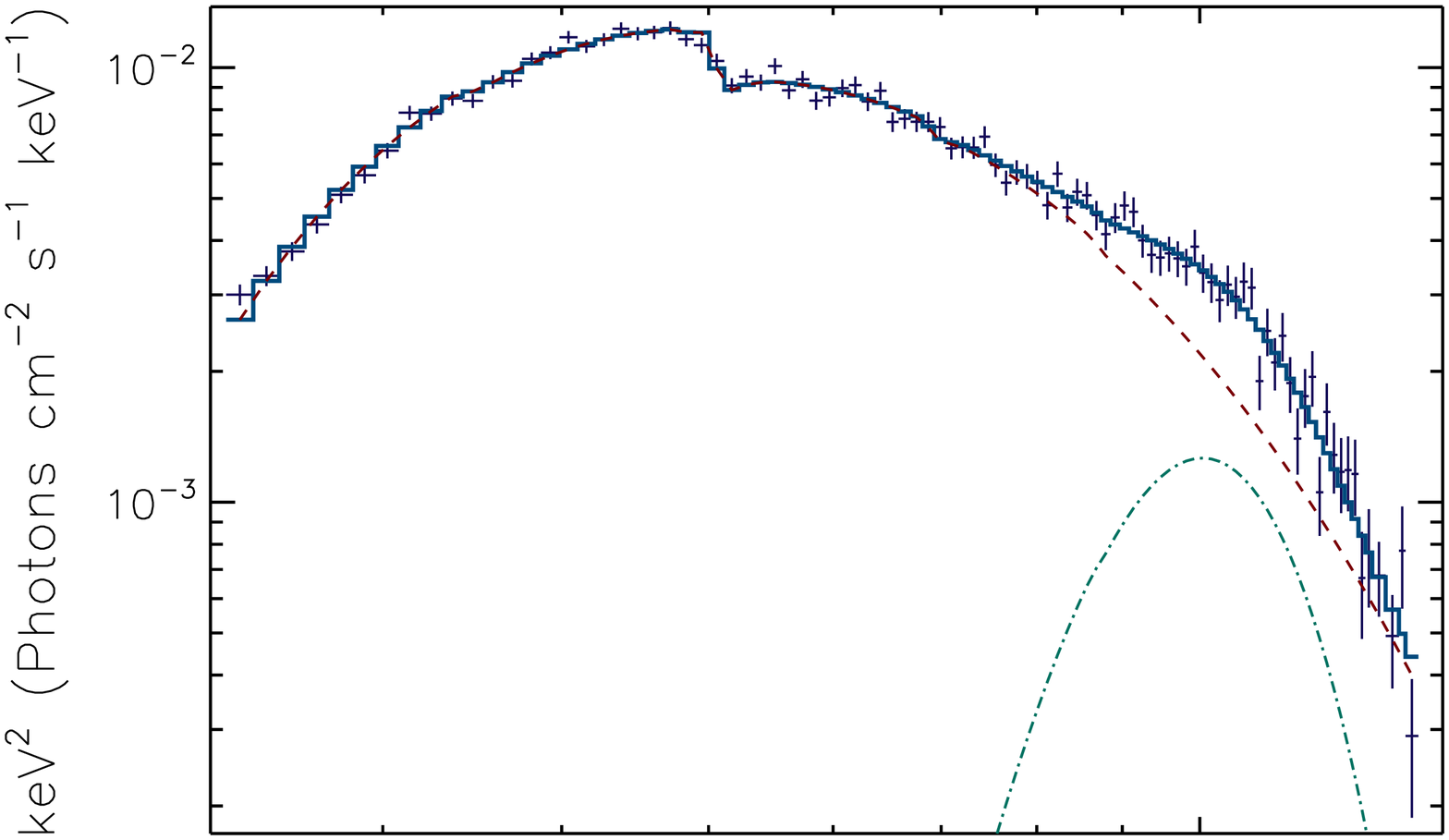}
\includegraphics[width=0.42\textwidth]{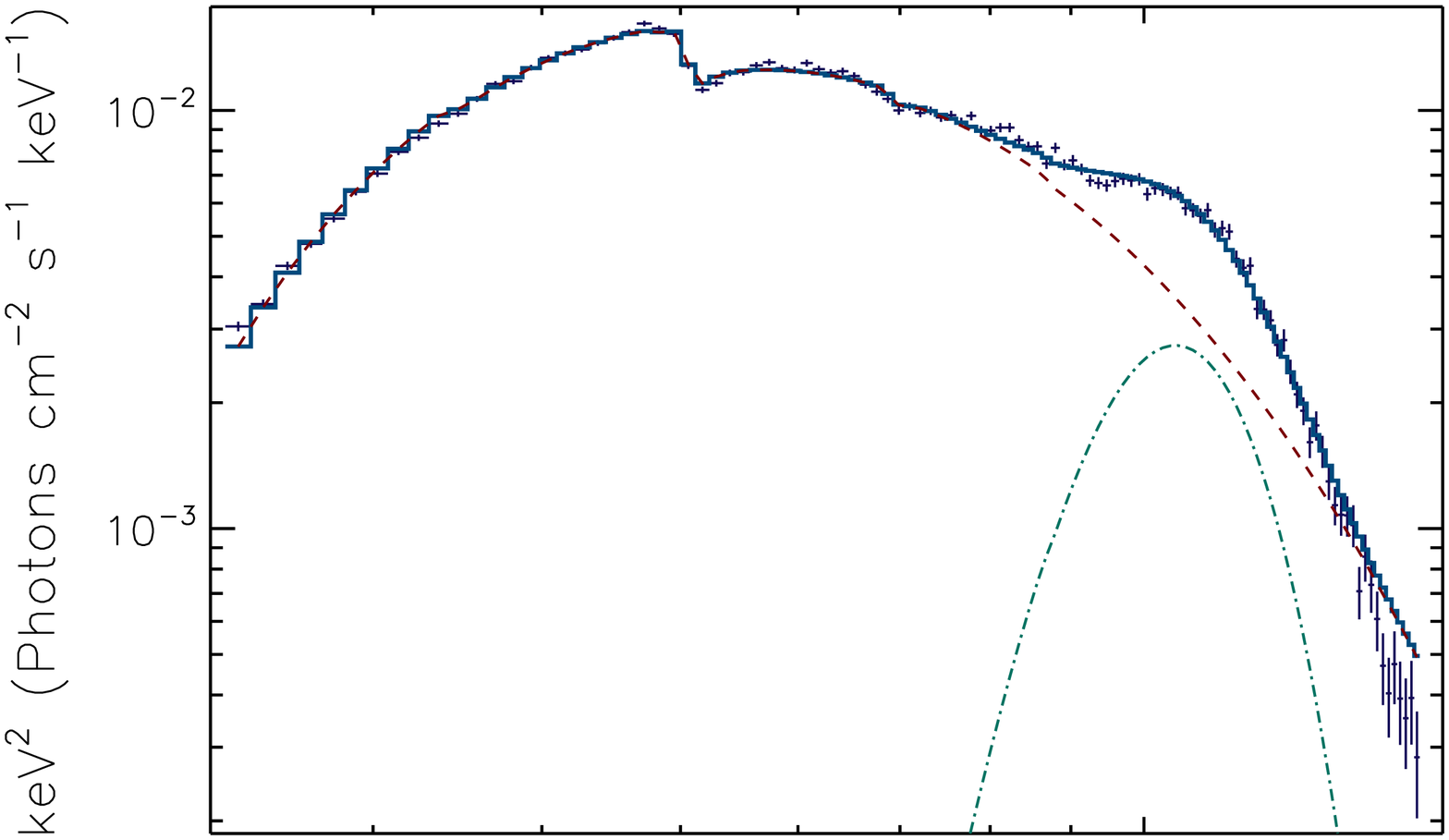}
\includegraphics[width=0.42\textwidth]{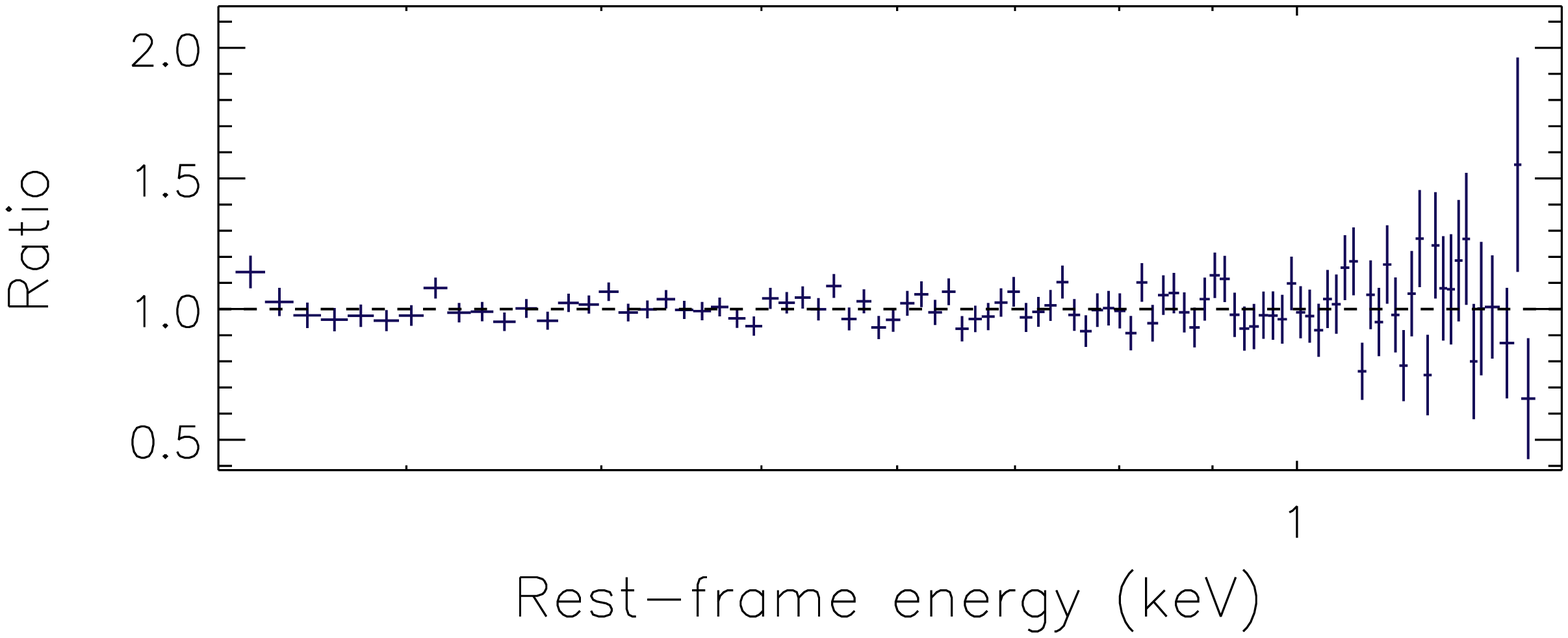}
\includegraphics[width=0.42\textwidth]{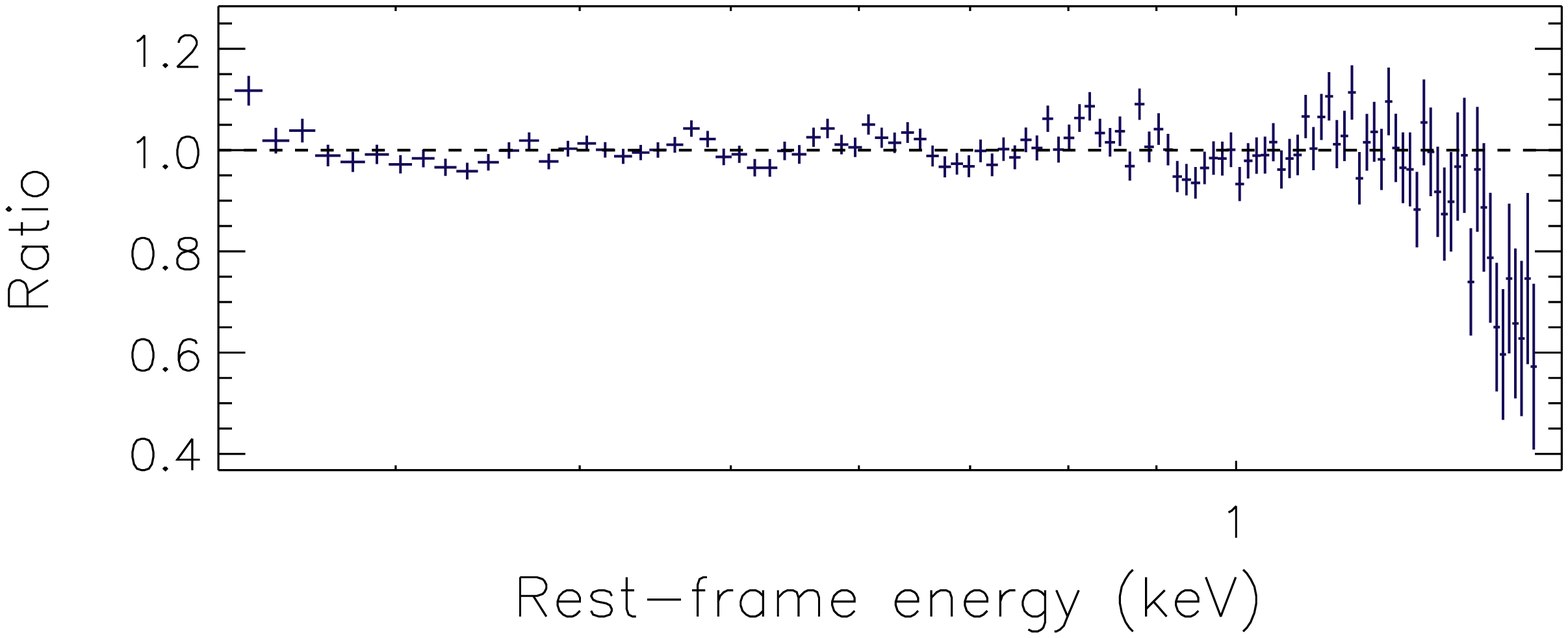}
\includegraphics[width=0.42\textwidth]{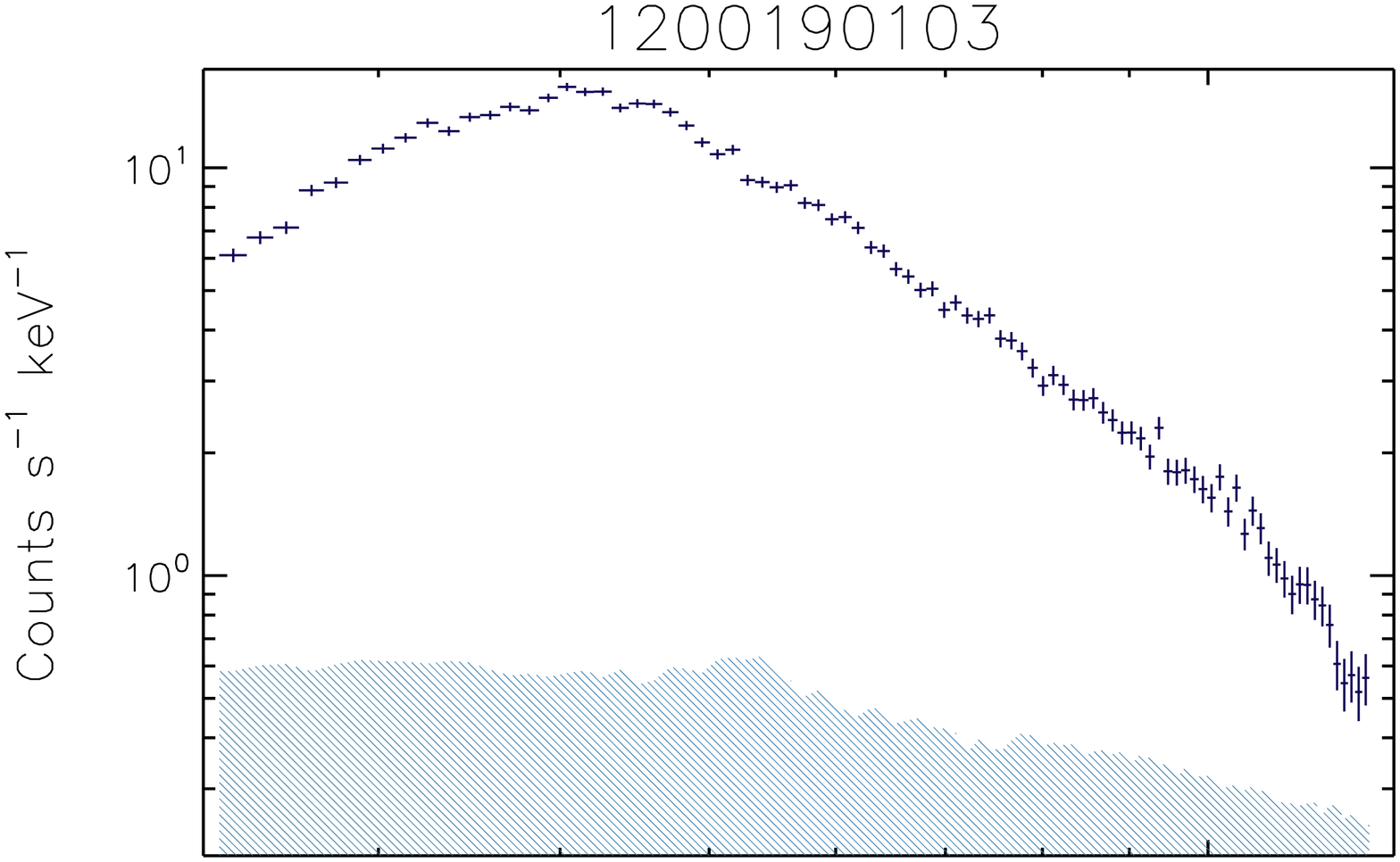}
\includegraphics[width=0.42\textwidth]{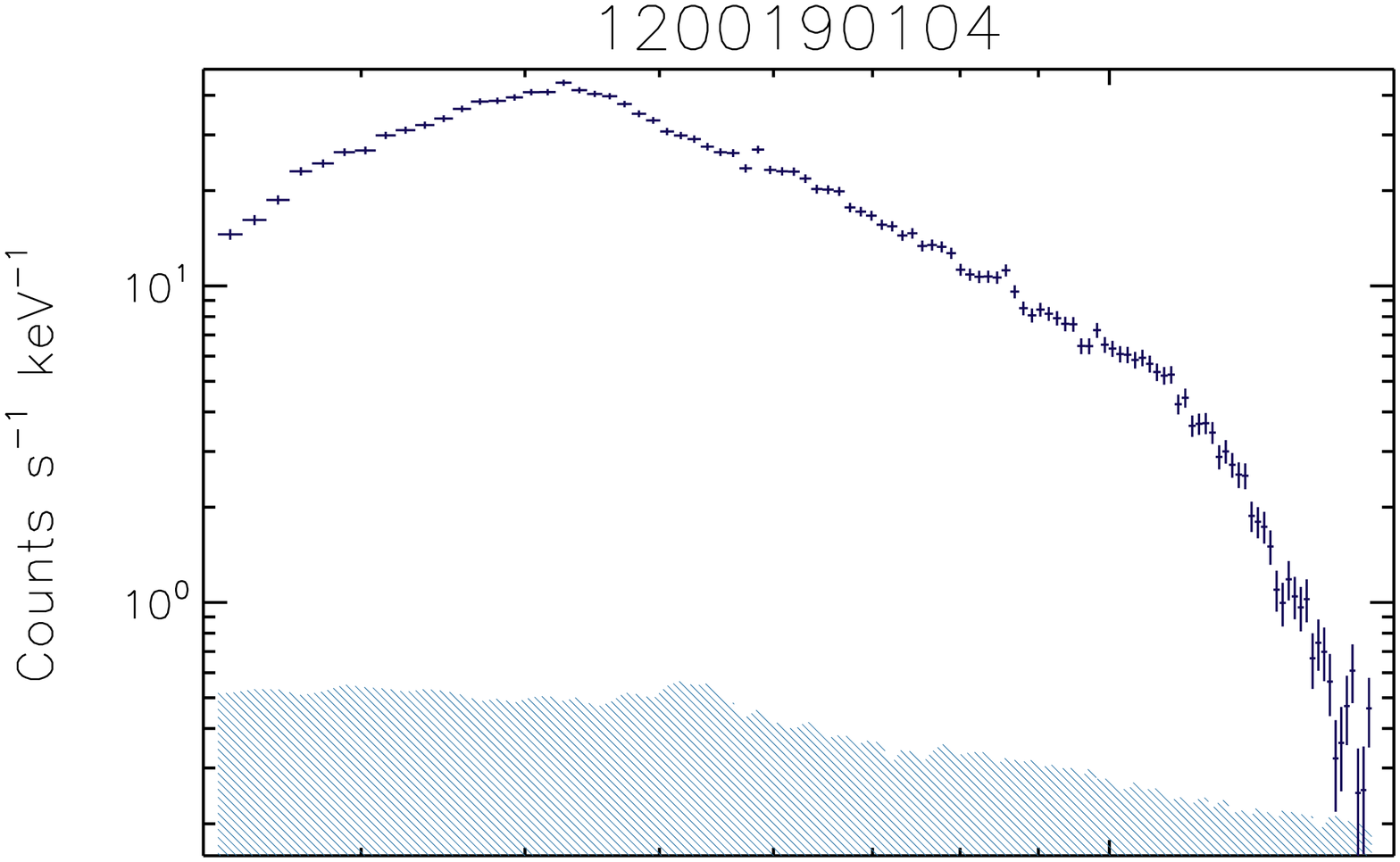}
\includegraphics[width=0.42\textwidth]{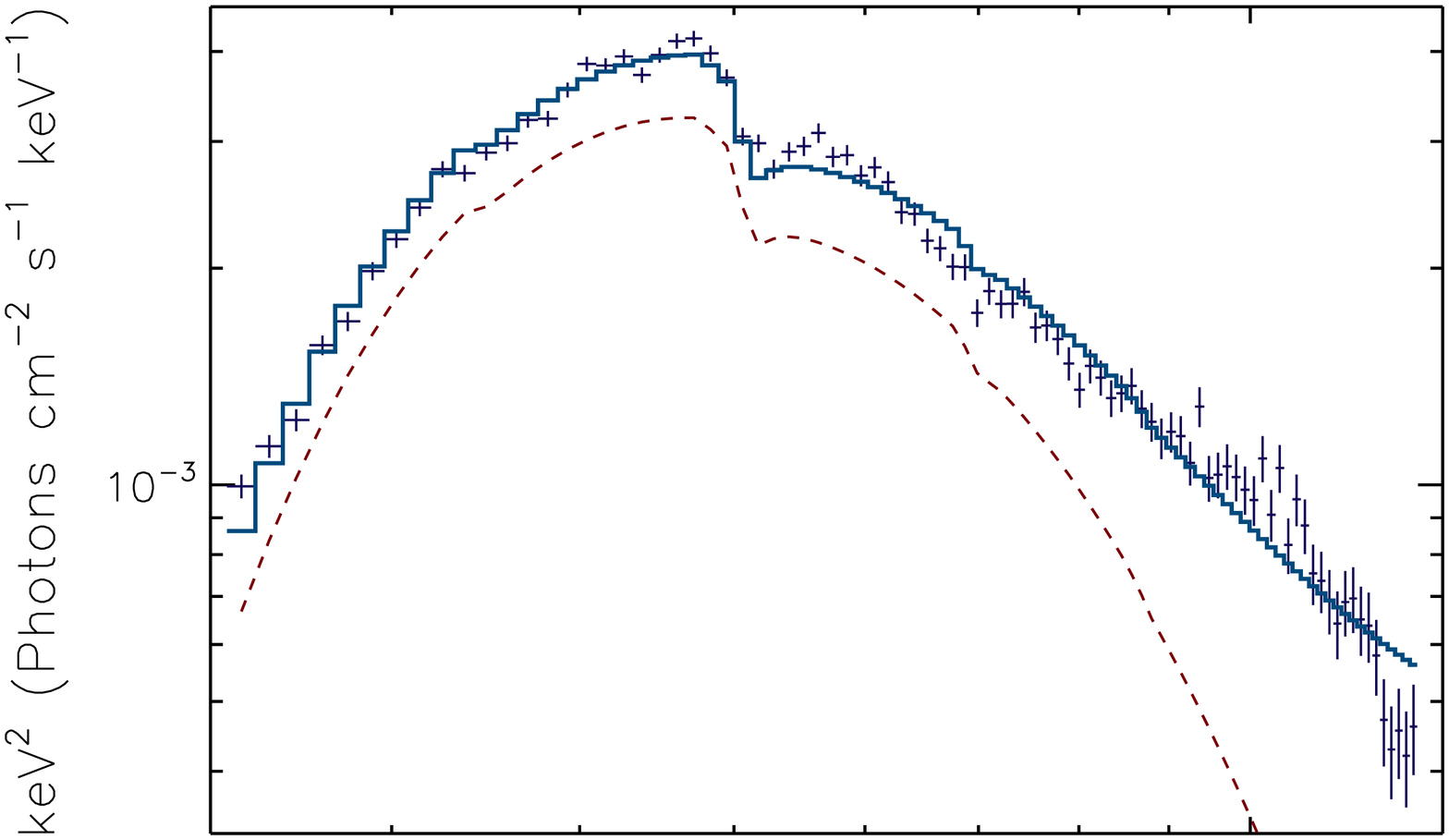}
\includegraphics[width=0.42\textwidth]{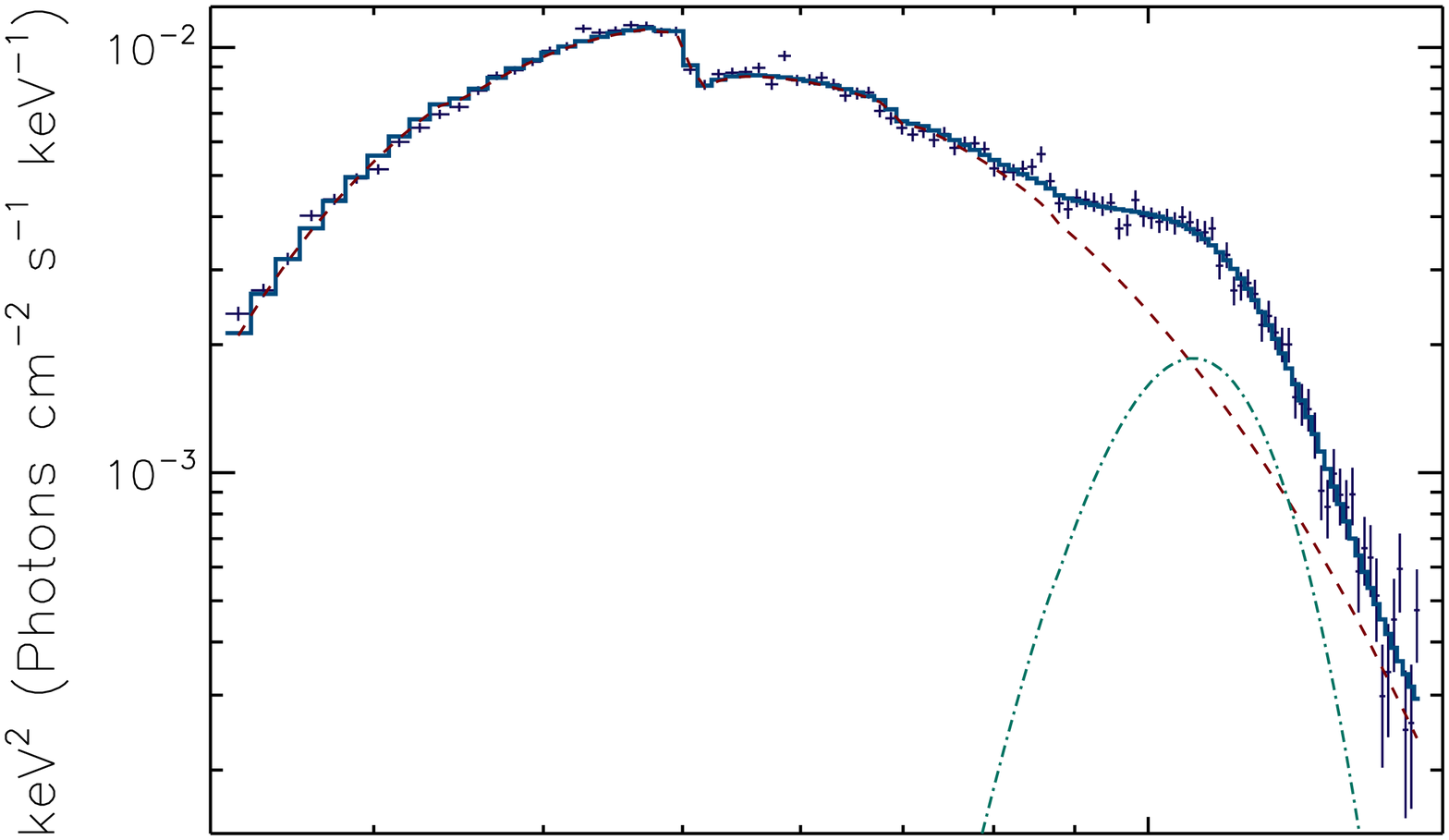}
\includegraphics[width=0.42\textwidth]{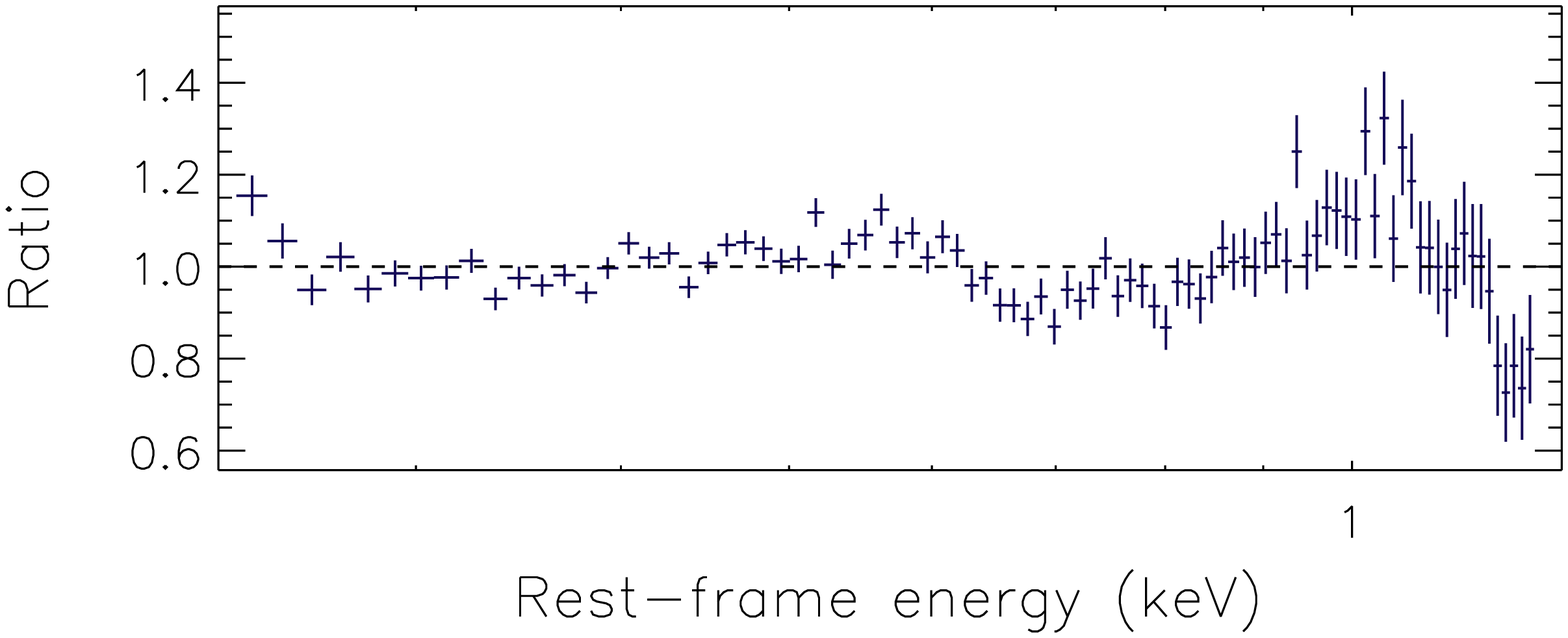}
\includegraphics[width=0.42\textwidth]{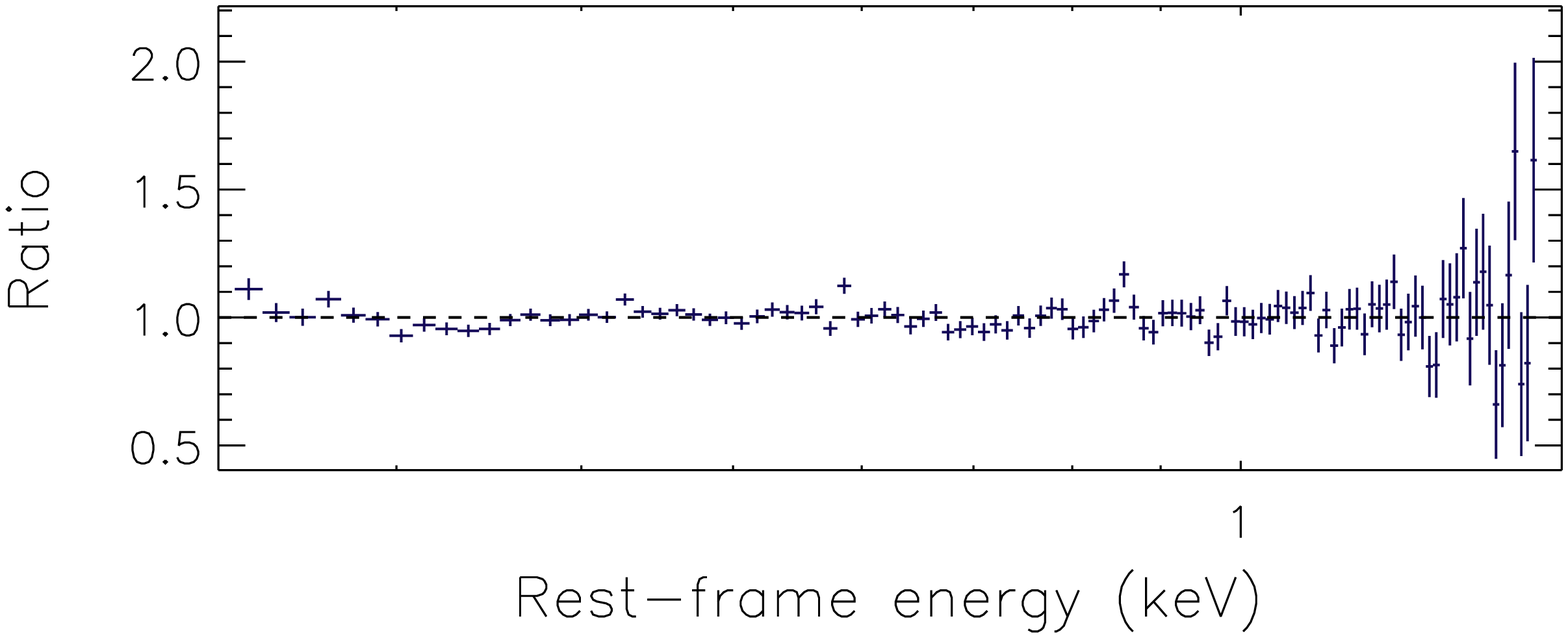}
\caption{{\it NICER} spectra of 1ES\,1927+654. The complete set of figures will be available on ApJS.}
\label{fig:NICERplots0}
\end{center}
\end{figure*}

\end{document}